\documentclass[twoside,12pt]{article}
\usepackage{epsfig}
\usepackage{axodraw}
\usepackage{colordvi}
\usepackage{graphicx}
\usepackage{bm}
\usepackage{amsmath}
\newcommand{\eep}{(e,e$'$p)}
\newcommand{\bra}[1]{\left\langle #1\right|}
\newcommand{\ket}[1]{\left| #1\right\rangle}

\newcommand{\be}{\begin{equation}}
\newcommand{\ee}{\end{equation}}
\newcommand{\bea}{\begin{eqnarray}}
\newcommand{\eea}{\end{eqnarray}}

\begin{document}


\title{ \vspace{1cm} Self-consistent Green's function method \\
for nuclei and nuclear matter }
\author{W.\ H.\ Dickhoff$^{1}$  and  C.\ Barbieri$^{2}$ \\
\\
$^1$Department of Physics, Washington University,
	 St.Louis, Missouri 63130, USA  \\ 
$^2$TRIUMF, 4004 Wesbrook Mall, Vancouver, 
          British Columbia, Canada V6T 2A3 }
\maketitle
\begin{abstract} 
Recent results obtained by applying the method of
self-consistent Green's functions to nuclei and nuclear matter are reviewed.
Particular attention is given to the description of
experimental data obtained from the (e,e$'$p) and (e,e$'$2N) reactions
that determine one and two-nucleon removal probabilities in nuclei since
the corresponding amplitudes are directly related to the imaginary parts of 
the single-particle and two-particle propagators.
For this reason and the fact that these amplitudes can now be calculated with
the inclusion of all the relevant physical processes, 
it is useful to explore the efficacy of the 
method of self-consistent Green's functions in describing these experimental
data.
Results for both finite nuclei and nuclear matter are discussed with
particular emphasis on clarifying the role of short-range correlations in
determining various experimental quantities.
The important role of long-range correlations in determining the structure
of low-energy correlations is also documented.
For a complete understanding of nuclear phenomena it is therefore essential
to include both types of physical correlations.
We demonstrate that recent experimental results for these reactions combined 
with the reported theoretical calculations yield a very clear 
understanding of the 
properties of \textbf{all} protons in the nucleus.
We propose that this knowledge of the properties of constituent fermions in
a correlated many-body system is a unique feature of nuclear physics.
\end{abstract}

\begin{flushleft}
TRIUMF preprint: TRI-PP-03-40
\end{flushleft}

\eject
\tableofcontents

\section{Introduction}

Various approaches exist to deal with the nuclear many-body problem.
A comparison of these approaches has recently been given in 
Ref.~\cite{mupo00}.
In this review only one specific method is singled out for discussion.
This method is most intimately connected to quantum field theory
since it corresponds to the nonrelativistic implementation of this approach
to the many-body problem.
While the propagator or Green's function method was the most important tool
in the formal development of many-body theory~\cite{AAA} - \cite{fetwa}, 
only in the last ten years has it been applied to many-body problems beyond 
its mean-field implementation (the Hartree-Fock approximation).
Examples of such applications have appeared in atomic~\cite{vnpw01}
and condensed matter physics~\cite{hoba98}.
In the atomic case, the essential new development involves taking the
self-consistent determination of the electron propagator from the
Hartree-Fock level to the fully
self-consistent inclusion of the second-order self-energy.
Important quantitative improvements over the Hartree-Fock results are obtained
for standard quantities like the binding energy but other experimental
data like spectroscopic factors are significantly improved as 
well~\cite{vnpw01}.
In the case of the electron gas, the self-consistent determination of the
electron propagator involves the summation of all ring diagrams in the
self-energy which is necessary to treat the well known divergence of the
Coulomb interaction in an infinite system.
Also in this case, one obtains a much improved description of the
energy per particle~\cite{hoba98} essentially in agreement with the
benchmark Monte Carlo results of Ref.~\cite{cepal}.

In the nuclear case, similar developments have taken place which have
focused on the self-consistent treatment of short-range correlations
(SRC) in nuclear matter by including ladder diagrams in the self-energy.
For finite nuclei, the self-consistent
treatment of
long-range correlations (LRC) has been considered at different levels
of sophistication.
The nuclear case is particularly unique due to the availability of a 
magnificent set of experimental data which have illuminated in great
detail the properties of the nucleon propagator in the medium.
These experiments involve nucleon knockout reactions induced by electrons
which are measured in coincidence with the ejected proton~\cite{frmo} - 
\cite{Paviabook}.
These experiments identify in an essentially unambiguous manner the
removal probability of protons near the Fermi energy from a variety
of target nuclei.
The results of these experiments have clearly identified the need to consider
nucleon properties beyond the mean-field since these removal probabilities
have consistently been about 35\% 
below the simple shell-model value of 1 for closed-shell systems.
By identifying these removal probabilities experimentally, one gains crucial
insight into those quantities which determine the proton propagator in the
medium.
It is then natural to approach the theoretical description of these
data by employing the propagator method. 

It is the purpose of this review to document the recent development of the
theoretical understanding of these (e,e$'$p) data over the last ten years.
The wonderful interplay between experimental and theoretical work
will be exploited and presented in some detail.
The currently emerging picture suggests that the properties of \textit{all}
protons in the nucleus are within the domain of experimental scrutiny
including those with high momenta.
The accompanying improved theoretical understanding of these properties
of protons in the nucleus has also acted as a catalyst for the renewed
study of the nuclear-matter saturation problem.

The paper starts in Sec.~\ref{sec:form} with the introduction of the 
relevant formalism for the
determination of the one and two-particle propagators.
By employing the equation of motion formulation it becomes
straightforward to introduce the concept of self-consistency
necessary for a proper description of the dynamics as it is encountered
in nuclei and nuclear matter.
Sec.~\ref{sec:exp} is devoted to a brief pedagogical 
exploration of the relation between
the experimental data accessible in (e,e$'$p) and (e,e$'$2N) reactions
and the relevant theoretical quantities introduced in Sec.~\ref{sec:form}.
A detailed analysis of Green's function calculations for nuclear matter is
presented in Sec.~\ref{sec:nucmat}.
We limit the discussion to zero-temperature results and
do not cover neutron matter.
This section starts with a brief introduction of the material
and gives a short review of the first-generation results for 
spectral functions to establish a framework for the assessment of the most
recent fully self-consistent results.
These early calculations solved the scattering problem in the medium by
propagating mean-field (mf) nucleons in the medium which is equivalent to
the so-called Galitski-Feynman treatment~\cite{Galitskii58}.
A similar approach is adopted to generate spectral functions for $\Lambda$
hyperons which are compared to the corresponding results for nucleons.
Particular attention is given to recent developments involving the fully
self-consistent inclusion of SRC.
The results for this second generation of spectral functions
are discussed together with an analysis of the scattering process in the 
medium
that underlies the properties of individual nucleons in nuclear matter.
A detailed discussion of healing properties of nucleons in the medium
is presented in order to resolve the paradox that arises when the original
discussion of healing~\cite{gww58,day67} is confronted with 
recent (e,e$'$p) data. 
The consequences for the nuclear matter saturation properties including
these recent developments concludes this section.

Section \ref{sec:finnuc} is devoted to a discussion of the results for the
single-particle (sp) propagator in finite nuclei.
After a brief introduction we start this section with an overview of
the early calculations for the spectral strength distribution.
The influence of LRC is illustrated by a direct comparison of the 
spectral strength distribution with the data.
The dual role of SRC in determining the distribution of the sp strength is 
then clarified in detail.
The important influence of both low-energy particle-particle (pp), hole-hole
(hh),
and particle-hole (ph) correlations in the form of collective behavior in 
determining the low-energy sp properties is emphasized to provide the
basis of a new development involving the Faddeev technique applied to the 
many-body problem.
This new technique is then reviewed and shown to incorporate
the correct way to include these correlations simultaneously.
State of the art results for ${}^{16}\textrm{O}$ demonstrate that 
further improvement of the description of LRC is required for a complete 
description of the experimental data.
This is further illustrated by analyzing the collective phonons in 
${}^{16}\textrm{O}$ that form the ingredients of the Faddeev summation.
A comparison with the experimental excitation spectrum of 
${}^{16}\textrm{O}$ demonstrates that further improvement of the theoretical 
description of 
this spectrum is necessary.
A first step in this direction is presented which takes the self-consistent 
coupling
of single phonons to two phonons into account.

A discussion of recent calculations of two-proton removal cross sections
follows including a comparison with experimental data.
Recent experimental developments will be briefly highlighted which
elucidate the properties of
all protons in the domain of the traditional nuclear mean-field.
These results are also discussed in Sec.~\ref{sec:finnuc} together
with recent attempts to determine the properties of high-momentum
protons in the nucleus.
In Sec.~\ref{sec:other} the results are summarized in as far as they
clarify the properties of protons in the nucleus.
A brief comparison with other correlated many-body systems is also presented
there.
A final summary and outlook for future work is given in Sec.~\ref{sec:sumout}.


\section{\label{sec:form}Formal development}

\subsection{Single-particle Green's function
\label{sec:spdef} }

The nuclear systems to be  discussed in this review will be described as 
a collection of nonrelativistic nucleons interacting by means of a two-body 
interaction $\hat{V}$ that describes nucleon-nucleon (NN) scattering data
up to pion production threshold.
One can generate a zero-order approximation of the 
system under study by introducing an appropriate mean-field
potential $\hat{U}$
and splitting the Hamiltonian into an unperturbed one-body
part $\hat{H}_0 = \hat{T} + \hat{U}$ and a residual interaction
$\hat{H}_1 = \hat{V} - \hat{U}$.
The total Hamiltonian can then be written as
 \begin{equation}
\hat{H}~=~\hat{H}_0 ~+~ \hat{H}_1
 ~=~\sum_{\alpha} \varepsilon^0_\alpha ~c^{\dag}_\alpha c_\alpha ~+~ 
   \left( {\textstyle \frac{1}{4}} \sum_{\alpha \beta \gamma \delta} 
              V_{\alpha \beta , \gamma \delta} 
                  ~c^{\dag}_\alpha ~c^{\dag}_\beta ~c_\delta ~c_\gamma 
                              ~-~ \sum_{\alpha \beta} U_{\alpha,\beta}
 ~c^{\dag}_\alpha c_\beta \right) \; ,
\label{eq:Hamiltonian}
\end{equation}
where we have chosen to use a set of sp states $\{\alpha\}$ that
diagonalize $H_0$  with eigenvalues $\{\varepsilon^0_\alpha\}$,
$c^{\dag}_\alpha$ ($c_\alpha$) are the
creation (destruction) operators
of a particle in the state $\alpha$, $V_{\alpha \beta , \gamma \delta}$
the antisymmetrized matrix elements of $\hat{V}$, and $U_{\alpha,\beta}$
correspond to the matrix elements of $\hat{U}$.
In general, the best choice of $\hat{H}_0$  (and therefore of the
basis $\{\alpha\}$) depends on the symmetry properties of the system under
study. For infinite nuclear matter translational invariance
suggests the use of momentum eigenstates without any need for choosing an
auxiliary potential, while for finite
nuclei one needs to employ a potential (\textit{e.g.} of the 
harmonic oscillator or Woods-Saxon type) to localize the nucleons.
Obviously, the spectrum $\{\varepsilon^0_\alpha\}$ can contain
both discrete and/or continuum states and 
we therefore use the summation symbols in Eq.~(\ref{eq:Hamiltonian})
to include both a summation over the discrete part of the spectrum 
as well as an integration over the continuum part.
 This convention will be used throughout this section.
 
Three-body forces
are known to produce important effects on details of nuclear
spectra~\cite{Pieper:2001mp,Navratil:2003ef}. These  forces can 
in principle be added to Eq.~(\ref{eq:Hamiltonian}) and included in the 
formalism.
Since the calculations to be discussed in this paper
do not explicitly include their effects, we will therefore not
consider them in the following but we will discuss some related issues in
Sec.~\ref{sec:satprop}.

The experimental information concerning sp properties
of a many-body nuclear system is usually obtained by probing
the ground state
$| \Psi^A_0 \rangle$  of a nucleus with $A$ particles by the 
addition or removal of a nucleon as discussed in more detail in 
Sec.~\ref{sec:exp}.
The one-body (sp) propagator
(or two-point Green's function) associated with
the state $| \Psi^A_0 \rangle$ embodies this information and is
defined according to~\cite{AAA} - \cite{fetwa}
\begin{equation}
g_{\alpha \beta}(t-t') ~=~ - i~\langle \Psi^A_0 | ~\mathcal{T}
 [c_\alpha(t)  
                      c^{\dag}_\beta(t') ]~| \Psi^A_0 \rangle \; , 
\label{eq:fullg_time}
\end{equation}
where $\mathcal{T}[...]$ represents the time-ordering operation and
$c^{\dag}_\alpha(t)$ and $c_\alpha(t)$ now correspond to
operators in the Heisenberg picture (with $\hbar = 1$).
 For $\tau=t-t' > 0$, Eq.~(\ref{eq:fullg_time}) incorporates the probability
amplitude for adding a particle in a state $\beta$ and removing it from
a state $\alpha$ after a time $\tau$. Similarly, a hole can be
generated in $\alpha$ and annihilated from  $\beta$  after a time $t'-t > 0$.

The information contained in the sp propagator
$g_{\alpha \beta}(\tau)$ becomes more transparent after 
Fourier transformation from the time to the energy formulation.
After separating the positive and negative time
contributions in Eq.~(\ref{eq:fullg_time})
one includes the completeness relations for $A\pm 1$ nucleons to obtain
\begin{eqnarray}
 g_{\alpha \beta}(\omega)  ~&=&~ 
   \frac{1}{2\pi}  \int_{-\infty}^{+\infty} d\tau   ~  e^{-i \omega \tau}
       g_{\alpha \beta}(\tau)
\nonumber  \\
 &=&~
 \sum_n  \frac{
   \langle \Psi^A_0     \vert  c_\alpha       \vert \Psi^{A+1}_n \rangle
   \langle \Psi^{A+1}_n \vert  c^{\dag}_\beta  \vert \Psi^A_0     \rangle
   }
                       {\omega - \varepsilon^{+}_n + i \eta }  ~+~
 \sum_k \frac{ 
   \langle {\Psi^{A-1}_k} \vert  c_\alpha        \vert {\Psi^A_0}     \rangle
   \langle {\Psi^A_0}     \vert  c^{\dag}_\beta \vert {\Psi^{A-1}_k} \rangle
   }
                       {\omega - \varepsilon^{-}_k - i \eta } \; ,
\label{eq:fullg_Leh}
\end{eqnarray}
which is know as the Lehmann representation of the sp 
propagator~\cite{Lehmann52}.
Here, and in the following, the indices $n$ and $k$ enumerate the 
eigenstates of the system with $A+1$ and $A-1$ particles, respectively.
As discussed above, the summations are intended to represent sums over the 
discrete spectrum and integrals over the continuum part.
 The poles $\varepsilon^{+}_n \equiv E^{A+1}_n - E^A_0$  and
$\varepsilon^{-}_k  \equiv  E^A_0 - E^{A-1}_k$ in Eq.~(\ref{eq:fullg_Leh}) 
correspond to the excitation
energies of the $A\pm1$-body systems with respect to the ground-state
energy $E^A_0$.
Equation~(\ref{eq:fullg_Leh}) includes information on the transition amplitudes
for the addition and removal of a nucleon in the numerator
and these excitation energies in the denominator.
This feature illustrates the wealth of information in the sp propagator
that can be compared to experimental data. Experimentally,
this information is typically obtained in the form
of the one-hole 
spectral function which is related to the imaginary part of
$g_{}(\omega)$ by
\begin{eqnarray}
  S^h_{\alpha}(\omega) &=& \frac{1}{\pi}  ~ {\rm Im} \; 
g_{\alpha \alpha}(\omega)
\nonumber \\
   &=& \sum_k 
   \left| \langle {\Psi^{A-1}_k} \vert  c_\alpha  
                  \vert {\Psi^A_0}     \rangle  \right|^2
               \delta(\omega - \varepsilon^{-}_k)  
                  ~ ~ ~ ~ ~ ~  ~ ~ ~ \omega < \varepsilon^{-}_F \; ,
\label{eq:Sh}
\end{eqnarray}
for nucleon removal.
The corresponding particle spectral function is given by
\begin{eqnarray}
  S^p_{\alpha}(\omega) &=& - \frac{1}{\pi}  ~ {\rm Im} \; g_{\alpha \alpha}
(\omega)
\nonumber \\
   &=& \sum_n 
   \left| \langle {\Psi^{A+1}_n} \vert  c^{\dag}_\alpha  
                  \vert {\Psi^A_0}     \rangle  \right|^2
               \delta(\omega - \varepsilon^{+}_n)  
                ~ ~ ~ ~ ~ ~  ~ ~ ~ \omega > \varepsilon^{+}_F \; ,
\label{eq:Sp}
\end{eqnarray}
and in this form is mostly of theoretical relevance.
$S^h(\omega)$ (~$S^p(\omega)$~) 
represent the probability for removing (adding)
a particle from (to)
an (in) orbital $\alpha$ while leaving the residual system in a state with
energy $-\omega$ ($\omega$) relative to the ground state $E^A_0$
of the system with $A$~particles.
The Fermi energies $\varepsilon_F^{+}$ and $\varepsilon_F^{-}$ denote
the minimum excitation energy needed to remove or add
a particle and are given by
\begin{subequations}
\label{eq:ef+ef-}
\begin{eqnarray}
   \varepsilon^{-}_F &\equiv& E^A_0 - E^{A-1}_0  \; 
\label{eq:ef+ef-a} \\
   \varepsilon^{+}_F &\equiv& E^{A+1}_0 - E^A_0  \; .
\label{eq:ef+ef-b}
\end{eqnarray}
\end{subequations}
For infinite systems with a ground state that does not exhibit pairing these
quantities are equal in the thermodynamic limit
and will be referred to as the Fermi
energy $\varepsilon_F=\varepsilon_F^{+}=\varepsilon_F^{-}$.
In finite nuclei a considerable difference exists between
$\varepsilon_F^{+}$ and~$\varepsilon_F^{-}$ for the closed-shell nuclei
considered in this review.

An important experimental result, when considering the knockout of nucleons
from a mf orbit, is the reduction of the observed cross section 
with respect to the prediction of the independent-particle model 
(IPM)~\cite{deWitt1} - \cite{Paviabook} as discussed in Sec.~\ref{sec:exp}. 
This is due to
the fact that these orbits are only partially occupied in the 
correlated system.
From the theoretical point of view, this effect can be quantified by
looking at the normalization
of the overlap integral between the initial
and the final states,%
~$\langle \Psi^{A-1}_k | c_\alpha | \Psi^A_0 \rangle$~\cite{Paviabook}
(see Sec.~\ref{sec:exp} for more details).
Considering the transition to a state $k$ of the $A-1$ particle system
one may define the theoretical spectroscopic factor as
\begin{equation}
 S_k = \sum_\alpha  \left| \langle {\Psi^{A-1}_k} \vert  c_\alpha  
                  \vert {\Psi^A_0}     \rangle  \right|^2 \; .
\label{eq:hole_Sk}
\end{equation}
Obviously, this quantity is also contained in the
hole spectral function given in Eq.~(\ref{eq:Sh}).

The information on nuclear structure that is contained
in Eqs.~(\ref{eq:Sh}) and~(\ref{eq:Sp})
can be fruitfully compared to the case of an
uncorrelated mf system.
When the residual interaction is neglected in Eq.~(\ref{eq:Hamiltonian}),
the wave function of the system is an eigenstate of
the unperturbed Hamiltonian $\hat{H}_0$.
The corresponding ground state, $| \Phi^A_0 \rangle$, is a Slater determinant
with all the lowest sp orbitals fully occupied up to the hole
Fermi energy~$\varepsilon_F^-$,
while those above are empty.
The unperturbed propagator associated with $| \Phi^A_0 \rangle$ 
is given by
\begin{equation}
 g^{(0)}_{\alpha \beta}(\omega) ~=~ \delta_{\alpha , \beta}   \left\{ 
 \frac{ \theta(\alpha - F) }{\omega - \varepsilon^0_\alpha + i \eta }  ~+~
 \frac{ \theta(F - \alpha) }{\omega - \varepsilon^0_\alpha - i \eta }
  \right\}   \; ,
\label{eq:g0}
\end{equation}
where $\theta(\alpha - F)$~(~$\theta(F - \alpha)$~) is equal to 0~(1) if
$\alpha$ is an occupied state and it is 1~(0) otherwise.
The poles of $g^{(0)}(\omega)$ correspond to
the unperturbed sp energies and the forward and backward going
contributions contain either orbitals
that are outside or inside the Fermi sea~$F$, respectively.
 We note that in order to consider the whole sp spectrum, one
needs to consider both the particle and hole states.
 The hole spectroscopic factors associated with Eq.~(\ref{eq:g0})
are exactly one for those states that belong to the Fermi sea
and zero otherwise,
reflecting the occupancy of the orbitals in $| \Phi^A_0 \rangle$.
A deviation from unity occurs in a finite system when the center-of-mass motion
is properly treated~\cite{diedef}.

This simple mf picture is modified when one includes many-body correlations,
generated by the term $\hat{H}_1$ in the full Hamiltonian.
Comparing Eqs.~(\ref{eq:fullg_Leh}) and~(\ref{eq:g0}),
one can still find a correspondence between the
unperturbed energies of particle orbitals and the excitation spectrum
$\{n\}$ of the
system with $A+1$ nucleons and between the hole energies 
and the spectrum $\{k\}$ of $A-1$ nucleons.
However, a larger number of states can appear in the correlated systems
due to the mixing of sp excitations with more complicated configurations.
 As discussed in Sec.~\ref{sec:EOM}, these more complicated states
 are at least of two-particle$-$one-hole (2p1h) or two-hole$-$one-particle
(2h1p) character.
In the case of nuclear systems, the mixing of these states is generated
by the presence of both collective modes of the system and 
the relevance of short-range effects associated with the nature of the
underlying two-body interaction $\hat{V}$.
This mixing generates a large number of sp fragments
(or poles in Eq.(\ref{eq:fullg_Leh})~). At the same time, it reduces the total
sp strength of each peak of the IPM and redistributes it among all
the fragments of both particle and hole type, over a wide range of energies.
The reduction of the spectroscopic factors~(\ref{eq:hole_Sk}) from 1
 can therefore be taken as a direct measure of correlations
in the system.

The occupation number of a given sp particle state
in the correlated ground state can be obtained from the hole
spectral function
\begin{equation}
 n_\alpha =
   \langle {\Psi^A_0} \vert  c^\dag_\alpha c_\alpha
    \vert {\Psi^A_0}     \rangle  =
  \sum_k  \left| \langle {\Psi^{A-1}_k} \vert  c_\alpha  
                  \vert {\Psi^A_0}     \rangle  \right|^2
  = \int_{-\infty}^{\varepsilon^-_F} \!\!\! 
d\omega \; S^h_\alpha(\omega) \; \; ,
\label{eq:occ_numb}
\end{equation}
where the integral ranges over all the fragments of the orbital 
$\alpha$ that are spread over different quasihole energies,
as discussed above.
Correlations have the effect of
reducing the occupation of the orbitals in the Fermi sea (which would
be 1 for the unperturbed state $| \Phi^A_0 \rangle$) and to partially
populate states that were originally empty.
A measure of the emptiness of the orbital is obtained by integrating the
particle spectral function
\begin{equation}
 d_\alpha =
   \langle {\Psi^A_0} \vert  c_\alpha c^\dag_\alpha
    \vert {\Psi^A_0}     \rangle  =
  \sum_n  \left| \langle {\Psi^{A+1}_n} \vert  c^\dag_\alpha  
                  \vert {\Psi^A_0}     \rangle  \right|^2
  = \int^{+\infty}_{\varepsilon^+_F} \!\!\! 
d\omega \; S^p_\alpha(\omega) \; \; .
\label{eq:depl_numb}
\end{equation}
Using the fermion anticommutation relations, one obtains
the sum rule $n_\alpha+d_\alpha=1$, valid for any orbital $\alpha$. This
is an important result since it connects information on
 the removal and the addition of a particle to the system which
refer to quite different experimental processes.
It also supports the suggestion that both quasiparticle and quasihole states,
together, form a natural space in which to study sp
properties~\cite{Escher:2002ud}.

It is also worth noting that the knowledge of the sp propagator allows
the evaluation of the expectation value of any one-body operator
$\hat{O}$ according to
\begin{equation}
\langle \Psi^A_0 | ~\hat{O}~| \Psi^A_0 \rangle =
 \sum_{\alpha \; \beta}  O_{\alpha , \beta}
  \int \frac{d\omega}{2 \pi i} e^{i \omega \eta} g_{\beta \alpha}(\omega)
  \; . 
  \label{eq:1b_expctval}
\end{equation}

Moreover, it is also possible
to compute the energy of the ground state by means of the 
Midgal-Galitski-Koltun (MGK) sum rule~\cite{MigGal58,Koltun}
\begin{equation}
 E^A_0 =
   \langle {\Psi^A_0} \vert  \hat{H} \vert {\Psi^A_0}     \rangle
  = \frac{1}{2}
   \int_{-\infty}^{\varepsilon^-_F} d\omega \; 
   \sum_{\alpha \beta}
     \left( \langle \alpha \vert \hat{T} \vert \beta \rangle
     ~+~
     \omega \; \delta_{\alpha \beta} \right)
       \frac{1}{\pi} {\rm Im} \; g_{\beta \alpha}(\omega)
    \; \; ,
\label{eq:MGKsum}
\end{equation}
which is an exact result when only two-body forces are considered.
Equation (\ref{eq:MGKsum}) is noteworthy since it includes 
the expectation value
of the potential energy, which is a two-body quantity, solely in terms of the
one-body propagator.
This contribution is obtained
by integrating over all possible separation energies the product of this
separation energy multiplied by the corresponding spectral strength (summed
over all relevant sp orbits).
 This result is very important because it shows how the details
of the hole spectral distribution strongly influence the
binding energy of the correlated ground state.
The occupation number $n_\alpha$ yields information
about the overall population of the state $\alpha$ but the energy integral
in Eq.~(\ref{eq:occ_numb}) loses track of the energies at which these nucleons 
are located inside the nucleus.
 In the case of nuclear systems a sizable amount of strength is shifted
down to very negative sp energies by short-range and tensor correlations,
with important consequences for the binding energy of the system
as discussed in Sec.~\ref{sec:satprop}.

\subsection{Equation of motion method and Dyson equation
\label{sec:EOM} }

\begin{figure}[tb]
\begin{center}
\begin{picture}(300,300)(20,0)
\SetColor{Black}
\ArrowLine(24,130)(24,170)
\ArrowLine(26,130)(26,170)
\Vertex(25,130){2}
\Vertex(25,170){2}
\Text(50,150)[l]{$=$}
\ArrowLine(75,130)(75,170)
\Vertex(75,130){2}
\Vertex(75,170){2}
\Text(95,150)[l]{$+$}
\Vertex(125,140){2}
\Vertex(125,160){2}
\BCirc(125,150){10}
\Text(120,150)[l]{$\Sigma^*$}
\ArrowLine(125,100)(125,140)
\ArrowLine(125,160)(125,200)
\Vertex(125,100){2}
\Vertex(125,200){2}
\Text(150,150)[l]{$+$}
\Vertex(175,130){2}
\Vertex(175,170){2}
\Vertex(175,190){2}
\Vertex(175,230){2}
\Vertex(175,110){2}
\Vertex(175,70){2}
\BCirc(175,120){10}
\Text(170,120)[l]{$\Sigma^*$}
\BCirc(175,180){10}
\Text(170,180)[l]{$\Sigma^*$}
\ArrowLine(175,130)(175,170)
\ArrowLine(175,190)(175,230)
\ArrowLine(175,70)(175,110)
\Text(200,150)[l]{$+$}
\Vertex(225,160){2}
\Vertex(225,140){2}
\Vertex(225,200){2}
\Vertex(225,220){2}
\Vertex(225,260){2}
\Vertex(225,80){2}
\Vertex(225,100){2}
\Vertex(225,40){2}
\BCirc(225,210){10}
\Text(220,210)[l]{$\Sigma^*$}
\BCirc(225,150){10}
\Text(220,150)[l]{$\Sigma^*$}
\BCirc(225,90){10}
\Text(220,90)[l]{$\Sigma^*$}
\ArrowLine(225,160)(225,200)
\ArrowLine(225,220)(225,260)
\ArrowLine(225,40)(225,80)
\ArrowLine(225,100)(225,140)
\Text(250,150)[l]{$+$}
\Vertex(275,170){2}
\Vertex(275,130){2}
\Vertex(275,190){2}
\Vertex(275,230){2}
\Vertex(275,250){2}
\Vertex(275,290){2}
\Vertex(275,110){2}
\Vertex(275,70){2}
\Vertex(275,50){2}
\Vertex(275,10){2}
\BCirc(275,240){10}
\Text(270,240)[l]{$\Sigma^*$}
\BCirc(275,180){10}
\Text(270,180)[l]{$\Sigma^*$}
\BCirc(275,120){10}
\Text(270,120)[l]{$\Sigma^*$}
\BCirc(275,60){10}
\Text(270,60)[l]{$\Sigma^*$}
\ArrowLine(275,130)(275,170)
\ArrowLine(275,190)(275,230)
\ArrowLine(275,70)(275,110)
\ArrowLine(275,10)(275,50)
\ArrowLine(275,250)(275,290)
\Text(290,150)[l]{$+ ...$}
\end{picture}
\begin{minipage}[t]{16.5 cm}
\caption{
 Illustration of how the Dyson equation generates the expansion in which the
 irreducible self-energy $\Sigma^\star(\omega)$ is summed to all orders.
\label{fig:Dysonb} }
\end{minipage}
\end{center}
\end{figure}
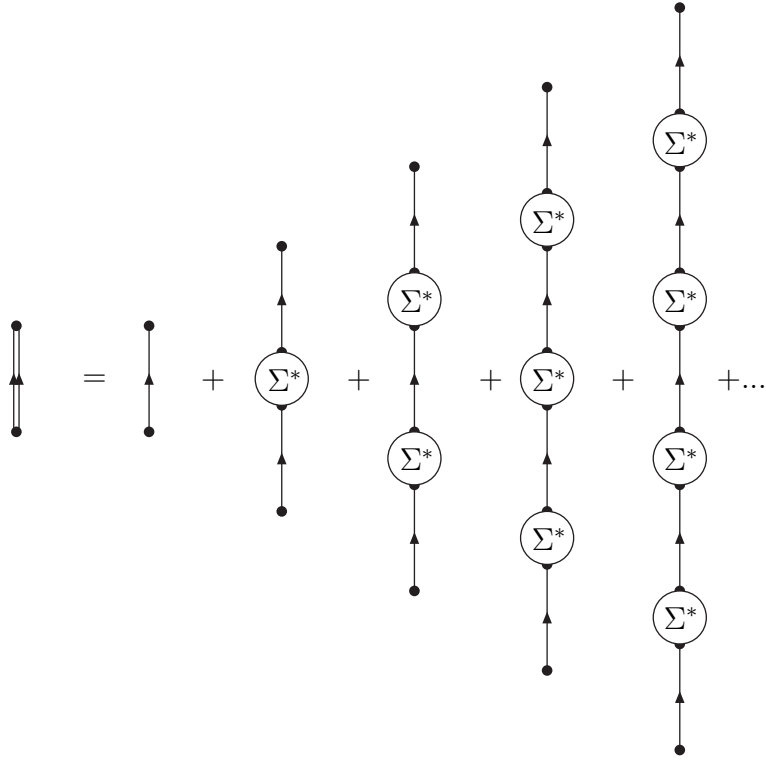

The Green's function~(\ref{eq:fullg_time}) can be computed by applying
the methods of quantum field theory.
This path was developed in the fifties
by Migdal and Galitski~\cite{MigGal58,Galitskii58} and
by Martin and Schwinger~\cite{masch} and makes use of a Feynman-diagram
expansion.
The method has found extensive application in several areas of
many-body physics including atomic, condensed matter, and nuclear physics.
One possible approach consists in choosing the propagator $g^{(0)}(\omega)$ 
as a starting point and to generate a perturbative expansion in the
interaction $\hat{H}_1$. The diagram rules for calculating the contributions
to $g(\omega)$ can be found in several text books (see for example
Refs.~\cite{AAA,Noz,fetwa}).
In general, the expansion will contain diagrams that describe all the
possible ways in which the unperturbed sp propagator interacts with
particle and hole excitations inside the system.
 Due to the strongly repulsive character of the nuclear force
a truncation of the perturbative expansion to a given power in
the interaction $\hat{H}_1$ is quite inadequate
and it becomes necessary to include the relevant physical
 processes by means of all-order summation techniques.
One therefore introduces the concept of the irreducible self-energy
$\Sigma^{\star}(\omega)$ as the collection 
of all connected diagrams that are one-particle irreducible, that is, they
cannot be separated by cutting a single propagator line. The sum of these
contributions
represents the effective interaction that a particle is subjected to
when interacting with the nuclear medium as will be shown explicitly
in Sec.~\ref{sec:exp}.
 These contributions can be summed to all orders by the Dyson equation
shown in Fig.~\ref{fig:Dyson} which generates the whole
perturbation series as shown in Fig.~\ref{fig:Dysonb}. 
This Dyson equation is given by
\begin{equation}
 g_{\alpha \beta}(\omega) ~=~ 
  g^{(0)}_{\alpha \beta}(\omega) ~+~ 
  \sum_{\gamma ,  \delta} g^{(0)}_{\alpha \gamma}(\omega) \; 
                        \Sigma^{\star}_{\gamma \delta}(\omega)  \;
                        g_{\delta \beta}(\omega) \; .
 \label{eq:dyson}
\end{equation}
\begin{figure}[bt]
\begin{center}
\begin{picture}(300,115)(0,-15)
\SetColor{Black}
\ArrowLine(75,25)(75,75)
\Text(80,75)[l]{$\alpha$}
\Text(80,25)[l]{$\beta$}
\ArrowLine(77,25)(77,75)
\Vertex(76,25){2}
\Vertex(76,75){2}
\Text(60,50)[l]{$g$}
\Text(90,50)[l]{$=$}
\ArrowLine(125,25)(125,75)
\Text(130,75)[l]{$\alpha$}
\Text(130,25)[l]{$\beta$}
\Vertex(125,25){2}
\Vertex(125,75){2}
\Text(105,50)[l]{$g^{(0)}$}
\Text(147,50)[l]{$+$}
\ArrowLine(180,60)(180,100)
\Text(185,100)[l]{$\alpha$}
\Text(185,0)[l]{$\beta$}
\Text(185,35)[l]{$\delta$}
\Text(185,65)[l]{$\gamma$}
\ArrowLine(179,0)(179,40)
\ArrowLine(181,0)(181,40)
\Vertex(180,0){2}
\Vertex(180,40){2}
\Vertex(180,60){2}
\Vertex(180,100){2}
\Text(188,84)[l]{$g^{(0)}$}
\Text(188,17)[l]{$g$}
\BCirc(180,50){10}
\Text(175,50)[l]{$\Sigma^*$}
\end{picture}
\begin{minipage}[t]{16.5 cm}
\caption{\label{fig:Dyson} Graphical representation of the Dyson equation
(Eq.~(\ref{eq:dyson})~). The double line represents the exact (dressed) sp 
propagator~$g(\omega)$, while the single line stands for the unperturbed 
propagator~$g^{(0)}(\omega)$.}
\end{minipage}
\end{center}
\end{figure}
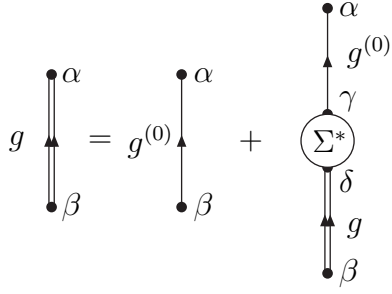

 An alternatve derivation of Eq.~(\ref{eq:dyson}) can be obtained
by either considering the equation of motion for $g(\omega)$~\cite{masch}
or working directly with the effective action functional~\cite{NegeleBook}.
The result is an exact formulation relating
each many-body Green's function to the exact propagators of higher order.
This generates a hierarchy of relations between the two-, three- and
$A$-body Green's functions, of which Eq.~(\ref{eq:g_derivative}) below
is the first example~\cite{masch,migdal67}.
Employing the equation of motion for a Heisenberg operator
$d c_\alpha(t) / dt = -i [c_\alpha(t),\hat{H}]$ one obtains the time derivative
of Eq.~(\ref{eq:fullg_time})~\cite{AAA},
\begin{eqnarray}
 i \frac{\partial}{\partial t}g_{\alpha \beta}(t - t') 
    ~&=&~ \delta(t - t') \delta_{\alpha \beta} ~+~
\varepsilon^0_\alpha g_{\alpha \beta}(t - t') ~-~
          \sum_\gamma U_{\alpha \gamma} g_{\gamma \beta}(t - t')
 \nonumber \\
 &+&~ \frac{1}{2} \sum_{\eta, \gamma, \zeta} V_{\alpha \eta , \gamma \zeta}
   \; (-i) \; \langle \Psi^A_0 |
             ~T [ c^{\dag}_\eta(t) c_\zeta(t) c_\gamma(t) 
                 c^{\dag}_\beta(t') ]
              ~| \Psi^A_0 \rangle \;   .
\label{eq:g_derivative}
\end{eqnarray}
This generates a term containing the 4-point 
Green's function
\begin{equation}
     g^{4-pt}_{\alpha \beta ,\gamma \delta}(t_1 , t_2 ; t_3 , t_4) ~=~
          - i ~ \langle \Psi^A_0 |
             ~T [c_\beta(t_2) c_\alpha(t_1) 
                  c^{\dag}_\gamma(t_3) c^{\dag}_\delta(t_4) ]
              ~| \Psi^A_0 \rangle \;   .
\label{eq:g_4pt}
\end{equation}
By contracting this with the matrix elements of the two-body interaction,
as in the last term of Eq.~(\ref{eq:g_derivative}), one can relate the
expectation value of $\hat{V}$ to the one-body
propagator~(\ref{eq:fullg_time}), whence the MGK sum rule~(\ref{eq:MGKsum})
follows.

\begin{figure}[tb]
\begin{center}
\begin{minipage}[t]{8 cm}
\includegraphics[origin=bl,angle=0,width=0.85\textwidth]{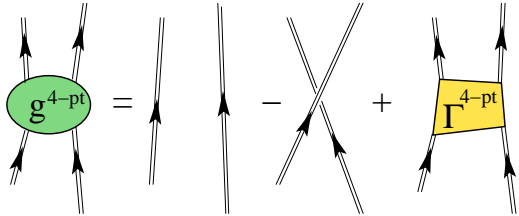}
\end{minipage}
\begin{minipage}[t]{16.5 cm}
\caption{
    Diagrammatic structure of the four-point Green's function in terms
    of the vertex function $\Gamma^{4-pt}$ (see Eq.~(\ref{eq:Gamma_4pt})~).
\label{fig:Gamma_4pt}}
\end{minipage} 
\end{center}
\end{figure}

In general, $g^{4-pt}$ can describe the propagation of either
two-particle (pp), two-hole (hh) or particle-hole (ph) excitations
depending of the ordering of its time arguments.
From a diagrammatical point of view, this propagator is the sum of
two contributions. The first is a diagram in which two different
particles propagate, fully correlated, without any interaction
among them.
The second group defines the four-point vertex function
$\Gamma^{4-pt}$ which includes
all the diagrams in which two dressed particles can interact with
each other, thereby generalizing the $T$ matrix for the
scattering of two free particles.
The corresponding contribution to $g^{4-pt}$ is simply obtained by 
adding external lines to this vertex. One therefore obtains
\begin{eqnarray}
 g^{4-pt}_{\alpha \beta ,\gamma \delta}(t_1 , t_2 ; t_3 , t_4) ~&=&~
     i \left[
     g_{\alpha \gamma}(t_1 - t_3) \;  g_{ \beta \delta}(t_2 - t_4) ~-~
     g_{\alpha \delta}(t_1 - t_4) \;  g_{ \beta \gamma}(t_2 - t_3) \right]
\nonumber \\
  & &-~ \int dt_5 \int dt_6 \int dt_7 \int dt_8
      \sum_{\mu, \nu, \lambda, \eta}
    g_{\alpha \mu}(t_1 - t_5) g_{\beta \nu}(t_2 - t_6) 
\nonumber \\
  & & ~ ~ \times 
    \; \Gamma^{4-pt}_{\mu \nu, \lambda \eta}(t_5 , t_6 ; t_7 , t_8) \;
    g_{\lambda \gamma}(t_7 - t_3) g_{\eta \delta}(t_8 - t_4) 
    \; ,
\label{eq:Gamma_4pt}
\end{eqnarray}
which is depicted in Fig.~\ref{fig:Gamma_4pt}.

\begin{figure}[t]
\begin{center}
\begin{picture}(220,100)(-40,0)
\SetColor{Black}
\Vertex(-30,60){2}
\Vertex(-30,40){2}
\ArrowLine(-30,60)(-30,70)
\ArrowLine(-29,30)(-29,40)
\ArrowLine(-31,30)(-31,40)
\BCirc(-30,50){10}
\Text(-35,50)[l]{$\Sigma^*$}
\ArrowLine(9,40)(9,50)
\ArrowLine(11,40)(11,50)
\ArrowLine(10,50)(10,60)
\Vertex(35,50){2}
\ZigZag(10,50)(35,50){3}{3}
\Vertex(10,50){2}
\ArrowLine(59,40)(59,50)
\ArrowLine(61,40)(61,50)
\ArrowLine(60,50)(60,60)
\Vertex(85,50){2}
\Text(-10,50)[l]{$=$}
\Text(43,50)[l]{$+$}
\Text(107,50)[l]{$+$}
\DashLine(60,50)(85,50){2}
\Text(70,0)[l]{$a)$}
\Vertex(60,50){2}
\ArrowArcn(91,50)(7,0,360)
\CArc(91,50)(7,0,360)
\ArrowArcn(91,50)(5,0,360)
\CArc(91,50)(5,0,360)
\ArrowLine(124,15)(124,25)
\ArrowLine(126,15)(126,25)
\ArrowLine(124,50)(124,90)
\ArrowLine(126,50)(126,90)
\ArrowLine(125,90)(125,100)
\Vertex(125,50){2} \Vertex(125,90){2}
\DashLine(125,90)(150,90){2}
\ArrowLine(149,50)(149,90)
\ArrowLine(151,50)(151,90)
\Vertex(150,50){2} \Vertex(150,90){2}
\Vertex(125,25){2} \Vertex(150,25){2}
\ArrowArcn(111,57.5)(50,40.54,319.46)
\ArrowArcn(113,57.5)(50,40.54,319.46)
\CBox(125,25)(150,50){Black}{White}
\Text(128,37)[l]{$\Gamma^{4pt}$}
\end{picture}
\\
\begin{picture}(220,100)(-40,0)
\SetColor{Black}
\Vertex(-30,60){2}
\Vertex(-30,40){2}
\ArrowLine(-31,60)(-31,70)
\ArrowLine(-29,60)(-29,70)
\ArrowLine(-30,30)(-30,40)
\BCirc(-30,50){10}
\Text(-35,50)[l]{$\Sigma^*$}
\ArrowLine(10,40)(10,50)
\ArrowLine(9,50)(9,60)
\ArrowLine(11,50)(11,60)
\Vertex(35,50){2}
\ZigZag(10,50)(35,50){3}{3}
\Vertex(10,50){2}
\ArrowLine(60,40)(60,50)
\ArrowLine(59,50)(59,60)
\ArrowLine(61,50)(61,60)
\Vertex(85,50){2}
\DashLine(60,50)(85,50){2}
\Text(-10,50)[l]{$=$}
\Text(43,50)[l]{$+$}
\Text(107,50)[l]{$+$}
\Text(70,0)[l]{$b)$}
\Vertex(60,50){2}
\ArrowArcn(91,50)(7,0,360)
\CArc(91,50)(7,0,360)
\ArrowArcn(91,50)(5,0,360)
\CArc(91,50)(5,0,360)
\ArrowLine(125,15)(125,25)
\ArrowLine(124,25)(124,65)
\ArrowLine(126,25)(126,65)
\ArrowLine(124,90)(124,100)
\ArrowLine(126,90)(126,100)
\Vertex(125,25){2} \Vertex(125,90){2}
\DashLine(125,25)(150,25){2}
\ArrowLine(149,25)(149,65)
\ArrowLine(151,25)(151,65)
\Vertex(150,65){2} \Vertex(150,90){2}
\Vertex(125,65){2} \Vertex(150,25){2}
\ArrowArcn(111,57.5)(50,40.54,319.46)
\ArrowArcn(113,57.5)(50,40.54,319.46)
\CBox(125,65)(150,90){Black}{White}
\Text(128,77)[l]{$\Gamma^{4pt}$}
\end{picture}
\\
\begin{picture}(270,100)(-40,0)
\SetColor{Black}
\Vertex(-30,60){2}
\Vertex(-30,40){2}
\ArrowLine(-30,60)(-30,70)
\ArrowLine(-29,30)(-29,40)
\ArrowLine(-31,30)(-31,40)
\BCirc(-30,50){10}
\Text(-35,50)[l]{$\Sigma^*$}
\ArrowLine(9,40)(9,50)
\ArrowLine(11,40)(11,50)
\ArrowLine(10,50)(10,60)
\Vertex(35,50){2}
\ZigZag(10,50)(35,50){3}{3}
\Vertex(10,50){2}
\ArrowLine(59,40)(59,50)
\ArrowLine(61,40)(61,50)
\ArrowLine(60,50)(60,60)
\Vertex(85,50){2}
\DashLine(60,50)(85,50){2}
\Text(-10,50)[l]{$=$}
\Text(43,50)[l]{$+$}
\Text(105,50)[l]{$+$}
\Text(175,50)[l]{$+$}
\Text(95,0)[l]{$c)$}
\Vertex(60,50){2}
\ArrowArcn(91,50)(7,0,360)
\CArc(91,50)(7,0,360)
\ArrowArcn(91,50)(5,0,360)
\CArc(91,50)(5,0,360)
\ArrowLine(124,10)(124,20)
\ArrowLine(126,10)(126,20)
\ArrowLine(124,20)(124,37)
\ArrowLine(126,20)(126,37)
\ArrowLine(124,63)(124,80)
\ArrowLine(126,63)(126,80)
\ArrowLine(125,80)(125,90)
\Vertex(125,20){2} \Vertex(125,80){2}
\DashLine(125,80)(155,80){2}
\DashLine(125,20)(155,20){2}
\ArrowLine(149,63)(154,80)
\ArrowLine(151,63)(156,80)
\ArrowLine(154,80)(159,63)
\ArrowLine(156,80)(161,63)
\ArrowLine(154,20)(149,37)
\ArrowLine(156,20)(151,37)
\ArrowLine(159,37)(154,20)
\ArrowLine(161,37)(156,20)
\Vertex(155,80){2}
\Vertex(155,20){2}
\CBox(120,37)(165,63){Black}{White}
\Text(130,50)[l]{$R^{2p1h}$}
\ArrowLine(194,20)(194,10)
\ArrowLine(196,20)(196,10)
\ArrowLine(194,37)(194,20)
\ArrowLine(196,37)(196,20)
\ArrowLine(194,80)(194,63)
\ArrowLine(196,80)(196,63)
\ArrowLine(195,90)(195,80)
\Vertex(195,20){2} \Vertex(195,80){2}
\DashLine(195,80)(225,80){2}
\DashLine(195,20)(225,20){2}
\ArrowLine(224,80)(219,63)
\ArrowLine(226,80)(221,63)
\ArrowLine(229,63)(224,80)
\ArrowLine(231,63)(226,80)
\ArrowLine(219,37)(224,20)
\ArrowLine(221,37)(226,20)
\ArrowLine(224,20)(229,37)
\ArrowLine(226,20)(231,37)
\Vertex(225,80){2}
\Vertex(225,20){2}
\CBox(190,37)(235,63){Black}{White}
\Text(200,50)[l]{$R^{2h1p}$}
\end{picture}
\begin{minipage}[t]{16.5 cm}
\caption{ Various possible expansions of the irreducible self-energy in terms 
of higher order many-body Green's functions.
   In all cases the first two terms represent the mean-field
   contributions $-\hat{U}$ 
   and the Hartree-Fock self-energy $\Sigma^{HF}$.
    The dynamic part of the self-energy can be expressed in terms of the 
   four-point vertex function $\Gamma^{4-pt}$.
    Part $a)$ represents the relation
   given in Eq.~(\ref{eq:selfen_Gamma}), while the part $b)$ corresponds
to an alternative derivation involving the time derivative of $g$ with respect
to $t'$.
 If $\Gamma^{4-pt}$ is approximated in
   such a way that the corresponding two diagrams in $a)$ and $b)$ are 
equivalent, the Dyson equation
   will satisfy appropriate sum rules.
    Part $c)$ gives the expansion in terms of the one-particle
   irreducible 2p1h/2h1p propagator $R(\omega)$, Eq~(\ref{eq:selfen_R}).
    Note that the full four-time dependence of $\Gamma^{4-pt}$ is needed in
    principle, while in the $R(\omega)$ formulation one
can specialize to a two-time quantity.
\label{fig:selfen}
}
\end{minipage}
\end{center}
\end{figure}
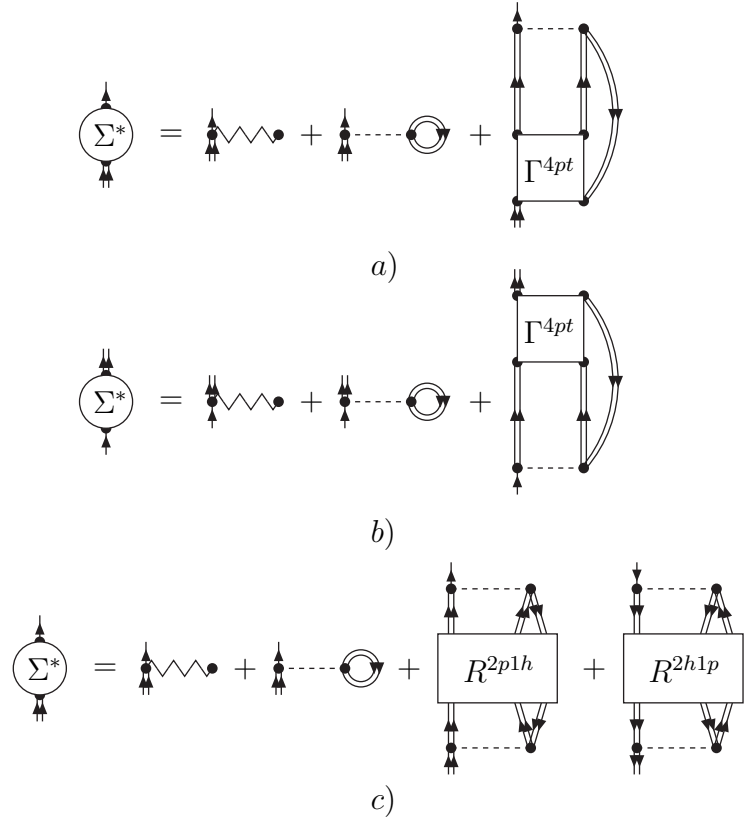

The Dyson equation~(\ref{eq:dyson}) can now
be obtained by Fourier transformation of 
Eq.(\ref{eq:g_derivative}). 
Employing Eq.~(\ref{eq:g0})
one then finds an exact expression for the self-energy
$\Sigma^{\star}(\omega)$ in terms
of the sp propagator $g(\omega)$ and the vertex $\Gamma^{4-pt}$.
By using Eq.~(\ref{eq:Gamma_4pt}) one obtains 
\begin{eqnarray}
 \Sigma^{\star}_{\gamma \delta}(\omega)  ~ &=& ~ 
    - ~ \langle \gamma \vert \hat{U} \vert \delta \rangle
  ~ + ~ \Sigma^{HF}_{\gamma \delta}
 \nonumber \\
  & & ~  - \frac{i^2}{2} \int \frac{d\omega_1}{2 \pi}
                      \int \frac{d\omega_2}{2 \pi}
        \sum_{\mu , \nu , \lambda}    \sum_{\alpha , \beta , \varepsilon}
   V_{\gamma \varepsilon , \alpha \beta} \;
       g_{\alpha \mu}(\omega_1)    g_{\beta \nu}(\omega_2)
 \nonumber \\
  & & ~ ~ \times  \Gamma^{4-pt}_{\mu \nu, \delta \lambda}
          (\omega_1 , \omega_2 ; \omega , \omega_1 + \omega_2 - \omega ) 
        g_{\lambda \varepsilon}(\omega_1 + \omega_2 - \omega )   
\label{eq:selfen_Gamma}
\end{eqnarray}
where the term $-\langle \gamma \vert \hat{U} \vert \delta \rangle$
subtracts the auxiliary potential which was added
in the unperturbed Hamiltonian $\hat{H}_0$.
This result is shown in part $a)$ of Fig.~\ref{fig:selfen}),
The second term in Eq.~(\ref{eq:selfen_Gamma}) is the Hartree-Fock
contribution to  the
self energy,
\begin{equation}
  \Sigma^{HF}_{\alpha \beta} =
\sum_{\gamma \delta}
    \int \frac{d \omega}{2 \pi i} ~ e^{+i \omega \eta^+}
  ~ V_{\alpha \gamma, \beta \delta}  ~  g_{\gamma \delta}(\omega)  \; \; .
\label{eq:selfen_HF}   
\end{equation}
which represents the (energy-independent) interaction of the nucleon
with the quasihole excitations inside the system.
 When all the higher-order terms of the self-energy are neglected,
the solution of the Dyson equation $g(\omega)$ takes the
same simple form as Eq.~(\ref{eq:g0}), which is equivalent to a 
Slater determinant description of the ground state. 
In this case the quasihole wave functions
contained in $g(\omega)$ refer to completely filled orbitals and
the iterative solution of
Eqs.~(\ref{eq:dyson}) and~(\ref{eq:selfen_HF}) generates the standard
Hartree-Fock approximation.

The many-body correlations beyond the nuclear mean field are included
in the energy-dependent part of the self-energy. A possible expansion
of this self-energy, in terms of $\Gamma^{4-pt}$, is given by the last term
of Eq.~(\ref{eq:selfen_Gamma}) and shown
diagrammatically in Fig.~\ref{fig:selfen}$a)$.
One should note that Eq.~(\ref{eq:selfen_Gamma}) is formally exact and
therefore the latter term contains implicitly the effect of both
pp, hh, and ph collective excitations and their coupling to the
sp motion. This is of course at the prize of having to
keep track of the different time arguments
in $\Gamma^{4-pt}$, which involves an awkward mathematical structure.
When only the pp (hh) effects or the ph ones need to be studied,
$\Gamma^{4-pt}$ can be conveniently approximated by the ladder or
the ring series, respectively (to be described in Sec.~\ref{sec:ring-ladd}).
In these cases $\Gamma^{4-pt}$ reduces to a two-time quantity, \textit{i.e.} 
depends on only one energy variable.
In the cases where this simplification is not possible,
Eq.~(\ref{eq:selfen_Gamma}) may no longer correspond to the best approach
and it becomes useful to go one step further in the hierachy 
of Green's functions by taking a second derivative in
Eq.~(\ref{eq:g_derivative}), with respect to the time argument~$t'$.
 This will result in introducing a 6-point Green's function
$R_{\alpha \beta \gamma , \mu  \nu  \lambda}(t - t')$ 
that includes explicitly the coupling of sp motion to 2p1h and 2h1p
propagation.
Only the two-time reduction of this propagator is needed to compute
the self-energy. After Fourier transformation one obtains,
\begin{equation}
 \Sigma^{\star}_{\gamma \delta}(\omega)  ~ = ~ 
    - ~ \langle \gamma \vert \hat{U} \vert \delta \rangle
  ~ + ~ \Sigma^{HF}_{\gamma \delta}   ~ + ~ 
    \sum_{\mu , \nu , \lambda}    \sum_{\alpha , \beta , \varepsilon}
   V_{\gamma \lambda , \mu \nu}  \; 
       R_{\mu  \nu  \lambda , \alpha \beta \varepsilon}(\omega) 
   \; V_{\alpha \beta , \delta  \varepsilon}  \; ,
\label{eq:selfen_R}
\end{equation}
which is depicted by part $c)$ of Fig.~\ref{fig:selfen}.
The propagator $R(\omega)$ contains all the diagrams 
that describe the propagation of 2p1h and 2h1p (and more complicated 
configurations) but is one-particle irreducible since these terms are
generated by the Dyson equation.

It is worth mentioning that the same development
of Eqs.~(\ref{eq:g_derivative}) to~(\ref{eq:selfen_Gamma}) can be carried
out by considering the time derivative of Eq.~(\ref{eq:fullg_time}) with
respect to~$t'$. This leads to an analogous result shown diagammatically
in part $b)$ of Fig.~\ref{fig:selfen}. Baym and Kadanoff have shown 
that an approximation chosen for  $\Gamma^{4-pt}$
should be such that parts  $a)$ and $b)$ in Fig.~\ref{fig:selfen}
generate the same self-energy~\cite{kaba1} - \cite{baym62}.
With this symmetry requirement it is assured that the solution of the Dyson 
equation $g(\omega)$ satifies basic conservation laws,
such as particle number,
total energy, total momentum, and total angular momentum.

\subsection{Self-consistent approach}
\label{sec:SCGF}

In Eqs.~(\ref{eq:selfen_Gamma}) and ~(\ref{eq:selfen_R}),
the term $- \langle \gamma \vert \hat{U} \vert \delta \rangle$
removes the auxiliary potential $\hat{U}$ included
in $g^{(0)}(\omega)$ (or equivalently in $\hat{H}_0$, as discussed in
Eqs.~(\ref{eq:2nd19})-(\ref{eq:2nd20}) below).
 This makes the solution of the Dyson equation
formally independent of the choice of $\hat{U}$. 
An implicit dependence on $\hat{H}_0 = \hat{T} + \hat{U}$
remains for the case of a standard perturbative expansion,
where the irreducible self-energy $\Sigma^{\star}(\omega)$
is expressed as a series of Feynman diagrams, in terms of $g^{(0)}(\omega)$
and the vertices of the residual interaction~\cite{AAA,fetwa}.
However, this is not the case when one uses the approach of the 
equation of motion discussed in the previous section since
this provides for an expansion of the self-energy in terms of 
the exact propagator $g(\omega)$ and  fully dressed higher-order
Green's functions.
 The use of Bethe-Salpeter-like equations to evaluate the 4- and 6-point
Green's functions results in an expansion of the self-energy only in
terms of $g(\omega)$.
While this $g$ is not exact for a given approximation, it is self-consistent
and for this reason a propagator obtained from this method can be referred to 
as a self-consistent Green's function (SCGF).
This self-consistency feature
has the advantage of generating a formalism in which the actual
excitations that propagate in the system are employed as the basic
degrees of
freedom, instead of a mf approximation of the nuclear orbitals.
Of course, this result is obtained at the price of an increased complexity in
practical calculations. In general, one seeks an adequate approximation of
the self-consistent propagator $g(\omega)$ in terms of the relevant
degrees of freedom that are important in the system under study.

 In the nuclear case, there exist strong couplings between the sp degrees
of freedom to both low-lying collective excitations and high-lying states,
the latter coupling being due to the strong short-range repulsion
in the nuclear force.
 These mechanisms
generate the sizable fragmentation of sp strengths which is observed 
experimentally.
 In order to obtain a detailed description
of the resulting nuclear structure it may be necessary to include 
the details of the fragmented one-body propagators directly
in the construction of the nucleon self-energy.

\begin{figure}[t]
 \begin{center}
  \begin{minipage}[t]{10 cm}
      \includegraphics[width=1.\textwidth]{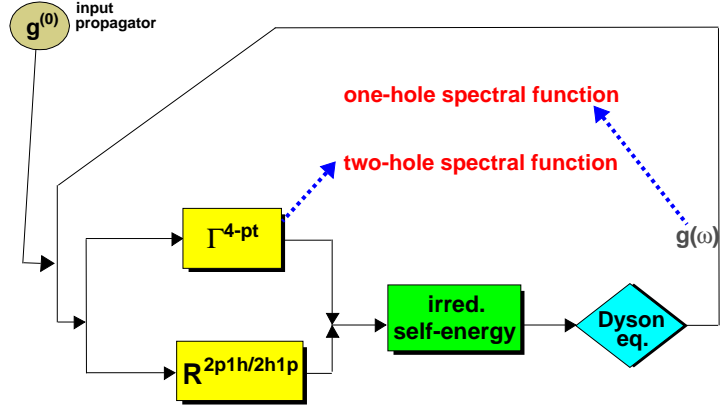}
   \end{minipage}      
\begin{minipage}[t]{16.5 cm}             
      \caption[SCGF scheme] {\label{fig:SCGF}
   \small   Iteration scheme used to reach self-consistency between the input
        propagator and the result of the Dyson equation. At 
each step in the iteration
 loop new approximations for the one-body propagator~(\ref{eq:fullg_time})
        and for $\Gamma^{4-pt}$~(\ref{eq:Gamma_4pt}) or $R(\omega)$ are
        obtained until self-consistency is obtained.
        }
      \end{minipage}
  \end{center}
\end{figure}

The practical approach of SCGF theory consists
in starting  with an approximation for the input Green's function (usually
a Hartree-Fock or other IPM propagator).
This can be used to construct an approximation for the irreducible
2p1h/2h1p propagator
$R_{\mu  \nu  \lambda , \alpha \beta \gamma}(\omega)$ or the four-point
Green's function $g^{4-pt}$ from which one derives the irreducible
self-energy
$\Sigma^{\star}(\omega)$.
Then, the solution of the Dyson Eq.~(\ref{eq:dyson}) will give an improved 
approximation to the self-consistent sp propagator that can,
subsequently, be employed in a next iteration step.
For infinite nuclear matter the self-energy should include both pp and hh
propagation to properly account for short-range effects and particle number 
conservation. This can be
done by summing the ladder equation for $\Gamma^{4-pt}$
and employing Eq.~(\ref{eq:selfen_Gamma}).
 In the case of finite nuclei both pp (hh) and ph collective motion can be
relevant and one is forced to seek a more complex expansion, based
on~Eq.~(\ref{eq:selfen_R}).
This may require the computation of both the pp (hh) and ph propagators
and then to couple them
through a Faddeev expansion~\cite{badi01}.
As shown in Fig.~\ref{fig:SCGF}, the whole procedure should be iterated 
until consistency is obtained for two successive solutions (\textit{i.e.}
between the input propagator and the solution that is generated).
 We note that the Hartree-Fock formalism, in which $\Gamma^{4-pt}$
and $R(\omega)$ are neglected, is the simplest realization
of this formulation.

 The attractive feature of the SCGF method is not only restricted to the fact
that the effects of fragmentation are included in the calculations.
Also, solutions for the one-body, two-body and higher-order propagators are
generated all at the same time while their mutual influence is
taken into account.
In addition, the comparison with experimental data can be performed for all
ingredients of the calculation yielding important clues for further 
improvements of the chosen approximation scheme.
 We stress again  that the final self-consistent solutions will not
depend in any way on the choice of $\hat{H}_0$.
Nevertheless, a good choice for the unperturbed
propagator $g^{(0)}$ can provide a useful starting point
 for the first iteration.

The self-consistent propagator  $g(\omega)$ enters the construction
of higher-order Green's functions through the use of Bethe-Salpeter
equations (BSE). We give a brief description of
this approach by considering the three-times polarization propagator,
\begin{eqnarray}
   \Pi_{\alpha \beta , \gamma \delta}(t_1 - t' , t_2 - t') &=&
   - i ~ 
   \langle \Psi^A_0     \vert ~ T [ 
    c^{\dag}_\beta(t_2)  c_\alpha(t_1)  c^{\dag}_\gamma(t')  c_\delta(t') ]
                       \vert \Psi^A_0 \rangle
\nonumber \\
   & &   + i ~ 
   \langle \Psi^A_0     \vert ~ T [ 
    c^{\dag}_\beta(t_2)  c_\alpha(t_1) ]  \vert \Psi^A_0 \rangle 
     \langle \Psi^A_0  \vert c^{\dag}_\gamma(t')  c_\delta(t') 
                       \vert \Psi^A_0 \rangle
  \; ,
\label{eq:Pi-3t}  
\end{eqnarray}
which governs the response of the system to an external perturbation.
Working with the equation of motion method, one finds that
Eq.~(\ref{eq:Pi-3t}) is the solution of the following Bethe-Salpeter
equation~\cite{kaba1,kaba2}.
\begin{eqnarray}
  \Pi_{\alpha \beta , \gamma \delta}(t_1 - t' , t_2 - t') &=&
       - i g_{\alpha \gamma}(t_1 - t') g_{\delta \beta}(t' - t_2)
\nonumber \\
  & &  + ~ \int dt_3 \int dt_4 \int dt_5 \int dt_6
       \sum_{\mu, \nu, \lambda, \eta}
   g_{\alpha \mu}(t_1 - t_3) g_{\nu \beta}(t_4 - t_2)
\nonumber \\
  & & ~ ~ \times 
      \; K^{ph}_{\mu \nu, \lambda \eta}(t_3 , t_4 ; t_5 , t_6) \;
      \Pi_{\lambda \eta , \gamma \delta}(t_5 - t' , t_6 - t')
  \; ,
\label{eq:Pi-3t_BSE}  
\end{eqnarray}
where the ph irreducible kernel $K^{ph}$, represents all the
basic interactions
that a quasiparticle and a quasihole can undergo. The latter
can be obtained from
the exact irreducible self-energy in terms of a functional derivative,
\begin{equation}
   K^{ph}_{\alpha \beta , \gamma \delta}(t_1 , t_2 ; t_3 , t_4 ) ~=~
   \frac{ \delta ~ \Sigma^{\star}_{\alpha \beta}(t_1 - t_2) }
        { \delta ~ g_{\gamma \delta}(t_3 - t_4) } \; .
\label{eq:Kph_vs_Sigma}  
\end{equation}
In the language of Feynman diagrams this corresponds to taking 
the diagrammatic expansion
of $\Sigma^{\star}(\omega)$ in terms of the propagator $g(\omega)$ and then 
generating the contributions to $K^{ph}$ by removing such a sp line
in every possible way.
Baym and Kadanoff have shown that a conserving approximation of the 
polarization propagator $\Pi$ is guaranteed if both $K^{ph}$ 
and $g(\omega)$ are derived from the same approximation of the
self-energy (by means of Eqs.~(\ref{eq:Kph_vs_Sigma}) and~(\ref{eq:dyson}),
respectively)~\cite{kaba1,baym62}.
From a practical point of view, this approach is not always possible
to implement
since relatively simple approximations to $\Sigma^{\star}(\omega)$
can generate a very large number of contributions to $K^{ph}$.
Fortunately, actual nuclear systems do not necessarily require
the inclusion of all of these terms to obtain reasonable results.
Nevertheless, the above construction can be used as
a guideline to help in choosing the approximation that is suitable for the
problem under study.
The resulting $K$ will involve an expansion in terms
 $g(\omega)$ and $\Gamma^{4-pt}$ or $R(\omega)$.
 The fact that this perturbative expansion is based on the fragmented
sp propagator ---and not on its unperturbed approximation
$g^{(0)}(\omega)$---, acts to renormalize the interaction kernel $K$
and drastically improves the convergence properties, while still allowing 
to keep track of the relevant physical ingredients in the calculation.
The simplest realization of this scheme consists in considering
only the bare interaction potential $\hat{V}$ in either the pp or ph 
channels. The result is the usual random phase approximation (RPA).
This approximation
is described in more detail in the next section since it is
relevant to most of the calculations discussed in this review.

\subsection{Ring and ladder approximations}
\label{sec:ring-ladd}

 By taking the limit $t_2 \rightarrow t_1^+ = t$ in Eq.~(\ref{eq:Pi-3t}),
one is left with the usual two-time polarization propagator
$\Pi_{\alpha \beta , \gamma \delta}(t - t')$. This can be Fourier 
transformed to its Lehmann representation,
\begin{equation}
  \Pi_{\alpha \beta , \gamma \delta}(\omega)  ~=~
   \sum_{n \ne 0}  \frac{
   \langle \Psi^A_0 \vert  c^{\dag}_\beta   c_\alpha  \vert \Psi^A_n \rangle
   \langle \Psi^A_n \vert  c^{\dag}_\gamma  c_\delta  \vert \Psi^A_0 \rangle
   }
      {\omega - \left( E^A_n - E^A_0 \right) + i \eta }  ~-~
   \sum_{n \ne 0}  \frac{
   \langle \Psi^A_0 \vert  c^{\dag}_\gamma  c_\delta  \vert \Psi^A_n \rangle
   \langle \Psi^A_n \vert  c^{\dag}_\beta   c_\alpha  \vert \Psi^A_0 \rangle
   }
      {\omega + \left( E^A_n - E^A_0 \right) - i \eta } 
  \; .
\label{eq:Pi_Leh}  
\end{equation}
where the poles correspond to the excitation
energies $\varepsilon^\pi_n \equiv E^A_n - E^A_0$ of the states 
of the system whereas the residues contain the corresponding
amplitudes to determine transition probabilities from the ground state to
these excited states. It is important to note that this
information is included for all those states that can be excited by means
of a one-body operator, therefore also those with important
two-particle$-$two-hole (2p2h) (or even more complex) admixtures.
In Eq.~(\ref{eq:Pi_Leh}), the forward- and the backward-going
contributions are identical apart from a time reversal transformation.

Analogously, one can define the
{\em free} ph propagator $\Pi^f(\omega)$ by coupling two 
excitations of particle and hole type, not interacting with each other.
 Graphically, this corresponds to two dressed sp lines
propagating with opposite direction in time. By applying Feynman diagram
rules~\cite{fetwa}, one obtains the following Lehmann representation
for~$\Pi^f(\omega)$,
\begin{eqnarray}
     \Pi^f_{\alpha \beta , \gamma \delta}(\omega) ~&=&~
            -i \int \frac{d \omega_1}{2 \pi} (-1) ~
          i g_{\alpha \gamma}(\omega+\omega_1) ~ i g_{\delta \beta}(\omega_1)
 \nonumber \\
     &=& 
 \sum_{n , k}  \frac{  
   \langle {\Psi^A_0}   \vert  c^{\dag}_\beta \vert \Psi^{A-1}_k \rangle
   \langle \Psi^A_0     \vert  c_\alpha       \vert \Psi^{A+1}_n \rangle
               \;
   \langle \Psi^{A+1}_n \vert  c^{\dag}_\gamma  \vert \Psi^A_0   \rangle
   \langle \Psi^{A-1}_k \vert  c_\delta         \vert \Psi^A_0   \rangle
                     }
     {\omega - \left( \varepsilon^{+}_n - \varepsilon^{-}_k \right) + i \eta }
 \nonumber \\
& &~-~ \sum_{n , k} \frac{
   \langle {\Psi^A_0}   \vert  c^{\dag}_\gamma \vert \Psi^{A-1}_k \rangle
   \langle \Psi^A_0     \vert  c_\delta        \vert \Psi^{A+1}_n \rangle
                    \; 
   \langle \Psi^{A+1}_n \vert  c^{\dag}_\beta   \vert \Psi^A_0   \rangle
   \langle \Psi^{A-1}_k \vert  c_\alpha         \vert \Psi^A_0   \rangle
                    }
  {\omega + \left( \varepsilon^{+}_n - \varepsilon^{-}_k \right) - i \eta }
  \; ,
\label{eq:Pif}
\end{eqnarray}
 where the forward- and the backward-going
contributions describe the propagation of a particle-hole and
hole-particle type excitation, respectively.

\begin{figure}[t]
 \begin{center}
 \begin{minipage}[t]{8 cm}
      \includegraphics[width=1.\textwidth]{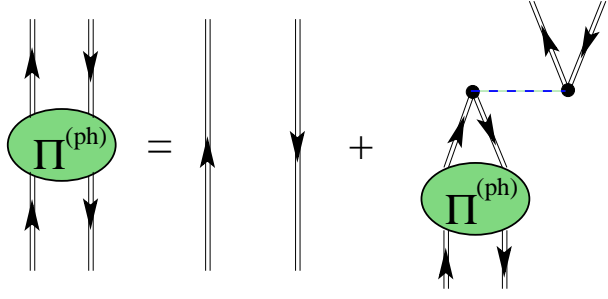}
   \end{minipage}      
\begin{minipage}[t]{16.5 cm}     
      \caption[ph DRPA equation]{ \small
       Feynman diagram representation of the DRPA equation
       for the polarization propagator $\Pi(\omega)$.
\label{fig:PiDRPA}  }
       \end{minipage}
  \end{center}
\end{figure}

\begin{figure}[t]
 \begin{center}
   \begin{minipage}[t]{9 cm}
      \includegraphics[width=1.\textwidth]{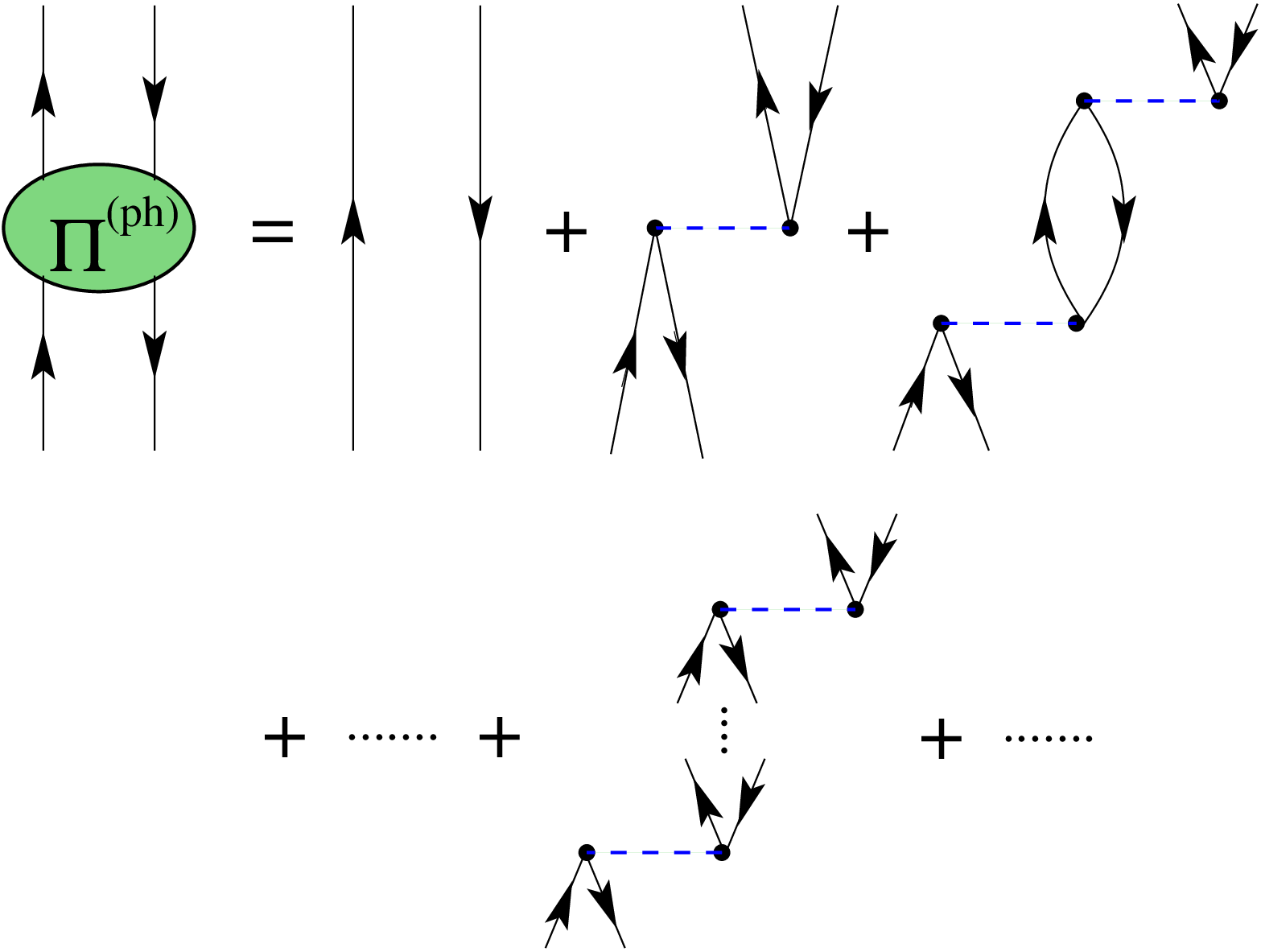}
           \end{minipage}      
\begin{minipage}[t]{16.5 cm}     
      \caption[ph TDA series]{\small
    Diagrammatic expansion of the ph propagator $\Pi(\omega)$ in TDA. 
    An implicit time ordering is intended for these diagrams.
    Dressing all the sp lines would result in the DTDA.
\label{fig:DTDAexp} }
       \end{minipage}
  \end{center}
\end{figure}

\begin{figure}[bt]
 \begin{center}
    \begin{minipage}[t]{9 cm}
      \includegraphics[width=1.\textwidth]{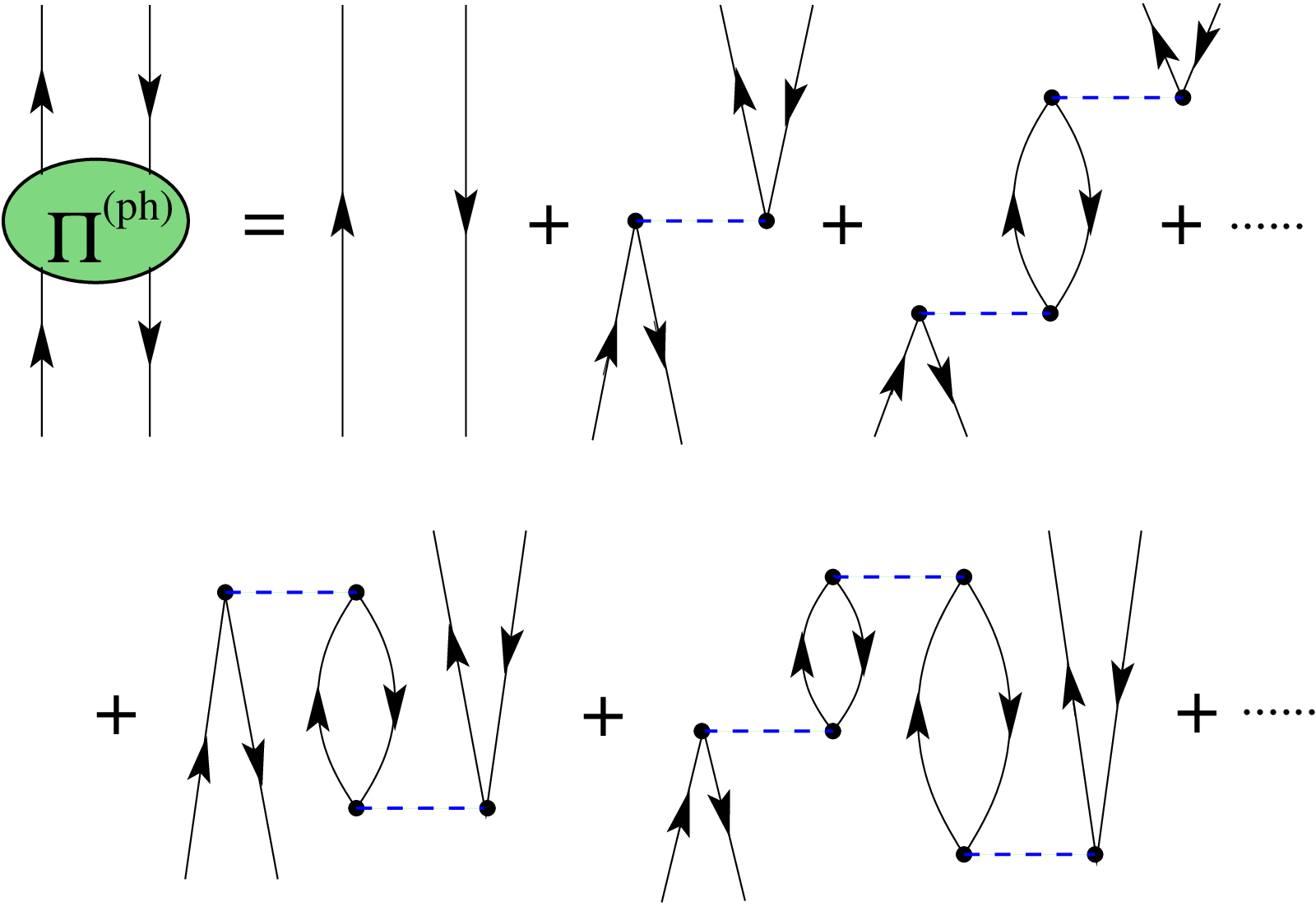}
    \end{minipage}      
\begin{minipage}[t]{16.5 cm}     
      \caption[ph RPA series]{\small
    Diagrammatic expansion of the ph propagator $\Pi(\omega)$ in RPA. 
     This contains the
    whole phTDA series plus diagrams involving the inversion of the
    propagation time. 
    Dressing all the sp lines results in the DRPA.
\label{fig:DRPAexp} }
       \end{minipage}
  \end{center}
\end{figure}

In order to account for the collective excitations present in the
$A$-body system, one needs to take into account the interactions
between the two lines propagating in $\Pi^f(\omega)$.
The simplest approximation consist in considering only the
contribution of the bare two-body potential $\hat{V}$.
 This corresponds to choosing 
$K^{ph}_{\alpha \beta , \gamma \delta} = V_{\alpha \delta , \beta \gamma}$
in the Bethe-Salpeter Eq.~(\ref{eq:Pi-3t_BSE}) and results in the
standard RPA approximation,
\begin{equation}
 \Pi_{\alpha \beta , \gamma \delta}(\omega) ~=~ 
        \Pi^f_{\alpha \beta , \gamma \delta}(\omega) ~+~ 
         \sum_{\mu , \nu , \rho , \epsilon}
           \Pi^f_{\alpha \beta , \mu \nu}(\omega) 
           ~ V_{\mu \epsilon ,  \nu \rho} 
       ~ \Pi_{\rho \epsilon , \gamma \delta}(\omega) \; ,
 \label{eq:PiDRPA}
\end{equation}
which is depicted in Fig.~\ref{fig:PiDRPA} in terms of Feynman diagrams.
Equation~(\ref{eq:PiDRPA}) implicitly generates a series of diagrams in 
which the
particle and the hole interact any number of
times.  The resulting expansion is depicted in Figs.~\ref{fig:DTDAexp}
and~\ref{fig:DRPAexp}, in terms of $g^{(0)}(\omega)$ propagators.

A simpler approximation consists in considering only
the forward-going propagation in Eq.~(\ref{eq:PiDRPA}), \textit{i. e.}
 to include only the first term 
of Eq.~(\ref{eq:Pif}).  This corresponds to the so-called dressed Tamm-Dancoff
approximation (DTDA), depicted in Fig.~\ref{fig:DTDAexp} (where an explicit
time ordering is assumed and the dressing of sp Green's functions is implied).
 When the full $\Pi^f(\omega)$ is employed in
Eq.~(\ref{eq:PiDRPA}), one obtains the dressed random phase
approximation~(DRPA). The relevant diagrammatic expansion is the one given in
Fig.~\ref{fig:DRPAexp}.
 Clearly, the whole TDA series is contained in the RPA one.
 Besides this contribution, the backward-going component of $\Pi^f(\omega)$
generates diagrams containing more that a single ph excitation at each time.
These refer to contributions in which additional intermediate ph states
are present and eventually annihilated by two-body interactions. 
 Consistent with the aim of performing a self-consistent calculation,
Eqs.~(\ref{eq:Pif}) and~(\ref{eq:PiDRPA}) have been formulated in terms
of the full, dressed, sp propagator $g(\omega)$ given in
 Eq.~(\ref{eq:fullg_time}).
When an unperturbed, IPM, Green's function
$g^0(\omega)$~(\ref{eq:g0}) is used as input, the approximations described here
reduce to the standard TDA and RPA results~\cite{schuckbook}.

 In general, the interparticle distance between nucleons inside the nucleus is
of the order of the nucleon size itself so that the nuclear medium
can be  considered a rather dense system. Besides, the nuclear 
effective interaction remains substantial.
 A natural consequence is that the ``vacuum fluctuation'' contributions
that are included in the RPA expansion play an important
role in the description of nuclear systems.
It must be realized, however, that the intermediate
{\it n}-particle--{\it n}-hole excitations depicted in Fig.~\ref{fig:DRPAexp}
do not represent all the possible diagrams
related by Pauli exchange of two sp lines.
The philosophy underlying the use of RPA relies on the assumption
that the corrections coming from Pauli correlations sum up in a random 
way and tend to cancel each other (whence the name
``random phase'')~\cite{PinesRPA} - \cite{PinesRPA2}.
As a result, the Pauli
principle may only be violated slightly and the 
improvement obtained by including RPA 
correlations is generally more
relevant. In such cases, the RPA is a useful
approximation.
  It must be kept in mind, however, that in systems containing strong
collective and/or pairing correlations the effects of Pauli violation may
sum up in a coherent way, invalidating the standard RPA approach. 
Typically this situation is signaled by the appearence of a pair of complex 
eigenvalues of Eq.~(\ref{eq:PiDRPA}) together with diverging solutions for
the ph spectroscopic amplitudes.

We also note that using dressed propagators $g(\omega)$ in
Eq.~(\ref{eq:Pif}) results in including implicitly diagrams that go beyond
the bare 1p1h configurations already at the level of $\Pi^f(\omega)$.
This is due to the self-energy insertions
in each particle and hole line. The fragmentation of $g(\omega)$ 
acts once again to renormalize the free propagator $\Pi^f(\omega)$
thereby adding stability to the DRPA approach, as compared the
the bare RPA one.

The saturation energy of nuclear matter, as well as the spectral distribution
at high missing energies, is governed by short-range effects and two-body
correlations. Moreover, some particular nuclei present strong 
pairing effects. 
 In these cases, one is faced with two-body correlations in the
system and a proper description of the physics
calls for a direct evaluation of the interaction between
two particles in the medium. From the experimental point of view, the
ideal approach to study such effects is represented by
addition or removal of the correlated pair (see Sec.~\ref{sec:exp}).
 The information required to study such processes are also contained in
the four-point Green's function, $g^{4-pt}$.
 One can define the two-particle propagator $g^{II}(t -t')$ as a
two-time reduction of Eq.~(\ref{eq:g_4pt})
\begin{equation}
 g^{II}_{\alpha \beta , \gamma \delta}(t - t') ~=~
      g^{4-pts}_{\alpha \beta ,\gamma \delta}(t , t^+ ; t'{}^+ , t') \; .
\label{eq:g2_time}
\end{equation}
Its Lehmann
representation,
\begin{equation}
 g^{II}_{\alpha \beta , \gamma \delta}(\omega) =
 \sum_n  \frac{  \langle {\Psi^A_0} \vert 
                c_\beta c_\alpha \vert {\Psi^{A+2}_n} \rangle \;
                 \langle {\Psi^{A+2}_n} \vert 
         c^{\dag}_\gamma c^{\dag}_\delta \vert {\Psi^A_0} \rangle }
            {\omega - \left( E^{A+2}_n - E^A_0 \right) + i \eta }
 - \sum_k  \frac{  \langle {\Psi^A_0} \vert 
    c^{\dag}_\gamma c^{\dag}_\delta \vert {\Psi^{A-2}_k} \rangle \;
                 \langle {\Psi^{A-2}_k} \vert 
                  c_\beta c_\alpha \vert {\Psi^A_0} \rangle }
            {\omega - \left( E^A_0 - E^{A-2}_k \right) - i \eta } \; .
\label{eq:g2}
\end{equation}
contains the spectroscopic amplitudes for a transition
from the ground state $\vert {\Psi^A_0} \rangle$ to the eigenstates
of the systems with $A \pm 2$ particles and the respective 
excitation energies with respect to $E^A_0$.

As for the polarization propagator,
the zero-order contribution consists of two noninteracting lines,
propagating in the same time direction.  The Lehmann
representation for the dressed case is given by
\begin{eqnarray}
   g^{II, f}_{\alpha \beta , \gamma \delta}(\omega)  &=&
     -i \int \frac{d \omega_1}{2 \pi}
        ~ i g_{\alpha \gamma}(\omega-\omega_1) 
        ~ i g_{\beta \delta}(\omega_1)    
 \nonumber \\
     &=&~ 
 \sum_{n_1 , n_2} ~ \frac{  
   \langle \Psi^A_0     \vert  c_\beta        \vert \Psi^{A+1}_{n_2} \rangle
   \langle \Psi^A_0     \vert  c_\alpha       \vert \Psi^{A+1}_{n_1} \rangle
               \;
   \langle \Psi^{A+1}_{n_1} \vert  c^{\dag}_\gamma  \vert \Psi^A_0   \rangle
   \langle \Psi^{A+1}_{n_2} \vert  c^{\dag}_\delta  \vert \Psi^A_0   \rangle
                }
    {\omega - \left( \varepsilon^{+}_{n_1} + \varepsilon^{+}_{n_2} \right)
        + i \eta }
 \nonumber \\
& & ~ ~ - ~ \sum_{k_1 , k_2} ~ \frac{ 
   \langle {\Psi^A_0}   \vert  c^{\dag}_\gamma \vert \Psi^{A-1}_{k_1} \rangle
   \langle {\Psi^A_0}   \vert  c^{\dag}_\delta \vert \Psi^{A-1}_{k_2} \rangle
            \; 
      \langle \Psi^{A-1}_{k_2} \vert  c_\beta     \vert \Psi^A_0   \rangle
      \langle \Psi^{A-1}_{k_1} \vert  c_\alpha    \vert \Psi^A_0   \rangle
          }
    {\omega - \left( \varepsilon^{-}_{k_1} + \varepsilon^{-}_{k_2} \right)
        - i \eta } \; ,
\label{eq:gIIf}
\end{eqnarray}
where the forward and backward parts refer to the (independent) propagation
of two particle and two hole lines, respectively.
The two-particle dressed RPA (ppDRPA) equation is given by
\begin{equation}
 g^{II}_{\alpha \beta , \gamma \delta}(\omega) ~=~ 
     \left( g^{II, f}_{\alpha \beta , \gamma \delta}(\omega)   - 
     g^{II, f}_{\beta \alpha , \gamma \delta}(\omega) \right) ~+~ 
       \frac{1}{2}  \sum_{\mu \nu \rho \varepsilon}
g^{II, f}_{\alpha \beta , \mu \nu}(\omega) \; 
           V_{ \mu \nu , \rho \epsilon} \; 
           g^{II}_{\rho \epsilon , \gamma \delta}(\omega) \; ,
 \label{eq:g2DRPA}
\end{equation}
where the symmetry factor $\frac{1}{2}$ is required by the Feynman
rules~\cite{fetwa}.

\begin{figure}[t]
 \begin{center}
  \begin{minipage}[t]{8 cm}
      \includegraphics[width=1\textwidth]{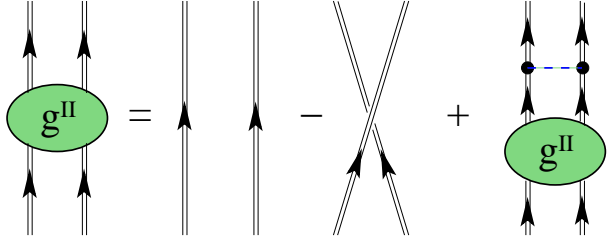}
        \end{minipage}      
\begin{minipage}[t]{16.5 cm}    
      \caption[pp DRPA equation]{\label{fig:gIIf}
   \small  DRPA (ladder)
equation for the two-particle propagator $g^{II}(\omega)$.}
         \end{minipage}
  \end{center}
\end{figure}

 Equation (\ref{eq:g2DRPA}) is depicted in terms of Feynman diagrams
in Fig.~\ref{fig:gIIf} and generates a series of ladder diagrams 
in a similar fashion as higher order terms are generated in
Fig.~\ref{fig:DRPAexp}.
We note that this approach is equivalent to computing the two-body
wave function of a pair propagating inside the medium, while
the Pauli blocking effects are being taken into
account in a similar fashion to the Bethe-Goldstone 
equation~\cite{brueckner,gost}.
The ppDRPA has two advantages over that approximation. 
First the details of the sp
particle fragmentation, included in Eq.~(\ref{eq:gIIf}), allows for
a partial propagation below the Fermi level. This is due to the fact that 
in the correlated system the Fermi sea is partially depleted.
The dressed result therefore corresponds to propagation with respect
to the correlated ground state and not the IPM as in the Bethe-Goldstone
equation.
Second, the advantage of an RPA approach over the TDA one is  that
it accounts for the propagation of hh
excitations in the ground state of the system.
Without this latter feature, the resulting self-energy would not yield
the fragmentation of the sp strength below the Fermi energy.
 As in the ph case, the corresponding ppDTDA approximation is obtained by
considering propagation only in one time direction.  This corresponds to
including only the first term from the left of Eq.(\ref{eq:gIIf}) 
into Eq.~(\ref{eq:g2DRPA}) when solving for the pp part of
$g^{II}$, therefore neglecting the hh contributions.

In general, both the two-particle and the polarization propagator 
represent 
different time orderings of the same 4-points Green's function, see 
Eqs.~(\ref{eq:Pi-3t}) and~(\ref{eq:g2_time}). 
 By making use of Eq.~(\ref{eq:g_derivative}) and~(\ref{eq:selfen_Gamma})
 one can include the effects of ph or pp(hh) motion directly in
the nuclear self-energy.
 

\section{\label{sec:exp}Relation to experimental data}

In this section we will explore the connection between the information 
contained in various propagators and experimental data.
Detailed work on this subject has been presented in Ref.~\cite{masa91}
and we refer to that paper for a discussion of the relation between
elastic nucleon scattering and the particle part of the sp propagator.
In this paper the focus is on the experimental properties that are
probed by the removal of nucleons.
Nevertheless, it will be necessary within the theoretical treatment to
simultaneously consider the addition and removal aspects of the propagators
as discussed in Sec.~\ref{sec:form}. 
Before making this connection it is useful to demonstrate that the
Dyson equation (Eq.~(\ref{eq:dyson})~) yields a Schr\"odinger-like
equation with corresponding interpretation of the removal or
addition amplitudes appearing in Eq.~(\ref{eq:fullg_Leh}).
We will discuss here the case when the spectrum for the $A \pm 1$-particle
system near the
Fermi energy involves discrete bound states which applies to a finite
system like a nucleus.
An appropriate form of the Lehmann representation in this case is given by
\begin{eqnarray}
\lefteqn{g_{\alpha \beta}(\omega)
= \sum_n \frac{\bra{\Psi^A_0} c_{\alpha}
\ket{\Psi^{A+1}_n} \bra{\Psi^{A+1}_n} c^\dagger_{\beta} \ket{\Psi^A_0}
}{ E - \varepsilon^{+}_n +i\eta }
+ \int_{\varepsilon^+_T}^\infty d\varepsilon^{+}_{\nu} 
\frac{\bra{\Psi^A_0} c_{\alpha}
\ket{\Psi^{A+1}_{\nu}} \bra{\Psi^{A+1}_{\nu}} c^\dagger_{\beta} \ket{\Psi^A_0}
}{ E - \varepsilon^{+}_{\nu} +i\eta } } \nonumber \\
& & + \sum_k \frac{\bra{\Psi^A_0} c^\dagger_{\beta} \ket{\Psi^{A-1}_k}
\bra{\Psi^{A-1}_k} c_{\alpha} \ket{\Psi^A_0} }{
E - \varepsilon^{-}_{k} -i\eta} 
+ \int^{\varepsilon^-_T}_{-\infty} d\varepsilon^{-}_{\kappa} 
\frac{\bra{\Psi^A_0} c^\dagger_{\beta} \ket{\Psi^{A-1}_{\kappa}}
\bra{\Psi^{A-1}_{\kappa}} c_{\alpha} \ket{\Psi^A_0} }{
E - \varepsilon^{-}_{\kappa} -i\eta}, \hspace{0.1cm}   
\label{eq:10.32}
\end{eqnarray}
where the continuum energy spectrum for the $A \pm 1$ systems has been
included and the corresponding energy thresholds are denoted by
$\varepsilon^{\pm}_T$. A change of integration variable was also used to
obtain this form of the Lehmann representation introducing the integration
variables for the continuum energies in the form $\varepsilon^{+}_{\nu}
= E^{A+1}_{\nu} - E^A_0$ and $\varepsilon^{-}_{\kappa}
= E^A_0 - E^{A-1}_{\kappa}$ , respectively.
For the unperturbed propagator one will encounter sp energies
associated with $H_0$ that are different from the poles of $g(\omega)$.
By exploring the equations of motion of the two-body propagator, it is possible
to show that a Lehmann representation also exists for the exact self-energy 
and has different poles from the ones for $g(\omega)$~\cite{migdal67}.
These features can be used to identify the residues of discrete solutions
in the Lehmann representation from the Dyson equation.
In practice, one may proceed with taking the following limit of the Dyson
equation for the case of the hole part of the propagator (without generating
contributions from the poles in the self-energy or the noninteracting 
propagator)
\begin{equation}
\lim_{\omega \to \varepsilon^-_k} (\omega-\varepsilon^-_k)\ 
\Big\{ g_{\alpha\beta}(\omega) = 
g_{\alpha \beta}^{(0)}(\omega) +
\sum_{\gamma \delta} g_{\alpha\gamma}^{(0)}(\omega)\ 
\Sigma_{\gamma \delta}^\star(\omega)\ 
g_{\delta \beta}(\omega)\ \Big\} .
\label{eq:10.33}
\end{equation}
This limit process generates an eigenvalue
equation of the following kind 
\begin{equation}
z^{k-}_{\alpha} = \sum_{\gamma,\delta}\ 
g^{(0)}_{\alpha \gamma}(\varepsilon^-_k)\
\Sigma^\star_{\gamma\delta}(\varepsilon^-_k)\ z^{k-}_{\delta} ,
\label{eq:2nd17}
\end{equation}
where 
\begin{equation}
z^{k-}_{\alpha} = \bra{\Psi^{A-1}_k}c_{\alpha} \ket{\Psi^A_0} .
\label{eq:10.35}
\end{equation}
Since the Dyson equation can be written in any sp basis, one may
choose the coordinate representation
with $sp$ quantum numbers ${\bf r},m$ for
the position and spin projection, respectively,
to solve Eq.~(\ref{eq:2nd17}).
In this basis one obtains
\begin{equation}
z^{k-}_{{\bf r}m} = \sum_{m_1,m_2} \int \!\! d^3r_1 \int \!\! d^3r_2\
g^{(0)}_{{\bf r}m ~{\bf r}_1 m_1}(\varepsilon^-_k)\
\Sigma^\star_{{\bf r}_1m_1 ~{\bf r}_2m_2}(\varepsilon^-_k)\
z^{k-}_{{\bf r}_2m_2}.
\label{eq:2nd18}
\end{equation}
Equation (\ref{eq:2nd18}) can be rearranged by inverting the unperturbed 
propagator according to
\begin{equation}
\sum_{m} \int \!\! d^3r \bra{{\bf r'}m'} \varepsilon^-_k - H_0 \ket{
{\bf r}m} g^{(0)}_{{\bf r}m ~{\bf r}_1m_1}(\varepsilon^-_k) =
\delta_{m',m_1} \delta({\bf r}' -{\bf r}_1) .
\label{eq:2nd19}
\end{equation}
The corresponding operation on $z^{k-}_{{\bf r}m}$ yields
\begin{equation}
\sum_{m} \int \!\! d^3r\ \bra{{\bf r'}m'} \varepsilon^-_k - H_0 \ket{
{\bf r}m} z^{k-}_{{\bf r}m} = \left\{\varepsilon^-_k + \frac{\hbar^2 \nabla'{}^2}
{2m} - U({\bf r}')\right\} z^{k-}_{{\bf r}'m'}, 
\end{equation}
where $U$ is assumed to be local and spin-independent for simplicity.
Combining these results yields the explicit
cancellation of the auxiliary potential $U$ and the 
following result
\begin{equation}
- \frac{\hbar^2 \nabla^2}{2m} z^{k-}_{{\bf r}m} + \sum_{m_1} \int \!\! d^3r_1\ 
\Sigma'{}^\star_{{\bf r}m ~{\bf r}_1m_1}(\varepsilon_k) z^{k-}_{{\bf r}_1m_1} = 
\varepsilon^-_k z^{k-}_{{\bf r}m} , 
\label{eq:2nd20}
\end{equation}
where the notation $\Sigma'{}^\star$ has been used to signify that the $U$ 
contribution has been removed.
This equation has the form of a Schr\"odinger equation with a nonlocal
potential which is represented by the self-energy $\Sigma'{}^\star$.
Note that an eigenvalue $\varepsilon^-_k$ can only be obtained when it 
coincides with the energy argument of the self-energy.
An important difference with the ordinary Schr\"odinger equation is related
to the normalization of the quasihole
``eigenfunctions'' $z^{k-}_{{\bf r}m}$.
The appropriate normalization condition is obtained by considering terms
beyond the pole contribution in the Dyson equation.
This result is most conveniently expressed in terms of the sp state (itself
normalized to 1)
which corresponds to the quasihole wave function $z^n_{{\bf r}m}$.
In other words, one can use the eigenstate which diagonalizes 
Eq.~(\ref{eq:2nd20}) with eigenvalue $\varepsilon^-_k$, 
to express the normalization condition.
Assigning the notation $\alpha_{qh}$ to this sp state,
one obtains with $z^{k-}_{\alpha_{qh}} = \bra{\Psi^{A-1}_k}c_{\alpha_{qh}}
\ket{\Psi^A_0}$
\begin{equation}
\mid z^{k-}_{\alpha_{qh}} \mid^2 = \bigg( {1 - 
\frac{\partial \Sigma'{}^\star_{\alpha_{qh},
\alpha_{qh}}(E)}{\partial E} \bigg|_{\varepsilon^-_k}}
\bigg)^{-1} , 
\label{eq:2nd21}
\end{equation}
where the subscript $qh$ refers to the quasihole nature of this state and the 
fact that for states very near to the Fermi energy with quantum numbers
corresponding to fully occupied mean-field states the normalization yields
a number of order 1.
 This result is equivalent to the definition given
in Eq.~(\ref{eq:hole_Sk}).

\subsection{Spectroscopic strength from the $(e,e'p)$ reaction}

In order to make the connection with experimental data obtained from
knockout reactions, it is useful to consider the response of a system to a weak
probe. The hole spectral function introduced in Sec.~\ref{sec:form} can be 
experimentally ``observed'' in so-called knockout reactions. The general 
idea is to transfer a large amount of momentum and energy to a 
proton of a bound nucleus in the ground state.
The proton is then ejected from 
the system, and one ends up with a fast-moving particle and a bound 
$(A-1)$-particle system. By observing the momentum of the ejected 
particle it is then possible the reconstruct the spectral function of the 
system, provided that the interaction between the ejected particle and the 
remainder is sufficiently weak or treated in a controlled fashion,
\textit{e. g.} by constraining this treatment with information from other
experimental data.  

We assume that the  $A$-particle system is initially in its ground 
state,
\begin{equation}
\ket{\Psi_i} =\ket{\Psi^A_0 },
\end{equation}
and makes a transition to a final $A$-particle eigenstate 
\begin{equation}
\label{dim:eep1}
\ket{\Psi_f}= a^{\dagger}_{\bm p}\ket{\Psi^{A-1}_{n}},
\end{equation}
composed of a bound $(A-1)$-particle eigenstate, 
$\ket{\Psi^{A-1}_{n}}$, and a particle with momentum $\bm{p}$.

For simplicity we consider the transition matrix elements for a scalar 
external probe 
\begin{equation}
\rho ({\bm q})=\sum_{j=1}^A \exp{(i{\bm q}\cdot{\bm r}_j)},
\label{eq:eep0}
\end{equation} 
which transfers momentum $\bm{q}$ to a particle.   
Suppressing other possible sp quantum numbers, like $\textit{e.g.}$ spin,
the second-quantized form of this operator is given by
\begin{equation} 
\hat{\rho}({\bm q}) = \sum_{\bm{p},\bm{p}'}\bra{\bm{p}}\exp{(i{\bm q}
\cdot{\bm r})}\ket{\bm{p}'}a^{\dagger}_{\bm p}a_{\bm{p}'}
=\sum_{\bm p} a^{\dagger}_{\bm p} a_{{\bm p}-{\bm q}}.
\end{equation}
The transition matrix element now becomes 
\begin{eqnarray}
\bra{\Psi_f }\hat{\rho}({\bm q})\ket{\Psi_i }&=&
\sum_{\bm{p}'} \bra{\Psi^{A-1}_{n}}a_{\bm{p}} a^{\dagger}_{\bm{p}'}
a_{\bm{p}'-\bm{q}}\ket{\Psi^A_0}
=
\sum_{\bm{p}'} \bra{\Psi^{A-1}_{n}}\delta_{\bm{p}',\bm{p}}
a_{\bm{p}'-\bm{q}}
+a^{\dagger}_{\bm{p}'}a_{\bm{p}'-\bm{q}}a_{\bm p} \ket{\Psi^A_0 }
\nonumber\\ 
&\approx
&\bra{\Psi^{A-1}_{n}}a_{{\bm p}-{\bm q}}\ket{\Psi^A_0 } .
\label{dim:eep2}
\end{eqnarray}
The last line is obtained in the so-called {\em Impulse Approximation}, 
where it is assumed that the ejected particle is the one that has absorbed 
the momentum from the external field. This is a very good approximation 
whenever the momentum ${\bm p}$ of the ejectile is much larger than typical 
momenta for the particles in the bound states; the neglected term in 
Eq.~(\ref{dim:eep2}) is then very small, as it involves the removal of a 
particle with momentum  ${\bm p}$ from $\ket{\Psi^A_0}$. 

There is one other assumption in 
the derivation: the fact that the final eigenstate of the $A$-particle system 
was written in the form of Eq.~(\ref{dim:eep1}), 
i.e.\ a plane-wave state for the ejectile on top of an $(A-1)$-particle 
eigenstate. This is again a good approximation 
if the ejectile momentum is large enough, as can be understood by rewriting 
the Hamiltonian in the $A$-particle system as 
\begin{equation} 
\label{dim:eep3}
H_A=\sum_{i=1}^A \frac{{\bm p}^2_i}{2m}+\sum_{i<j=1}^A V(i,j) 
=H_{A-1}+\frac{{\bm p}^2_A}{2m}+\sum_{i=1}^{A-1} V(i,A) .
\end{equation}
The last term in Eq.~(\ref{dim:eep3}) represents the 
{\em Final State Interaction}, or the 
interaction between the ejected particle $A$ and the other 
particles $1\dots A-1$. If the relative momentum between particle $A$ and the 
others is large enough their mutual interaction can be neglected, and 
$H_A \approx H_{A-1} + \bm{p}^2_A/2m$. 
The result given by Eq.~(\ref{dim:eep2}) is called the 
{\em Plane Wave Impulse Approximation} or 
PWIA knock-out amplitude, for obvious reasons, and is precisely 
a removal amplitude (in the momentum representation) appearing in the Lehmann
representation of the sp propagator (see Eq.~(\ref{eq:fullg_Leh})~).

The cross section of the knock-out reaction, where the momentum and energy of
the ejected particle and the probe are either measured or known, is according 
to Fermi's golden rule proportional to
\begin{equation}
\label{dim:eep4}
d\sigma \sim
\sum_n \delta (\omega 
+E_i -E_f) |\bra{\Psi_f}\hat{\rho}({\bm q})\ket{\Psi_i}|^2,
\end{equation}
where the energy-conserving $\delta$-function contains the energy transfer 
$\omega$ of the probe, and the initial and final energies of the system 
are $E_i = E^A_0$ and $E_f =E^{A-1}_n +{\bm p}^2/2m$, respectively. Note that 
the internal state of the residual $A-1$ system is not measured, hence the 
summation over $n$ in Eq.~(\ref{dim:eep4}). 
Defining the missing momentum $\bm{p}_{miss}$ and missing energy $E_{miss}$ 
of the 
knock-out reaction as\footnote{We will neglect here the recoil of the residual
$A-1$ system, i.e.\ we assume the mass of the $A$ and $A-1$ system to be much 
heavier than the mass $m$ of the ejected particle.}
\begin{equation}
\bm{p}_{miss}={\bm p} - {\bm q}
\label{eq:eepx}
\end{equation}
and
\begin{equation}
E_{miss} ={\bm p}^2/2m -  \omega = E^A_0-E^{A-1}_n ,
\label{eq:eepy}
\end{equation} 
respectively,
the PWIA knock-out cross section can be rewritten as 
\begin{eqnarray} 
d\sigma &\sim&
\sum_n \delta (E_{miss}-E^A_0+E^{A-1}_n ) |\bra{\Psi^{A-1}_{n}}
a_{\bm{p}_{miss}}\ket{\Psi^A_0 }|^2
\nonumber\\
& = & S_h (\bm{p}_{miss},E_{miss}) .
\label{dim:eep6}
\end{eqnarray}
The PWIA cross section is therefore exactly proportional to the hole spectral 
function defined in Eq.~(\ref{eq:Sh}). 
This is of course only true in the PWIA, but when the 
deviations of the impulse approximation and the effects of the final state 
interaction are under control, it is possible to obtain precise 
experimental information on the hole spectral function of the system under 
study. Although the actual (e,e$'$p) experiments involve more complicated
one-body excitation operators than the one considered here in this simple
example, the basic conclusions are not altered~\cite{frmo}.

An even more realistic description of the proton knockout reaction
again identifies the external electron probe and the corresponding
virtual photon as represented by a one-body excitation operator.
\begin{equation}
\hat{O} = \sum_{\alpha \beta} \bra{\alpha} O \ket{\beta} a^\dagger_\alpha
a_\beta .
\label{eq:op}
\end{equation}
The response of the system to such an external excitation operator is described
by the polarization propagator which has a Lehmann representation
given by Eq.~(\ref{eq:Pi_Leh}).
The transition probability induced by $\hat{O}$
from the ground state to an excited state is given by
\begin{equation}
\left| \bra{\Psi^A_n} \hat{O} \ket{\Psi^A_0} \right|^2 =
\sum_{\alpha \beta} \sum_{\gamma \delta} 
 \bra{\gamma} O \ket{\delta} \bra{\Psi^A_n} a^\dagger_\gamma 
a_\delta \ket{\Psi^A_0} 
 \bra{\alpha} O \ket{\beta}^* \bra{\Psi^A_n} a^\dagger_\alpha 
a_\beta \ket{\Psi^A_0}^* .
\label{eq:14.6}
\end{equation}
This result demonstrates that the numerator of the first term in
Eq.~(\ref{eq:Pi_Leh}) contains the relevant transition amplitudes for a
given state $n$ to evaluate this transition probability.
In general, there are important correlations between the ph states,
in particular at low energy where collective surface vibrations and
giant resonances occur.
At higher excitation energy and momentum transfer, these collective
coherence effects tend to disappear.
In this domain, one may therefore write the polarization propagator
as the product of two dressed sp propagators as in Eq.~(\ref{eq:Pif}).
Replacing $\Pi$ by this dressed $\Pi^{f}$ one neglects contributions
where the particle and hole interact but includes all other correlations
associated with the full dressing of the removed particle in terms
of the corresponding removal amplitude (or spectral function) and
the corresponding dressing of the particle that will ultimately be detected
in the (e,e$'$p) experiment.
The wave function of this dressed particle corresponds to the addition
amplitude in coordinate space and can be associated with an optical model
wave function~\cite{masa91}.
From this analysis it becomes therefore clear that one may use empirical
information associated with the elastic scattering of protons in terms
of optical potentials to describe this wave function of the outgoing
proton.
Clearly, it is this interpretation that clarifies the assumptions that
underly the standard analyis of the (e,e$'$p) reaction~\cite{deWitt1} - 
\cite{Paviabook}.
In the practical analysis of an (e,e$'$p) experiment it is conventional to find
a local potential well (mostly of Woods-Saxon type) which will generate a sp
state at the removal energy for the transition that is studied.
This state is further required to provide the best possible fit to
the experimental momentum dependence of the cross section (with proper
inclusion of complications due to electron and proton 
distortion)~\cite{deWitt1} - \cite{Paviabook}.
The overall factor necessary to bring the resulting calculated cross section
into agreement with the experimental data, can then be interpreted as the
spectroscopic factor corresponding to the ``experimental''
quasihole wave function
according to Eq.(\ref{eq:2nd21}).

\begin{figure}[tbb]
  \begin{center}
    \includegraphics[height=0.65\textheight,angle=-90]{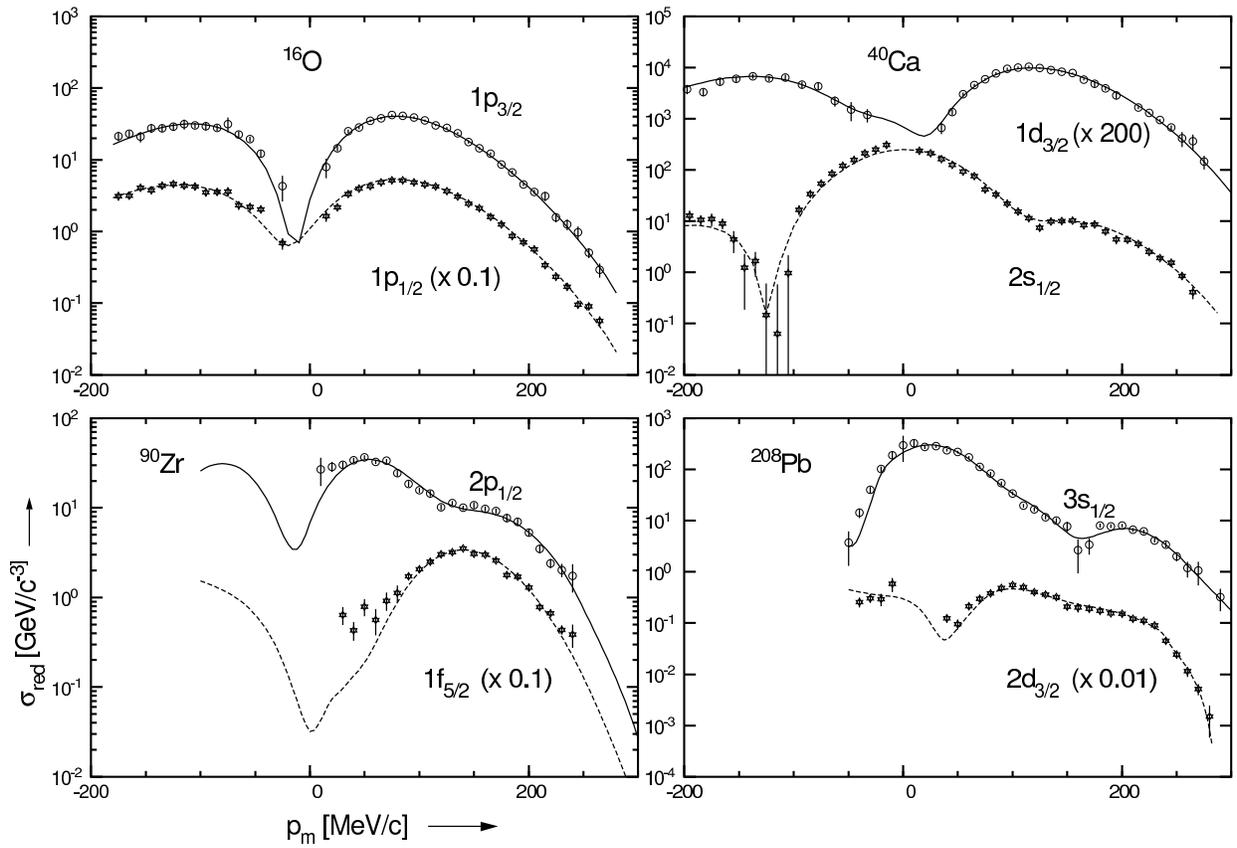}
\begin{minipage}[t]{16.5 cm}
    \caption{\label{fig:louk1}
      Momentum distributions for various nuclei obtained from the
\eep\ reaction performed at NIKHEF~\cite{lap93}.}
\end{minipage}
    \end{center}
\end{figure}

The resulting cross sections obtained at 
the NIKHEF facility are shown for four different nuclei
in Fig.~\ref{fig:louk1}~\cite{lap93}.
It is important to realize that the shapes of the wave functions
in momentum space correspond closely to the ones expected on
the basis of a standard Woods-Saxon potential well (or more involved
mf wave functions).
This is itself an important observation since the (e,e$'$p) reaction
probes the interior of the nucleus, a feat not available with hadronically
induced reactions.

\begin{figure}[!pbt]
\vspace*{.5cm}       
 \begin{center}
    \includegraphics[height=0.30\textheight]{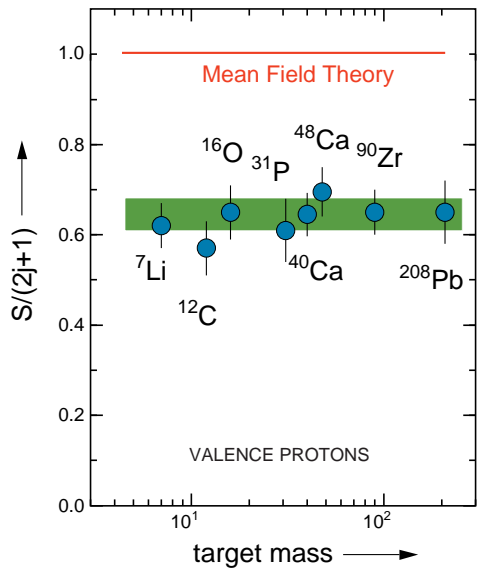}
\begin{minipage}[t]{16.5 cm}
\caption{Spectroscopic factors from the (e,e$'$p) reaction as a function
of target mass. Data have been obtained at the NIKHEF facility~\cite{lap93}.
\label{fig:louk2} }
\end{minipage}
\end{center}
\end{figure}

While the shapes of the valence nucleon wave functions correspond to the basic
ingredients expected on the basis of years of nuclear structure physics
experience, there is a significant departure with regard to the integral
of the square of these wave functions.
This quantity is of course the spectroscopic factor
and is shown in Fig.~\ref{fig:louk2} for the data obtained at 
NIKHEF~\cite{lap93}.
The results shown in Fig.~\ref{fig:louk2} indicate that there is an
essentially global reduction of the sp strength of about
35 \% 
which needs to be explained by the theoretical calculations.
This depletion is somewhat less for the strength associated with slightly
more bound levels.
An additional feature obtained in the (e,e$'$p) reaction is the fragmentation
pattern of these more deeply bound orbitals in nuclei.
This pattern is such that single isolated peaks are obtained only
in the immediate vicinity
of the Fermi energy whereas for more deeply bound states a stronger
fragmentation of the strength is obtained with larger distance from
$\varepsilon_F$.
This is beautifully illustrated by the \eep\ data from 
Quint~\cite{qui} (see Sec.~\ref{sec:finnuc}).
Whereas the $3s_{1/2}$ orbit exhibits a single peak, there is a
substantial fragmentation of the $1f$ strength as indicated in 
this figure.
Additional information about the occupation number of the former orbit 
is also available and can be 
obtained by analyzing elastic electron scattering cross sections of 
neighboring nuclei~\cite{wag}.
The actual occupation number for the $3s_{1/2}$ proton orbit obtained
from this analysis is about 10\% 
larger than the quasihole spectroscopic
factor~\cite{grab,grabx} and therefore corresponds to 0.75.
All these features of the strength need to be explained theoretically.
This will be attempted in the material covered in later sections.

\subsection{Information from the (e,e$'$2N) reaction}

Suggestions to explore SRC in two-nucleon emission
reactions go back to the work of Gottfried~\cite{GO58}.
More recently, theoretical work has focused on the possibility to utilize
the (e,e$'$2N) reaction to probe nucleon-nucleon
correlations~\cite{CI90} - \cite{BO90}.
Practical descriptions of this reaction have been developed by 
the Pavia group~\cite{GP91} - \cite{GP95}.
Proceeding in a similar vein as in the analysis of the (e,e$'$p) reaction,
which yields information about the one-nucleon (removal) spectral function, 
one may hope to learn about the two-nucleon (removal) spectral function
in two-nucleon emission processes.
The emission of two protons is particularly promising for studies of
SRC since the effect of meson-exchange currents and
isobars is not expected to dominate the cross section
under suitable kinematic conditions~\cite{GP92}.

Experiments have been carried out for $^{12}$C~\cite{KHK95} and
$^{16}$O (discussed in more detail in Sec.~\ref{sec:ee1-2N}) 
to explore the feasibility of gaining insight 
into nucleon-nucleon correlations in finite nuclei using the (e,e$'$pp)
reaction. 
Triple coincidence measurements involving protons with large initial
momenta seem particularly suitable to provide information on SRC. 
The scattered electron is then expected to 
transfer a virtual photon to one of these two protons which have large
and opposite momenta and therefore a relatively small center-of-mass momentum.
This strong correlation results from hard collisions due to the strong 
repulsive core of the NN interaction.
When one of the protons is removed by the absorption of the virtual photon,
its partner will also leave the nucleus under the assumption that the 
energy transfer is mainly to the hit pair (the residual nucleus stays at a 
low excitation energy)~\cite{GP91}.
It is therefore hoped that, if the coupling of the virtual photon to one
nucleon is the dominant mechanism, the
(e,e$'$pp) process may be exploited as a useful tool to
investigate these short-range correlation effects (see also Ref.~\cite{Ry94}).

One of the goals of triple coincidence measurements is to illuminate
the features of the interaction between two nucleons before their knockout
of the nucleus and their subsequent detection.
Early experiments on $^{12}$C employing the (e,e$'$pp) reaction
already assumed in the analysis of the data obtained at 
NIKHEF~\cite{Kes93,GP91}
that the virtual photon is coupled to
one of the detected protons. 
This approximation can be understood by considering the transition matrix
element of the nuclear charge operator in momentum space
	\begin{equation}
		\hat{\rho} (\bm{q}) = e \sum_{\bm{ p}'} 
		a^\dagger_{\bm{p}'+\bm{ q}} a_{\bm{p}'} 
	\label{eq:rhoq}
	\end{equation}
(with spin implicit in the summation)
between the initial state $\ket{\Psi^A_0}$ and an approximate
final state of the form
	\begin{equation}
	\ket{\Psi_{final} } =
a^\dagger_{\bm{p}_{1'}} a^\dagger_{\bm{p}_{2'}} | \Psi^{A-2}_n \rangle .
	\label{eq:finst}
	\end{equation}
This final state contains two plane wave protons and an exact
state $\ket{\Psi_n^{A-2}}$ for the system with two protons removed. 
In calculating the matrix elements of the transition charge operator
one obtains
	\begin{equation}
	\langle \Psi_{final} | {\hat \rho} (\bm{q}) | \Psi^A_0 \rangle =
\langle \Psi^{A-2}_n | a_{\bm{p}'_2} a_{\bm{p}'_1-\bm{q}}
| \Psi^A_0 \rangle + \langle \Psi^{A-2}_n |
a_{\bm{p}'_2-\bm{q}} a_{\bm{p}'_1}| \Psi^A_0 \rangle ,
	\label{eq:finstme}
	\end{equation}
assuming that one of the detected protons absorbs the momentum $\bm{q}$.
To obtain the contribution to the cross section one requires the square of
this matrix element.
This yields four terms, each representing a particular term of 
Eq.~(\ref{eq:SpecFunc}) for the two-nucleon spectral function.
An important ingredient in the description of the two-nucleon knock-out 
reaction is the two-hole spectral function defined by
	\begin{equation}
		S^{hh}(\bm{p}_1, \bm{p}_2, \bm{p}_{1'}, \bm{p}_{2'}, 
		\omega )
	= \sum_n
\langle\Psi^A_0| a^\dagger_{\bm{p}_{1'}} a^\dagger_{\bm{p}_{2'}}|%
		\Psi^{A-2}_n \rangle
		\langle\Psi^{A-2}_n | a_{\bm{p}_1} a_{\bm{p}_2}|%
		\Psi^A_0 \rangle
		\delta( \omega - (E^{A}_0 - E^{A-2}_n) )
	\;,
	\label{eq:SpecFunc}
	\end{equation}
where $\ket{\Psi^A_0}$ may denote
 the ($0^+$) ground state of the target system 
(for example $^{16}$O) and $\ket{\Psi^{A-2}_n}$ the $n$-th excited state
of the residual nucleus ($^{14}$C).
In Eq.\ (\ref{eq:SpecFunc}), 
 $a^\dagger_{{\bf p}}$ ($a_{{\bf p}}$) represent the addition (removal)
operators of a nucleon with momentum ${\bf p}$ (spin and isospin are implicit).
It is therefore clear that the appropriate expression
for the transition probability is to consider the quantity
	\begin{eqnarray}
		\hat{S}(\bm{p}'_1, \bm{p}'_2, E)
& = &
		 S^{hh}( \bm{p}'_1-\bm{q}, \bm{p}'_2, 
		    \bm{p}'_1-\bm{q}, \bm{p}'_2, E )
		+S^{hh}( \bm{p}'_1-\bm{q}, \bm{p}'_2, 
		    \bm{p}'_1, \bm{p}'_2-\bm{q}, E )
	\nonumber \\
	&+&
		 S^{hh}( \bm{p}'_1, \bm{p}'_2-\bm{q}, 
		    \bm{p}'_1-\bm{q}, \bm{p}'_2, E )
		+S^{hh}( \bm{p}'_1, \bm{p}'_2-\bm{q}, 
		    \bm{p}'_1, \bm{p}'_2-\bm{q}, E ) 	\;,
\label{eq:Shat}	
\end{eqnarray}
where the vector $\bm{q}$ is the momentum of the virtual photon. 

More realistic implementations of this reaction model naturally require
the consideration of the distortion effects of the outgoing particles.
Nevertheless, this simple representation of the reaction clarifies
the intrinsic importance of two-nucleon spectral functions in
describing two-nucleon knockout reactions.
More details related to actual comparison of calculations with
experimental data will be presented in Sec.~\ref{sec:ee1-2N}.

\section{\label{sec:nucmat}Theoretical calculations for nuclear 
matter}

The study of nuclear matter has remained a prominent field of study
in recent years. We will focus in this section on recent developments
related to self-consistent Green's functions and put these results in 
perspective with regard to other many-body approaches to the relevant
quantities of interest where appropriate.
All many-body techniques that depart from the free nucleon-nucleon interaction
are required to treat the short-range repulsion of this interaction in
an appropriate manner.
In perturbative methods one must sum the infinite set of ladder diagrams
that fulfills this requirement.
This was realized by Brueckner long time ago~\cite{brueckner}.
Since then many different implementations have been studied with perturbative
methods to treat short-range correlations.
Some of this work has been reviewed in Refs.~\cite{dimu1,mupo00}.
The main emphasis in the present review is on the progress that has been made
recently in implementing the self-consistency condition on the single-particle 
(sp) propagators in consort with the summation of ladder diagrams for the
effective interaction.
According to the Lehmann representation of the sp propagator 
(see Eq.~(\ref{eq:fullg_Leh})~) knowledge of the propagator is equivalent to
knowledge of the particle and hole spectral functions given in 
Eqs.~(\ref{eq:Sh}) and (\ref{eq:Sp}).
The first nuclear-matter spectral functions in SCGF theory
were obtained for a semirealistic
interaction by employing mean-field (mf) propagators in the two-body scattering
equation~\cite{anthes,rpd1}.
Self-consistency was limited to the sp spectrum obtained from the real part
of the on-shell self-energy.
The corresponding ladder 
equation includes both particle-particle (pp) and
hole-hole (hh) propagation and is sometimes called the Galitski-Feynman
equation~\cite{Galitskii58}.
Hole spectral functions using correlated basis function theory (CBF)
were first obtained in Ref.~\cite{bff89} while particle spectral functions
were reported in Ref.~\cite{bff92}.
Spectral functions for the full Reid potential~\cite{reid} were obtained 
in SCGF by employing a self-consistent gap in the sp 
spectrum~\cite{vond1} - \cite{vond3} to avoid pairing instabilities in the
${}^1S_0$ and ${}^3S_1$-${}^3D_1$
channels~\cite{vond4}.
The first solution of the ladder equation using fully dressed sp 
propagators~\cite{christh} was obtained by employing a parametrization of the
spectral functions~\cite{gdpr}.
Important consequences of treating this dressing of the sp propagators
include a strong reduction of the in-medium cross section and the
disappearance of the pairing instabilities at normal nuclear-matter
density~\cite{dgrpr}.
The latter conclusion was recently confirmed in detail in Ref.~\cite{bozek0}.

Several approaches towards the goal of self-consistent calculations of
nuclear-matter Green's functions have 
recently been reported in 
Refs.~\cite{djl} - \cite{rd}.
In Ref.~\cite{fgmc} the average effect of the spectral strength distribution
was included in the determination of the sp spectrum thereby avoiding
pairing instabilities. 
In addition several realistic interactions were employed to study the
dependence of spectral functions on the choice of the nucleon-nucleon (NN)
interaction. 
The Ghent group has concentrated on a discrete representation of the
the sp propagator which yields a scattering equation containing several
discrete poles but otherwise similar to the usual one based on mean-field
propagators~\cite{dw1,dwvnw1,dwvnw2}.
This type of approach will likely generate the same results
for the energy per particle as the continuous version implemented in
Refs.~\cite{libth} and \cite{bozek0,bozek2}
for an appropriate choice of the discretization scheme. 
Initial results from the various groups indicate
an important change of the saturation properties with respect to Brueckner
Hartree Fock (BHF)
calculations with a continuous choice for the sp spectrum.
These results have been obtained
for an updated version of the Reid potential~\cite{reid93} in 
Ref.~\cite{dwvnw2}.
In Refs.~\cite{bozek1,bozek2} a separable representation of
the Paris interaction~\cite{hp1,hp2} has been used to generate self-consistent
propagators including short-range correlations by employing a direct
discretization of the spectral functions and self-energy as a function
of energy.
Also in this work~\cite{bozek2} a repulsive effect is observed for the energy
per particle leading to less binding as compared to the conventional
BHF approach with a continuous sp spectrum. These results therefore also
generate a corresponding substantial reduction of the saturation density.
Several realistic interactions were studied in Ref.~\cite{ddsvw03}.
In this recent work a new argument about the importance of
separating long and short-range contributions to binding in nuclear matter
was introduced.
The hope is that such new developments will lead to new insights
into the long-standing problem of nuclear-matter saturation.

A discussion of the formalism necessary to accomodate a self-consistent
treatment of short-range correlations in nuclear matter is presented
in Sec~\ref{sec:formalismnm}.
We begin the discussion of results by reviewing some early calculations of 
first-generation spectral functions in Sec.~\ref{sec:brian}.
A useful digression is made in Sec.~\ref{sec:neil} where spectral functions
for the $\Lambda$ hyperon in nuclear matter~\cite{neilth} are discussed.
These spectral functions are also based on the solution of the corresponding 
ladder equation in the medium employing mf intermediate propagators.
A useful comparison between nucleon and $\Lambda$ spectral functions
is then possible, illuminating the simililarities and differences between
these nuclear constituents.
The difficulty of propagating fully dressed particles in the ladder equation
is associated with the necessity to treat all off-shell aspects of the
scattering process in the medium.
In Sec.~\ref{sec:healing} a discussion of the scattering process
of dressed particles~\cite{dic98} is presented
that has wider implications for the conceptual understanding of the
properties of nuclei in terms of reconciling
their violently interacting constituents with aspects of the simple 
shell-model picture.
Results that compare typical quantities that characterize scattering
for dressed, mf, and free particles are presented in Sec.~\ref{sec:scatt}.
Results that illustrate the properties of fully self-consistent
Green's functions in compariosn with those of the first generation
are presented in Sec.~\ref{sec:selfnmr}.
The last subsection~\ref{sec:satprop} concerned with nuclear matter properties 
 is devoted to the discussion of the consequences of
self-consistent treatments of short-range correlations for the understanding
of nuclear saturation properties.

\subsection{\label{sec:formalismnm}Formalism for self-consistent 
Green's functions in nuclear matter}

The formalism for the improved determination of the
sp propagator in nuclear matter using a self-consistent scheme which
includes the full treatment of short-range correlations will be outlined below.
In symmetric nuclear matter one may take advantage of the invariance properties
of the system to write the sp propagator in the following way
\begin{equation}
\label{eq:nmg}
g(k,\omega )=\int _{\varepsilon _{F}}^{\infty }d\omega'
\frac{S_{p}(k,\omega' )}{\omega -\omega' +i\eta }+
\int ^{\varepsilon _{F}}_{-\infty }d\omega'
\frac{S_{h}(k,\omega' )}{\omega -\omega' -i\eta },
\end{equation}
where $S_{p}(k,\omega )$ ($S_{h}(k,\omega )$) represents the probability
distribution to add (remove) a nucleon with momentum $k$ to (from)
the system leaving it at an energy $\omega$. 
The latter functions are also referred to as the particle and hole spectral 
functions, respectively, and were introduced in Sec.~\ref{sec:spdef} by 
considering 
the Lehmann representation of the sp propagator in Eq.~(\ref{eq:fullg_Leh}).
The propagator can be obtained diagrammatically by considering its 
perturbation expansion~\cite{AAA,fetwa} or, alternatively, by considering the
equation of motion which relates it to the two-body and, subsequently,
higher-body propagators~\cite{masch} as discussed in Sec.~\ref{sec:EOM}.
In the latter approach a natural introduction of self-consistent propagators 
is obtained.
The link between one-body and two-body propagators is made through the
self-energy which can be studied at various levels of approximation.
The one including SRC is obtained by describing the two-body propagator
in the same way as one would proceed in free space, {\textit{i. e.}} sum all 
ladder diagrams.
This procedure leads to the in-medium equivalent of the $T$ matrix in free
space and therefore properly accounts for short-range correlations.
Using a notation with ${\bm q}, {\bm q}'$ for relative and ${\bm K}$ for
the conserved total momentum,
the resulting two-body propagator can be written as
\begin{subequations}
\label{eq:g2gam}
\begin{eqnarray}
g^{II}({\bm q},{\bm q}'; {\bm K}, \Omega )  & = &
g^{II}_f({\bm q},{\bm q}'; {\bm K}, \Omega ) 
+\ g^{II}_f({\bm q};{\bm K}, \Omega )
 \int d^3q''\ \bra{{\bm q}} V \ket{ {\bm q}'' }\
g^{II}({\bm q}'',{\bm q}'; {\bm K},\Omega )
\label{eq:g2gama} \\
& = &  g^{II}_f({\bm q},{\bm q}'; {\bm K},\Omega )
+ g^{II}_f({\bm q};{\bm K},\Omega ) 
 \bra{ {\bm q} } \Gamma({\bm K},\Omega ) \ket{ {\bm q}' }\
g^{II}_f({\bm q}';{\bm K},\Omega ) ,
\label{eq:g2gamb}
\end{eqnarray}
\end{subequations}
where
\begin{equation}
g^{II}_f({\bm q},{\bm q}'; {\bm K},\Omega )
~ = ~ \delta({\bm q}-{\bm q}')\
g^{II}_f({\bm q};{\bm K},\Omega)
\label{eq:g2f}
\end{equation}
is the noninteracting but dressed two-particle propagator which 
conserves the relative momentum as expressed by the
$\delta$ function in Eq.~(\ref{eq:g2f}).
The presence of exchange terms in Eqs.~(\ref{eq:g2gam}) and (\ref{eq:g2f})
and possible summations over spins and isospins
is hereby acknowledged but suppressed in the presentation.
Eq.~(\ref{eq:g2gamb}) links the two-particle
propagator with the four-point vertex function 
$\Gamma$ shown explicitly in Fig.~\ref{fig:Gamma_4pt}.

\begin{figure}[tpb]
\begin{center}
\begin{minipage}[t]{8 cm}
\includegraphics[origin=bl,angle=0,width=0.8\textwidth]{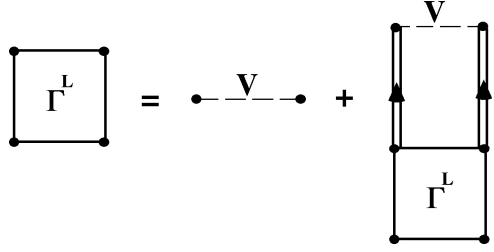}
\end{minipage}
\begin{minipage}[t]{16.5 cm}
\caption{\label{fig:gammalast}
Diagrammatic representation of the ladder equation [Eq.~(\ref{eq:gaml})]
in the medium. Note that fully dressed propagators are iterated between
successive interactions. Only in this diagram the subscript ``L'' is used
to identify the ladder summation.
}
\end{minipage}
\end{center}
\end{figure}

The four-point vertex function or effective interaction $\Gamma$ then contains 
the summation of all ladder diagrams in the following way
\begin{subequations}
\label{eq:gaml}
\begin{eqnarray}
\bra{ {\bm q} } \Gamma({\bm K},\Omega ) \ket{ {\bm q}' } & = &
\bra{ {\bm q} } V \ket{ {\bm q}' } + \int d^3q''\ \bra{ {\bm q} } V
\ket{ {\bm q}'' }\ g^{II}_f({\bm q}'';{\bm K},\Omega)
\bra{ {\bm q}'' } \Gamma({\bm K},\Omega ) \ket{ {\bm q}' } 
\label{eq:gamla} \\
& \equiv & \bra{ {\bm q} } V \ket{ {\bm q}' } + 
\bra{ {\bm q} } \Delta \Gamma({\bm K},\Omega ) \ket{ {\bm q}' } .
\label{eq:gamlb}
\end{eqnarray}
\end{subequations}
The ladder equation is shown diagrammatically in Fig.~\ref{fig:gammalast}.
It is convenient to decompose the effective interaction in two
parts as in done in Eq.~(\ref{eq:gamlb}) which separates the 
energy-dependent part from the energy-independent NN interaction $V$.
Employing this separation, one obtains the nucleon self-energy in 
the form shown diagrammatically in Fig.~\ref{fig:selfen}$a)$ with $U$ discarded
and the full vertex function replaced by the corresponding ladder 
approximation.
The diagram with the closed loop 
corresponds to the energy-independent contribution
involving the NN interaction
\begin{equation}
\Sigma_V(k) = \int \frac{d^3k'}{(2\pi)^3}
\bra{ {\bm k} {\bm k}' } V \ket{ {\bm k} {\bm k}' } n(k') ,
\label{eq:sighf}
\end{equation}
with
\begin{equation}
n(k) = \int_{-\infty}^{\varepsilon_F} d\omega\ S_h(k,\omega)
\label{eq:momdis}
\end{equation}
representing the correlated momentum distribution.
Note that explicit summation over spin and isospin projections
have been suppressed and a notation with individual particle
momenta ${\bm k}$ and ${\bm k}'$ is used for the two-body interaction
in Eq.~(\ref{eq:sighf}).
Both the effective interaction and the self-energy fulfill important
dispersion relations which are helpful in devising strategies to achieve
a numerically tractable self-consistency procedure.
For the energy-dependent part of the effective interaction one has
\begin{subequations}
\label{eq:gamdis}
\begin{eqnarray}
\bra{ {\bm q} } \Delta \Gamma({\bm K},\Omega ) \ket{ {\bm q}' } 
& = & \frac{-1}{\pi} \int _{2\varepsilon _{F}}^{\infty } \!\!\! d\Omega'\
\frac{ \textrm{Im}\
\bra{ {\bm q} } \Delta \Gamma({\bm K},\Omega' ) \ket{ {\bm q}' } 
}
{\Omega -\Omega' +i\eta } 
 +  \frac{1}{\pi}
\int ^{2\varepsilon _{F}}_{-\infty } \!\!\! d\Omega'\
\frac{ \textrm{Im}\
\bra{ {\bm q} } \Delta \Gamma({\bm K},\Omega') \ket{ {\bm q}' } 
}
{\Omega -\Omega' -i\eta } 
\label{eq:gamdisa} \hspace{0.6cm} \\
& \equiv & \bra{ {\bm q} } \Delta \Gamma_\downarrow ({\bm K},\Omega )
\ket{ {\bm q}' }
+ \bra{ {\bm q} } \Delta \Gamma_\uparrow ({\bm K},\Omega )
\ket{ {\bm q}' }
.
\label{eq:gamdisb}
\end{eqnarray}
\end{subequations}
The decomposition of the effective interaction into forward and
backward-going contributions as exhibited in Eq.~(\ref{eq:gamdis})
is essential to obtain a proper construction of the self-energy~\cite{hhpap}.
Indeed, using Eqs.~(\ref{eq:nmg}) and (\ref{eq:gamdisa}) one
obtains the remaining contributions to the self-energy in the
following way
\begin{subequations}
\label{eq:delsig}
\begin{eqnarray}
\Sigma_{\Delta \Gamma} (k,\omega)  & = &
\int \frac{d^3k'}{(2\pi)^3} \int_{-\infty}^{\varepsilon_F} d\omega'\
\bra{ {\bm k} {\bm k}' } \Delta \Gamma_\downarrow (\omega + \omega')
\ket{ {\bm k} {\bm k}' } S_h(k',\omega') \nonumber \\
& - & 
\int \frac{d^3k'}{(2\pi)^3} \int_{\varepsilon_F}^{\infty} d\omega'\
\bra{ {\bm k} {\bm k}' } \Delta \Gamma_\uparrow (\omega + \omega')
\ket{ {\bm k} {\bm k}' } S_p(k',\omega') \label{eq:delsiga} \\
& \equiv & \Delta \Sigma_\downarrow (k,\omega) +
\Delta \Sigma_\uparrow (k, \omega) . \label{eq:delsigb}
\end{eqnarray}
\end{subequations}
As for Eq.~(\ref{eq:sighf}) the notation for the
two-body matrix elements of $\Delta \Gamma$ involve individual momenta.
The result of Eq.~(\ref{eq:delsig})
also implies the following dispersion relation for the self-energy
\begin{subequations}
\label{eq:selfdis}
\begin{eqnarray}
\Sigma(k,\omega ) 
& = & \Sigma_V(k) - \frac{1}{\pi} \int _{\varepsilon _{F}}^{\infty }d\omega'
\frac{ \textrm{Im }\
\Sigma( k,\omega' )  } 
{\omega -\omega' +i\eta } 
 +  \frac{1}{\pi}
\int ^{\varepsilon _{F}}_{-\infty }d\omega'
\frac{ \textrm{Im}\
\Sigma( k,\omega )  }
{\omega -\omega' -i\eta } 
\label{eq:selfdisa} \\
& = & \Sigma_V(k) + \Delta \Sigma_\downarrow (k,\omega )
+ \Delta \Sigma_\uparrow (k,\omega ) .
\label{eq:selfdisb}
\end{eqnarray}
\end{subequations}
Using this self-energy in the Dyson equation one obtains
\begin{subequations}
\label{eq:dysonx}
\begin{eqnarray}
g(k,\omega) & = & g^{(0)}(k,\omega)+
g^{(0)}(k,\omega) \Sigma(k,\omega) g(k,\omega)
\label{eq:dysona} \\
& = & \frac{\omega - \varepsilon(k)- \textrm{Re}\ \Sigma(k,\omega)
+i \textrm{Im}\ \Sigma(k,\omega) }
{\left(\omega - \varepsilon(k) - \textrm{Re}\ \Sigma(k,\omega) \right)^2
+ \left( \textrm{Im}\ \Sigma(k,\omega) \right)^2 } ,
\label{eq:dysonb}
\end{eqnarray}
\end{subequations}
where the explicit form of the noninteracting propagator
\begin{equation}
g^{(0)} (k,\omega) = \frac{\theta(k-k_F)}
{\omega-\varepsilon(k)+i\eta} +  \frac{\theta(k_F-k)}
{\omega-\varepsilon(k)-i\eta}
\label{eq:nmg0}
\end{equation}
with
\begin{equation}
\varepsilon(k) = \frac{k^2}{2m} 
\label{eq:spsp}
\end{equation}
is used to obtain the result of Eq.~(\ref{eq:dysonb}).
The diagrammatic version of the Dyson equation is shown in
Fig.~\ref{fig:Dyson}.
Although it is now possible to solve the ladder equation with mean-field
propagators without an angle-averaging 
procedure~\cite{smc99a,sokn00}, it is necessary
for a practical implementation using fully dressed propagators
to make this approximation.
To obtain the ladder equation in a partial wave basis one therefore proceeds
by an angle-averaging procedure for the
noninteracting but dressed two-particle propagator
\begin{eqnarray}
g^{II}_f({\bm q};{\bm K},\Omega) & = &
\frac{i}{2\pi} \int d\omega\ g({\bm K}/2+{\bm q},
\Omega/2 + \omega)
g({\bm K}/2-{\bm q}, \Omega/2 -\omega) \nonumber \\
& = &
\int_{\varepsilon_F}^{\infty} d\omega_1 \int_{\varepsilon_F}^\infty 
d\omega_2\ \frac{ S_p({\bm K}/2+{\bm q}, \omega_1)
S_p({\bm K}/2-{\bm q}, \omega_2) }
{\Omega -\omega_1 -\omega_2 +i\eta} \nonumber \\
& - &
\int_{-\infty}^{\varepsilon_F} d\omega_1 \int_{-\infty}^{\varepsilon_F}
d\omega_2\ \frac{ S_h({\bm K}/2+{\bm q}, \omega_1)
S_h({\bm K}/2-{\bm q}, \omega_2) }
{\Omega -\omega_1 -\omega_2 -i\eta} .
\label{eq:g2fdis}
\end{eqnarray}
Eq.~(\ref{eq:g2fdis}) shows that the angle-averaging is confined to the
numerators. 
It is therefore possible to consider the imaginary part of 
$g^{II}_f$ for this purpose.
For $\Omega > 2\varepsilon_F$ one has
\begin{equation}
\textrm{Im}\ g^{II}_f({\bm q};{\bm K},\Omega) = 
- \frac{1}{\pi} \int_{2\varepsilon_F}^{\infty} d\omega\ 
 S_p({\bm K}/2+{\bm q}, \Omega/2 + \omega)\
S_p({\bm K}/2-{\bm q}, \Omega/2 -\omega) 
\label{eq:img2fa}
\end{equation}
and for $\Omega < 2\varepsilon_F$ the corresponding result is given by
\begin{equation}
\textrm{Im}\ g^{II}_f({\bm q};{\bm K},\Omega) = 
\frac{1}{\pi} \int_{-\infty}^{2\varepsilon_F} d\omega\ 
 S_h({\bm K}/2+{\bm q}, \Omega/2 + \omega)\
S_h({\bm K}/2-{\bm q}, \Omega/2 -\omega) 
\label{eq:img2fb}
\end{equation}
The convolution integrals in Eqs.~(\ref{eq:img2fa}) and (\ref{eq:img2fb})
also provide a practical route to obtain the real part of 
$g^{II}_f$ by means of the following dispersion relation
\begin{equation} 
g^{II}_f({\bm q};{\bm K}, \Omega) =
- \frac{1}{\pi} \int_{2\varepsilon_F}^\infty d\Omega'\
\frac{ \textrm{Im}\ g^{II}_f({\bm q};{\bm K}, \Omega') }
{\Omega -\Omega' +i\eta}
+ \frac{1}{\pi} \int_{-\infty}^{2\varepsilon_F} d\Omega'\
\frac{ \textrm{Im}\ g^{II}_f({\bm q};{\bm K}, \Omega') }
{\Omega -\Omega' -i\eta} .
\label{eq:g2fdisa}
\end{equation}
The cycle of steps necessary to perform a self-consistent calculation of the
sp propagator is now complete.
Starting from a reasonable set of spectral functions, the construction of
the noninteracting but dressed two-particle propagator first requires
the convolution integrals given in Eqs.~(\ref{eq:img2fa}) and 
(\ref{eq:img2fb}). In a next step the dispersion integrals in
Eq.~(\ref{eq:g2fdisa}) can be used to obtain the real part of the
this propagator. After an angle-averaging procedure the resulting
propagator only depends on the absolute values of the relative and
total momenta.
It is therefore permissible to solve for the ladder equation in
a partial wave basis
\begin{eqnarray}
\bra{q\ell}\Gamma^{JST}(K,\Omega)\ket{q'\ell'}
& = & \bra{q\ell}V^{JST}\ket{q'\ell'} \label{eq:ladpw} \\
& + & \sum_{\ell''} \int_0^\infty dp\ p^2
\bra{q\ell}V^{JST}\ket{p\ell''} g^{II}_f(p;K,\Omega)
\bra{p\ell''}\Gamma^{JST}(K,\Omega)\ket{q'\ell'} . \nonumber
\end{eqnarray}
According to Eq.~(\ref{eq:delsigb}) one only needs the diagonal
elements of the imaginary part of $\Gamma$ (properly adding the
different partial wave contributions) and the original input
spectral functions to obtain the imaginary part of the self-energy.
The dispersion integrals in Eq.~(\ref{eq:selfdis}) together with
Eq.~(\ref{eq:sighf}) then allow the construction of the complete real
part of the self-energy.
The solution of the Dyson equation is then straightforward.
One finally obtains the new spectral functions from
\begin{equation}
S_p(k,\omega) = \frac{-1}{\pi}
\frac{ \textrm{Im}\ \Sigma(k,\omega) }
{\left(\omega - \varepsilon(k) -\textrm{Re}\ \Sigma(k,\omega) \right)^2
+ \left( \textrm{Im}\  \Sigma(k,\omega) \right)^2 } 
\label{eq:specp}
\end{equation}
for energies above $\varepsilon_F$ and from
\begin{equation}
S_h(k,\omega) = \frac{1}{\pi}
\frac{ \textrm{Im}\ \Sigma(k,\omega) }
{\left(\omega - \varepsilon(k) - \textrm{Re}\ \Sigma(k,\omega) \right)^2
+ \left( \textrm{Im}\ \Sigma(k,\omega) \right)^2 } 
\label{eq:spech}
\end{equation}
for energies below $\varepsilon_F$.
At this point the cycle can be repeated until by some numerical criteria the
input sp propagator is sufficiently close to the output sp propagator.
One of the most important quantities to emerge from the spectral functions
is the energy per particle~\cite{MigGal58}
\begin{equation}
\frac{E}{A} = \frac{2}{\rho} \int \frac{d^3k}{(2\pi)^3}
\int_{-\infty}^{\varepsilon_F} d\omega\ \left( \frac{k^2}{2m} +
\omega \right) S_h(k,\omega) ,
\label{eq:epp}
\end{equation}
where $\rho$ is the density both related to the Fermi momentum by
\begin{equation}
\rho = \frac{2k_F^3}{3\pi^2}
\label{eq:rho1}
\end{equation}
and the self-consistent momentum distribution
\begin{equation}
\rho = \frac{2}{\pi^2} \int_0^\infty dkk^2\ n(k)
\label{eq:rho2}
\end{equation}
since self-consistency guarantees particle number 
conservation~\cite{kaba1,kaba2}.
One may also separately consider the kinetic energy per particle
\begin{equation}
\frac{T}{A} = \frac{2}{\pi^2} \int_0^\infty dkk^2\ \frac{k^2}{2m} n(k) .
\label{eq:kpp}
\end{equation}
The potential energy can then be obtained by considering the difference
between Eqs.~(\ref{eq:epp}) and (\ref{eq:kpp}).
This general outline requires further details related to a suitable numerical
implementation. Some of these aspects will be considered in 
Sec.~\ref{sec:selfnmr} where results for fully self-consistent spectral
functions are discussed.

\subsection{\label{sec:brian}Spectral functions obtained from
mean-field input}

The first generation spectral functions were based on the solution of the
scattering equation (Eq.~(\ref{eq:ladpw})~) employing mf propagators in the
construction of $g^{II}_f$ with corresponding $\delta$ function spectral
functions.
The corresponding result for the self-energy given in Eq.~(\ref{eq:delsig})
also simplifies and is given by
\begin{eqnarray}
\Sigma_{\Delta\Gamma}(k,\omega)& = & \int\frac{d^3 k'}{(2\pi )^3}
\bra{\bm{k}\bm{k}'} \Delta\Gamma_\downarrow^{(0)}(\omega+\varepsilon(k'))
\ket{\bm{k} \bm{k}'}\ \theta(k_F- k')  \nonumber \\
& - & \int \frac{d^3 k'}{(2\pi )^3}
\bra{\bm{k}\bm{k}'} \Delta\Gamma_\uparrow^{(0)}(\omega+\varepsilon(k'))
\ket{\bm{k}\bm{k}'}\
 \theta(k' -k_F) , 
\label{eq:selfgo}
\end{eqnarray}
where a superscript has been attached to $\Gamma$ to identify that it
was obtained from a scattering equation with mf propagators.
A serious difficulty arises when propagating mf propagators corresponding
to a continuous sp spectrum in the ladder equation
when a realistic NN interaction is employed~\cite{vond2}.
Indeed, so-called pairing instabilities arise in the ${}^3S_1$-${}^3D_1$
and ${}^1S_0$ partial wave channels in certain density regimes~\cite{vond4}.
Self-consistency with unperturbed propagators in the ladder 
equation can be achieved while solving the above difficulties 
and still maintaining a connection between the sp potential 
and the full $\Sigma$ by making the following choices~\cite{vond1}: 
Below $k_F$, let $\varepsilon(k)$ be the 
first moment of $S_h(k,\omega)$ with respect to energy (the 
contribution to the total energy at that momentum) divided by 
the number of particles in the system with that momentum.  One 
cannot extend this intuitive prescription to $k>k_F$ by replacing 
$S_h$ with $S_p$  since $S_p$ 
contains a significant fraction of strength up to GeV energies and 
would result 
in unrealistically large values for $\varepsilon (k)$ when $k>k_F$.
Therefore, above $k_F$, $S_p$ is replaced with the quasiparticle (qp)
contribution to $S_p$.  This is equivalent to
setting $\varepsilon(k)$ equal to the qp energy 
(discussed in more detail below).
One then obtains
\begin{equation}
\varepsilon (k)= \frac{\int d\omega\ \omega\ S_h(k,\omega)}{
\int d\omega\ S_h(k,\omega)} \hspace{5.5cm} k<k_F 
\label{eq:spenh}
\end{equation}
and
\begin{equation}
\varepsilon (k)= \frac{\int d\omega\ \omega\ S_{qp}(k,\omega)}{
\int d\omega\ S_{qp}(k,\omega)} 
=\frac{k^2}{2m}+ \textrm{Re}\ \Sigma (k,\varepsilon_{qp}(k)) \hspace{1.cm} 
k>k_F. 
\label{eq:spenp}
\end{equation}
The above prescription for the $sp$ spectrum requires calculation of the full
energy dependence of $\Sigma$ in each iteration step and is therefore more
computationally intensive.
However, it naturally contains a gap large enough to prevent 
the pairing instabilities.  More details related to this prescription
of self-consistency can be found in Ref.~\cite{vond2}.
The qp energy
alone has no appreciable gap, but near the 
Fermi momentum there is considerable spreading of hole strength to 
energies below the qp energy.  

\begin{figure}[tbp]
\begin{center}
\includegraphics[origin=bl,angle=0,width=0.4\textwidth]{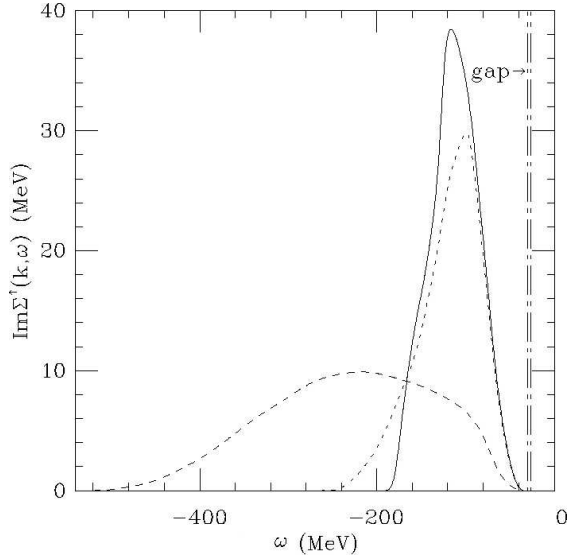}
\begin{minipage}[t]{16.5 cm}
\caption{\label{fig:imselfb}
Imaginary part of the self-energy as a function of
energy below the gap for three momenta, $k=0.01$ (solid), $0.51$ (dotted),
and  $2.1~\textrm{fm}^{-1}$ (dashed).
}
\end{minipage}
\end{center}
\end{figure}

Some results from this scheme will be reviewed now to establish a benchmark
for the second-generation spectral functions discussed in 
Sec.~\ref{sec:selfnmr}.
One characteristic feature of the resulting imaginary part of the self-energy
is shown in Fig.~\ref{fig:imselfb}. 
Comparing different values of $k$ for these energies below the Fermi energy
indicates that Im$\Sigma^\uparrow$ becomes weak 
while its energy range enlarges in 
a smooth manner as $k$ increases. 
The energy range for this Im$\Sigma^\uparrow$ is determined by
using the analysis of Ref.~\cite{anthes}.
This analysis shows that for each $k$ there is a minimum energy above
which 2h1p states (of mf type) can mix in the self-energy.
So momentum conservation and the constraint on the location of 2h1p energies
due to their mf character are responsible for the energy range observed
in  Fig.~\ref{fig:imselfb}. 
This wide range of energies will 
be available to the hole spectral functions discussed below.

\begin{figure}[tbp]
\begin{center}
\includegraphics[origin=bl,angle=0,width=0.5\textwidth]{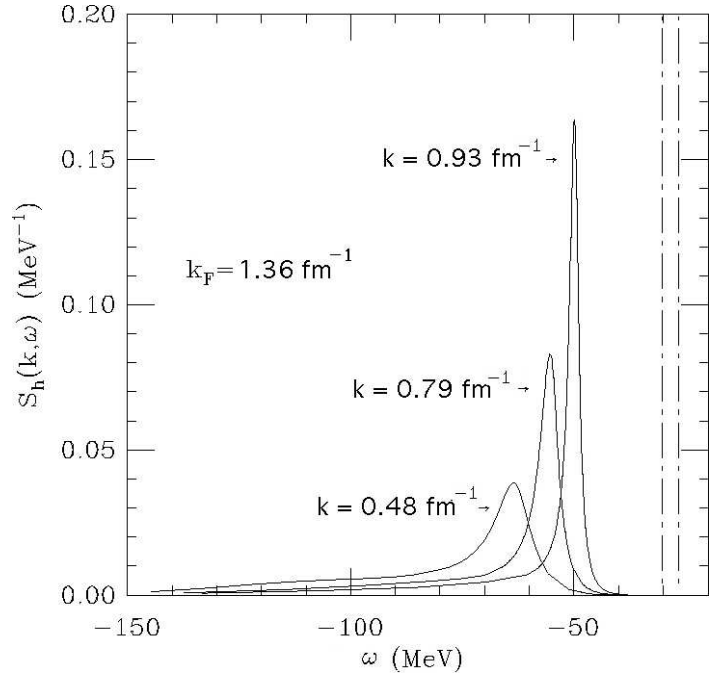}
\begin{minipage}[t]{16.5 cm}
\caption{\label{fig:spechmf}
Illustration of the decreasing width of the quasiparticle peak in the 
spectral function for three 
momenta below $k_F$ given by $k=0.48$, 0.79, and 
$0.93~\textrm{fm}^{-1}$. The vertical lines indicate the position of the gap
in the sp spectrum that results from employing Eqs.~(\ref{eq:spenh}) and 
(\ref{eq:spenp}).
}
\end{minipage}
\end{center}
\end{figure}

For momenta near $k_F$ one finds that the energy dependence of $S_h$ and $S_p$
is dominated by the qp peak characterizing the extent to which noninteracting
features are maintained.
Each momentum has an 
associated qp energy which is the solution of
\begin{equation}
\varepsilon_{qp}(k)=\frac{k^2}{2m} + \textrm{Re} \Sigma(k,\varepsilon_{qp}(k)).
\label{eq:Eqp}
\end{equation}   
The spectral function displays a peak at $\varepsilon_{qp}$ 
because of the vanishing term in the denominator of Eq.~(\ref{eq:specp})
or (\ref{eq:spech}).
The qp-peak itself is represented by 
\begin{equation}
S_{qp}(k,\omega)= \frac{1}{\pi} \frac{Z^2
(k) \mid W(k) \mid }{ (\omega-\varepsilon_{qp}(k))^2 +(Z(k)W(k))^2}
\label{eq:qpspec}
\end{equation}
where $W(k)= \textrm{Im}\ \Sigma(k,\varepsilon_{qp}(k))$ and 
\begin{equation}
Z(k)=\left\{1-\left(\frac{\partial \textrm{Re}\ \Sigma(k,\omega)}{
\partial\omega} \right)_{\omega=\varepsilon_{qp}(k)} \right\}\sp{-1} 
\label{eq:zfac}
\end{equation}
is the strength contained in the peak.
These first-generation SCGF calculations show that near $k_F$ the qp
peak contains around 70\% 
of the total sp strength, while an extra 13\% 
is contained in the background composing the remainder of hole strength.
The background is uniformly distributed across several hundred MeV below
$\varepsilon_F$ corresponding to the range of Im~$\Sigma$
but depends significantly on the value of $k$ as shown in
Fig.~\ref{fig:spechmf}.   
The final 17\% 
of the strength has moved to energies greater than $\varepsilon_F$
including a significant high-energy tail in $S_p$ discussed
momentarily.  Farther below $k_F$ this picture breaks down as the
qp peak melts into the background resulting in hole strength which is
spread over a much wider range of energies.
To the extent that spectral functions are described by
the qp approximation, the excitations in nuclear matter are
like those of a Landau Fermi-liquid as illustrated in
 Fig.~\ref{fig:spechmf}.
This figure contains several hole
spectral functions for momenta below the Fermi momentum
at $k_F=1.36~\textrm{fm}^{-1}$.
Notice that as $k\rightarrow  k_F$ this peak becomes extremely 
sharp due to the vanishing Im$\Sigma$ in Eq.~(\ref{eq:spech}).
The infinite-lifetime character of such excitations is made 
possible by the loss of phase-space available to the states in 
$\Sigma$ near the Fermi energy.  This is essentially the same 
argument used by Landau in more general terms to 
develop the microscopic foundations of Fermi-liquid theory~\cite{lan}.

\begin{figure}[tbp]
\begin{center}
\includegraphics[origin=bl,angle=0,width=0.5\textwidth]{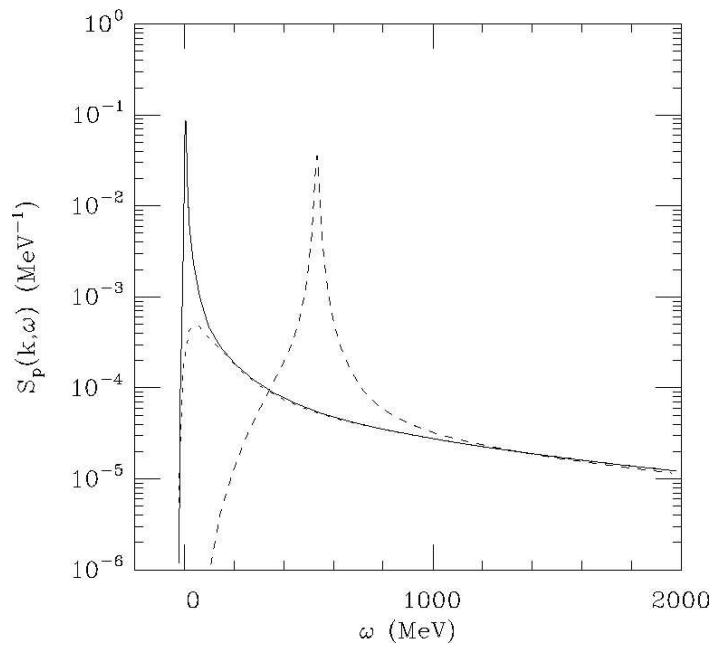}
\begin{minipage}[t]{16.5 cm}
\caption{\label{fig:specpmf}
Particle spectral functions at $k=0.79$
(dotted), $1.74$ (solid), $5.04~\textrm{fm}^{-1}$ (dashed).  All three spectral
functions converge to the same tail at high energy.
}
\end{minipage}
\end{center}
\end{figure}

In Fig.~\ref{fig:specpmf}
 the particle spectral function is plotted for three different
momenta, $k= 0.79,~ 1.74,$ and $5.04~\textrm{fm}^{-1}$ as a function of energy.
All momenta below $k_F$ have the same high-energy tail as the dotted curve
for $k=0.79~\textrm{fm}^{-1}$ in Fig.~\ref{fig:specpmf}.
For momenta larger than $k_F$ a quasiparticle peak, which broadens with
increasing momentum, can be observed on top of the same high-energy tail.
The results therefore display a common, essentially momentum independent,
high-energy tail.
The location of sp strength at high energy simply means that the
interaction has sufficiently large matrix elements to compensate
energy denominators encountered in the ladder equation.
For this particular interaction a significant amount of strength is found at
high energy. This result was of course already anticipated a long
time ago~\cite{bro69}.

A quantitative characterization of the missing $sp$ strength 
for $k= 0.79~\textrm{fm}^{-1}$ shows that
the integrated strength accounts for 17\% 
of the sp strength.
This is in agreement with the sum rule since the integrated
hole strength provides 83\% 
of sp strength. 
The strength in the interval from 100 MeV above the Fermi energy to infinity
amounts to 13\% 
with 7\% 
residing above 500 MeV.
To understand the influence of the tensor force on this distribution,
a calculation of the ladder equation was performed in which the tensor coupling
in the $^3S_1$-$^3D_1$ coupled channel was switched off.
In this case,
the integrated sp strength amounts to 10.5\% 
and should be regarded as
resulting from pure short-range correlations.
In Ref.~\cite{vond1} it is shown that the tensor force moves the additional 
6.5\%  
of strength to the first 1000 MeV above the Fermi energy.
This is consistent with CBF calculations of the momentum
distribution which show depletions of a similar size due to tensor
correlations~\cite{fantp}.

\begin{figure}[tbp]
\begin{center}
\includegraphics[origin=bl,angle=0,width=0.4\textwidth]{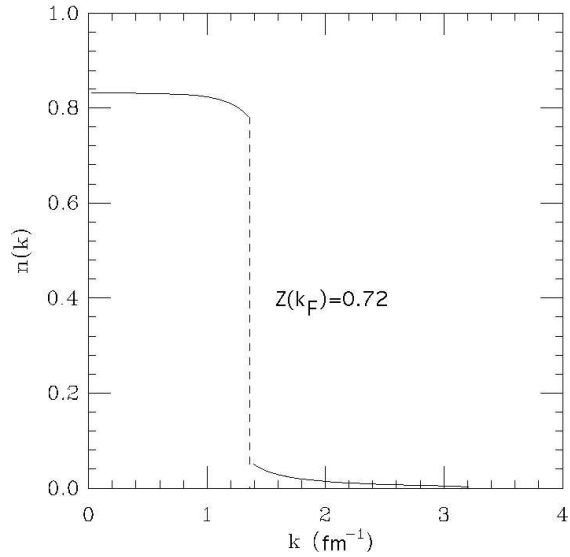}
\begin{minipage}[t]{16.5 cm}
\caption{\label{fig:nkreidb}
Occupation probability for nuclear matter at 
equilibrium density obtained by integrating hole spectral functions
obtained with mean-field propagators as input.
}
\end{minipage}
\end{center}
\end{figure}

Figure \ref{fig:nkreidb} exhibits the graph of the occupation probability, 
$n(k)$, or the number of particles 
in the ground state of the system with sp quantum number $k$.  
Near $k=0$, 
$n(k)$ becomes fairly constant with a value of 0.83.  From the discussion of
$S_p$, roughly $1/3$ of the 17\% 
depletion is due to the effect
of tensor correlations in the ladder equation.  
Another $1/3$ is due the to
high-energy tail in $S_p$ at energies above 500~MeV.  Results from other
many-body methods such as Brueckner theory~\cite{gcl87,bal90}
and CBF theory~\cite{bff89} using other realistic interactions
give very similar occupation near $k=0$.  
In Refs.~\cite{gcl87} and \cite{bal90}, 0.82 is
reported for the Paris potential.  
Older CBF calculations
for the Urbana $v_{14}$ interaction give 0.87 whereas more recent CBF
results~\cite{fantp} give 0.83.  All these calculations for different
interactions using different methods give a strikingly similar result for
$n(0)$.
This was encouraging since it implies that non-relativistic many-body
calculations are under control in the region where one would like to compare to
finite nuclei.  

In contrast to $n(0)$,
the occupation at $k_F$ varies significantly between methods.  This 
variability is most easily expressed in terms of the discontinuity in $n(k)$
at $k_F$, $Z(k_F)\equiv n(k^-_F)-n(k^+_F)$.  For the result shown in
Fig.~\ref{fig:nkreidb},
$Z(k_F)=0.72$ and 
for the CBF calculation $Z(k_F)=0.70$.  However, for the Paris interaction, 
$Z(k_F)=0.35$ and $Z(k_F)=0.47$ has been obtained in Refs.~\cite{gcl87} and
\cite{bal90},
respectively.  The extra depletion in $n(k)$ as $k\rightarrow k_F$
arises from the enhanced ability of the sp state to couple to low lying 2p1h
excitations as its energy approaches the 2p1h states. 
Also, the discontinuity depends on the level at
which pairing correlations are included in the calculation.  In the case of
a paired system $Z(k_F)$ would be zero.

\subsection{\label{sec:neil}Spectral functions for $\Lambda$-hyperons
obtained from mean-field input}

Hypernuclei, especially those with one $\Lambda$ hyperon,
have been studied for a long time~\cite{gal:77,povh:78}.
When a $\Lambda$ hyperon is placed in a nucleus or nuclear matter it will 
interact with the nucleons in its environment.  
As a result of these strong interactions, the $\Lambda$ becomes
correlated with nucleons in the medium.
The study of the properties of the $\Lambda$ hyperon in an environment of
nucleons aims to answer a number of fundamental questions related to
the properties of strange particles in the nuclear medium.
Considerable attention has been given to the potential energy the $\Lambda$
experiences in the nucleus and the corresponding single-particle (sp) energies.
From this experimental work it becomes clear that the $\Lambda$ hyperon
is less strongly bound to nucleons than either a proton or a neutron.
Such sp properties have been studied theoretically for finite nuclei 
by several groups~\cite{doga:83} - \cite{hald:00}.
The general conclusion from the
experimental work is that the $\Lambda$ hyperon experiences
a potential well in the nucleus that has the familiar Woods-Saxon shape with
a depth of about 30 MeV for a wide range of heavier nuclei.
Global sp properties of the $\Lambda$ hyperon can be studied in nuclear matter.
Results of such calculations have also been reported by several 
groups~\cite{rozy:79} - \cite{dabro:01}.

\begin{figure}[tbp]
\begin{center}
\includegraphics[width=0.5\textwidth]{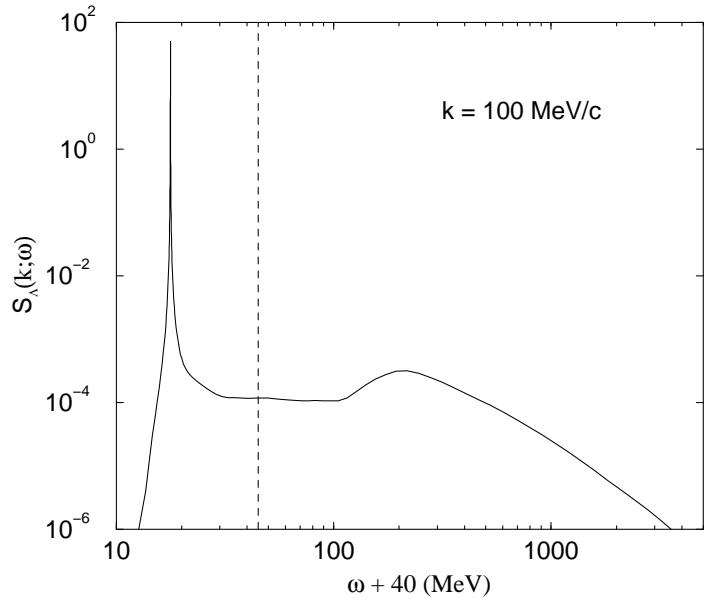}
\begin{minipage}[t]{16.5 cm}
\caption[Lambda spectral function]
{Spectral function for a $\Lambda$ with $k$ = 100 MeV/c~\cite{neilth,nd04}.  
The vertical dashed line indicates the position of a delta-function 
spectral distribution for the limiting case of a free particle.
Because of the 30 MeV binding for a $\Lambda$ in nuclear matter, it
is convenient to shift the horizontal axis by 40 MeV for plotting
on a log scale.
\label{fig:lspf}}
\end{minipage}
\end{center}
\end{figure}

The correlations of the $\Lambda$ hyperon in nuclear matter 
have typically been studied at the level of its average binding or sp energy.
Full propagation of the $\Lambda$ including the determination of its complex
self-energy has not been reported sofar.
In view of the relevance of the properties of a strange particle in a nuclear
system, it seems timely to study the properties of a $\Lambda$ hyperon
when it is embedded in such a nuclear system.
As in the case of NN interactions, typical hyperon-nucleon
(YN) interactions~
\cite{nagels:75} - \cite{rijken:99}
 incorporate substantial repulsion at short distance.
As discussed above, the consequences of this strong interaction
can be accounted for in the framework of
the Green's function formalism by including the
proper treatment of these short-range correlations (SRC)
in the form of ladder-diagram summation for the hyperon-nucleon interaction in
the medium ($G$ matrix).
The effects on the dynamical single-particle properties of the $\Lambda$
can then be explored by evaluating the complex self-energy of the $\Lambda$
in nuclear matter.
The solution of the Dyson equation for the $\Lambda$ then yields again 
information on the net binding of the $\Lambda$ in nuclear
matter, but also determines the distribution of spectral strength
for its addition to the nuclear-matter ground state
as a result of SRC.
Such a spectral function for a $\Lambda$ hyperon is shown in 
Fig.~\ref{fig:lspf}.
The addition of a strange test particle opens the door for
quantitative comparisons
with spectral functions obtained for nucleons discussed in the
previous section.
The weaker YN interaction is expected to result in similar but less
extreme modifications to the spectral distribution.
However, the presence of the $\Lambda$ hyperon also requires 
consideration of its heavier sibling, the isospin one $\Sigma$ hyperon.
The two hyperons have a small enough mass difference that a coupled-channel
problem must be solved.
In the case of the $\Lambda$ propagator some qualitative changes occur in 
comparison with the case of the nucleon propagator.
The result corresponding to Eq.~(\ref{eq:nmg}) now becomes
\begin{equation}
g_{\Lambda}(k;\omega) = \int_{\varepsilon_{T}^{\Lambda}}^{\infty}
d{\omega}' 
\frac{S_{\Lambda}(k;{\omega}')}{\omega - {\omega}' + i \eta} .
\label{lehmannl}
\end{equation}
Since no Fermi surface for $\Lambda$ hyperons is considered here,
it is not possible to remove a $\Lambda$ from the NM ground state.
As a result, the propagator only contains the probability amplitude for
adding a $\Lambda$ particle with momentum $k$ at an energy ${\omega}'$.
The energy threshold at which it becomes possible to add a $\Lambda$ to the NM
ground state is denoted by $\varepsilon_{T}^{\Lambda}$.
One expects this energy to be accessible only for a $\Lambda$ 
at rest ($k = 0$).  
If the $\Lambda$ were added to a free Fermi gas NM ground state
and would not be correlated otherwise, this threshold energy would simple be 
the kinetic energy of a $\Lambda$ with zero momentum and would therefore
correspond to zero energy.
Based on previous work by other groups~\cite{yama:90,schulze:95}
one expects the actual value of the threshold energy at normal NM density to
be around $-30\ \textrm{MeV}$ indicating the 
substantial attraction a $\Lambda$ experiences in NM.
Since the $\Lambda$ can only propagate as a particle, the spectral strength
above the threshold energy must integrate to 1.
This condition on the $\Lambda$ spectral function is given by
\begin{equation}
\int_{\varepsilon_{T}^{\Lambda}}^{\infty} d{\omega}' 
S_{\Lambda}(k;{\omega}') = n_{p,\Lambda}(k) = 1.
\label{eq:sumrulel}
\end{equation}
This result is quite different from the corresponding nuclear case where
one has to consider the strength in the particle domain to supplement
the contribution given by
Eq.~(\ref{eq:momdis}).
As in the case of nucleons one can obtain the spectral function from the
imaginary part of the propagator
\begin{equation}
S_{\Lambda}(k;\omega) = -\frac{1}{\pi} \textrm{Im}~g_{\Lambda}(k;\omega)
\hspace{15 mm} \omega > \varepsilon_{T}^{\Lambda} .
\label{spfalgl}
\end{equation}

The sp spectral function as shown for example in Fig.~\ref{fig:lspf}
involves the overlap between the simple physical state given by
\begin{equation}
a_{\Lambda\bm{k}}^{\dag} \ket{\Psi_0} 
\label{eq:rightside}
\end{equation}
and the complicated eigenstate 
\begin{equation}
\ket{ {}^{\Lambda}\Psi(E) } .
\label{eq:leftside}
\end{equation}
which includes all
interactions between the $\Lambda$ and the nuclear medium.  
The extent to which there is overlap illustrates how well
the $\Lambda$ sp state survives intact in the medium.
For the case of no interactions between the $\Lambda$ and the nucleons,
the overlap is perfect, since the state of Eq.~(\ref{eq:rightside}) is
an eigenstate in this situation.  This is evidenced by the 
$\delta$ function spectral distribution appropriate for a free 
particle, as indicated by a dashed line at the kinetic energy 
in Fig.~\ref{fig:lspf}.
Interactions between the $\Lambda$ and nucleons 
are responsible for the transition from the simple $\delta$ function 
structure to the more complex distribution of sp strength realized in NM.
The mechanism behind the spreading of sp strength can be understood as
the mixing of a sp state at a given energy with 2p1h states which span a 
continuum of energies.

\begin{figure}[tbp]
\begin{center}
\includegraphics[angle=-90,width=0.5\textwidth]{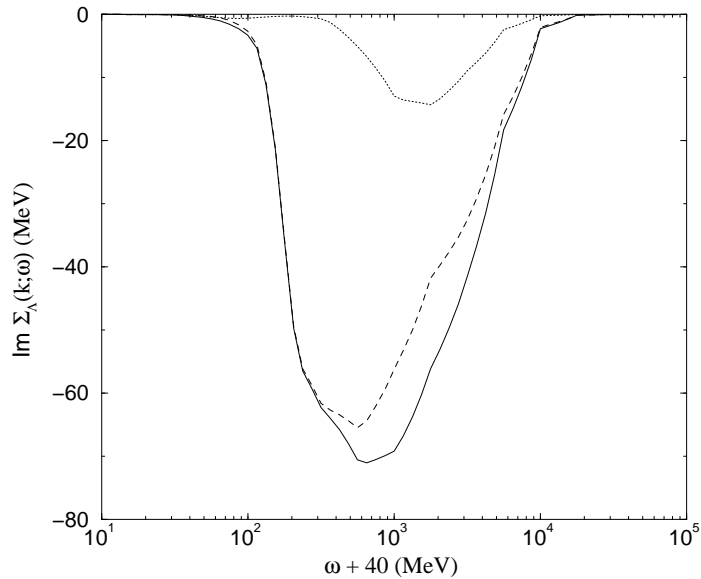}
\begin{minipage}[t]{16.5 cm}
\caption[Imaginary part of lambda self-energy]
{Imaginary part of the $\Lambda$ self-energy for $k$ = 100 MeV/c.
The broken curves represent contributions to the overall
self-energy from the $^{3}S_{1}$ (dash) and $^{1}S_{0}$ (dot)
partial wave channels.
\label{fig:lims}}
\end{minipage}
\end{center}
\end{figure}

\begin{figure}[tbp]
\begin{center}
\includegraphics[width=0.5\textwidth]{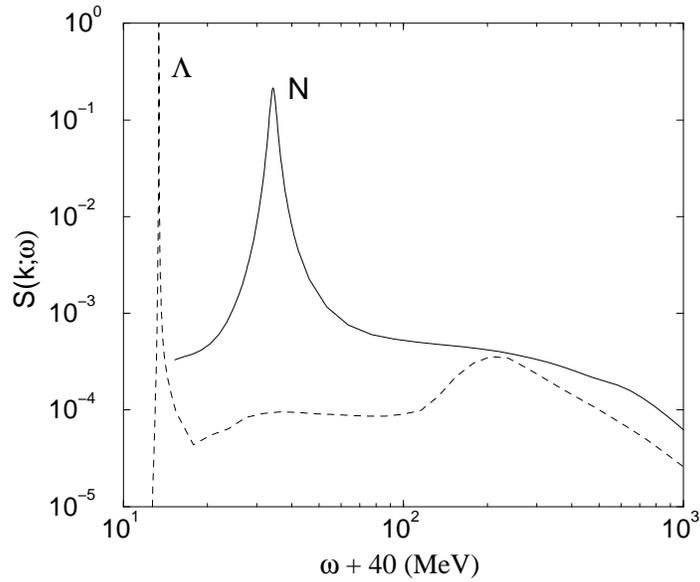}
\begin{minipage}[t]{16.5 cm}
\caption[Nucleon and lambda spectral functions]
{Nucleon particle spectral function (solid) for $k = 316$ MeV/c
with lambda spectral function (dashed) at $k = 60$ MeV/c for comparison.
\label{fig:nspf}}
\end{minipage}
\end{center}
\end{figure}

Although the sp state is no longer an eigenstate of the many-body
Hamiltonian, its quantum numbers are still conserved by the interaction.
The total strength associated with the original sp state, though 
fragmented, is fixed.  This is reflected in the sum rule of 
Eq.~(\ref{eq:sumrulel}).
Details of the strength distribution are determined by the density of
2p1h states which increases with energy and the strength with which 
the interaction couples these states to the unperturbed 
sp state.
This information is summarized in the imaginary part of the 
self-energy, shown for example in Fig.~\ref{fig:lims}.
Note that the decomposition in partial wave contributions emphasizes the
dominance of the ${}^3S_1$ channel.
A nuclear matter calculation for nucleons similar to this 
one~\cite{vond1} yields a 
particle spectral function shown in Fig.~\ref{fig:nspf}, for a momentum
just above $k_F$.
The $Z$-factor obtained from this calculation is 
$Z_{N}(k_F) = 0.72$, which is substantially reduced compared
to $Z_{\Lambda}(0) = 0.87$ for a similar $\Lambda$ qp state.
These two momentum values are compared because each qp
sits at the lowest possible excitation energy for a qp
in the respective systems.
In Ref.~\cite{vond1}, this depletion of the qp strength
is explained in terms of couplings to 2h1p states, which moves
approximately 10\% 
of the sp strength to energies below $\varepsilon_F$, and coupling
to 2p1h states, which distributes another 18\% 
to higher energies in the particle domain.
The corresponding fraction of sp strength in the particle domain is 13\% 
for the $\Lambda$, compared to 18\% 
for nucleons.
Turning off the $\Lambda N$-$\Sigma N$ coupling 
in the Nijmegen soft core (NSC)
potential reveals~\cite{maessen:89} that tensor effects are responsible for 
almost half
of the reduction in the $\Lambda$ qp strength consistent with the nucleon
case discussed in Sec.~\ref{sec:brian}.
A value of $Z_{\Lambda}(k_{F}^{\Lambda}) = 0.94$ is obtained when 
coupling to $\Sigma N$ states is cut off.

\begin{figure}[tbp]
\begin{center}
\includegraphics[angle=-90,width=0.5\textwidth]{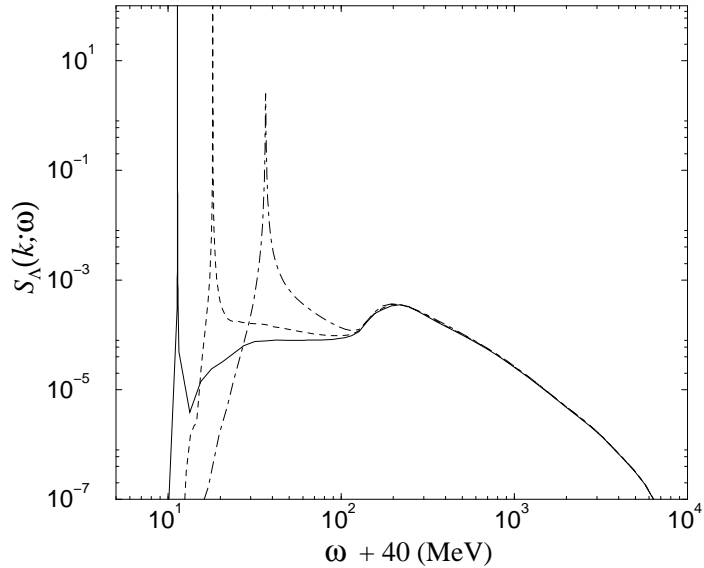}
\begin{minipage}[t]{16.5 cm}
\caption[Spectral function tail: $k$-independence]
{Spectral function for three values of $k_{\Lambda}$
show the $k$-independence of the high energy tail for
$k$ = 10 MeV/c (solid curve), 110 MeV/c (dashed),
and 210 MeV/c (dash-dot).
\label{fig:spfhie}}
\end{minipage}
\end{center}
\end{figure}

Away from the qp peak, at high-energy, the size and structure of the 
spectral function is primarily determined by two factors.  The 
density of 2p1h states increases like $\omega^{1/2}$ at high energy.
This growth in spectral strength with energy is moderated by
the strength of the coupling to these high-energy states.
A $\Lambda$ with a reasonably low momentum couples to a nucleon hole
state only with a low relative momentum.  The high-energy $\Lambda N$ 
two-particle states couple most strongly to high relative momentum and the
strength of the potential matrix elements between these two states
depends on the short-range characteristics of the two-body 
interaction.  
A harder core allows a stronger coupling between
states and correspondingly more spectral strength at high 
energy.
The fact that structure in the high-energy region of the spectral 
function is primarily determined by the short-range behavior of
the two-body interaction should be tempered by the knowledge that
the short-range part of baryon-baryon interactions are poorly known.  
Typical potentials are designed, within whatever
model, to fit only low-energy experimental data which does little
to constrain the details of the repulsive core.  
This situation can be taken in two ways.  
On the one hand, the high-energy tail of the 
spectral function is just as uncertain in detail as the core of the 
interaction from which it is derived.  
On the other hand, it is also just as experimentally inaccessible and
any observable which can be related to the detail of the tail
in the spectral strength distribution could be used to gain insight
into the behavior of the bare two-body interaction at short-range.
In Fig.~\ref{fig:spfhie} the similarities of the tail of the 
spectral strength for different momenta is illustrated.
We note that this common feature of the strength is identical to the case
of nucleons illustrated in Fig.~\ref{fig:specpmf}.
The only difference is in the distribution of the strength which reflects
the hardness of the underlying interaction.
In conclusion, one may state that the properties of a $\Lambda$ hyperon in
nuclear matter are qualitatively quite similar to those of a nucleon.
The physical consequences of the underlying interactions with their strong
repulsive cores are also quite similar.
Naturally, there are quantitative differences which indicate that a nucleon
is somewhat stronger correlated with the nuclear medium than a $\Lambda$.
Nevertheless, a common high-energy tail for the sp strength and a similar
role for the tensor force in reducing the qp strength also emphasize
the common features of these strongly interacting particles in the medium.

\subsection{\label{sec:healing}Healing properties of nuclear wave functions}
 
The conceptual understanding of strongly interacting nucleons yielding
a mf shell-model (Fermi-gas) picture has relied heavily
on the concept of the healing of the relative wave function to the
noninteracting one as discussed in Refs.~\cite{gww58,day67}.
Experimental evidence based on the (e,e$'$p) reaction~\cite{lap93}
has demonstrated that nucleon sp motion must be understood in
terms of Landau's quasiparticle description~\cite{lan}.
In turn, this requires a substantial modification of the simple shell-model
or Fermi-gas picture.
The conventional Bethe-Goldstone propagator used to determine the effective
interaction in the medium leads to the healing property of the
relative two-nucleon wave function and takes proper care of short-range 
correlations~\cite{gww58,day67}.
However, the corresponding effective interaction is 
not sufficient to generate a nucleon
self-energy which realistically describes the sp strength
distribution below the Fermi energy both in nuclear matter and finite
nuclei.
Inclusion of additional terms involving hole-hole propagation as in a
Galitski-Feynman propagator is essential to achieve a realistic spectral
function~\cite{dimu1,di94}.
The main problem with the Bethe-Goldstone propagator is that it generates
no self-energy diagrams which couple to 2h1p states.
As a result, the experimentally observed fragmentation of the sp strength
can not be described in this approach.
Inclusion of hole-hole propagation in the ladder equation solves
this problem of describing the fragmentation of the sp strength below
the Fermi energy.
However, this inclusion destroys the healing property of the
relative function since it produces a nonvanishing phase shift in the
relative wave function for energies
below twice the Fermi energy~\cite{bish} when mf propagators are employed.
The resolution of this puzzle requires the consideration of the consequences
of the dressing of the nucleons for the description of the scattering process
in the medium~\cite{dic98}.
In order to clarify this issue it is therefore necessary to delve in to the
details of this scattering process and elucidate the essential
differences with the corresponding process involving free or mf particles.

For the purpose of the present discussion it is sufficient to consider the
two-time two-particle propagator as given by Eq.~(\ref{eq:g2gam}).
The main objective of this development is to study
the consequences of propagating
dressed particles for the description of the scattering process.
All subsequent discussion will be based on the solution of a
Lippmann-Schwinger-type scattering equation for the effective interaction
which is equivalent to summing the ladder diagrams for a particular choice
of noninteracting two-body propagator and two-body interaction.
In order to clarify the difference between the conventional discussion of
scattering in free space and the one necessary for the medium, it will
be useful to cast the description completely in the language of the two-body
propagator.
While this is certainly not necessary for the scattering of free particles,
it will provide a simple way to clarify the changes that are required to
extend the description to dressed particles in the medium.

It is important to realize that the usual results from scattering theory
are obtained in the coordinate representation.
The relevant double Fourier transform of the two-particle propagator
in Eq.~(\ref{eq:g2gam}) is given by
\begin{equation}
g^{II}(\bm{r},\bm{r}';\Omega ) ~ = ~
\frac{1}{(2\pi)^3} \int \!\! d^3q\
\int \!\! d^3q'\ e^{i \bm{q} \cdot \bm{r}}
g^{II}(\bm{q},\bm{q}';\Omega )\ e^{-i\bm{q}'\cdot \bm{r}'} ,
\label{eq:dft}
\end{equation}
where without loss of generality we consider the case
that the total momentum $K$ is zero.
The transform of the noninteracting propagator only involves one integration
due to the presence of the $\delta$ function in Eq.~(\ref{eq:g2f})
\begin{equation}
g^{II}_f(\bm{r},\bm{r}';\Omega ) = 
\frac{1}{(2\pi)^3} \int \!\! d^3q\
\int \!\! d^3q'\ e^{i\bm{q} \cdot \bm{r}} 
g^{II}_f(\bm{q},\bm{q}';\Omega )\
e^{-i\bm{q}' \cdot \bm{r}'} 
= \frac{1}{(2\pi)^3} \int \!\! d^3q\ 
e^{i\bm{q} \cdot(\bm{r}-\bm{r}')}
g^{II}_f(\bm{q};\Omega) .
\label{eq:ftg2f}
\end{equation}
The result for Eq.~(\ref{eq:g2gam}) can then be transformed to yield
\begin{subequations}
\label{eq:g2gamr}
\begin{eqnarray}
g^{II}(\bm{r},\bm{r}';\Omega ) & = &
g^{II}_f(\bm{r},\bm{r}';\Omega ) 
+ \int \!\!  d^3r_1\ \int \!\! d^3r_2\ g^{II}_f(\bm{r},
\bm{r}_1;\Omega )\
\langle \bm{r}_1 | V |\bm{r}_2
\rangle\ g^{II}(\bm{r}_2,\bm{r}';\Omega )
\label{eq:g2gamra} \\
& = & g^{II}_f(\bm{r},\bm{r}';\Omega )
+  \int \!\! d^3r_1\ \int \!\! d^3r_2\ g^{II}_f(\bm{r},
\bm{r}_1;\Omega )\
\langle \bm{r}_1 | \Gamma (\Omega ) |
\bm{r}_2
\rangle\ g^{II}_f(\bm{r}_2,\bm{r}';\Omega ) .
\label{eq:g2gamrb}
\end{eqnarray}
\end{subequations}
Employing these equations one can arrive at an asymptotic analysis and 
resulting
definition of the cross section which is equivalent to a standard analysis
involving the equation for the wave function in the case of scattering in
free space or mf particles in the medium.
Before this result is developed, it is useful to summarize the propagator
equations in the partial-wave representation.
These allow the introduction of the phase shift which contains the relevant
information to obtain the asymptotic
propagator or wave function in this representation.

A partial wave decomposition of the two-body propagator in 
Eq.~(\ref{eq:g2gam}) yields the corresponding integral equation and the
relation between the propagator and the vertex function
\begin{subequations}
\label{eq:prop}
\begin{eqnarray}
g^{II}_{JST}(q\ell,q'\ell';\Omega) & = & \frac{\delta(q-q')}{q^2}
\delta_{\ell,\ell'} g^{II}_f(q;\Omega) 
+  g^{II}_f(q;\Omega) \sum_{\ell''} \int \!\! dp p^2 \langle q\ell | 
V^{JST} |p \ell''
\rangle g^{II}_{JST}
(p\ell'',q'\ell';\Omega) \hspace{0.9cm} \label{eq:propa} \\
& = & \frac{\delta(q-q')}{q^2} \delta_{\ell,\ell'} g^{II}_f(q;\Omega) 
+  g^{II}_f(q;\Omega) \langle q\ell|\Gamma^{JST}(\Omega)|q'\ell'\rangle
g^{II}_f(k';,\Omega) ,
\label{eq:propb}
\end{eqnarray}
\end{subequations}
where the vertex function or effective interaction $\Gamma$ can be obtained 
from the numerical solution of the ladder equation in a partial-wave momentum
representation as given in Eq.~(\ref{eq:ladpw}).
 The energy $\Omega$ is conserved and must
be viewed as a variable upon which the propagator depends (it also depends
on the total momentum in the case of the medium).
The noninteracting propagator is again denoted by $g^{II}_f$ and may include 
the dressing of the individual particles when the scattering takes place in
matter.

The coordinate space version of Eq.~(\ref{eq:prop}) is obtained by a double
Fourier-Bessel transform
\begin{equation}
g^{II}_{JST}(r\ell,r'\ell';\Omega)  =
\frac{2}{\pi} \int_{0}^{\infty} \!\!\! dq~q^2\int_{0}^{\infty} \!\!\!
dq'~q'{}^2
~j_{\ell}(qr) j_{\ell'}(q'r')\ g^{II}_{JST}(q\ell,q'\ell';\Omega) .
\label{eq:propr}
\end{equation}
The corresponding result for the noninteracting part of the propagator,
represented by the first term in Eq.~(\ref{eq:prop}), reduces to one integral
on account of the $\delta$ function which conserves relative momentum
\begin{equation}
g^{II}_{f,\ell}(r,r';\Omega)  =
\frac{2}{\pi} \int_{0}^{\infty}\!\!\! dq~q^2~
j_{\ell}(qr) j_{\ell}(qr') g^{II}_{f}(q;\Omega) .
\label{eq:nipropr}
\end{equation}
The Fourier-Bessel transform of Eq.~(\ref{eq:prop}) then has the following
form
\begin{eqnarray}
g^{II}_{JST}(r\ell,r'\ell';\Omega) \!\!\! & = &  
\!\! \delta_{\ell,\ell'}  g^{II}_{f,\ell}(r,r';\Omega)  
 + \!\! \sum_{\ell''} \int_0^{\infty}
\!\!\!\! \int_{0}^{\infty} \!\!\!\! dr_1 r_1^2\
  dr_2 r_2^2
g^{II}_{f,\ell}(r,r_1;\Omega)
  \langle r_1\ell|V^{JST}|r_2\ell''\rangle
g^{II}_{JST}(r_2\ell'',r'\ell';\Omega) \hspace{0.3cm} 
\nonumber  \\
& =& \!\! \delta_{\ell,\ell'}  g^{II}_{f,\ell}(r,r';\Omega)  
 +  \!\! \int_{0}^{\infty}\!\!\!\! \int_{0}^{\infty} \!\!\! dr_1 r_1^2
 dr_2 r_2^2\ 
g^{II}_{f,\ell}(r,r_1;\Omega)
  \langle r_1\ell|\Gamma^{JST}(\Omega)|r_2\ell'\rangle
g^{II}_{f,\ell'}(r_2,r';\Omega) .
\label{eq:propas}
\end{eqnarray}
When the bare two-body interaction $V$ is local in the
relative coordinate, only one integral in the first equality remains.
The second equality can be used to study the asymptotic behavior of the
propagator outside the range of the interaction.

The subsequent discussion for the scattering of dressed particles in the
medium requires the consideration of the two-body propagator in the medium.
For this reason it is advantageous to collect the conventional results for
scattering in free space using this language.
In the case of free particles the noninteracting propagator in momentum
space is given by
\begin{equation}
g^{II}_f(q;\Omega) = \frac{1}{\Omega-\hbar^2q^2/m+i\eta} .
\label{eq:nimom}
\end{equation}
Defining the on-shell momentum by
\begin{equation}
\Omega = \frac{\hbar^2q_0^2}{m} ,
\label{eq:onshell}
\end{equation}
one can perform the relevant Fourier-Bessel transform of the noninteracting
propagator in Eq.~(\ref{eq:nipropr}) analytically (see {\em e.g.}
\cite{got89}) with the well-known result
\begin{equation}
g^{II}_{f,\ell}(r,r';q_0) =  -iq_0\frac{m}{\hbar^2}
j_{\ell}(q_0r_<)h_{\ell}(q_0r_>) .
\label{eq:grr'}
\end{equation}
The coordinate argument in the spherical Hankel function must be the larger
of $r$ and $r'$ and is denoted by $r_>$ while the argument of the spherical
Bessel function is the smaller and denoted by $r_<$.
For the current analysis it will be assumed that the interaction 
has a finite range, $\langle r | V | r' \rangle = 0$ for $r, r'$ larger than
some $r_0$.
Substituting Eq.~(\ref{eq:grr'}) in the second part of Eq.~(\ref{eq:propas})
in the case of an uncoupled channel for $r' > r$ and $r' > r_0$ yields
\begin{eqnarray}
g^{II}_{\ell JST}(r,r';q_0) & = &  
 -iq_0\frac{m}{\hbar^2}j_{\ell}(q_0r)h_{\ell}(q_0r') \nonumber \\
& + &   \int_{0}^{\infty} \!\!\!\!
\int_{0}^{\infty} \!\!\!\! dr_1r_1^2\ dr_2r_2^2 ~g^{II}_{f,\ell}(r,r_1;q_0)
  \langle r_1|T^{JST}_{\ell}(q_0)|r_2\rangle
 \left( -iq_0\frac{m}{\hbar^2} \right) j_{\ell}(q_0r_2)h_{\ell}(q_0r') 
 \nonumber \\
 & = & -iq_0\frac{m}{\hbar^2} \psi^{JST}_{\ell}(r;q_0)h_{\ell}(q_0r') ,
\label{eq:propsep}
\end{eqnarray}
where
\begin{equation}
\psi^{JST}_{\ell}(r;q_0) = j_{\ell}(q_0r) 
+  \int_{0}^{\infty} \!\!\!\!  \int_{0}^{\infty} dr_1r_1^2\
dr_2r_2^2 ~g^{II}_{f,\ell}(r,r_1;q_0)
  \langle r_1|T^{JST}_{\ell}(q_0)|r_2\rangle
  j_{\ell}(q_0r_2) ,
\label{eq:feq}
\end{equation}
and the conventional notation $T$ instead of $\Gamma$ has been introduced
together with the replacement of $\Omega$ by $q_0$.
This result demonstrates that under the given conditions the propagator
separates as a product of a function of $r$ and a function of $r'$.
This result can be substituted into the first part of Eq.~(\ref{eq:propas})
to obtain the relevant integral equation for the wave function $\psi$
(under the condition that $r' > r_0$)
\begin{equation}
\psi^{JST}_{\ell}(r;q_0) = j_{\ell}(q_0r) 
+  \int_{0}^{\infty} \!\!\!\! \int_{0}^{\infty} \!\!\!\! dr_1r_1^2\
dr_2r_2^2 
~ g^{II}_{f,\ell}(r,r_1;q_0)
  \langle r_1|V^{JST}_{\ell} | r_2\rangle
\psi^{JST}_{\ell}(r_2;q_0) ,
\label{eq:finteq}
\end{equation}
which can be found in standard textbooks 
(see {\em e.g.} \cite{got89} for the case of a local potential).
It is derived here to demonstrate the relation between the propagator and
the wave function since for the case of dressed particles one has to start
with the formulation in terms of propagators.

The asymptotic analysis of the propagator can be performed by using
Eq.~(\ref{eq:grr'}) in Eq.~(\ref{eq:propas}) under the assumption that
the propagator will be considered for $r < r'$ while both
these coordinates are larger than $r_0$, the range of the interaction.
Values of $r_1$ and $r_2$ in Eq.~(\ref{eq:propas}) larger than $r_0$
yield no contributions to the integral.
As a result, the effective interaction, $T$, has a range similar
to the one of the bare interaction $V$.
Using the relation between spherical Bessel and Hankel functions 
\begin{equation}
j_{\ell}(\rho) = \frac{1}{2} \left( h_{\ell}(\rho) + h^*_{\ell}(\rho) \right) ,
\label{eq:bess}
\end{equation}
one obtains the asymptotic behavior of the propagator for the case of an
uncoupled partial wave channel from the second part of Eq.~(\ref{eq:propas})
in the following form
\begin{eqnarray}
g^{II}_{\ell,JST}(r,r';q_0)  \rightarrow & &
-i\left( \frac{m}{2\hbar^2} \right) q_0h_{\ell}(q_0r')
\bigg\{ h_{\ell}^*(q_0r) + h_{\ell}(q_0r) \times  \nonumber \\
& & \left[ 1- 2i\frac{m}{\hbar^2}q_0 
\int_0^{\infty} \!\!\!\! \int_0^{\infty} \!\!\!\! dr_1~r_1^2~
dr_2~r_2^2~ \langle r_1 |T^{JST}_{\ell} (q_0) |r_2 \rangle
j_{\ell}(q_0r_1) j_{\ell}(q_0r_2)\right] \bigg\} \nonumber \\
 =  & & -i\frac{m}{2\hbar^2}q_0h_{\ell}(q_0r')
 \left\{ h_{\ell}^*(q_0r) + h_{\ell}(q_0r)
\left[ 1-2\pi i \left( \frac{mq_0}{2\hbar^2} \right)
\langle q_0 | T^{JST}_{\ell}(q_0) | q_0 \rangle \right] \right\} . \nonumber \\
\label{eq:asymni}
\end{eqnarray}
In the last step of Eq.~(\ref{eq:asymni}) one can return to the on-shell
matrix element of the $T$-matrix in momentum space which completely
determines the outcome of the scattering process.
The term in square brackets corresponds to the
$S$-matrix element in terms of which one can define the phase shift
\begin{equation}
\langle q_0 | S^{JST}_{\ell}(q_0)|q_0\rangle = \left[ 1-2\pi i
\left( \frac{mq_0}{2\hbar^2} \right)
\langle q_0 | T^{JST}_{\ell}(q_0) | q_0 \rangle \right] \equiv
e^{2i\delta_{\ell}^{JST}} .
\label{eq:phaseni}
\end{equation}
This result can be represented by
\begin{equation}
\tan \delta^{JST}_{\ell} 
= \frac{\textrm{Im}\ \langle q_0 | T^{JST}_{\ell}(q_0) | q_0 \rangle}
{\textrm{Re}\ \langle q_0 | T^{JST}_{\ell}(q_0) | q_0 \rangle} ,
\label{eq:tanph}
\end{equation}
which explicitly shows that a nonzero imaginary part of the effective
interaction is required to obtain a nonvanishing phase shift.
In turn, this imaginary part of the interaction only appears for energies
where the noninteracting propagator has a nonvanishing imaginary part.
For the scattering of free particles this corresponds to all positive
energies.
By substituting the explicit form of the spherical Hankel functions for
$\ell = 0$ in Eq.~(\ref{eq:asymni}) one can construct the asymptotic propagator
for the ${}^1S_0$ channel explicitly 
\begin{equation}
g^{II}_{\ell,{}^1S_0}(r,r';q_0) \rightarrow -i \frac{m}{2q_0\hbar^2}
\frac{1}{rr'} e^{i(q_0r' + \delta_{{}^1S_0})} \sin (q_0r+
\delta_{{}^1S_0}) .
\label{eq:gasym}
\end{equation}
The standard result for the asymptotic wave function is contained 
in this equation and the imaginary part of Eq.~(\ref{eq:gasym}) is simply
the product of these wave functions as a function of $r$ and $r'$,
respectively.

To obtain the relation between the cross section and the propagator it is
necessary to return to Eq.~(\ref{eq:g2gamr}) and perform the Fourier transform
of the noninteracting propagator (Eq.~(\ref{eq:nimom})) in 
Eq.~(\ref{eq:ftg2f}).
This Fourier transform is given by the well-known result (again replacing
the energy $\Omega$ by the on-shell momentum $q_0$)
\begin{equation}
g^{II}_f(\bm{r},\bm{r}';q_0 ) = 
-\frac{m}{4\pi \hbar^2} \frac{
e^{iq_0
|\bm{r}-
\bm{r}'|}
}{|\bm{r}-\bm{r}'|} .
\label{eq:g2fvec}
\end{equation}
A similar procedure as used for the asymptotic analysis in the partial wave
basis can be employed to obtain the corresponding result for 
Eq.~(\ref{eq:g2gamr}).
Whereas in the former analysis the separable form of the noninteracting
propagator in Eq.~(\ref{eq:grr'}) is valid without constraint on $r$ and
$r'$, this is not the case here since Eq.~(\ref{eq:g2fvec}) only
becomes separable in the case $r' \gg r$ or vice versa.
In the former case one can write Eq.~({\ref{eq:g2fvec}) as
\begin{equation}
g^{II}_f(\bm{r},\bm{r}';q_0 ) \rightarrow
-\frac{m}{4\pi \hbar^2} \frac{e^{iq_0r'}}{r'}
e^{-iq_0\hat{\bm{r}}' \cdot \bm{r} } .
\label{eq:g2fvas}
\end{equation}
Substituting this result in the second part of Eq.~(\ref{eq:g2gamr}) for
both $r' \gg r$ and $r' \gg r_2$ demonstrates that $g^{II}$ is separable
and can be written as
\begin{equation}
g^{II}({\mbox{\boldmath $r$}},{\mbox{\boldmath $r$}}';q_0 ) =
-\frac{m}{4\pi \hbar^2} \frac{e^{iq_0r'}}{r'} \psi(\bm{r};q_0)
\label{eq:g2as}
\end{equation}
in the asymptotic domain.
By substituting this result in turn in Eq.~(\ref{eq:g2gamr})
one obtains the standard integral equation for the wave function
and the appropriate formulation for the asymptotic wave function to obtain
the scattering amplitude
\begin{eqnarray}
\psi(\bm{r};q_0) & = &
e^{-iq_0\hat{\bm{r}}' \cdot
\bm{r} }  
+  \int \!\!\! d^3r_1 \int \!\!\! d^3r_2 ~g^{II}_f(\bm{r},
\bm{r}_1;q_0 )
\langle\bm{r}_1 | V |\bm{r}_2
\rangle \psi(\bm{r}_2;q_0)
\nonumber \\
& = & e^{-iq_0\hat{\bm{r}}' \cdot \bm{r} }  
 +  \int \!\!\! d^3r_1 \int \!\!\! d^3r_2 ~g^{II}_f(\bm{r},
\bm{r}_1;q_0 )
\langle\bm{r}_1 | T (q_0 ) |
\bm{r}_2 \rangle
e^{-iq_0\hat{\bm{r}}' \cdot
\bm{r}_2 }  .
\label{eq:wfasym}
\end{eqnarray}
One may identify the origin of the motion in the direction of
the negative $z$-axis,
meaning that $\hat{\bm{r}}'$ points in that direction,
so that $\bm{q} \equiv -q_0\hat{\bm{r}}'$
points into the positive $z$-direction.
If one assumes that $r$ is also much larger than the range of the potential,
and, therefore much larger than any contributing value of $r_1$, one
can use Eq.~(\ref{eq:g2fvas}) again in the second part of
Eq.~(\ref{eq:wfasym}) to identify the coefficient multiplying the outgoing
spherical wave $e^{iq_0r}/r$ as the scattering amplitude (while double Fourier
transforming the $T$-matrix element back to momentum space)
\begin{equation}
f_{q_0}(\theta , \phi ) = -\frac{2m\pi^2}{\hbar^2} \langle 
\bm{q}' | T(q_0 ) |\bm{q} \rangle ,
\label{eq:scamp}
\end{equation}
where $\theta ,\phi$ denote the angles associated with the direction of
$\hat{\bm{r}}$ and 
$\bm{q}' \equiv q_0 \hat{\bm{r}}$
corresponds to the momentum of the detected motion in the direction
$\hat{\bm{r}}$ with the same absolute value $q_0$ as the
initial state.
The differential cross section for the direction $(\theta,\phi)$
is then simply the square of the scattering amplitude as given by
Eq.~(\ref{eq:scamp})
\begin{equation}
\frac{d\sigma}{d\Omega} = \left| f_{q_0}(\theta,\phi) \right|^2 .
\label{eq:cross}
\end{equation}
The present formulation is closely tailored to the conventional
experimental situation where a collimated beam propagates along the $z$-axis
characterized by a given energy or momentum toward a target situated at the
origin. Detection then takes place in
a particular direction away from the origin characterized by angles
$\theta$ and $\phi$.
The only difference is that the present formulation is appropriate for the
corresponding center-of-mass system ($K = 0$).

To obtain the phase shifts and cross sections for particles propagating in the
medium with mf sp energies one can proceed in similar fashion.
A useful reference is the work of Bishop {\em et al.}~\cite{bish} where
the introduction of the phase shift for the case of hole-hole propagation
is discussed.
While some details are different in this case, the main points of the previous
analysis remain valid when both particle-particle and hole-hole propagation
is taken into account.
The corresponding mf propagator in the medium, also known as the
Galitski-Feynman propagator, is given in
momentum space by
\begin{equation}
g^{II}_{mf}(q;\Omega) = \frac{\theta(q-k_F)}{\Omega-2\varepsilon(q)+i\eta} 
 - \frac{\theta(k_F-q)}{\Omega-2\varepsilon(q)-i\eta} ,
\label{eq:nimmom}
\end{equation}
where, again, the case of zero center-of-mass
momentum is considered.
The sp energy $\varepsilon(q)$ can deviate from the simple
kinetic energy spectrum and therefore yield a different relation between
the energy $\Omega$ and the on-shell momentum $q_0$
\begin{equation}
\Omega \equiv 2\varepsilon(q_0) ,
\label{eq:mfonshell}
\end{equation}
nevertheless the uniqueness of $q_0$ for a given energy is still preserved.
Although one can no longer evaluate the noninteracting propagator in
coordinate space completely analytically from Eq.~(\ref{eq:nipropr}),
the separability of the propagator is maintained for the contribution of the
pole term as in Eq.~(\ref{eq:grr'}) (with a different constant prefactor), 
while the remaining term vanishes asymptotically for $r$ sufficiently
different from $r'$.
A discussion of a similar result for the Fourier transform of the mf
propagator given in Eq.~(\ref{eq:nimmom}) can be found in Ref.~\cite{brro}
for the Bethe-Goldstone propagator.
As a result, one preserves the integral equations for the wave
function either in a partial wave basis as in Eq.~(\ref{eq:feq}) or for the
wave function in coordinate space as in Eq.~(\ref{eq:wfasym}) in the case of
mf propagators. 
The only difference with the free scattering case involves the use of the
mf equivalents of the noninteracting propagators in coordinate
space in Eqs.~(\ref{eq:finteq}) and (\ref{eq:wfasym}).
This result is due to the uniqueness of the on-shell momentum at a given
energy which guarantees that the noninteracting wave function is a plane wave
or spherical Bessel function (in a partial wave basis).
One can therefore proceed with a similar asymptotic analysis as for free
particles yielding a corresponding definition of the phase shifts as in
Eq.~(\ref{eq:phaseni}}) in terms of the on-shell scattering matrix.
The result of Eq.~(\ref{eq:tanph}) also remains valid in this case.
The presence of a nonvanishing phase shift is therefore continued to be
linked to the nonvanishing of the imaginary part of the noninteracting
propagator. In the case of mf scattering the corresponding energy
domain resides above $2\varepsilon(q=0)$ which corresponds to the lowest
energy of two hole states.

These modest modifications of the quantities that characterize the
scattering process as compared to the case of free-particle scattering,
are related to the continued one-to-one relation of the energy
with a unique relative momentum for which the noninteracting propagator has
an imaginary part. This on-shell momentum emerges as the momentum
that characterizes the plane wave (or spherical Bessel function) function
describing the relative motion.
The plane-wave character of the wave function allows for a conventional
interpretation of the scattering process as in free space.

Since the correlated wave function does not heal to the uncorrelated one
when hole-hole propagation is included, 
the usual discussion of healing must be indeed be modified as indicated above.
The standard interpretation of the validity of the shell model 
is couched in terms of the healing of the relative wave
function to the noninteracting one.
This interpretation was put forward in Ref.~\cite{gww58} and is based on the
use of the Bethe-Goldstone propagator in describing the effective interaction.
Since this propagator excludes the propagation of two holes in
Eq.~(\ref{eq:nimmom}), the correction to the relative wave function in
Eq.~(\ref{eq:finteq}) due to the strong interaction potential 
heals within a characteristic healing distance to the spherical Bessel
function provided that the energy is less than twice the Fermi energy.
In scattering language this simply states that there is no phase shift
for energies less than $2\varepsilon_F$ when the Bethe-Goldstone propagator is
employed since no corresponding imaginary part of the propagator exists
\cite{day67}.

An apparent contradiction then arises with this interpretation when one 
realizes
that it is not permitted to neglect the propagation of the hole-hole term in
Eq.~(\ref{eq:nimmom}) since it is essential for the understanding of the
fragmentation of the sp strength below the Fermi
energy~\cite{rpd1}.
Inclusion of hole-hole propagation yields a nonvanishing phase shift below
$2\epsilon_F$~\cite{bish} which is at odds with the healing interpretation
of the relative wave function.
On the other hand this contribution to the effective interaction is
responsible for the presence of an imaginary part of the nucleon self-energy
which is required in order to describe the experimental situation in nuclei
as obtained from the (e,e$'$p) reaction~\cite{lap93}.
Evidently it is not possible to propagate mf nucleons which
can generate a realistic sp strength distribution, and,
at the same time, obtain the healing of the relative wave function
to the noninteracting one which supposedly underlies the success of the
mf picture.

While recent (e,e$'$p) experiments have sharpened the range of the
validity of the sp picture in terms of the more appropriate
Landau quasiparticle description~\cite{lan} which is adequately described by
microscopic theory~\cite{dimu1,di94}, the paradox at the level of the relative
wave function remains.
A clue to the solution of this puzzle is provided by noting the inconsistency
of the description of the sp strength in terms of a realistic
spectral function and the construction of the effective interaction by means
of a mf propagator.
Clearly if the dressing effect of the nucleon is substantial and
experiment~\cite{lap93} indicates it is, then one is forced to consider the
construction of the effective interaction in terms of dressed nucleons.
The consequences of this extension for the description of the scattering
process in matter and the resolution of the healing paradox will be taken up
in the following.

The propagation of dressed nucleons requires a different treatment of
the description of the scattering process. 
The main ingredient for this change is the form of the noninteracting
propagator which is given in the medium by Eq.~(\ref{eq:g2fdis}) and 
for sp momenta $\bm{k}$ and $\bm{k}'$ reads
\begin{equation}
g^{II}_f(\bm{k},\bm{k}';\Omega) = 
\int_{\varepsilon_F}^{\infty} \!\!\! d\omega 
\int_{\varepsilon_F}^{\infty} \!\!\! d\omega'
\frac{S_p(\bm{k},\omega)
S_p(\bm{k}',\omega ')}
{\Omega -\omega -\omega ' +i\eta} 
- \int_{-\infty}^{\varepsilon_F} \!\!\! d\omega 
\int_{-\infty}^{\varepsilon_F} \!\!\! d\omega '
\frac{S_h(\bm{k},\omega) S_h(\bm{k}',\omega ')}
{\Omega -\omega -\omega ' -i\eta} .
\label{eq:gnidr}
\end{equation}
The particle and hole spectral functions, $S_p$ and $S_h$ respectively,
describe the distribution of the sp strength for a given momentum over the
energy. 
These distributions are continuous and have sizable peaks either above or
below the Fermi energy, corresponding to a momentum state above or below
$k_F$, at the so-called quasiparticle energy.
For $k_F = 1.36~\textrm{fm}^{-1}$, corresponding to normal density,
the strength contained in the peak for momenta close to $k_F$ is typically
only 70\%~\cite{vond1,bff92}.
From the rest of the strength we refer for more details to the discussion 
in Sec.~\ref{sec:brian}.

\begin{figure}[tbp]
\begin{center}
\includegraphics[angle=0,width=0.5\textwidth]{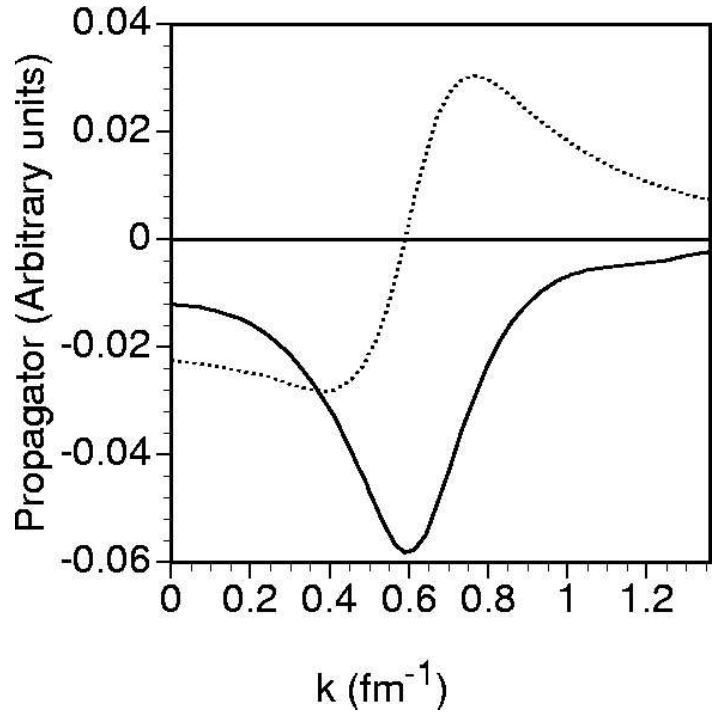}
\begin{minipage}[t]{16.5 cm}
\caption{Real (dotted) and imaginary part (full line)  of the
noninteracting propagator of
dressed particles at an energy corresponding to an on-shell momentum
of 0.5 $\textrm{fm}^{-1}$ for the mf propagator.
\label{fig:irdr} }
\end{minipage}
\end{center}
\end{figure}

It should be noted that the noninteracting propagator in Eq.~(\ref{eq:gnidr})
becomes the familiar mf Galitski-Feynman propagator (see
Eq.~(\ref{eq:nimmom})) when mf spectral functions
are inserted which are characterized by a $\delta$ function peak of strength
1 at a sp energy either above the Fermi energy ($k > k_F$) or below
($k < k_F$).
The difference between the Galitski-Feynman propagator and the dressed
propagator is qualitatively different for the imaginary part
and quantitatively for the real part.
In Fig.~\ref{fig:irdr} both the real and imaginary part of the dressed
propagator (Eq.~(\ref{eq:gnidr})) are shown as a function of the relative
momentum for zero total momentum while the energy corresponds to an on-shell
momentum $q_0 = 0.5~\textrm{fm}^{-1}$ in the Galitski-Feynman case using a sp
spectrum from Ref.~\cite{vond2}.
In the latter case the imaginary part of the propagator contains a
$\delta$ function only at a single momentum corresponding to
0.5 $\textrm{fm}^{-1}$.
It arises on account of the vanishing of the denominator 
in the hole-hole term at the on-shell momentum.
The full curve in Fig.~\ref{fig:irdr} corresponds to the imaginary
part of the dressed propagator shown here as a function of momentum up to 
$k_F$.
\begin{figure}[tbp]
\begin{center}
\includegraphics[angle=0,width=0.5\textwidth]{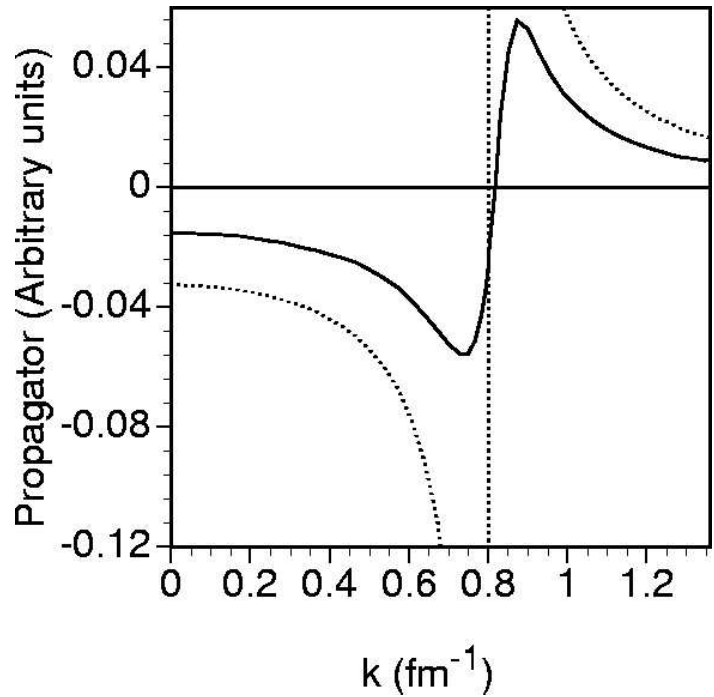}
\begin{minipage}[t]{16.5 cm}
\caption{Comparison of the real part of the mf (dotted) and dressed
(full line) propagator at an on-shell momentum of 0.8 
$\textrm{fm}^{-1}$ indicated by the
vertical dotted line.
\label{fig:mfdr} }
\end{minipage}
\end{center}
\end{figure}
There is a substantial spreading
in the imaginary part containing even small high-momentum components (not
shown in Fig.~\ref{fig:irdr}).
This spreading is a critical feature which completely alters the conventional
picture of the scattering process.
In Fig.~\ref{fig:mfdr} the real part of the dressed and mf
propagator are compared for momenta below $k_F$ for an energy corresponding
to an on-shell momentum of 0.8 $\textrm{fm}^{-1}$.
The dotted line corresponds to the mf propagator and also indicates
the pole present at 0.8 $\textrm{fm}^{-1}$.
The dressed propagator shows a less dramatic momentum dependence and is in
general substantially reduced from the mf propagator except for
high momenta where both coincide.

In order to perform the analysis of the scattering process it will be
illustrative to use an analytical approximation to the noninteracting
propagator of dressed particles (Eq.~(\ref{eq:gnidr}))
which contains the essential new features.
In addition, only the case of zero total momentum
of the propagating pair will be considered in the following.
As a result, the noninteracting propagator contains equal and opposite
momenta for the two particles (holes).
Since the spectral functions do not depend on the direction of the momentum
one can rewrite Eq.~(\ref{eq:gnidr}) for the present purposes as
\begin{equation}
g^{II}_f(k,\Omega) =
\int_{\varepsilon_F}^{\infty} \!\!\! d\omega 
\int_{\varepsilon_F}^{\infty} \!\!\! d\omega '
\frac{S_p(k,\omega)
S_p(k,\omega ')}
{\Omega -\omega -\omega ' +i\eta}
- \int_{-\infty}^{\varepsilon_F} \!\!\! d\omega 
\int_{-\infty}^{\varepsilon_F} \!\!\ d\omega '
\frac{S_h(k,\omega) S_h(k,\omega ')}
{\Omega -\omega -\omega ' -i\eta} .
\label{eq:gnidra}
\end{equation}
The momentum $k$ not only corresponds to the absolute value of the
sp momenta but also represents the relative momentum for
the case of zero total momentum.

\begin{figure}[pbt]
\begin{center}
\includegraphics[angle=0,width=0.5\textwidth]{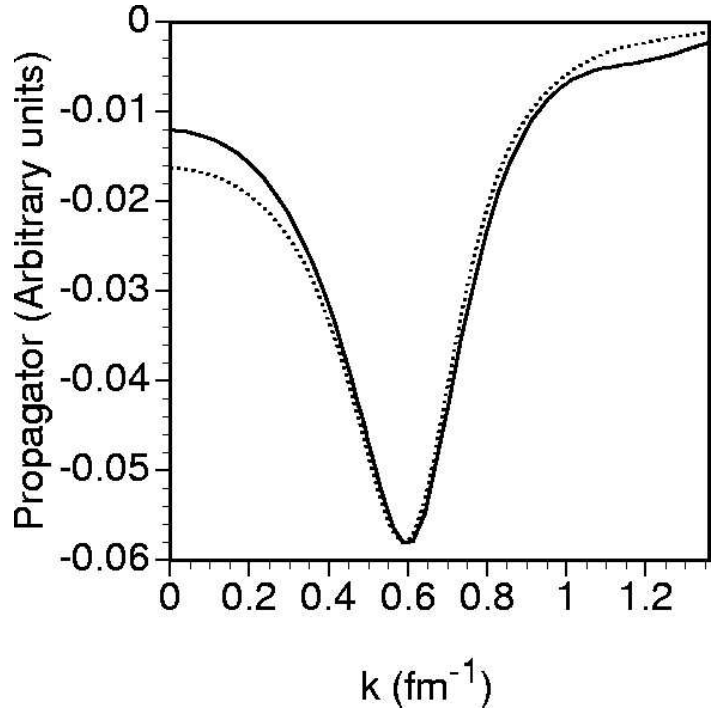}
\begin{minipage}[t]{16.5 cm}
\caption{Comparison for the imaginary part of the noninteracting propagator
of dressed nucleons between the complete (numerical) result given by the
full line and the simple CPA given by the dotted line for momenta below
$k_F$.
\label{fig:cpa} }
\end{minipage}
\end{center}
\end{figure}

Introducing a two-body self-energy term for purely practical reasons, one can
attempt to write this noninteracting propagator as
\begin{equation}
g^{II}_f(k,\Omega) =  \frac{\pm 1}{\Omega-\Sigma^{II}(k,\Omega)} ,
\label{eq:sigII}
\end{equation}
where the sign is determined by whether the energy $\Omega$ is 
above (+) or below (-) 2$\varepsilon_F$.
By assuming that this ad hoc self-energy $\Sigma^{II}$ has a slowly varying
imaginary part as a function of the relative momentum $k$ one can expand the
self-energy at the momentum $k_0$ for which
\begin{equation}
\Omega \equiv \textrm{Re}\ \Sigma^{II}(k_0,\Omega) .
\label{eq:q0}
\end{equation}
Noting that the expansion is in the square of the momentum, one obtains a
complex pole approximation (CPA) to the propagator by only keeping the real
and imaginary part
of $\Sigma^{II}$ at $k_0^2$ and the first derivative of the real part.
The resulting propagator has the form
\begin{equation}
g^{II}_{f,CPA}(k,\Omega) = \frac{m}{\hbar^2} \frac{\pm c}{k_0^2-k^2 \pm
i\gamma} ,
\label{eq:cpa}
\end{equation}
where the constant $c$ is obtained from
\begin{equation}
c = \frac{\hbar^2}{m} \left( \left. \frac{\partial {\rm Re} \Sigma^{II}}{
\partial k^2} \right\vert_{k_0^2} \right)^{-1}
\label{eq:const}
\end{equation}
and $\gamma$ from
\begin{equation}
\gamma = \left\vert {\rm Im} \Sigma^{II}(k_0,\Omega) \right\vert
\left(
\left. \frac{\partial {\rm Re} \Sigma^{II}}{\partial k^2} \right\vert_{k_0^2}
\right)^{-1} .
\label{eq:gamm}
\end{equation}
Typical values of $c$ at low energies correspond to 0.5 whereas it rises
slowly to 1 for higher energies. This feature is closely related to the
pattern of the distribution of the sp strength. The quasiparticle pole
strength at $k_F$ = 1.36 $\textrm{fm}^{-1}$ is about 0.7 
(see Sec.~\ref{sec:brian}) so for a two-particle
propagator close to these energies a factor of $(0.7)^2$ is expected.
For higher momenta the strength in the peak grows back to 1 yielding
a propagator which is more of the mf or even free-particle kind.
It is also apparent that this factor of about 0.5 can be identified 
from Fig.~\ref{fig:mfdr}.
It should be noted that the CPA is obtained after first numerically
calculating the noninteracting propagator of the dressed 
particles.
In Fig.~\ref{fig:cpa} the quality of this CPA to the
propagator can be judged by comparing it to the numerically calculated
result for the imaginary part of the propagator at an energy below
2$\varepsilon_F$.

\begin{figure}[pbt]
\begin{center}
\includegraphics[angle=0,width=0.5\textwidth]{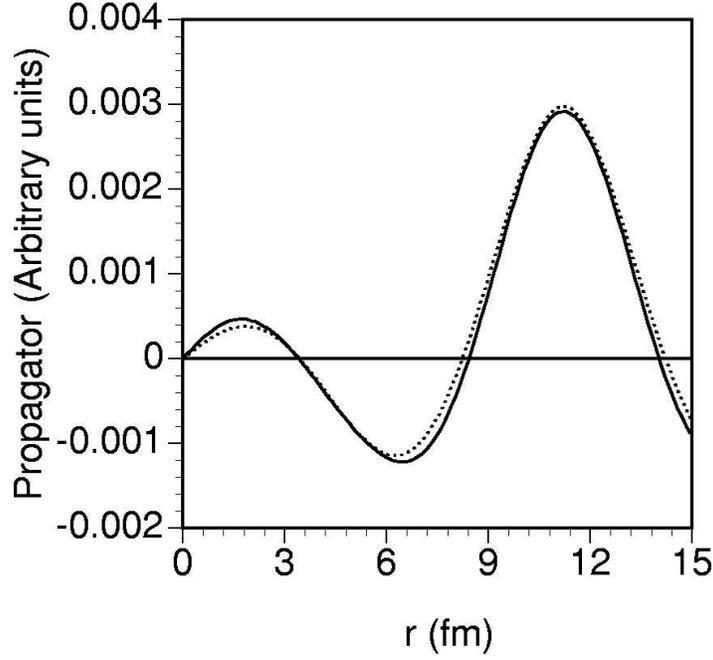}
\begin{minipage}[t]{16.5 cm}
\caption{Comparison between the analytical approximation and the
complete numerical result for the dressed noninteracting propagator
in coordinate space.
Shown is the imaginary part for an energy below 2$\varepsilon_F$ for the
CPA propagator (dotted line) and the complete result (full line).
The corresponding propagators in momentum space are shown in
Fig.~\ref{fig:cpa}.
\label{fig:cpaim} }
\end{minipage}
\end{center}
\end{figure}

\begin{figure}[pbt]
\begin{center}
\includegraphics[angle=0,width=0.5\textwidth]{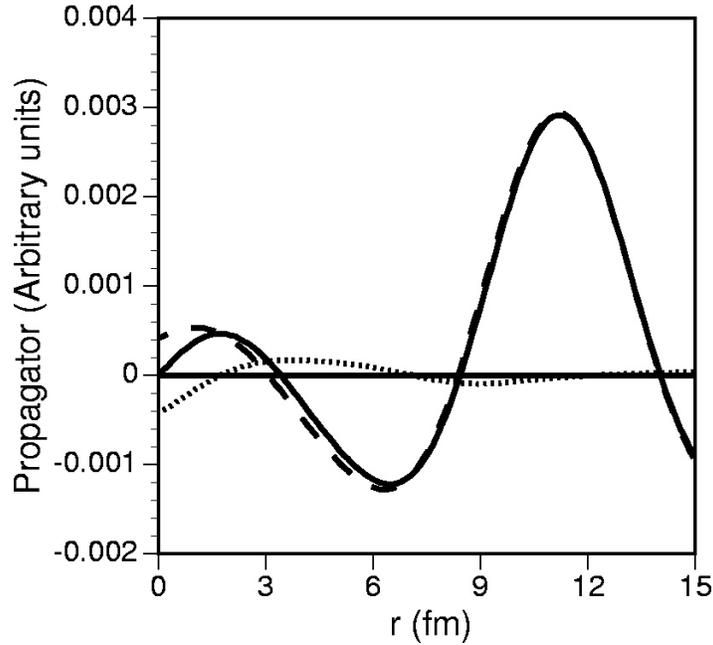}
\begin{minipage}[t]{16.5 cm}
\caption{
Decomposition of the numerical propagator in coordinate space into
in- (dashed) and outgoing wave (dotted line).
Also shown is the sum of both contributions (full line).
\label{fig:decom} }
\end{minipage}
\end{center}
\end{figure}

The CPA result for the propagator cannot be used to solve any of
the integral equations for the effective interaction
for example since it is only a good approximation to the noninteracting
propagator close to the peak of the imaginary part.
The full solution also requires an accurate
representation of the high-momentum components of the propagator in order
to properly include the effect of short-range correlations in the
interaction or wave function.
The CPA result does provide a reasonable representation of the long-range
part of the
propagator and therefore can be profitably used to discuss the asymptotic
analysis of the scattering process.
Indeed, using the CPA to the dressed propagator 
one can repeat the steps involved in the Fourier-Bessel transform
leading to Eq.~(\ref{eq:grr'}). 
For free or mf particles the integral in Eq.~(\ref{eq:nipropr})
yields the product of a spherical Bessel function and one of the spherical
Hankel functions with as argument the real on-shell momentum $k_0$ (see
Eqs.~(\ref{eq:onshell}) and (\ref{eq:mfonshell})).
This on-shell momentum is real since the corresponding noninteracting
propagators (Eqs.~(\ref{eq:nimom}) and (\ref{eq:nimmom})) can only have
a vanishing denominator for a real momentum.
Since Eq.~(\ref{eq:nipropr}) can be calculated by a contour integral for
Eqs.~(\ref{eq:nimom}), (\ref{eq:nimmom}) (at least for the long-range part),
as well as for Eq.~(\ref{eq:cpa}), it is clear that the presence of
a nonvanishing imaginary part for the pole of Eq.~(\ref{eq:cpa}), due to a
the nonvanishing of $\gamma$, will lead to a complex on-shell momentum
which will be denoted by $\kappa_0$.
Using the CPA (Eq.~(\ref{eq:cpa})) for $\Omega < 2\epsilon_F$ one obtains from
Eq.~(\ref{eq:nipropr})
\begin{equation}
g^{II}_{\ell,CPA}(r,r';\Omega) =
-ic\frac{m}{\hbar^2}\kappa_0j_{\ell}(\kappa_0r_<)h_{\ell}^*(\kappa_0r_>) .
\label{eq:cparr'}
\end{equation}
The momentum argument of the spherical Bessel and Hankel functions, $\kappa_0$,
is now complex, its real and imaginary part, $\kappa_0^R$ and $\kappa_0^I$,
respectively, are easily obtained from $k_0$ and
$\gamma$ (see Eqs.~(\ref{eq:q0}) and (\ref{eq:gamm})) by determining the
zeros of the denominator of Eq.~(\ref{eq:cpa}).
Eq.~(\ref{eq:cparr'}) contains the Hankel function $h_{\ell}^*$ due to the
different boundary condition associated with hole-hole propagation for
energies below $2\epsilon_F$.
This leads to a pole in the upper half of the complex $k$-plane in contrast
to the case of particle-particle or free-particle scattering.
As a result, $\kappa_0^I$ is negative for $\Omega < 2\epsilon_F$ and 
its magnitude can become as large as 0.2 to 0.3 $\textrm{fm}^{-1}$.
The resulting propagator for $\ell = 0$ can be written as (for $r < r'$)
\begin{equation}
g^{II}_{\ell=0,CPA}(r,r';\Omega) = \frac{-icm}{2\hbar^2(\kappa_0^R+i
\kappa_0^I)} \left(\frac{e^{i\kappa_0^Rr}e^{-\kappa_0^Ir}}{r}-
\frac{e^{-i\kappa_0^Rr}e^{\kappa_0^Ir}}{r}\right)
\frac{e^{-i\kappa_0^Rr'}e^{\kappa_0^Ir'}}{r'} .
\label{eq:l0cpa}
\end{equation}
A comparison between this analytical result and the numerical Fourier-Bessel
transform of the dressed noninteracting propagator which it approximates, is
very successful in the domain where the propagator has its maximum
as shown in Fig.~\ref{fig:cpaim}.
This is also true for the real part.
This demonstrates that the propagator for dressed particles
is radically different from the noninteracting or mf
one (see Eq.~(\ref{eq:grr'})) due to presence of damping terms related to
the nonzero value of $\kappa_0^I$.
As noted before, there is no longer a unique on-shell momentum.
Indeed, the complex pole at $\kappa_0$ in the CPA propagator is just a
simple (and approximate) representation of this feature.
As a consequence, the relative wave function of the dressed particles
contains a spread in momentum states.
This, in turn, must yield a localization of the corresponding wave function
in coordinate space.
This is exhibited in the propagator Eq.~(\ref{eq:l0cpa}) which represents
the probability amplitude for removing a pair with relative distance $r'$
and adding the pair after propagation at $r$ (without interaction between
the pair included yet). This amplitude peaks at $r = r'$ (see
Fig.~\ref{fig:cpaim})
and is exponentially damped with the decay constant $\vert \kappa_0^I \vert$.
This feature has interesting physical consequences since it means that
if the separation distance between the scatterers is too large there is
little probability that they will actually interact because this requires a
small relative distance.
Indeed, taking $r'$ to much larger values than in Fig.~\ref{fig:cpaim}
yields a negligible contribution to the noninteracting propagator near small
$r$ where the interaction will act to modify the wave function.
Clearly this effect is governed by the size of $\kappa_0^I$
the imaginary part of the pole of the CPA.
It should be noted that this value will only approach zero when the
scattering energy approaches 2$\varepsilon_F$.
Just as in the case of the sp motion, this means that the
noninteracting wave function will tend to a plane wave again only in this 
limit.
The corresponding result for Fig.~\ref{fig:cpaim} then yields a simple
sine wave characterized by $\kappa_0^R$ which approaches $k_F$.
For all other energies damping does occur sufficiently rapidly to warrant
the following observation:
since only the part of the wave which returns from the scattering
can be affected and this part always decreases with increasing $r$,
only that part of the noninteracting wave can be influenced
by the scattering which is exponentially damped in $r$.
This is illustrated in Fig.~\ref{fig:decom} where a numerical
decomposition of the
propagator is presented in terms of the
incoming and outgoing wave.
For values of $r'$ outside the range of the interaction as in
Fig.~\ref{fig:decom} this implies that even a substantial modification
of the outgoing wave will hardly affect the total propagator
and the wave function must automatically heal, according to the 
equivaqlent value
of $\kappa_0^I$ in CPA approximation, to the noninteracting one.

The above observations allow for the resolution of the paradox related to the
healing properties of wave functions in the medium.
This property has been considered the physical justification of the
mf-like properties observed in nuclei in the presence of strong
short-range interactions.
There is overwhelming experimental evidence~\cite{lap93} that
sp motion in nuclei must be described in terms of dressed
nucleons with substantial fragmentation of the strength.
The original discussion of the healing properties of the relative wave
function of particles in the medium~\cite{gww58} used a Bethe-Goldstone
propagator involving mf nucleons to arrive at the healing property
of the relative wave function.
The present discussion explores 
the consequences of scattering dressed particles
and demonstrates that this dressing of nucleons automatically leads
to a localization of the relative wave functions in coordinate space.
The results for the CPA propagator analysis
indicates that even with sizable phase shifts the localization of the wave
function leads to the desired healing property of the wave function since
the part of the wave function affected by the scattering event is
exponentially damped.
Also for the complete numerical propagator the same features are
observed.
Even in the presence of strong interaction processes the resulting picture
of the nuclear
medium is a tranquil one in which the dressed particles no longer remember
their scattering event beyond some finite distance
and their wave functions heal to the corresponding noninteracting ones.
This appears to be a satisfactory picture of a correlated medium in which
particles do not carry the information of their interaction indefinitely
around unlike a description of scattering using a mf Galitski-Feynman
propagator. 

\subsection{\label{sec:scatt}In-medium cross sections and phase shifts}

For various reasons the study of the interaction between nucleons in the
nuclear medium has retained a considerable amount of interest over the years.
Of particular interest is the exploration of the ``in-medium''
interaction in the context of transport-theory descriptions of
heavy-ion reactions~\cite{bdg88}.
Typical analyses simulate the dynamics of a heavy-ion reaction on the basis
of kinetic equations like the Boltzmann-Uehling-Uhlenbeck (BUU)
equation~\cite{bgp92}.
An essential ingredient in these BUU calculations is the nucleon-nucleon
cross section in the medium.
Attempts to include medium-modified cross sections in such calculations have
been described in Ref.~\cite{arbds95}.
Calculations of the cross section between nucleons in nuclear matter
have been reported in 
Refs.~\cite{hm87} - \cite{srls98}.
Some recent issues that have emerged from this work include the enhancement
of the cross section at finite temperature due to the vicinity of a pairing
instability~\cite{ars94}, the sensitivity of the cross section to the choice
of the single-particle (sp) spectrum at zero temperature~\cite{gls96}, the
density and energy dependence~\cite{ssrl97}, and temperature dependence
of the cross sections~\cite{srls98}.

All results obtained in 
Refs.~\cite{hm87} -\cite{srls98}
have been generated under the assumption that the sp
motion of the nucleon in the medium is that of a mean-field (mf) particle.
Under this assumption the scattering process in the medium takes place
between nucleons which at most have a sp spectrum different from free space
but are otherwise unaffected by the presence of other nucleons except for
the Pauli principle related to a mf Fermi gas.
This assumption has been contradicted unequivocally in recent years for
finite nuclei by the careful analysis of the (e,e$'$p) reaction.
It is therefore fair to say that both on the basis of experimental results
as well as theoretical calculations for nuclear matter, it is prudent to
consider the sp dressing of the nucleons in matter and the effect this has
on the scattering process.
One of the consequences of employing dressed particles is the localization of 
the two-body propagator in coordinate space as discussed in 
Sec.~\ref{sec:healing}, severely limiting the range of the
propagation for most energies.
This feature is due to the presence, at a given energy, of a range of 
momenta which determine the relative wave function of the propagating 
particles.
While for mf particles a unique (on-shell) momentum characterizes the
relative wave function which therefore corresponds to a plane wave (or
spherical wave), the presence of different momentum components implies that
the determination of phase shifts and cross sections requires some kind of
folding procedure over these momenta in the case of dressed particles.
In Ref.~\cite{dic98} a set of expressions have been introduced to
characterize the scattering event of dressed particles involving such a
folding procedure.
Results for phase shifts and cross sections obtained from these
expressions can therefore be compared with the corresponding results
propagating free and mf particles using the Reid soft core
potential~\cite{reid}. 

As proposed in Ref.~\cite{dic98} and followed up in
Ref.~\cite{dgrpr} this folding of the effective interaction
takes the following form for the $S$-matrix element in the case of coupled 
channels
\begin{equation}
S^{JST}_{\ell,\ell'}(\Omega) = 1 + 2i \int_0^\infty dk~k^2 ~{\rm Im}
\left\{g^{II}_f(k;\Omega)\right\} \langle k\ell|\Gamma^{JST}
(\Omega ) |k\ell'\rangle .
\label{eq:smatd}
\end{equation}
This result reduces to the conventional results
for free or mf particles.
In the case of coupled channels Eq.~(\ref{eq:smatd}) can be used to follow
the procedure to obtain phase shifts by diagonalization of the $S$-matrix.
For an uncoupled channel, the matrix element in Eq.~(\ref{eq:smatd})
directly yields the phase shift in the form
\begin{equation}
S^{JST}_{\ell,\ell}(\Omega) = 1 + 2i \int_0^\infty dk~k^2 ~{\rm Im}
\left\{g^{II}_f(k;\Omega)\right\} \langle k\ell|\Gamma^{JST} |k\ell\rangle
\equiv e^{2i\delta^{JST}_\ell} .
\label{eq:phadr}
\end{equation}
A consequence of the present
approximation is that the phase shifts $\delta^{JST}_\ell$ remain 
real~\cite{dic98}.
As a result, the phase shifts can be fruitfully compared with results
for mf or free particles.
Equation (\ref{eq:smatd}) is exact for noninteracting or mf
particles and for dressed particles includes the
physically reasonable expectation that the distribution over the momenta
as contained in the imaginary part of 
the propagator will feature in determining the scattering process.
An example of such a distribution was shown in Figs.~\ref{fig:irdr}
and \ref{fig:mfdr}.
While this approximation does not make sense at large distance scales
as discussed in the previous section, it
provides, locally, a very reasonable generalization of the phase shift. 
The corresponding ``short-distance'' approximation to the scattering amplitude
yields the following result~\cite{dic98}
\begin{eqnarray}
f^S_{m_s'm_s}(\theta,\phi) & = & 4\pi \sum_{\ell\ell'J} \sum_{mm'M} i^{\ell'} 
(-i)^\ell
Y_{\ell m_\ell}(\hat{\bm{r}})Y^*_{\ell'm_\ell'}(\hat{\bm{z}})  
(\ell m_\ell~Sm_s | JM) (\ell'm'_\ell~Sm'_s|JM)
\nonumber \\
& & \int_0^\infty dk~k ~{\rm Im}\left\{ g^{II}_f(k;\Omega) \right\}
\langle k(\ell S)J | \Gamma(\Omega) | k(\ell'S)J\rangle ,
\label{eq:scampd}
\end{eqnarray}
where a coupling to total spin $S$ and projections $m_s, m'_s$ for initial
and final spin states has been included together with the usual
decomposition in partial waves.
For the total cross section (in the neutron-proton case) one obtains
\begin{equation}
\sigma_{tot} = \pi \sum_{S\ell \ell'J} (2J+1)
\left| \int_o^\infty dk~k ~{\rm Im} \left\{ g^{II}_f(k;\Omega) \right\}
\langle k(\ell S)J | \Gamma(\Omega) | k(\ell'S)J\rangle \right|^2 .
\label{eq:tcrd}
\end{equation}
Eq.~(\ref{eq:tcrd}) demonstrates that a sensible cross section will be
obtained in the case of dressed particles at all energies for which a
nonvanishing imaginary part of the propagator exists.
For two particles deep in the Fermi sea, for example, Eq.~(\ref{eq:tcrd})
avoids the divergence associated with the $k_0^{-2}$ term in the corresponding
expression for free or mf particles.
The formulation of the cross section in terms of Eq.~(\ref{eq:tcrd})
provides a reasonable way to
assess the strength of the interaction between dressed particles in the
medium in terms of the square of the relevant transition matrix element
($\Gamma$) multiplied by an appropriate measure of the density of states
represented by the imaginary part of the noninteracting propagator.

\begin{figure}[ptb]
\begin{center}
\includegraphics[angle=0,width=0.8\textwidth]{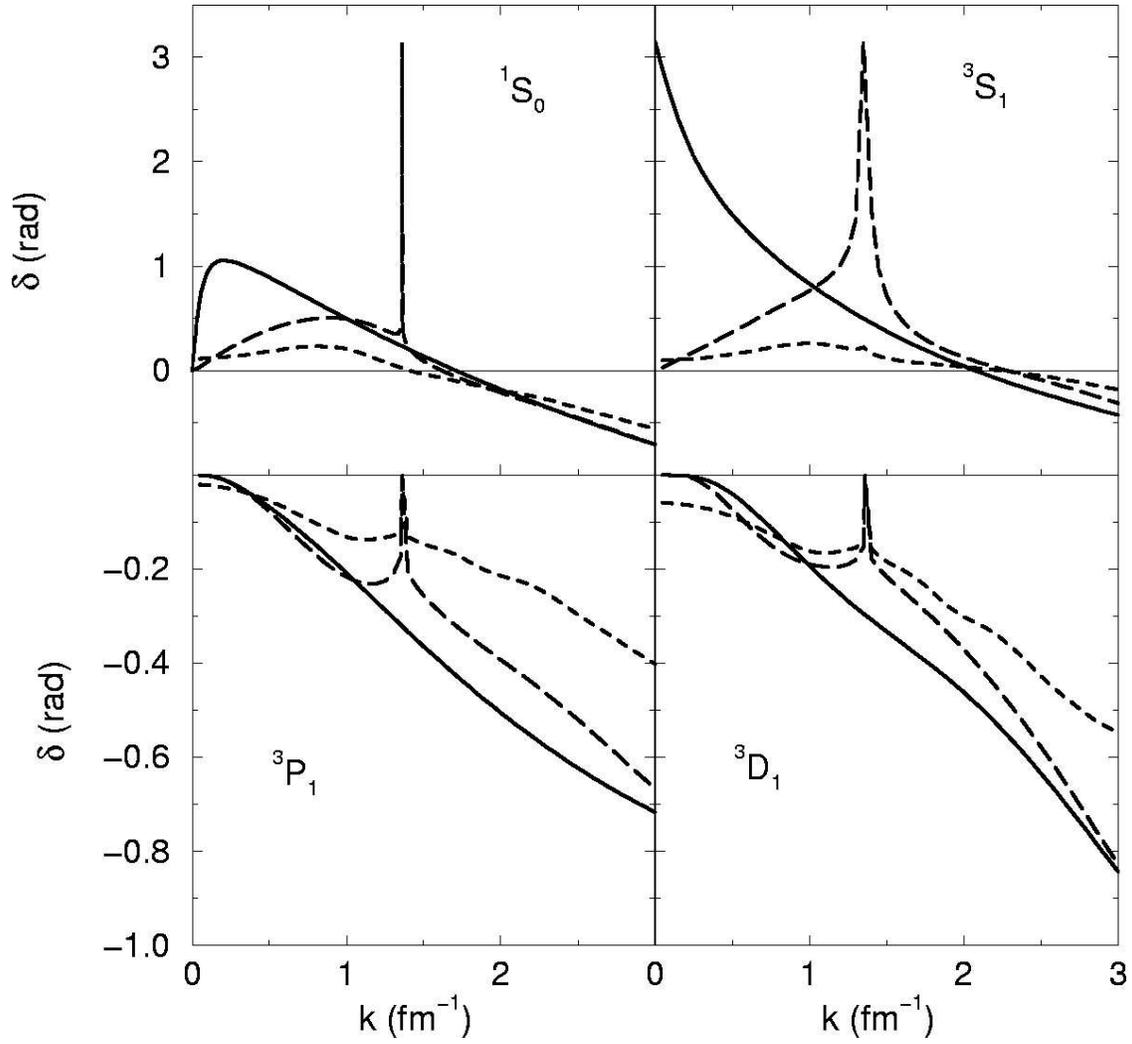}
\begin{minipage}[t]{16.5 cm}
\caption{Comparison of phase shifts for free particles (solid),
mf particles (dashed), and dressed particles (short-dashed lines)
for different partial waves. The density of the medium corresponds to
$k_F = 1.36~{\rm fm}^{-1}$.
\label{fig:drcompmf} }
\end{minipage}
\end{center}
\end{figure}

The main results for the phase shifts for some of the more important partial
wave channels are summarized in Fig.~\ref{fig:drcompmf}. A comparison
is made between phase shifts for free particles (solid line), mf particles at
$k_F = 1.36~{\rm fm}^{-1}$ (dashed line), and dressed particles
(short-dashed line) at the same
density for the ${}^1S_0$, ${}^3S_1$, ${}^3P_1$, and ${}^3D_1$ channels
(corresponding to the different panels in Fig.~\ref{fig:drcompmf})
as a function of the on-shell momentum.
In general one finds that the dressed phase shifts suggest weaker
interactions since in essentially all cases they are either less repulsive
or less attractive than the mf result.
By studying the individual contributions of the real and imaginary part of
the effective interaction one may gain more insight into this issue (see
below).
For the two $S$-wave channels the most striking feature of the dressed phase
shift is the disappearance of the pairing signature for the ${}^1S_0$
channel and the enormous reduction of the signal in the ${}^3S_1$ case.
While the dressed ${}^1S_0$ phase shift is essentially zero at $k_F$,
it is still clearly attractive at this momentum for the ${}^3S_1$ channel.
The actual calculation of the phase shift
for this channel displays a slight kink
close to $k_F$ suggesting that the phase shift may actually rise very
rapidly to $\pi$ very close to $k_F$.
This implies a tremendous reduction in the strength of the pairing
correlations in this coupled channel as compared to a mf treatment.
Gaps of the order of 10 MeV have been obtained for this channel in
Refs.~\cite{vond4,ars90,bbl92}.
Clearly, the dressing of the nucleons has a strong influence on pairing.
While one would expect to obtain
a gap using dressed nucleons based on the attractive effective interaction at
the Fermi surface, its magnitude is drastically reduced as
suggested by the phase shift calculation shown in Fig.~\ref{fig:drcompmf}.
This latter feature has recently been confirmed in Ref.~\cite{bozek0}.
The main ingredient in this reduction is the decrease in the 
density of states at $2\epsilon_F$ when dressed nucleons are propagated.
This reduction is
essentially the square of the strength of the quasiparticle pole at $k_F$
leading to a reduction factor of about 0.5.
Since pairing correlations are particularly sensitive to this density of
states, it is not surprising that the strength of the pairing is
substantially diminished when dressing is taken into account.
It is also noteworthy that one observes a smaller negative phase shift for
both $S$-waves at higher energy as compared to the mf result.
A similar conclusion may be drawn by inspecting the phase shifts for the
${}^3P_1$ and ${}^3D_1$ channel in the bottom panels of
Fig.~\ref{fig:drcompmf}. Also for these partial waves which represent
repulsive effective interactions, one observes a reduction of the magnitude
of the phase shift when dressing is considered.
It is also important to note that in the case of mf propagation the results
in Fig.~\ref{fig:drcompmf} show that the corresponding results tend to those
of free particles at high energy, whereas this is not the case for dressed
particles.
This latter result indicates that the effect of the dressing extends to a
large energy domain.
This observation is not too surprising since the spreading of the sp
strength due to short-range and tensor correlations takes place in a very
large energy domain~\cite{vond1,bff92,vond2}
and is quite different from a local (in energy)
spreading of the strength as would be obtained by a complex quasiparticle
energy.
\begin{figure}[ptb]
\begin{center}
\includegraphics[angle=0,width=0.4\textwidth]{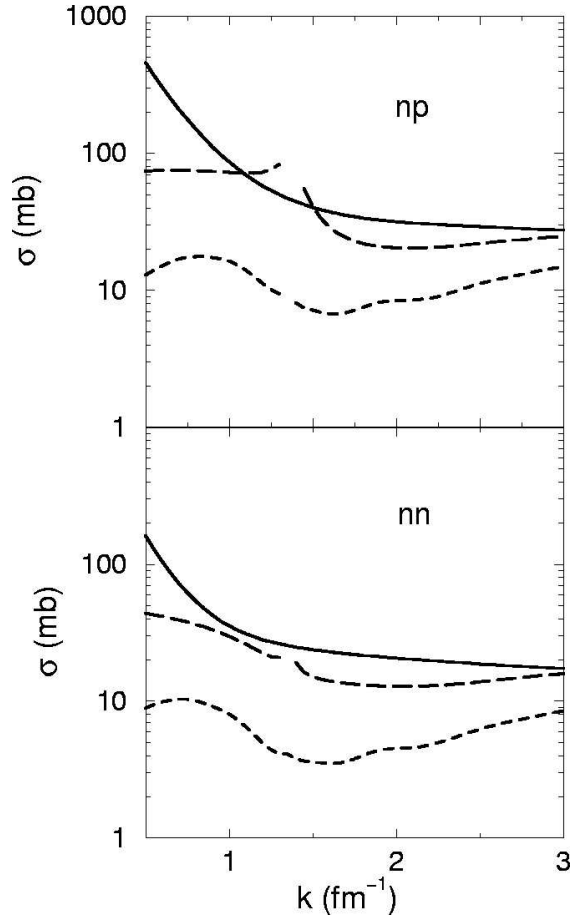}
\begin{minipage}[t]{16.5 cm}
\caption{Total neutron-proton (top) and neutron-neutron (bottom panel)
cross sections for free (solid), mf (dashed), and dressed particles 
(short-dashed line) as a function of the on-shell momentum.
\label{fig:tcross} }
\end{minipage}
\end{center}
\end{figure}
The results for the phase shifts obtained for dressed particles lead to the
expectation that the corresponding total cross sections are substantially
reduced compared to the mf results.
The results for neutron-proton (np)
and neutron-neutron (nn) total cross
sections displayed in Fig.~\ref{fig:tcross} confirm this expectation.
Results have been obtained for free (solid), mf (dashed), and dressed
particles (short-dashed line) by including all partial wave channels of the 
Reid potential with $J\leq 2$.

Results for mf particles were generated with a realistic sp energy spectrum
and are similar to the corresponding results obtained {\em e.g.} in
Ref.~\cite{gls96}.
The effect of the pairing correlations on the cross sections yields a
cusp-like behavior around $k_F$ reminiscent of the enhancement of the cross
sections obtained by the Rostock group at finite
temperature~\cite{arbds95,ars94}.
As the phase shifts for mf particles suggest, the corresponding cross sections
in the medium become essentially identical to the one in free space at high
energy.
Both for the np and nn total cross sections the effect of dressing the
nucleons is quite dramatic leading to a substantial reduction of the total
cross section at all energies.
Indeed, on average a cross section of only about 10 mb is obtained.
While this may seem a small number, it should be kept in mind that this by
no means implies that the effective interaction in the medium has become
insignificant~\cite{dgrpr}.
In addition, one should recall that the concept of asymptotic flux in the
medium representing preserved information of a scattering event deep in the
medium is not a realistic consideration when the dressing of the nucleons is
significant~\cite{dic98}, as it is at $k_F = 1.36~{\rm fm}^{-1}$.
The main ingredient representing the dressing is the two-particle
density of states, its reduction for dressed particles
is to a large extent responsible for the reduction of the cross section.
In this respect it should be noted that the results of Ref.~\cite{ars94}
also show a substantial reduction of the total cross section at high energy
with increasing temperature.
This result implies that with increasing temperature which also means a
larger depletion of the Fermi sea due to thermal excitations, one obtains
a reduction of the total cross section similar to the one obtained here at
zero temperature when the dressing of the particles is incorporated
(and therefore the depletion of the Fermi sea due to correlations is included).
While no results are shown in Fig.~\ref{fig:tcross} below 0.5 
${\rm fm}^{-1}$ in order to avoid the large value of the total cross
sections for free particles, it should be noted that the cross sections for
dressed particles smoothly go to zero when expression (\ref{eq:tcrd})
is used at lower energies.
This expression avoids the problem associated with the expression for
free or mf particles which
would yield an infinite cross section for the on-shell momentum going to zero.
In addition, the latter expression does not yield a cross section for energies
that do not yield a solution for the on-shell momentum, 
{\em i.e.} for energies deep in the hole-hole continuum.

\begin{figure}[tbp]
\begin{center}
\includegraphics[angle=-90,width=0.5\textwidth]{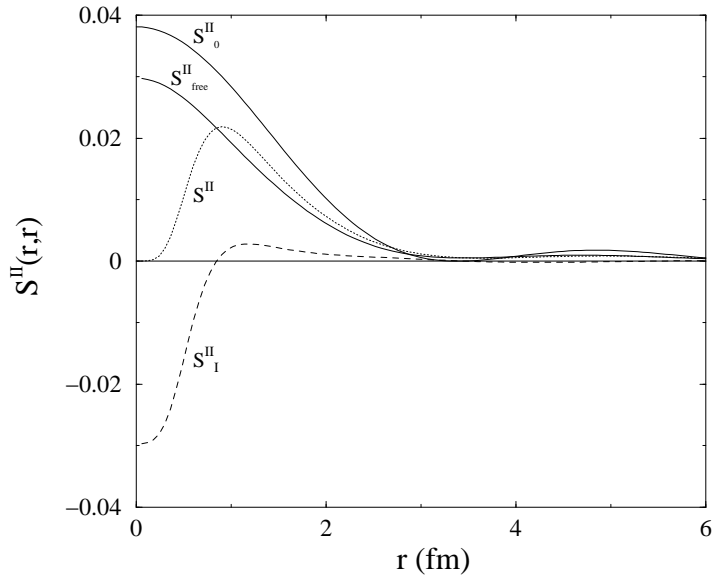}
\begin{minipage}[t]{16.5 cm}
\caption{Two-nucleon spectral function for the ${}^1S_0$ channel in nuclear
matter for zero total momentum and a removal energy corresponding to $- 100$
MeV.
\label{fig:SIIrr} }
\end{minipage}
\end{center}
\end{figure}

\begin{figure}[ptb]
\begin{center}
\includegraphics[angle=-90,width=0.5\textwidth]{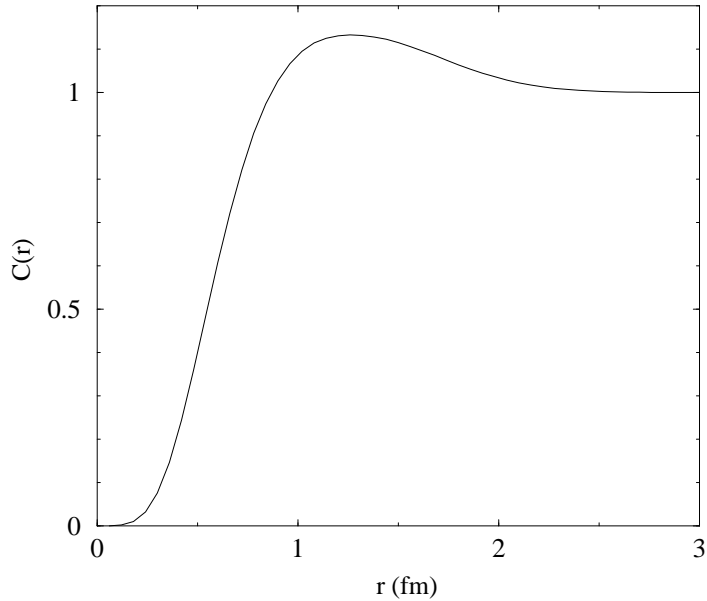}
\begin{minipage}[t]{16.5 cm}
\caption{Correlation function for $^1S_0$ nucleons with total momentum zero.
\label{fig:corrfunc} }
\end{minipage}
\end{center}
\end{figure}

A related quantity to the two-body interaction is the two-body propagator
given by Eq.~(\ref{eq:prop}) in momentum space and Eq.~(\ref{eq:propr}) in
coordinate space.
Since a Lehmann representation exists for this two-particle propagator,
it is possible to define a spectral density for this quantity either
in momentum or coordinate space.
We will consider here
\begin{equation}
S^{II}_{JST}(r\ell,r'\ell';\Omega) = - \frac{1}{\pi} 
g^{II}_{JST}(r\ell,r'\ell';\Omega) 
\label{eq:specII}
\end{equation}
which represents the removal or addition amplitude for a pair of particle
(for zero total momentum) depending on whether the energy $\Omega$ is
below or above $2\varepsilon_F$, respectively.
In Fig.~\ref{fig:SIIrr} this quantity is plotted for $r = r'$ in the case of
the ${}^1S_0$ partial wave for an energy corresponding to $- 100$ MeV.
This diagonal form of Eq.~(\ref{eq:specII}) corresponds to the
two-particle spectral function in this chosen representation.
The curves in this figure correspond to the full propagator (labeled $S^{II}$),
the contribution of the first term from Eq.~(\ref{eq:prop}) representing
two dressed but noninteracting nucleons ($S^{II}_{free}$), and the contribution
of the second term of Eq.~(\ref{eq:prop}) representing the vertex function
$\Gamma$ ($S^{II}_I$).
In addition, the result for two mf particles is shown by the curve labeled
$S^{II}_0$. 
Notice that $S^{II}_{free}$ has a large positive value at the origin
and $S^{II}_{I}$ has a large negative value at the origin.  There is a
precise cancellation which is illustrated in Fig.~\ref{fig:SIIrr}.
This cancellation is of course associated with the effect of the short-range
core of the Reid interaction which suppresses the relative wave function
at these disctances.
Integrating Eq.~(\ref{eq:specII}) over the allowed two-particle removal
energies yields the corresponding contribution to the two-particle
density matrix
\begin{equation}
\rho^{II}(r\ell,r'\ell';JST) = \int_{-\infty}^{2\epsilon_F}d\Omega\
 S^{II}_{JST}(r\ell,r'\ell';\Omega) 
\label{eq:rhoII}
\end{equation}
This quantity also characterizes the behavior at short-range for a chosen
partial wave and can be compared to the corresponding quantity using
dressed but noninteracting propagators.
It is attractive to cast this comparison in the form of the ratio
\begin{equation}
\label{eq:corrfunc}
C(r) = \frac{\rho^{II}(r\ell,r\ell;JST)}
{\rho^{II}_{free}(r\ell,r\ell;JST)} .
\end{equation}
In Fig.~\ref{fig:corrfunc} we plot the correlation function calculated for
$^1S_0$ nucleons with center of mass momentum zero.
Note that this correlation function shows the characteristic depression at
$r=0$ and an enhancement somewhat above $r=1$ fm.
This correlation function has been employed in the analysis of two-proton
removal experiments as discussed in Sec.~\ref{sec:ee1-2N}.
In Ref.~\cite{bffila} a related study has been reported for the probability
for the removal of pairs with the actual properties of the deuteron
yielding information about the deuteron distribution in nuclear matter.

\subsection{\label{sec:selfnmr}Results for self-consistent 
Green's functions in Nuclear Matter}

The solution to the scheme outlined in Sec.~\ref{sec:formalismnm} has 
been studied by different groups. Aspects
of self-consistency have been treated in 
Refs.~\cite{djl} - \cite{rd}.
Complete self-consistency has been implemented for a realistic interaction
in terms of a discrete representation of the propagator in Ref.~\cite{dwvnw2}.
A representation of the spectral functions on a momentum and energy grid
was made self-consistent in Re.~\cite{bozek1} for a separable version
of the Paris potential.
Both of these strategies work better when softer interactions are used.
Indeed, it appears that yet another method is required to deal with
interactions like the Reid potential~\cite{reid}.
The main reasons for potential problems is
the need to consider a very large energy domain since the peaks in the
imaginary part of the self-energy and effective interaction occur around
10 GeV~\cite{vond1,vond2} for a potential like Reid's.
This feature puts stringent demands on the sizes of grids
to describe the spectral strength as has been found in Ref.~\cite{bozek1}.

The essential difficulty in the self-consistency procedure is 
therefore the handling
of the information which describes the complete energy and momentum
dependence of the spectral functions over the relevant domains.
For interactions like the Reid potential the energy and momentum
range needed to store the relevant information precludes a
straightforward numerical discretization of the spectral functions.
For this reason earlier attempts by the St. Louis
 group have involved a parametrization
of the spectral functions themselves~\cite{christh,gdpr}.
Although manageable for the calculation of the effective interaction
at zero total momentum~\cite{dgrpr}, it is quite difficult to manage
the strong energy dependence of these functions in a complete self-consistency
scheme.
In particular it is necessary to represent the spectral strength by at least
two functions, one of which maybe sharply peaked.
This leads to some difficulties in generating the real parts of the
propagators from dispersion relations.
As a result, the self-energy that can be obtained from the propagator by
inverting the Dyson equation does not behave sufficiently smoothly
at all energies for a given momentum.

We therefore outline here another method that is able to deal with
such interactions as the Reid potential~\cite{libth,rd} implying that it will 
certainly work for softer interactions.
First, it is noted that the sp spectral function is completely determined
by the $\textrm{Im }\Sigma$.
The spectral function can be written in terms of the self-energy as
in Eqs.~(\ref{eq:specp}) and (\ref{eq:spech}).
The self-energy can be written according to Eq.~(\ref{eq:selfdis})
which demonstrates that 
the spectral function can be completely expressed in terms of
$\textrm{Im }\Sigma (k,\omega )$,
$\textrm{Re }\Delta\Sigma_\downarrow (k,\omega )$,
$\textrm{Re }\Delta\Sigma_\uparrow (k,\omega )$,
and $\Sigma_V(k)$.
The contributions 
$\textrm{Re }\Delta\Sigma_\downarrow (k,\omega )$ and
$\textrm{Re }\Delta\Sigma_\uparrow (k,\omega )$
can be obtained from
the $\textrm{Im }\Sigma (k,\omega )$ by performing the dispersion
integrals given in Eq.~(\ref{eq:selfdisa}).
The correlated Hartree-Fock contribution, 
$\Sigma_V(k)$, can be determined in a given iteration step as follows.
The quasi-particle (qp) spectrum usually identifies the location of the
peak of the spectral function and is determined by Eq.~(\ref{eq:Eqp}).
The momentum distribution given by
Eq.~(\ref{eq:momdis}) from a previous iteration, say the $(i-1)$th one,
can be used to determine the correlated Hartree-Fock self-energy from
Eq.~(\ref{eq:sighf}).
This allows the construction of a new Fermi energy and allows the calculation
of the $i$th iteration for the spectral functions closing the self-consistency
loop.
This analysis indicates that it is sufficient to accurately
represent the imaginary part of the self-energy for a complete
determination of the spectral functions.
Since this imaginary part is very well-behaved as a function of energy
and changes also smoothly as a function of energy, a strategy has been
developed to represent $\textrm{Im }\Sigma$
in terms of a limited set of gaussians~\cite{libth,rd}.
Two gaussians below the Fermi energy and three above appear to provide
sufficient flexibility for a complete representation of the imaginary
part of the self-energy.

\begin{figure}[ptb]
\begin{center}
\includegraphics[angle=-90,width=0.5\textwidth]{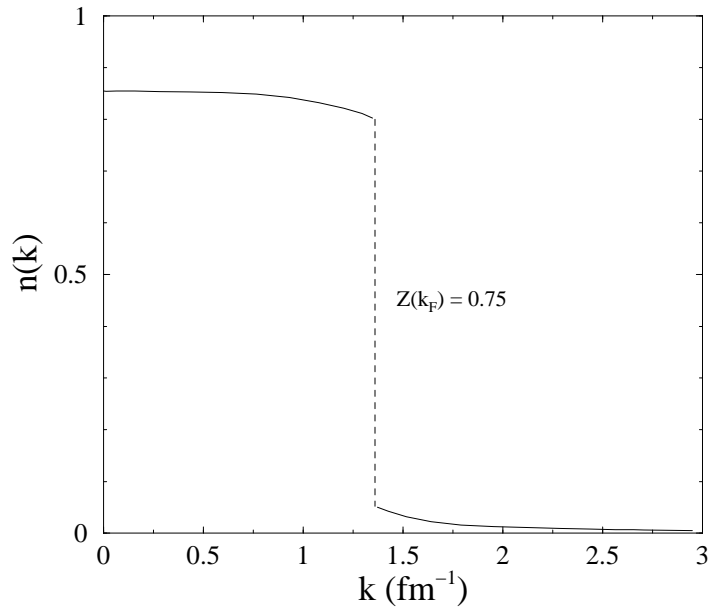}
\begin{minipage}[t]{16.5 cm}
\caption{Momentum distribution in nuclear matter obtained for self-consistent
Green's functions using the Reid potential.
\label{fig:nkself} }
\end{minipage}
\end{center}
\end{figure}

Some results of this type of self-consistent calculation will now be presented
here.
We start by discussing the results for the momentum 
distribution at a density corresponding to $k_F = 1.36~\textrm{fm}^{-1}$.
The corresponding result is shown in Fig.~\ref{fig:nkself}.
This figure should be compared to the first generation results shown in
Fig.~\ref{fig:nkreidb} for a calculation that can be considered as
a first iteration step in this self-consistency procedure since it is
based on mf propagators as input for the spectral functions.
One notices immediately the similarity of the occupation of momenta below
$k_F$. Indeed, only a slight increase in the occupation of at most 3\%
is observed.
This implies that the corresponding depletion due to short-range correlations
is still about 15\% 
for nucleons deep in the Fermi sea.
The slightly increased occupation below $k_F$ results in a slight reduction
of momentum states above $k_F$.
The occupation of momenta just above $k_F$, however, hardly changes.
This is further corroborated by
the result for the quasiparticle pole at the Fermi momentum since it changes
from 0.72 for the first-generation result to 0.75 in the self-consistent case.

\begin{figure}[ptb]
\begin{center}
\includegraphics[angle=-90,width=0.5\textwidth]{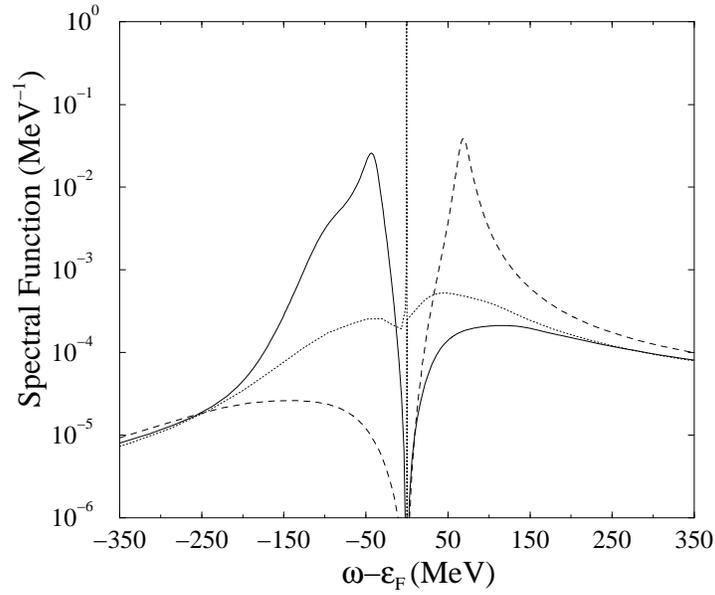}
\begin{minipage}[t]{16.5 cm}
\caption{Self-consistent spectral functions for three different momenta
at $k_F = 1.36~\textrm{fm}^{-1}$ corresponding to 0 (full), 1.36 (dotted),
and 2.1 $\textrm{fm}^{-1}$ (dashed) as a function of $\omega - \varepsilon_F$.
\label{fig:selfsf} }
\end{minipage}
\end{center}
\end{figure}

\begin{figure}[ptb]
\begin{center}
\includegraphics[angle=-90,width=0.5\textwidth]{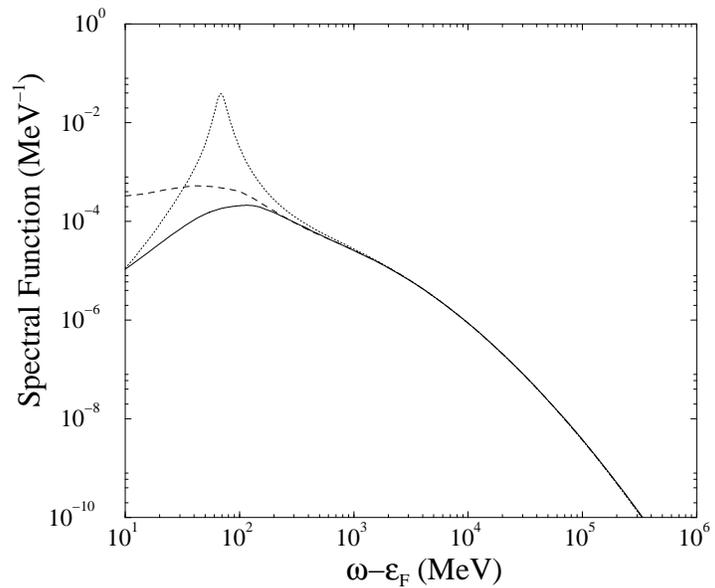}
\begin{minipage}[t]{16.5 cm}
\caption{Self-consistent spectral functions for three different momenta
at $k_F = 1.36~ \textrm{fm}^{-1}$ corresponding to 0 (full), 1.36 (dashed),
and 2.1 $\textrm{fm}^{-1}$ (dotted) but plotted on a logarithmic energy scale
as a function of $\omega-\varepsilon_F$.
\label{fig:selfsfa} }
\end{minipage}
\end{center}
\end{figure}

Some results for spectral functions are shown in Fig.~\ref{fig:selfsf}
around the Fermi energy as a function of $\omega - \varepsilon_F$.
A comparison with the results reviewed in Sec.~\ref{sec:brian} yields a
striking difference for the strength distribution below the Fermi energy.
The spectral strength in Fig.~\ref{fig:selfsf}
is essentially identical at large negative energies for all 
relevant momenta.
This feature reflects an essentially similar energy distribution of 
the imaginary part of the self-energy for these momenta.
This could have been expected since the constraint imposed by mf sp energies
no longer applies and all momenta have similar energy domains associated
with the imaginary part of the self-energy.
In the latter case, this imaginary part of the self-energy has a
fixed lower bound depending on the momentum as discussed in 
Sec.~\ref{sec:brian} and shown in Fig.~\ref{fig:imselfb}.
The appearance of sp strength at large negative energies has important 
implications for the binding energy of nuclear matter as will be discussed
shortly.
The results for spectral functions at high energy
are shown in Fig.~\ref{fig:selfsfa}. 
One continues to encounter a common tail
of the sp strength at high energies above $\varepsilon_F$ for
all momenta which is quantitatively similar to the one shown
in Fig.~\ref{fig:specpmf} for the first-generation result.

\begin{figure}[ptb]
\begin{center}
\includegraphics[angle=-90,width=0.6\textwidth]{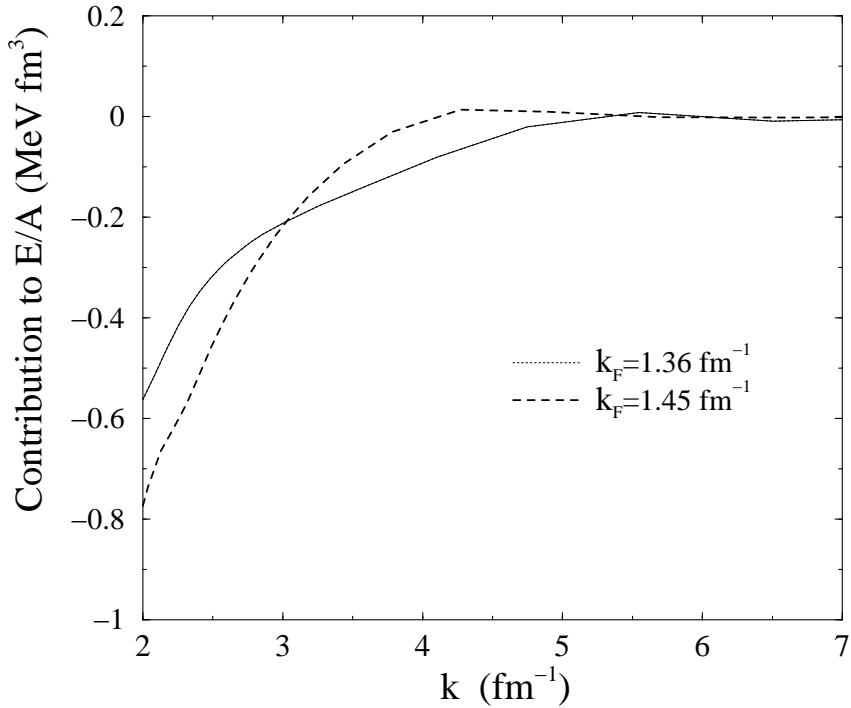}
\begin{minipage}[t]{16.5 cm}
\caption{The high-momentum contribution to the energy per particle for
$k_F =$ 1.36 ${\rm fm}^{-1}$ (solid) and 
1.45 ${\rm fm}^{-1}$ (dashed).
This result illustrates the source of the saturation process when 
short-range correlations are considered self-consistently.
\label{fig:Image_2}}
\end{minipage}
\end{center}
\end{figure}

The special role of short-range correlations in obtaining saturating
behavior of nuclear matter is illustrated in Fig.~\ref{fig:Image_2}.
In this figure we plot the integrand corresponding to both terms
in Eq.~(\ref{eq:epp}) as a function
of momentum after performing the energy integral over the spectral function
for densities corresponding to  $k_F =$ 1.36 ${\rm fm}^{-1}$ and 
$k_F =$ 1.45 ${\rm fm}^{-1}$, respectively.
At $k_F =$ 1.36 ${\rm fm}^{-1}$ the 
high-momentum components still provide attractive contributions whereas
for $k_F =$ 1.45 ${\rm fm}^{-1}$ a changeover occurs suggesting
that at an even higher density these high-momentum terms will provide
only repulsion.
From this analysis it is clear that the expected relevance of SRC in obtaining
reasonable saturation properties of nuclear matter is fully confirmed.
It remains to relate this observation to
the vast body of work on the nuclear-matter saturation
problem.

\subsection{\label{sec:satprop}Saturation of nuclear matter from
short-range correlations}

A correct description of the saturation properties of nuclear matter
has remained an unresolved issue for a very long time.
The Brueckner-Bethe-Goldstone (BBG) expansion~\cite{day67} supplies a converged
result for the energy per particle in the relevant density range,
for a given realistic interaction, at the level of three hole-line
contributions~\cite{day2,bald}.
Such calculations fail to reproduce the empirical saturation properties
which require a minimum in the equation of state at a density corresponding
to a Fermi momentum, $k_F$, of about 1.33 fm$^{-1}$ with a binding energy
of about 16 MeV.
The authors of Ref.~\cite{bald} obtain for the Argonne $v_{14}$ 
interaction~\cite{wsa84} a saturation density corresponding
to 1.565 fm$^{-1}$ with about the correct amount of binding.
This corresponds to an overestimation of the empirical density by about
60\% 
but appears completely consistent with corresponding variational 
calculations~\cite{dayw} for the same interaction.

Several different remedies for this serious problem have been proposed
over the years. The intrinsic structure of the nucleon and its related
strong coupling to the $\Delta$-isobar inevitably requires the consideration
of three-body (or more-body) forces. When three-body forces are considered
in variational calculations it is possible to achieve better saturation
properties only when an adhoc repulsive short-range component of this
three-body force is added~\cite{cpw,wff}.
It has also been suggested that a relativistic treatment of the nucleon
in the medium using a Dirac-Brueckner approach provides the necessary
ingredients for a better description of saturation~\cite{acps} - \cite{amtj}. 

All many-body methods developed for nuclear matter have focused on a
proper treatment of short-range correlations (SRC) without
the benefit of experimental information on the influence of these
correlations on the properties of the nucleon in the medium.
This influence can now be clearly identified by considering recent results
from (e,e$'$p) reactions~\cite{deWitt1} - \cite{lap93} and theoretical 
calculations
of the nucleon spectral function in nuclear matter~\cite{rpd1} - 
\cite{bff92,vond1}.

A recent analysis of the \eep\ reaction on ${}^{208}$Pb
in a wide range of missing energies and for missing momenta below 270 MeV/c
yields information on the occupation numbers of all the
deeply-bound proton orbitals (see also Sec.~\ref{sec:Pb208deep}).
These data indicate that all these orbitals are depleted by the same
amount of about 15\%~\cite{lap2,bat}.
These occupation numbers are associated with
the orbits which yield an accurate fit to the \eep\ cross section.
The properties of these occupation numbers suggest that the main effect of
the global depletion of these mean-field orbitals is due to SRC.
Indeed, the effect of the coupling
of hole states to low-lying collective excitations only affects occupation
numbers of states in the immediate vicinity of the Fermi energy~\cite{rijs}.
In addition, nuclear matter momentum distributions display such
an overall global depletion due to short-range and tensor
correlations~\cite{vond1,vond2,fantp} as discussed in Sec.~\ref{sec:brian}
and confirmed in Sec.~\ref{sec:selfnmr} for the fully self-consistent results.
The latter results formed the basis of the now corroborated 
prediction~\cite{dimu1,papa,benf90}
for the occupation numbers in ${}^{208}$Pb~\cite{lap2}.

Most of this depleted sp strength is located at 
energies more than 100 MeV
above the Fermi energy~\cite{vond1,vond2,dimu1}.
This appearance of strength at high energy is another important aspect of
the influence of short-range and tensor correlations.
Yet another characteristic feature of these SRC is that
this depletion of the sp strength
must be compensated by the admixture of a corresponding number
of particles with high-momentum components.
These high-momentum components have not yet been unambiguously identified
but are currently studied experimentally~\cite{rohe}
as discussed in Sec.~\ref{sec:Pb208deep}.
Solid theoretical arguments~\cite{ciofi} and calculations clearly
pinpoint this strength at high excitation energy in the hole 
spectrum both for nuclear matter~\cite{vond1} and finite
nuclei~\cite{mpd1} (see Sec.~\ref{sec:finiteSRC2}).
Indeed, experiment confirms that no substantial admixture of these 
high-momentum
components is observed in the vicinity of the Fermi energy~\cite{bobel}
as discussed in Sec.~\ref{sec:finiteSRC}.

We now present an argument showing that short-range correlations
are the dominant
factor in determining the empirical saturation density of nuclear matter.
We recall that elastic electron scattering from
${}^{208}$Pb~\cite{froi} accurately determines
the value of the central charge density in this nucleus.
By multiplying this number by $A/Z$ one obtains the relevant central density
of heavy nuclei, corresponding to 0.16 nucleons/fm${}^3$ or $k_F = 1.33$
fm${}^{-1}$.
Since the presence of nucleons at the center of a heavy nucleus is confined
to $s$-wave nucleons, and, as discussed above, their depletion is
dominated by SRC, one may therefore conclude that the same is true for 
the actual value of the empirical saturation density of nuclear matter.
While this argument is particularly appropriate for the deeply bound
$1s_{1/2}$ and $2s_{1/2}$ protons, it continues to hold to
a large extent for the $3s_{1/2}$
protons which are depleted predominantly by short-range effects (up to 15\%
)
and by at most 10\% 
due to long-range correlations~\cite{deWitt3,grabx}.
These considerations demonstrate clearly that one may expect
SRC to have a decisive influence on the actual value
of the nuclear-matter saturation density.

High-momentum components due to SRC also have a 
considerable impact on the binding energy of nuclear matter.
This result can be inferred from the energy sum rule given in 
Eq.~(\ref{eq:epp}).
Eq.~(\ref{eq:epp}) illustrates the link between the energy of the system and
the hole spectral function, $S_h(k,\omega)$ illustrated in 
Fig.~\ref{fig:Image_2}.
Results for the momentum distribution and true potential energy
based on the spectral function show that enhancements
as large as 200\% 
for the kinetic and 
potential energy over the mean-field values can be obtained
for both nuclear matter~\cite{vond2} and finite nuclei~\cite{mpd1} (see
Sec.~\ref{sec:finiteSRC2}).
These large attractive contributions to the potential energy
of nuclear matter are mainly from weighting the 
high-momentum components in the spectral function with 
large negative energies in Eq.~(\ref{eq:epp}).
The location of these high-momentum components as a function of energy
is therefore an important ingredient in the determination of the energy
per particle as a function of density.
So far, the determination of this location has relied only on
quasiparticle properties in the construction of the self-energy.
A self-consistent determination of the spectral function
including the location of these high-momentum 
components therefore includes the dominant physics of SRC in 
the description of nuclear matter and is consistent with the
experimental observations of the nucleon spectral function in nuclei.

Such a determination
requires the solution of the ladder equation for the effective interaction
in the medium as discussed above.
The implementation of this self-consistency scheme is numerically
quite involved and has been attempted by several groups~\cite{djl} - \cite{rd}
as discussed above.
As reported in Ref.~\cite{ddsvw03} 
two different approaches have been used to generate
results for different interactions.
In the continuous scheme, 
a representation of the imaginary part of the self-energy
in terms of four gaussians is used to completely describe the imaginary
part of the self-energy and thereby the sp
propagator as discussed in the previous section. 
The parameters of these gaussians are then determined
self-consistently ~\cite{rd} for the Reid potential~\cite{reid} by solving
Eq.~(\ref{eq:ladpw}) with the convolution of spectral functions
in Eq.~(\ref{eq:g2fdis}) as input, constructing the self-energy,
and then solving the Dyson equation (\ref{eq:dysonx}).
In the discrete scheme a representation of the propagator
in terms of three discrete poles~\cite{dwvnw1,dwvnw2} is used
avoiding a full continuum solution of Eq.~(\ref{eq:ladpw}).
The latter approach is equivalent to a continuous version as far as the 
energy per particle is concerned, since it 
requires  a reproduction of the relevant energy-weighted moments of the 
hole and particle spectral function~\cite{dwvnw1,dwvnw2}.
This is substantiated by comparing the results of this discrete scheme with 
the results of the continuous self-consistency scheme used in 
Ref.~\cite{bozek0} for the Mongan-type separable interaction \cite{mon} 
and in Ref.~\cite{bozek2} for the separable Paris interaction \cite{hp1,hp2}.
It is found that the binding energies correspond to 
within 5\%  
over the relevant $k_F$ range around the minimum, and moreover 
that the location of the minimum agrees to within 3\%.

\begin{figure}[ptb]
\begin{center}
\vspace*{-1.0cm}       
\includegraphics[width=0.5\textwidth,height=0.5\textwidth] {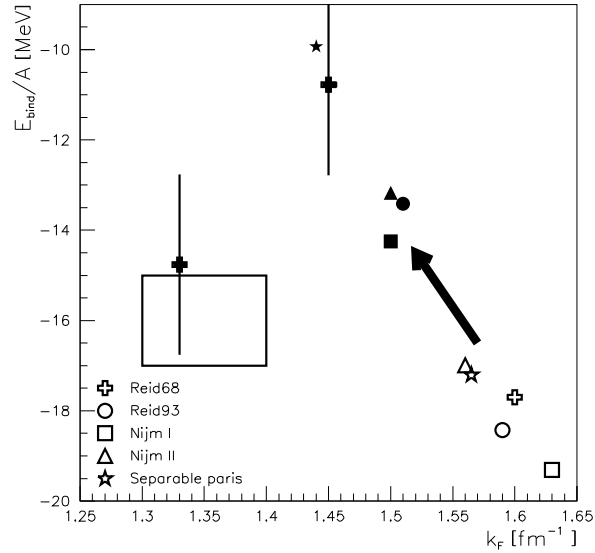}
\vspace*{-.8cm}       
\begin{minipage}[t]{16.5 cm}
\caption{\label{fig:coester} Nuclear matter
saturation points calculated with various realistic
NN interactions. The open symbols refer to continuous choice 
Brueckner-Hartree-Fock results. The filled symbols refer to
self-consistent results and represent saturation points calculated in the 
discrete scheme, except for the old Reid (Reid68)
interaction where the binding energy at two densities is shown in the 
continuous scheme. 
 }
\end{minipage}
\end{center}
\end{figure}

In Fig.~\ref{fig:coester} the saturation points obtained within the 
discrete scheme of Ref.~\cite{dwvnw1,dwvnw2}
for the updated Reid potential (Reid93), the NijmI
and NijmII interaction ~\cite{reid93} and the separable Paris interaction
\cite{hp1,hp2} are shown. 
The results demonstrate 
an important and systematic change of the saturation properties with respect 
to continuous choice Brueckner-Hartree-Fock (ccBHF) calculations, 
leading to about 4-6 MeV less binding, 
and reduced values of the saturation density, closer to the empirical one.   
Such a trend is entirely consistent with the observations in 
Ref.~\cite{bozek2} made for a separable NN interaction, and is now extended 
to more realistic (non-separable) interactions.   

As discussed in the previous section, the discrete scheme~\cite{dwvnw1,dwvnw2}
could not be used with the original 
Reid (Reid68) potential because of its slow decay in momentum space,  
but some results are available in the continuous scheme of  
Ref.~\cite{libth,rd}. In Fig.~\ref{fig:coester} the binding energy is shown 
at two densities ($k_F = 1.33$ and 1.45 fm$^{-1}$); 
the error bars are an estimate of the remaining uncertainty due to incomplete 
convergence and the non-selfconsistent treatment of some higher
order partial waves~\cite{libth,rd}. 
The results again seem to indicate a substantial shift in the 
saturation density for the Reid68 potential, from the ccBHF value of about 
1.6 fm$^{-1}$, to a value below 1.45 fm$^{-1}$, without seriously 
underbinding nuclear matter.

The present self-consistent treatment of SRC (scSRC) differs in two main 
aspects from the ccBHF approach,
the latter giving saturation results essentially equivalent to converged
three hole-line calculations~\cite{bald}. 
Firstly, hole and particle lines are treated 
on an equal footing, thereby ensuring thermodynamic consistency~\cite{bozek0}. 
Intermediate hole-hole propagation in the ladder diagrams is included to all 
orders. This feature provides, compared to ccBHF, a substantial repulsive 
effect in the $k<k_F$ contribution to Eq.~(\ref{eq:epp}), and 
 comes primarily from an upward shift of the quasi-particle 
energy spectrum as a result of including $\omega < \epsilon_F$ 
contributions to the imaginary part of the self-energy.
The effect increases with density, and is the dominant factor in the 
observed shift of the saturation point. 
Secondly, the realistic spectral functions, generated from
the self-consistent procedure outlined in Sec.~\ref{sec:formalismnm} 
 and used in the evaluation of the
in-medium interaction $\Gamma$ and self-energy $\Sigma$, are in agreement 
with experimental information obtained from \eep\ reactions.  
For the Reid93 interaction at $k_F=1.37$ fm$^{-1}$ we find $z=0.74$ for 
the quasiparticle strength at the Fermi momentum, whereas the hole 
strength for $k=0$, integrated up to 100 MeV missing energy, equals 83\%;
similar values are found for the other interactions
and confirmed by the results discussed in the previous section for
the Reid potential.  
The depletion of the quasiparticle peaks is primarily important to 
suppress unrealistically large pairing instabilities around normal density.
The improved treatment of the high-momentum components does 
affect the binding energy, through the $k>k_F$ contribution to 
Eq.~(\ref{eq:epp}).  
This feature, studied in \cite{dwvnw1,dwvnw2}, provides a sizeable 
attraction, but is smaller than the afore-mentioned repulsive effect.  

The inclusion of $hh$-propagation in scSRC also leads 
to a somewhat stiffer equation-of-state than in ccBHF. A recent analysis of 
the giant monopole resonance in heavy nuclei 
\cite{blaizot} yields an experimental estimate  
$K_{nm}=210\pm 30$MeV for the 
nuclear matter compression modulus,
\begin{equation}
K_{nm}=\left. k_F^2\frac{d^2 E/A}{dk_F^2} \right|_{k_F=k_{F,0}}.
\end{equation}
At the saturation points in Fig.~\ref{fig:coester} we find ccBHF values  
$K_{nm}=154$ MeV for Reid93 and $K_{nm}=148$ MeV for the 
separable Paris interaction, which are enhanced to $K_{nm}=177$ MeV 
and $K_{nm}=216$ MeV, respectively, in our scSRC calculation.  
These values agree reasonably well with the experimental estimate. 
Note that reasonable values for $K_{nm}$ imply that the Reid68 energies in 
Fig.~\ref{fig:coester} 
may still deviate by 1-1.5 MeV from numerically exact scSRC 
values, as indicated by the error bars.

The present results indicate that a sophisticated treatment 
of SRC lowers the ccBHF saturation densities, bringing them closer to the 
empirical one. 
It remains to be understood why apparently converged hole-line 
calculations~\cite{bald} yield higher saturation densities.
The three hole-line terms obtained in Ref.~\cite{bald} 
indicate reasonable convergence properties compared to
the two hole-line contribution. One may therefore assume that
these results provide an accurate representation of the energy per particle
of nuclear matter as a function of density for the case of nonrelativistic
nucleons and two-body forces.
At this point it is useful to identify an underlying assumption when the 
nuclear-matter problem is posed~\cite{rd,libth}.
This assumption asserts that the influence of long-range correlations
in finite nuclei and nuclear matter are commensurate. 
It has been suggested in Ref.~\cite{ddsvw03} that this underlying assumption 
is questionable. 

Three
hole-line contributions include a third-order ring diagram characteristic
of long-range correlations.
The effect of long-range correlations on nuclear saturation properties
is sizeable, as shown by the results for three-
and four-body ring diagrams calculated in Ref.~\cite{dfm}
(see also Refs.~\cite{day2,bald}).
The agreement of three hole-line calculations with advanced variational
calculations~\cite{dayw} further supports the notion that important
aspects of long-range correlations are included in both these calculations.
This conclusion can also be based on the observation that hypernetted chain
calculations effectively include ring-diagram contributions
to the energy per particle although averaged over the Fermi sea~\cite{jls}. 
The effect of these long-range correlations on nuclear saturation properties
is \textit{not} small and can be quantified by quoting explicit results for 
three- and four-body ring diagrams~\cite{dfm}.
These results for the Reid potential,
including only nucleons, demonstrate that such ring-diagram terms
are dominated by attractive
contributions involving pion quantum numbers propagating
around the rings.
Furthermore, these contributions increase in importance with increasing
density.
Including the possibility of the coupling of these pionic excitation modes
to $\Delta$-hole states in 
these ring diagrams leads to an additional large increase in the binding
with increasing density~\cite{dfm}.
\begin{figure}[ptb]
\begin{center}
\includegraphics[angle=0,width=0.5\textwidth]{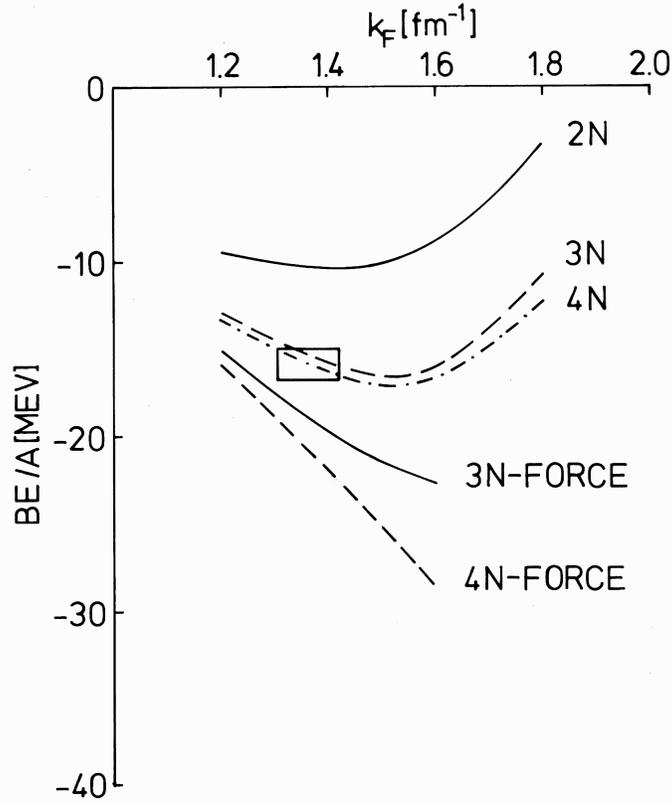}
\begin{minipage}[t]{16.5 cm}
\caption{
The curves labeled 2N, 3N, and 4N correspond to the 2-, 3-, and 4-hole-line
results from Ref.~\cite{day2} for the Reid interaction.
Added to the 4N results are the contribution of three- and four-body ring 
diagrams involving at least
one $\Delta$-isobar configuration. 
These terms can be considered as three- and four-body force contributions
when only nucleons are considered are therefore labeled with 3N- and 4N-force,
respectively.
\label{fig:delt}}
\end{minipage}
\end{center}
\end{figure}
This result is illustrated in Fig.~\ref{fig:delt} where these contributions
are added to the hole-line expansion results from Ref.~\cite{day2}.
Alternatively, these terms involving $\Delta$-isobars
can also be considered as contributions due to three- and four-body
forces in the space of only nucleons.
These $\Delta$-isobar terms have a very different behavior from the
phenomenological three-body forces that have been used to describe
the properties of nuclear matter~\cite{cpw,wff}.
The latter three-body forces turn repulsive at higher density.
The importance of the pionic long-range contributions to the binding energy
shown in Fig.~\ref{fig:delt} is related to the possible appearance of pion 
condensation at higher nuclear density~\cite{mig78,di81}.
These long-range pion-exchange dominated contributions to the
binding energy appear because of conservation of momentum in nuclear
matter. For a given momentum $q$ carried by a pion around a ring diagram,
one is able to sample coherently the attractive interaction that exists
for values of $q$ above 0.7 ${\rm fm}^{-1}$.
All ring diagrams contribute coherently when the interaction is attractive
and one may therefore obtain huge contributions at higher densities
which reflect the importance of this collective pion-propagation 
mode~\cite{dickdel}.
This is clearly illustrated by the results shown in Fig.~\ref{fig:delt}.

No such collective pion-degrees of freedom are actually observed
in finite nuclei.
A substantial part of the explanation of this fact is provided by the
observation that in finite nuclei both the attractive and repulsive parts of
the pion-exchange interaction are sampled before a build-up
of long-range correlations can be achieved.
Since these contributions very nearly cancel each other, which is
further facilitated by the increased relevance of exchange terms~\cite{cz1},
one does not see any marked effect on pion-like excited states in nuclei
associated with long-range pion degrees of freedom
even when $\Delta$-hole states are included~\cite{cz2}.
It seems therefore reasonable to call into question the relevance
of these coherent long-range pion-exchange contributions
to the binding energy per particle since their behavior is so markedly 
different in finite and infinite systems.
One may consider the salient difference of the ratio of spin-longitudinal
and spin-transverse response function in nuclear matter and finite
nuclei as another indication of the relevance of this suggestion~\cite{alber}.
We also like to point out that experimental information of these response 
functions~\cite{carey} - \cite{wakas}
suggests no characteristic enhancement of the (pionic) spin-longitudinal
response as expected on the basis of nuclear matter calculations.

Clearly, the assertion that long-range pion-exchange contributions
to the energy per particle need not be considered in explaining nuclear
saturation properties, needs to be further investigated.
In practice, this means that one must establish whether pion-exchange
in heavy nuclei already mimics the corresponding process in nuclear matter.
If this does not turn out to be the case, the arguments for considering
the nuclear-matter saturation problem only on the basis of the contribution
of SRC will be strenghtened considerably.
Furthermore, one would then also expect that the contribution of three-body
forces~\cite{ppwc}
to the binding energy per particle in finite nuclei continues to be
slightly attractive when particle number is increased substantially beyond 
ten particles~\cite{wpcp}.
This point and the previous discussion also suggest that there
would be no further need for the ad-hoc repulsion added to three-body forces
used to fit nuclear-matter saturation properties~\cite{cpw}. 

\section{\label{sec:finnuc}Theoretical calculations for finite nuclei}

The structure of finite nuclei presents different features from the case 
of nuclear matter due to the fact that nucleons are localized inside
the system.
 This requires the consideration of single particle basis states that
have good total angular 
momentum (and parity) and correspond to discrete set of levels, as in the IPM.
The coupling to low-lying collective excitations in the system tends to 
fragment the sp strength of
these orbitals and to spread it over a wide range of energies associated
with the energy scales of these excitations.
 Along with these effects, short-range correlations can move a sizable
fraction of nucleons to high-momentum states as in the case of nuclear
matter discussed in the previous section.

The influence of SRC in finite nuclei has been
studied theoretically by calculating the momentum distribution 
in the ground state of selected nuclei~\cite{zabol} - \cite{pwp92}.
These results clearly show that for momenta above 400 MeV/c short-range 
and tensor correlations completely dominate the momentum distribution.
Although the momentum distribution is a sp quantity, 
and thus allows only an indirect measurement of
correlations,
a careful study can give some indications on the importance of 
these effects.
More information is contained in the hole spectral function
$S_{\ell j\tau}(k,E)$, which gives the distribution
of nucleons both in momentum and energy  for particles with
orbital angular momentum $\ell$, total
angular momentum $j$, and isospin $\tau$.
%
%
%
In nuclei, substantial correlations
beyond the IPM approach are present
and a full understanding of the many-body system
requires, both in experimental as well as theoretical studies,
the knowledge of $S_{\ell j\tau}(k,E)$
for a large range of  energies $E$ as will be discussed in 
Sec.~\ref{sec:finiteSRC}.

Microscopic calculations of the spectral function and the momentum 
distribution of finite nuclei based on realistic nuclear hamiltonians
have been performed for very light nuclei ($A\leq 4$)
in Refs.~\cite{morita} - \cite{benpa}.
Results
for heavier nuclei are typically derived from investigations of nuclear
matter assuming a local density
approximation~(LDA)~\cite{string,sffb94} -\cite{neck94}.
Other variational calculations for $^{16}{\rm O}$ yield the momentum
distribution
\cite{pwp92} and the $p$-shell quasihole wave functions
\cite{rad94,Fabr01} but not the complete energy dependence
of the hole spectral functions.
The latter was studied by Benhar et al.~\cite{bffs94}, for different nuclei 
using nuclear matter results and LDA, and in Refs.~\cite{md94,mpd1,pmd2} for
$^{16}{\rm O}$  using
the Green's function formalism.
 These calculations suggest that most
of this strength is found along a ridge in the momentum-energy plane
($k$-$E$) which spans several hundreds of MeV/c (and MeV).
Moreover, although the distribution of this strength accounts only for a small
fraction of the total number of nucleons, it is responsible for 
as much as 60\% 
of the total binding energy of the system~\cite{mpd1}. 
As a consequence, the present understanding of nuclear correlations would
gain important information from the experimental knowledge
of this distribution.
 Such experiments require a high energy
transfer from the probe to the nucleus 
and a careful choice of the kinematics in order to
minimize the final state interactions~\cite{E97proposal}. For this reason 
it has only recently been possible to obtain reliable experimental
data in this correlated region which are still being 
analyzed~\cite{danielaElba,danielaPavia}.

\begin{figure}[!pbt]
 \begin{center}
    \includegraphics[width=0.65\textwidth,height=0.45\textheight,angle=-90]{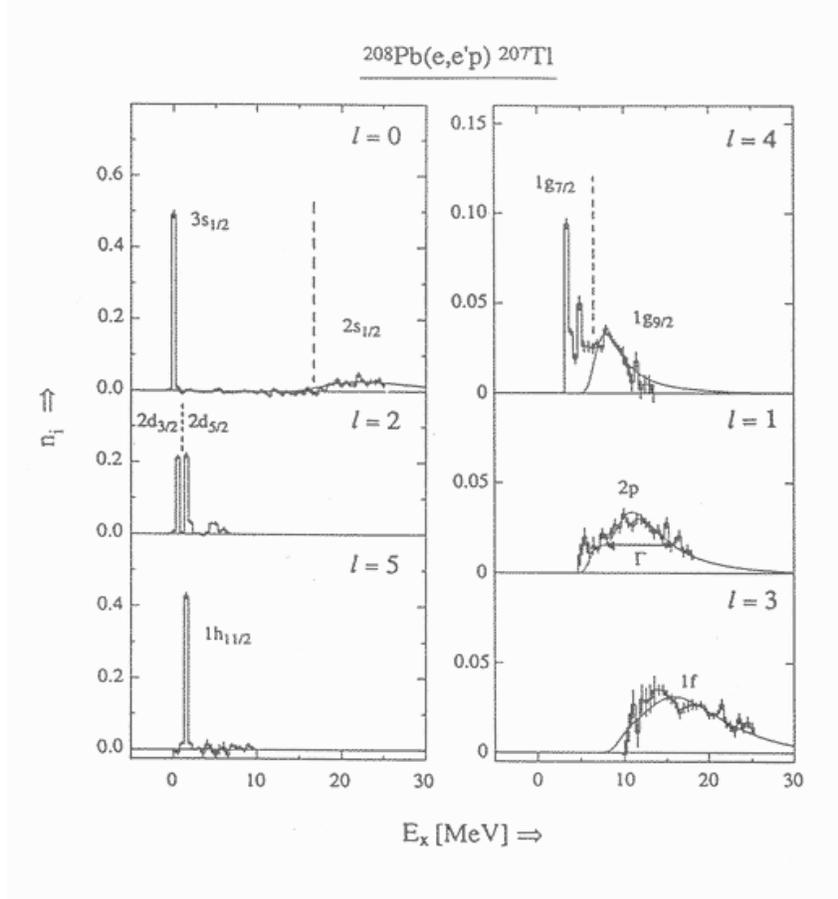}
\begin{minipage}[t]{16.5 cm}
\caption{Spectroscopic factors from the \eep\ reaction on ${}^{208}\textrm{Pb}$
for different valence hole orbits. The spectroscopic factor for the main
fragment of the $3s_{1/2}$ orbit must be changed to 0.65 from the values in
the figure to reflect
the analysis given in Ref.~\cite{deWitt3}.
\label{fig:quint} }
\end{minipage}
\end{center}
\end{figure}

Several studies have been carried out to obtain nuclear spectral
functions at small and medium missing energies, mostly by computing the
many-body Green's
function~\cite{rijs}, \cite{brand91} - \cite{badi02}.
%
The qualitative features of the strength distribution
at small and medium missing energies can be understood by
realizing that a considerable mixing occurs between hole states
and 2h1p configurations.
This mixing leads to different strength patterns depending on the
location of the orbital under consideration.
An orbital in the immediate vicinity of the Fermi energy
will still yield a large fragment near its original position
since the 2h1p states are quite far in energy.
For the same reason, only small components of this orbit appear at 2h1p
energies.
These features are observed for the $3s_{1/2}$ proton orbit in 
${}^{208}\textrm{Pb}$ as shown in Fig.~\ref{fig:quint}.
An orbit which is surrounded by many 2h1p states will yield a strongly
fragmented pattern, the width depending on the strength of the mixing
interaction.
These features are all observed experimentally as shown in 
Fig.~\ref{fig:quint}.
The actual values of the spectroscopic strength shown in this figure have not
been corrected for the effect of the Coulomb distortion on the electron.
This effect is sizable in heavy nuclei and leads to an increase of the 
spectroscopic factor for the $3s_{1/2}$ strength to 0.65 from the values of
0.5 shown in this figure~\cite{deWitt3}.

Calculations for medium-heavy nuclei like ${}^{48}$Ca and ${}^{90}$Zr
gave a fairly good description of the fragmentation pattern of valence hole
states~\cite{rijs} in terms of the coupling between quasiparticles
and TDA/RPA collective modes.
 The spectroscopic factors obtained in that work typically
overestimates the experimental summed strength by about a 15\% 
if only these long-range correlations are included. The remaining
surplus of strength is presumed to be moved to very high missing energies by 
the effects of short-range and tensor correlations as discussed in
 Sec.~\ref{sec:nucmat}.
 This is consistent with the reduction of the occupation of low
momentum states in nuclear matter also discussed in that section.
 This result from 
Ref.~\cite{vond1}
is supported by the observation that all the removed sp strength for
such small momenta in nuclear matter ends up at very high energies in the
particle domain associated with the repulsive core and tensor components
of the nucleon-nucleon interaction~\cite{vond1} 
(see also Sec.~\ref{sec:brian}).
Applying a corresponding 
adhoc reduction of the theoretical strength in medium-heavy
nuclei like ${}^{48}\textrm{Ca}$ yields an overall satisfactory description
of the sp strength distribution as discussed in Sec.~\ref{sec:48Ca90Zr}.

The conclusion that all is well is, however, premature. 
The effects of SRC on the spectroscopic factors of  $^{16}{\rm O}$
have been computed in Refs.~\cite{rad94} - \cite{md94}. All these 
works yield spectroscopic factors of 90\% 
for the knockout of a 
proton for the $p_{1/2}$ and $p_{3/2}$ shells in the (e,e$'$p) reaction
before the effects of the spurious motion of the center-of-mass
are taken into account.
 Such center-of-mass corrections are known to raise the theoretical 
spectroscopic
factor by about 7\%~\cite{diedef,O16com}.
 These results are in disagreement with the experimental value of 
about 60\% for
these orbitals~\cite{leus}.
The inclusion of the coupling to 2p1h and 2h1p states at low energy has been 
considered in Refs.~\cite{muetherO16,GeurtsO16} without leading to
an explanation of this discrepancy.
  In particular, in Ref.~\cite{GeurtsO16} a successful
attempt to combine the treatment of low-energy
long-range and short-range (and tensor) correlations was achieved.
The final result yielded a 
$p_{3/2}$ spectroscopic factor of 76\%, while
still neglecting center-of-mass corrections. This leaves a
discrepancy with the experiment of about 15-20\% (taking into
account the quoted uncertainties).
The calculations of Refs.~\cite{muetherO16,GeurtsO16}
take into account the interaction between the hole and particles
propagating in the system at the TDA level and therefore cannot
go beyond the 2h1p level.
 The calculation of Ref.~\cite{rijs} can account for the
coupling of quasiparticles to RPA collective modes and therefore
goes beyond 2h1p contributions,
 although it is limited by the difficulties in including
the coupling to 
collective excitations in \textit{both} the pp(hh) and ph channels.
The effects of fragmentation (or self-consistency) were only taken into account
in Ref.~\cite{vanneckCa48} were a fully self-consistent calculation
was obtained\ with a satisfactory representation of the continuum.
However, this calculation does not contain the
coupling to collective excitations and uses a phenomenological effective
interaction.
A procedure to include both ph and pp(hh) excitations in the 
construction of the self-energy was recently proposed and implemented in
Refs.~\cite{badi01,badi02}.
It was pointed out in this work that the RPA description of
the excitation spectrum for $^{16}{\rm O}$ is quite inadequate at present
and may be partly responsible for the unsatisfactory situation for
${}^{16}$O.

We begin this section by reviewing some early calculations for finite nuclei in
Sec.~\ref{sec:48Ca90Zr}. These results show that the properties of the
spectral fragmentation can be understood in terms of the 
coupling of single particle motion to collective pp(hh) and ph
motion, except for the reduction of the spectroscopic 
factors that is due to SRC. 
 A method that allows the evaluation of the effects of SRC on the
spectral function is reviewed in Sec.~\ref{sec:finiteSRC}.
Calculations for the particular case of $^{16}{\rm O}$ confirm
the expected reduction of spectroscopic factors due to short-range effects.
This calculation also gives quantitative information on the distribution
of the correlated spectral strength. The location of this
strength and the effects on the binding energy of nuclear systems
are discussed on section~\ref{sec:finiteSRC2}.
The formalism necessary to include the effects of both pp(hh) and ph
phonons in the nuclear self-energy, while describing them at least 
at the RPA level, is discussed in Sec.~\ref{sec:Faddform}.
  We present recent results for the
spectral function of $^{16}{\rm O}$ at low missing energy
in Sec.~\ref{sec:Faddres} and compare them to the 
available experimental data.
The relation between the details of the spectral distribution and the
results for the excitation spectrum
allows one to identify relevant missing ingredients
necessary to understand the spectroscopic factors of this nucleus.
 A  first attempt to improve the description of the spectrum~\cite{badi03}
is discussed in Sec.~\ref{sec:BSEO16}.

The two-hole spectral function can be computed, including the 
fragmentation effects, by solving the pp(hh) DRPA equation. The high
momentum components --excluded form the model space-- can be included
afterwards by computing the relevant defect functions.
Some results for two-nucleon emission, that clearly
show the signature of SRC are discussed in Sec.~\ref{sec:ee1-2N}.
The issue of how to deal with final state interactions
in reactions at high momentum transfer with regard to the extraction
of spectroscopic factors is discussed in Sec.~\ref{sec:highQ}.
We conclude this section by discussing two recent experiments
that give the first direct experimental information of the occupation numbers
of deep-lying orbitals in $^{208}{\rm Pb}$
and on the spectral strength distribution at high momenta,
Sec.~\ref{sec:Pb208deep}.

\subsection{Early results for the spectral distribution
\label{sec:48Ca90Zr} }

\begin{figure}[!tb]
\begin{center}
\includegraphics[angle=0,width=0.6\textwidth]{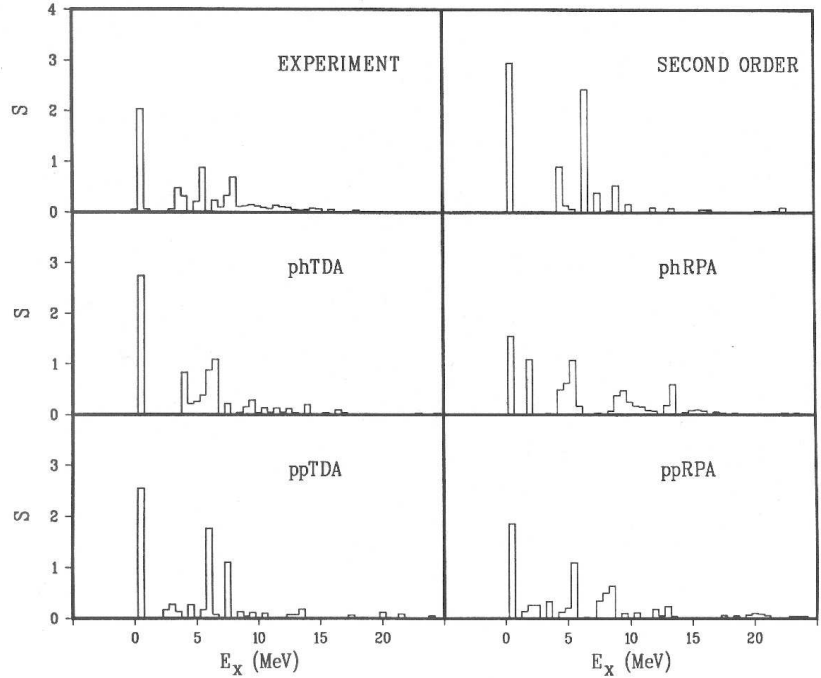}
\begin{minipage}[t]{16.5 cm}
\caption{Distribution of $sp$ strength for $\ell =2$ proton removal from
 $^{48}{\rm Ca}$. Experimental data are from Ref.~\cite{kra90}. 
 Theoretical results are displayed for various approximations to the
 self-energy which are discussed in the text. 
\label{fig:rijs_Sh} }
\end{minipage}
\end{center}
\end{figure}

The earliest calculations of the spectral strength in nuclei were reported
in Ref.~\cite{brand91,vanneckCa48}.
The study of the spectral strength distribution starting from a realistic NN
interaction is only possible when
short-range correlations are properly included.
These early calculations either relied on the construction of a local
$G$ matrix interaction obtained from nuclear matter~\cite{wim83} to represent
SRC~\cite{brand91} or relied on phenomenological forces~\cite{vanneckCa48}.
The latter approach actually generated the first self-consistent calculation
of the spectral strength in nuclei based on the inclusion of the second-order
self-energy diagram.
The former calculations also used a second-order approach but included the
$G$ matrix effective interaction without achieving self-consistent results.
Useful insights were obtained from these calculations which described certain
features of the spectral distribution obtained from the ${}^{48}$Ca(e,e$'$p)
reaction~\cite{kra90}.
The inclusion of SRC on the distribution of the strength was not considered
in this work (see Sec.~\ref{sec:finiteSRC}).
As a general conclusion of this work on $^{48}$Ca,
it is found that the total strength in the experimentally accessible domain
in the (e,e$'$p) reaction is overestimated by about 10-15\% 
although the shape of the strength distribution is already well described
as shown in Fig.~\ref{fig:rijs_Sh}.
These results for the sp strength distribution
for proton removal from $^{48}{\rm Ca}$ in $\ell=2$ states
are compared to the corresponding experimental data~\cite{kra90}.
The theoretical results are labeled by ``2nd Order'' in the figure.
The normalization in the figure is such that if a single state carried all
the strength, the peak height at the corresponding energy would be $2j+1$.
The first peak corresponds to $d_{3/2}$ removal whereas most of the rest of
the strength corresponds to $d_{5/2}$ removal.
The difference between the fragmentation of these two $sp$ orbitals
characterizes the basic features which are found in experiment.
For an orbital which in mean field is very near the Fermi energy, like the
$d_{3/2}$ orbital, fragmentation of strength to all 2p1h and 1p2h-like
states takes place in such a way that the distribution is characterized by a
single large fragment and hundreds of tiny contributions spread out over the
whole energy domain covered by the 2p1h and 1p2h states.
This background contribution to the hole strength, $i.e.$ the strength which
is not contained in the area of the main peak, is of the order of 10\%.

\begin{figure}[!tb]
\begin{center}
\includegraphics[angle=0,width=0.6\textwidth]{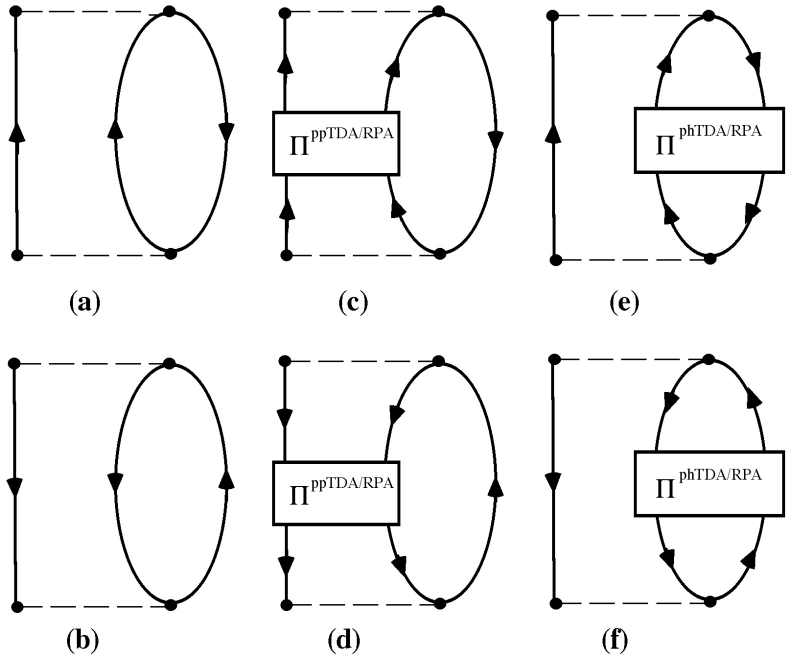}
\begin{minipage}[t]{16.5 cm}
\caption{Second-order self-energy terms represented by Goldstone diagrams
 $a$ and $b$.  Extensions are made by including TDA or RPA correlations
 in the particle-particle (diagrams $c$ and $d$) or particle-hole channel 
 (diagrams $e$ and $f$).
\label{fig:rijs_selfen} }
\end{minipage}
\end{center}
\end{figure}

The comparison between the second-order results and experiment
in Fig.~\ref{fig:rijs_Sh} also shows that
stronger fragmentation in the theoretical calculation at low energy
is required.
Improvement of the description of the intermediate states in the self-energy
was investigated in Ref.~\cite{rijs}.
In this paper various possible descriptions of the intermediate 2p1h/2h1p
in the self-energy were considered.
This approach can be summarized by considering Fig.~\ref{fig:rijs_selfen}
which exhibits the diagrammatic content of these approximations.
These different approximations correspond to the standard versions of the TDA
and RPA for both ph and pp(hh) forms as discussed in Sec.~\ref{sec:form}
and are represented by diagrams $c$ through $f$ in Fig.~\ref{fig:rijs_selfen}.
These approximations include the effect of long-range (low-enery)
correlations on the fragmentation pattern at low energy.
It should be noted that three independent types of possible collective 
excitations occur, i.e. neutron
particle - proton hole, proton particle - neutron hole, and a mixture of
neutron particle - neutron hole and proton particle - proton hole
with isospin components $T_z$=1, -1, and 0, respectively.
The $T_z$=0 collective low-lying $2^+$ and $3^-$
phonons are especially relevant as their coupling to the sp motion
is an important source of fragmentation of spectral functions at low energies.
While collective effects are stronger in the RPA, it is possible to encounter
serious problems with this method.
They arise, especially in phRPA, when the RPA correlations are so strong that
the lowest solution for some angular momentum and parity, has imaginary
eigenvalues and the amplitudes cannot be
properly normalized. This happens for the 3$^-$ phonon, which is crucial for
the fragmentation of the $\frac{7}{2}^-$ strength in $^{48}$Ca.
As the imaginary solution
can not be included it has to be discarded and
the phRPA method is unsatisfactory in this case~\cite{rijs}.
For the same reason the lowest
1$^+$ states in $^{46}$K en $^{50}$Sc become unstable in ppRPA and
therefore were also discarded in that approach.
These problems are expected to 
disappear when a self-consistent approach would be
adopted, which has not been implemented for this nucleus sofar.
The comparison of the various approximations in Fig.~\ref{fig:rijs_Sh}
to the experimental data shows that increased fragmentation at low energy
is obtained when collective effects are considered.
The sp strength farther away from the Fermi energy is largely unaffected.

\begin{figure}[!tb]
\begin{center}
\includegraphics[angle=0,width=0.5\textwidth]{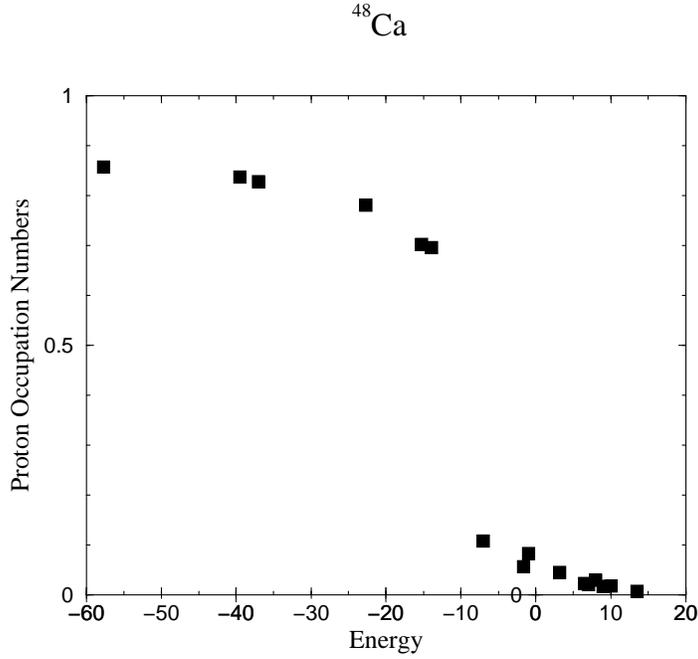}
\begin{minipage}[t]{16.5 cm}
\caption{Proton shell occupation probabilities deduced from a comparison
 of the present calculation and the (e,e$'$p) data~\cite{kra90}.
 The figure displays the calculated values with the ppTDA correlations in the
 self-energy, multiplied by a factor 0.9 to simulate the effect of
 short-range correlations. 
\label{fig:rijs_occ} }
\end{minipage}
\end{center}
\end{figure}

The summed sp strength below the Fermi energy generates the occupation
numbers for the orbits considered as given by Eq.~(\ref{eq:occ_numb}).
Experimentally, the hole strength has been
determined only within a limited energy region. For the valence $\ell=0$ and
$\ell=2$ shells 
the main portion of the total strength falls within this range
of energies. In all theoretical approaches discussed here the calculated extra
hole strength at higher energies is only 5-10$\%$ of the $2j+1$ sum rule for
the 2s shell and only 10-15$\%$ for the 1d shell.
For the major shells above the Fermi level the occupation probability is
about 5-10$\%$ and for more remote shells only 1-2$\%$.
A meaningful quantity is the jump in occupation numbers at the Fermi level.
This jump in occupation number for the ppTDA result is 0.74.
In Fig.~\ref{fig:rijs_occ} these occupation numbers in the ppTDA
approximation are multiplied by a factor of 0.9 to simulate the effect of
SRC.
The resulting occupation numbers show a flat behavior for deeply bound shells
while they bend down when the orbits approach the Fermi energy.
This feature also emphasizes that LRC are most important in the immediate
vicinity of the Fermi energy.

\subsection{Depletion due to short-range and tensor correlations
\label{sec:finiteSRC} }

\begin{figure}[!tb]
\vspace*{0.cm}       
\begin{center}
\includegraphics[width=0.5\textwidth]{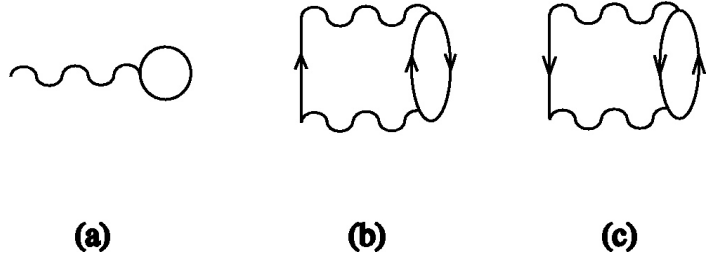}
\vspace*{0.cm}   
\begin{minipage}[t]{16.5 cm}    
\caption{\label{fig:diag}
Graphical representation of the Hartree-Fock (a), the 2-particle
1-hole (2p1h, b) and the 2-hole 1-particle contribution (2h1p, c) to the
self-energy of the nucleon.
}
\end{minipage}
\end{center}
\end{figure}

The treatment of short-range correlations requires the consideration
of the details of the two-particle wave function at small internucleon 
distances (or high relative momenta). 
This suggests that it is
appropriate to work directly in coordinate
or momentum space.
 The approach introduced in Refs.~\cite{borro,md94,mpd1} computes the
self-energy for finite nuclei in terms of a $G$-matrix
which is obtained as a solution of the Bethe-Goldstone equation for
nuclear matter
\begin{eqnarray}
\lefteqn{
\left\langle k \ell \right| G_{S J_S K L T} \left| k' \ell' \right\rangle
 =
  \left\langle k \ell   \right| V \left| k' \ell' \right\rangle  
  } \hskip .5in   & &
\nonumber\\
 & &  + \sum_{\ell''} \int dk'' (k'')^2 \; 
             \left\langle k \ell \right| V \left| k'' \ell'' \right\rangle
\frac{Q (k,K)}{\omega_{NM}- \frac{K^2}{4m} - \frac{k^2} {2m}} 
  \left\langle k'' \ell'' \right| G_{S J_S K L T} \left| k' \ell' \right\rangle
   \; ,
\label{eq:betheg}
\end{eqnarray}
where $k$, $k'$, and $k''$ denote the relative momenta 
between the two nucleons,
$\ell$, $\ell'$, and $\ell''$ the orbital angular momenta 
for the relative motion,
$K$ and $L$ the corresponding quantum numbers for the center-of-mass
motion, $S$ and $T$ the total spin and isospin, and $J_S$ is obtained
by coupling the orbital angular momentum of the relative motion to the spin
$S$.
 We note that Eq.~(\ref{eq:betheg}) is analogous to the ladder equation
for the vertex function $\Gamma$, Eq.~(\ref{eq:gamla}), except that it 
describes
the intermediate propagation of two free nucleons (instead of dressed
propagators), taking into account only the effects of Pauli blocking
with regard to the free Fermi gas.
This equation
differs from the usual $T$-matrix of the two-body scattering problem only
because of the inclusion of the Pauli operator $Q$ that prevents the
two nucleons to scatter into states inside the Fermi sea (which are already
occupied).  
Equation (\ref{eq:betheg}) therefore generates an appropriate solution
of two-body short-range dynamics.
 The choices for the density
of nuclear matter and the starting energy $\omega_{NM}$ 
are rather arbitrary. 
The calculation of the corresponding Hartree-Fock term is not very
sensitive to this choice~\cite{bm89}. Furthermore, this
nuclear-matter approximation is corrected by calculating
the 2p1h term displayed in Fig.\ \ref{fig:diag}(b)
directly for the finite system. This second-order correction, which 
assumes harmonic oscillator states for the occupied (hole) 
states and plane waves for the intermediate unbound particle 
states, incorporates the correct energy and density dependence 
characteristic of a finite nucleus $G$-matrix.
 To evaluate the 
diagrams in Fig.~\ref{fig:diag}, one requires matrix elements in a mixed 
representation of one particle in a bound harmonic oscillator 
while the other is in a plane wave state.
 Using vector bracket 
transformation coefficients~\cite{vecbr}, one can transform matrix 
elements from the representation in coordinates of relative and 
center-of-mass momenta to the coordinates of sp momenta in the 
laboratory frame in which the two particle state is described by
\begin{equation}
\left| k_1 \ell_1 j_1 k_2 \ell_2 j_2 J T \right\rangle ,  \label{eq:bra2p}
\end{equation}
where $k_i$, $\ell_i$ and $j_i$ refer to the momentum and angular momenta of 
particle $i$ whereas $J$ and $T$ define the total angular momentum and
isospin of the two-particle state.
Performing an integration over one of
the $k_i$, one obtains a two-particle state in a mixed representation
\begin{equation}
\left| n_1 \ell_1 j_1 k_2 \ell_2 j_2 J T \right\rangle =
 \int_0^\infty dk_1 \, k_1^2
R_{n_1, \ell_1}(k_1) \, 
\left| k_1 \ell_1 j_1 k_2 \ell_2 j_2 J T\right\rangle
 \; .
\label{eq:branp}
\end{equation}
Here $R_{n_1, \ell_1}$ stands for the radial oscillator function and the
oscillator length is chosen to be $b = 1.72 $ fm$^{-1}$ to achieve an
 appropriate description of the bound states
of $^{16}{\rm O}$.
Using the results of
Eqs.~(\ref{eq:betheg}) - (\ref{eq:branp}) the Hartree-Fock approximation
for the self-energy can be obtained in the momentum representation
\begin{equation}
\Sigma^{HF}_{\ell_1j_1} (k_1,k'_1) =
\frac{1}{2(2j_1+1)} \sum_{n_2 \ell_2 j_2} 	\sum_{J T} (2J+1) (2T+1)
\left\langle k_1 \ell_1 j_1 n_2 \ell_2 j_2 \right| G_{J T} \left|
k'_1 \ell_1 j_1 n_2 \ell_2 j_2 \right\rangle .
\label{eq:selhf}
\end{equation}
The summation over the oscillator quantum numbers is restricted to the
states occupied in the IPM of $^{16}$O. This
Hartree-Fock part of the self-energy is real and does not depend on the
energy.

The terms of lowest order in $G$ which give rise to an imaginary part
in the self-energy are represented by the diagrams displayed in Figs.\ 
\ref{fig:diag}(b) and \ref{fig:diag}(c),
refering to intermediate 2p1h and 2h1p
states, respectively. The 2p1h contribution to the imaginary part is given
by
\begin{eqnarray}
\lefteqn{
{W}^{2p1h}_{\ell_1 j_1} (k_1 , k'_1 ; E)
= \frac{-1}{2(2j_1+1)}  \sum_{n_2 \ell_2 j_2} \sum_{\ell L}
\sum_{J  S T} \int k^2 dk \int K^2 dK \; (2J+1) (2T+1) 
} ~ & &
 \label{eq:w2p1h} \\
& & \times \left\langle k_1 \ell_1 j_1 n_2 \ell_2 j_2 \right| G_{J T} \left|
k \ell S K L \right\rangle
  \left\langle k \ell S K L  \right| 
G_{J T} \left| k'_1 \ell_1 j_1 n_2 \ell_2 j_2 
  \right\rangle 
 \; \pi  \;  \delta\left(E +\varepsilon_{n_2 \ell_2 j_2}
                     -\frac{K^2}{4m}-\frac{k^2}{m}\right)  ,
\nonumber
\end{eqnarray}
where the average experimental quasihole energies 
$\varepsilon_{n_2 \ell_2 j_2}$
are used for the hole states ($-47$ MeV, $-21.8$ MeV, $-15.7$ MeV for 
$s_{1/2}$,  $p_{3/2}$ and $p_{1/2}$ states, 
respectively), while the energies of the particle states are 
given in terms of the kinetic energy only.
 The plane waves
associated with the particle states in the intermediate states 
are properly orthogonalized to the bound sp states following the 
techniques discussed in Ref.~\cite{borro}. The 2h1p 
contribution to the imaginary part 
${W}^{2h1p}_{\ell_1 j_1}(p_1,p'_1; E)$ can be calculated in a 
similar way~\cite{borro}.

The choice of pure kinetic energies for the particle states in
calculating the imaginary parts of $W^{2p1h}$ (Eq.~(\ref{eq:w2p1h})) and 
$W^{2h1p}$ may not be very realistic for the excitation modes at low energy.
Indeed a sizeable imaginary part in $W^{2h1p}$ is obtained only for
energies $E$ below -40~MeV. However, when the primary interest 
is in the effects of short-range correlations,
 the choice appears appropriate since these involve excitations of particle
states with high momenta. 
A different approach is required to treat the coupling
to the very low-lying 2p1h and 2h1p
states in an adequate way, as
discussed in Secs.~\ref{sec:48Ca90Zr} and~\ref{sec:Faddform}.
The 2p1h contribution to the real part of the self-energy can be calculated
from the imaginary part $W^{2p1h}$ using a dispersion relation 
\cite{mbbd}
\begin{equation}
V^{2p1h}_{\ell_1j_1}(k_1,k_1';E)=\frac{\mathcal{P}}{\pi} 
\int_{-\infty}^{\infty} \frac
{W^{2p1h}_{\ell_1j_1}(k_1,k_1';E')}{E'-E} dE', \label{eq:disper1}
\end{equation}
where $\mathcal{P}$ stands for a principal value integral. A similar 
dispersion relation holds for $V^{\rm 2h1p}$ and $W^{\rm 2h1p}$.
 
Since the Hartree--Fock contribution $\Sigma^{\rm HF}$ has been 
calculated in terms of a nuclear matter $G$-matrix, it already 
contains 2p1h terms of the kind displayed in Fig.~\ref{fig:diag}(b). 
In order to avoid this overcounting of the 
pp ladder terms, one subtracts from the real part of 
the self-energy a correction term ($V_{\rm c}$), which just 
contains the 2p1h contribution calculated in nuclear matter. 
Summing up the various contributions one obtains the following expression
for the self-energy
\begin{equation}
\Sigma  = \Sigma^{\rm HF} + \Delta\Sigma = \Sigma^{\rm HF} + 
\left( V^{\rm 2p1h} - V_{\rm c} + V^{\rm 2h1p} \right)
+ \left( W^{\rm 2p1h} + W^{\rm 2h1p} \right) \; .
\label{eq:defsel}
\end{equation}

The Dyson equation~(\ref{eq:dyson}) for this self-energy can be solved in
momentum space. In Ref.~\cite{mpd1} the integrals were discretized 
by  considering a complete basis within a spherical 
box of a radius $R_{\rm box}$. The calculated observables are 
independent of the choice of $R_{\rm box}$, if it is chosen to be 
around 15 fm or larger. 
By taking into account the different normalization of these
basis functions from the case of plane waves in the continuum, one can
express the matrix elements of nucleon self-energy in the ``box''
basis.
Energies and wave functions of 
the quasihole states can be determined by diagonalizing the 
Hartree-Fock Hamiltonian plus $\Delta\Sigma$ in this ``box basis''
[see Eq.~(\ref{eq:2nd20})]:
\begin{equation}
\sum_{n=1}^{N_{\rm max}} \left\langle k_i \right| 
\frac{k_i^2}{2m} \delta_{in} + \Sigma^{\rm HF}_{\ell j} + 
\Delta\Sigma_{\ell j} (E=\varepsilon^{qh}_{n- \ell j})
\left| k_n \right\rangle \left\langle k_n \vert z^{n-} 
\right\rangle_{\ell j} = \varepsilon^{qh}_{n- \ell j} \  
\left\langle k_i \vert z^{n-} \right\rangle_{\ell j} \; .
\label{eq:qhequ}
\end{equation}
In this approach $\Delta\Sigma$ only contains a sizable 
imaginary part for energies $E$ below the quasihole
eigenvalues $\varepsilon^{qh}_{n- \ell j}$ ($n-$ identifying discrete
solutions). Therefore the
solutions of Eq.~(\ref{eq:qhequ}) are separated in energy
from the continuum contribution to the spectral function.
The eigenvector corresponding to these discrete states yields
the quasihole wave function
$\left\langle k \vert z^{n-} \right\rangle_{\ell j}$ in momentum space,
which still needs to be normalized by the spectroscopic factor
$Z^{qh}_{n- \ell j}$ by means of Eq.~(\ref{eq:2nd21}).
The diagonal part of the quasihole contribution to the spectral
function is given, in the box basis, by
\begin{equation}
S^{qh}_{n- \ell j} (k_n; E) = Z^{qh}_{n- \ell j}
 \left|
\left\langle k_n \vert z^{n-} \right\rangle_{\ell j} 
  \right|^2
 \, \delta (E - \varepsilon^{qh}_{n- \ell j}) \; .
 \label{eq:skoqh}
\end{equation}

 In the calculations described in Refs.~\cite{mpd1,pmd2},
the Bethe-Goldstone equation~(\ref{eq:betheg}) was solved by employing
for $V$ 
the one-boson-exchange potential Bonn-$B$ developed by Machleidt
in Ref.~\cite{bonnABC} (Tab.~A.2) and the Pauli operator $Q$ was approximated 
by the so-called angle-averaged approximation for nuclear matter with a 
Fermi momentum $k_F = 1.4\ \textrm{fm}^{-1}$. This roughly corresponds
to the saturation density of nuclear matter. The starting energy
$\omega_{NM}$ for computing the $G$-matrix, Eq.~(\ref{eq:betheg}),
was chosen to be -10 MeV.

\begin{figure}[!tb]
\vspace*{.5cm}       
\begin{center}
\includegraphics[width=0.6\textwidth]{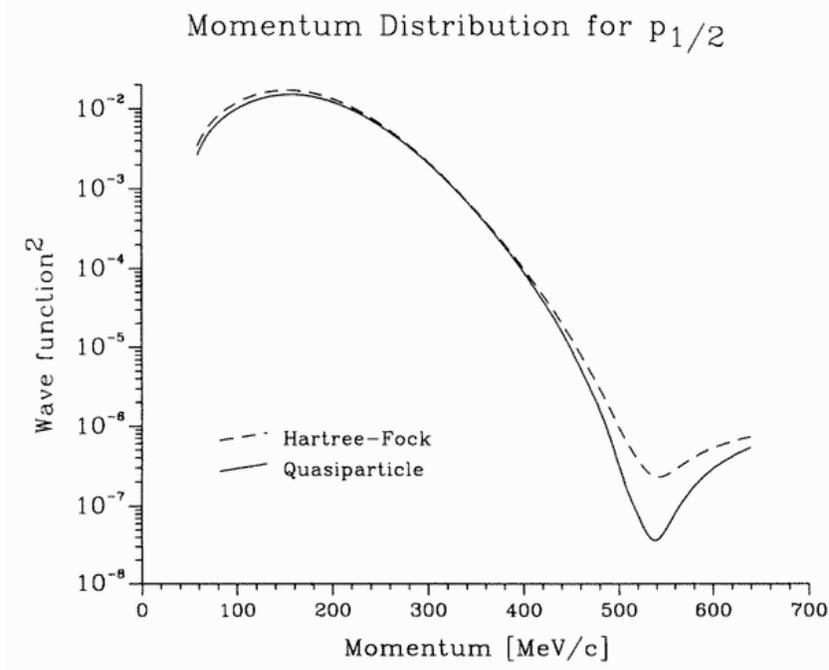}
\begin{minipage}[t]{16.5 cm} 
\caption{ \label{fig:qpwf}
 Square of the quasihole wave function for the $p_{1/2}$ state
 in $^{16}{\rm O}$
(full curve), normalized to the spectroscopic factor according to 
Eq.~(\ref{eq:2nd21}), compared to the Hartree-Fock result (dashed curve).
}
\end{minipage}
\end{center}
\end{figure}
The square of the quasihole wave function for the $p_{1/2}$ state 
(normalized to the spectroscopic factor, $Z^{\rm qh}_{0p1/2}$) is
shown in Fig.~\ref{fig:qpwf} as a full line. 
For comparison the result for the Hartree-Fock wave function is shown
by the dashed line.
From the comparison one can infer that at the quasihole energies
no substantial change in the wave functions occurs and that the Hartree-Fock
wave function is a good approximation.
It should be further noted that the wave function of a Woods-Saxon
potential, which is constructed as the local equivalent of the
Hartree-Fock potential~\cite{borro}, is indistinguishable from the  
Hartree-Fock wave function.
This suggests that the explicit inclusion of short-range correlations 
does not lead to the strong suppression of the wave function
in the interior of the nucleus as has been suggested by 
Refs.~\cite{mawam,masa91}. 
These findings are in line with the observation that the quasiholes
representing mean-field orbitals can be adequately described by
a local potential~\cite{Escher:2002ud}.
 It should be emphasized that this is not necessarily
the case for other overlap functions~\cite{Escher:2002ud}, since
the coupling to collective excitations can lead to significant changes 
to the quasihole wave function~\cite{Escher:2002ud,Al-Khalili:2003di}.
 These effects are unlikely to generate changes in the
high-momentum components discussed here.
It is important to note that these results also show that 
the influence of high-momentum components in the quasihole
wave function is of minor importance, so the enhancement of the
momentum distribution has to come from excitations at higher missing energies.

\begin{figure}[!tb]
\vspace*{0.cm}       
\begin{center}
\includegraphics[width=0.6\textwidth]{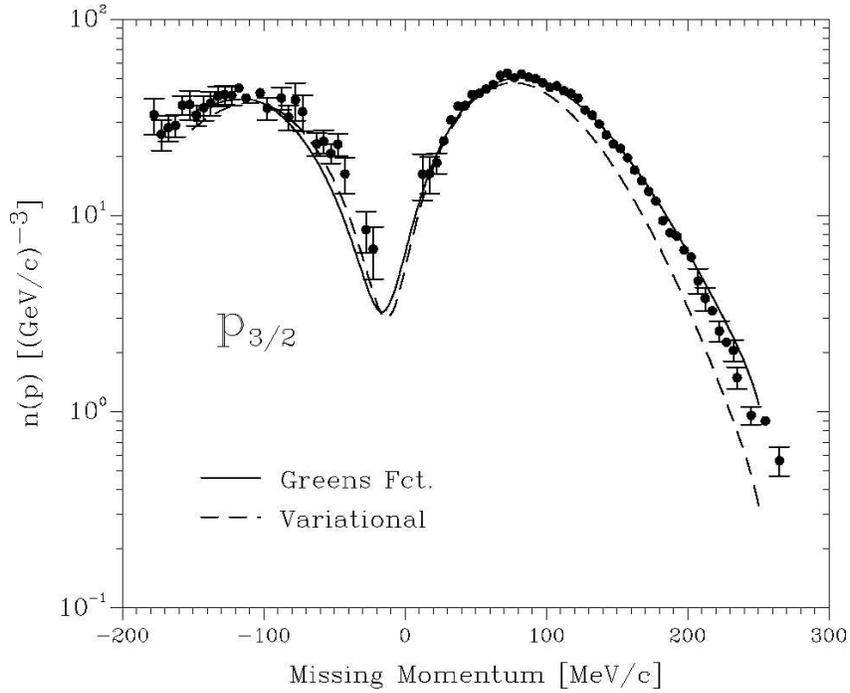}
\vspace*{0.cm}       
\begin{minipage}[t]{16.5 cm}
\caption{\label{fig:N15p32}
Reduced cross section for the $^{16}$O$(e,e'p)$ reaction 
in parallel kinematics leading to the ${3/2}^-$ 
state at $-6.32$ MeV of the residual nucleus $^{15}$N. Results of  
the Green's function approach (solid line) are compared 
to those obtained in the variational calculation of 
Ref.~\cite{rad94} (dashed line) and the experimental 
data~\cite{leus}. A spectroscopic 
factor of 0.537 was required for the Green's function result, 
while $Z_{0p3/2} = 0.459$ has been used to 
for the variational calculation.
}
\end{minipage}
\end{center}
\end{figure}
We show in Fig.~\ref{fig:N15p32} an example of the reduced cross section
for the (e,e$'$p)
reaction on $^{16}$O leading to bound quasihole state of
$^{15}$N at an excitation energy of $-6.32$ MeV~\cite{rad97}.
 In this picture the effects of FSI have been computed in the
distorted wave impulse approximation (DWIA)~\cite{Paviabook} 
and the data points have been obtained at NIKHEF
for the so-called parallel kinematics~\cite{leus}.
 Using the quasihole part 
of the spectral function computed from Eq.~(\ref{eq:qhequ})
but adjusting the spectroscopic factor for 
the quasihole state contribution $Z^{qh}_{0p3/2}$ to fit the 
experimental data, one obtains the solid line of 
Fig.~\ref{fig:N15p32}. Comparing this result with the experimental 
data one finds that the calculated spectral function reproduces 
the shape of the reduced cross section as a function of the 
missing momentum quite well.
 The absolute value for the reduced 
cross section can only be reproduced by employing a spectroscopic 
factor $Z^{qh}_{0p{3/2}} = 0.537$, a value considerably below 
the one of 0.914 calculated using Eq.~(\ref{eq:qhequ})~\cite{mpd1}.
 An analogous result holds for transition to the $p_{1/2}$ ground state
of $^{15}$N for which the same analysis yields a spectroscopic factor
$Z^{qh}_{0p{1/2}} = 0.644$.
For comparison, one may note that the use of
phenomenological Woods-Saxon wave functions adjusted to fit 
the shape of the reduced cross section require spectroscopic 
factors ranging from 0.61 to 0.64 for the lowest 
$0p_{1/2}$ state and from 0.50 to 0.59 for the 
$0p_{3/2}$ state, respectively, 
depending upon the choice of the optical potential for the 
outgoing proton~\cite{leus}.

Figure~\ref{fig:N15p32} also contains the results for the reduced 
cross section derived from
the variational result for overlap wave function of 
Pieper et al.~\cite{rad94}, 
who employed the Argonne $v_{14}$ potential for the NN 
interaction~\cite{wsa84}. Also in this case the shape of the 
experimental data is globally reproduced with a slightly better 
agreement for small negative values of $p_{m}$ but with a 
clear underestimation at larger $p_{m}$. The overall quality 
of the fit is somewhat worse than for the Green's function approach 
and the required adjusted spectroscopic factor is 
$Z^{qh}_{0p_{3/2}} = 0.459$, below the value of 0.537 for the latter approach.
 It is not clear, however, whether the
differences in the calculated reduced cross section are due to the
use of different interactions or to the methods employed in
calculating the spectral function.

Both the variational calculations of Refs.~\cite{rad94,Fabr01} and
the Green's function
approach give spectrocsopic factors for the $p$-shell orbitals of
about 0.90. 
 This value should be compared to the $\sim$0.63 obtained form
the NIKHEF experimenf for these shells.
 The additional contribution due to the proper consideration of
the center-of-mass motion raises the theoretical value to 0.98,
worsening the agreement with data~\cite{O16com}.
Nevertheless, the observation that the same results are
obtained with two independent many-body
methods but including the same short-range physics suggests that the
effects of SRC on the quasihole strength are well under control.
At the same time,
the discrepancy with the experiment is partly due to the emphasis
on the accurate treatment of short-range correlations only and
one should view the quasihole strength that has been discussed here
to be due only to the influence of short-range
correlations~\cite{borro}.
 It is clear that a considerable renormalization
of the strength is to be expected due the coupling of
the quasihole states to the low-lying collective excitations,
as will be discussed in Sec.~\ref{sec:Faddres}.

\subsection{High-momentum components at high missing energies
\label{sec:finiteSRC2} }

The continuum part of the hole spectral strength is found at higher
energies and stems from the coupling to the continuum of 2h1p states.
 In this region it is useful to proceed from the Hartree-Fock propagator
with states $\ket{\alpha}$ that diagonalize the corresponding self-energy
\begin{equation}
g_{\ell j}^{(0)} (\alpha; E) = 
\frac{1}{E-\varepsilon^{HF}_{\alpha \ell j} \pm i\eta} , 
\label{eq:green0}
\end{equation}
where the sign in front of the infinitesimal imaginary quantity 
$i\eta$ is positive (negative) depending on whether 
$\varepsilon^{HF}_{\alpha \ell j}$ 
is above or below the Fermi energy.
 The Dyson equation can then be solved by iterating the $\Delta\Sigma$ 
component in Eq.~(\ref{eq:defsel})
of the self-energy to generate the reducible self-energy
\begin{equation}
\left\langle \alpha \right| \Sigma^{red}_{\ell j}(E) \left| 
\beta \right\rangle =
\left\langle \alpha \right| \Delta\Sigma_{\ell j}(E) \left| \beta 
\right\rangle
+ \sum_\gamma
\left\langle \alpha \right| \Delta\Sigma_{\ell j}(E) \left| \gamma 
\right\rangle
g_{\ell j}^{(0)} (\gamma ; E) \left\langle \gamma \right| 
\Sigma^{red}_{\ell j}(E) \left| \beta \right\rangle
\end{equation}
and obtain the propagator from
\begin{equation}
g_{\ell j} (\alpha ,\beta ;E ) = \delta_{\alpha ,\beta} \  
g_{\ell j}^{(0)} (\alpha ;E ) + g_{\ell j}^{(0)} (\alpha ;E )
\left\langle \alpha \right| \Sigma^{red}_{\ell j}(E) \left| 
\beta \right\rangle g_{\ell j}^{(0)} (\beta ;E ) .
\end{equation}
Using this representation of the Green's function one can 
calculate the spectral function in the ``box basis'' from
[see Eq.(\ref{eq:Sh})]
\begin{equation}
S_{\ell j}^{c} (k_n; E) = \frac{1}{\pi} \  \mbox{Im} 
\left( 
 \sum_{\alpha, \beta} \left\langle k_n \vert \alpha 
\right\rangle_{lj} \;  g_{lj} (\alpha ,\beta ;E ) \,
\left\langle \beta \vert k_n \right\rangle_{lj}  \right) . 
\label{eq:skob}
\end{equation}
This spectral function is different from zero
for energies $E$ below the lowest sp energy of a
given Hartree-Fock state (with $\ell j$)
only due to the imaginary part in $\Sigma^{red}$.
This contribution involves the coupling to the continuum of 2h1p 
states and is therefore nonvanishing only for energies at which 
the corresponding irreducible self-energy $\Delta\Sigma$ has a 
non-zero imaginary part. 
The 2p1h contribution to the self-energy is responsible for the depletion
of strength, which in mean
field is located below the Fermi energy, to high energy.
The 2h1p term instead, is essential for the accumulation of sp
strength below the Fermi energy from states (in particular those with high
momenta) which are empty in the mean field.
%
The continuum contribution of Eq.~(\ref{eq:skob}) and 
the quasihole parts of Eq.~(\ref{eq:skoqh}), which are obtained 
in the basis of box states, can be added and renormalized to 
obtain the complete spectral function in the continuum representation at 
the momenta defined by the box basis from
\begin{equation}
S_{\ell j} (k_i;E) = \frac{2}{\pi} \  \frac{1}{N_{i \ell}^2}
\bigl( S^{c}_{\ell j} (k_i;E) + 
\sum_{n-}  S^{qh}_{n- \ell j} (k_i;E)
\bigr) ,
\label{eq:renor}
\end{equation}
where the $N_{i \ell}$ correspond to appropriate normalization 
constants~\cite{mpd1}.

\begin{figure}[!tb]
\vspace*{.5cm}       
\begin{center}
\includegraphics[width=0.6\textwidth]{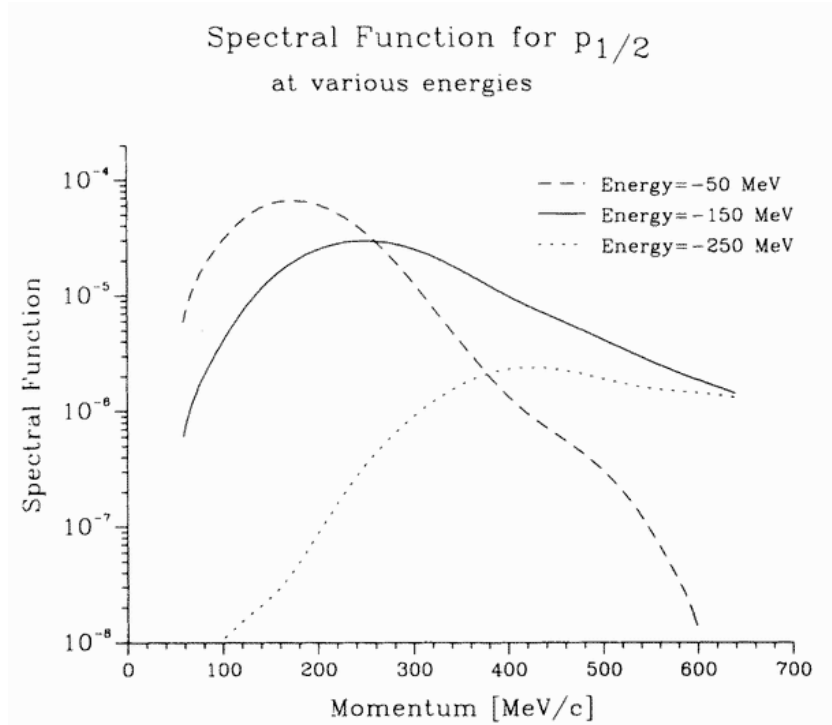}
\begin{minipage}[t]{16.5 cm}
\caption{\label{fig:npe12}
 The $p_{1/2}$ spectral strength as a function of momentum at fixed
energies corresponding to -50, -150, and -250 MeV. 
The results demonstrate the increasing importance of high-momentum components
with higher excitation energy in the $A-1$ system.
}
\end{minipage}
\end{center}
\end{figure}
\begin{figure}[!tb]
\vspace*{0.5cm}       
\begin{center}
\parbox[b]{.45\linewidth}{
\includegraphics[width=0.43\textwidth]{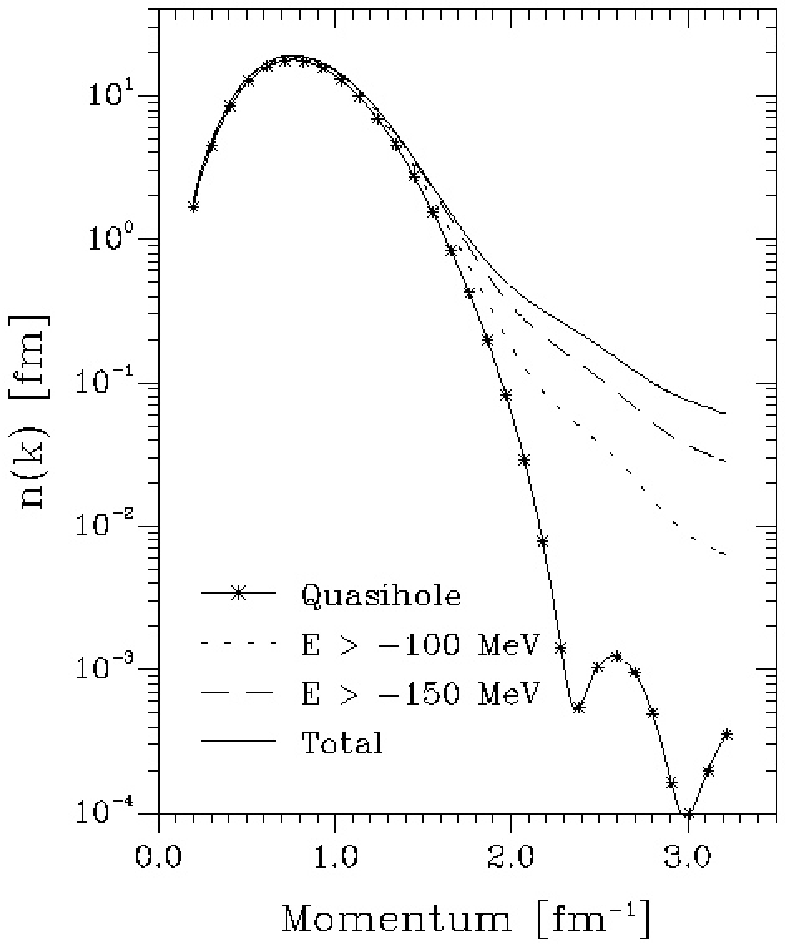}
    }
    \parbox[b]{7mm}{~}
    \parbox[b]{.45\linewidth}{
\includegraphics[width=0.4\textwidth]{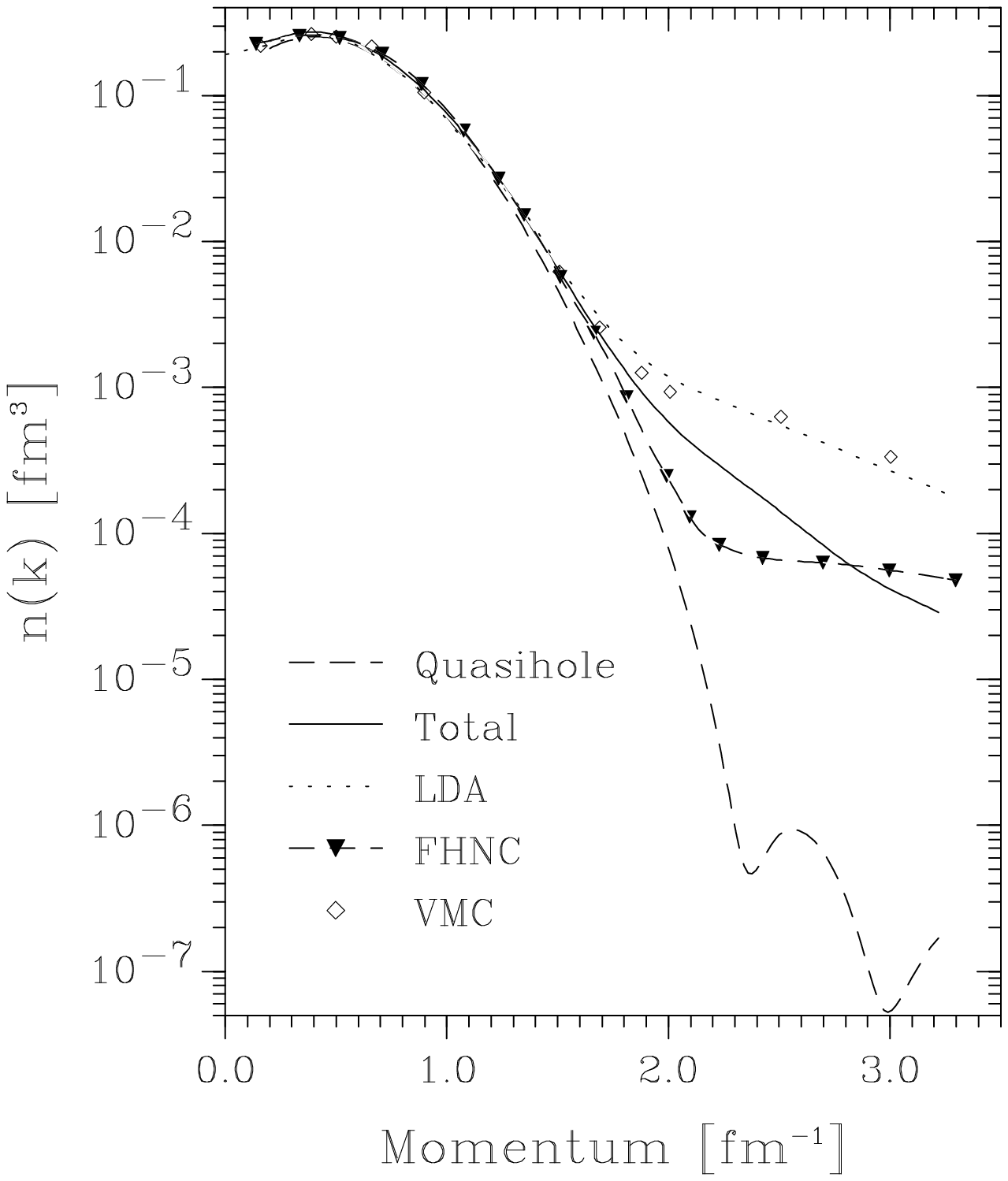}
     }
%
\begin{minipage}[t]{16.5 cm}
\caption{\label{fig:nto1}
The total momentum distribution of $^{16}$O. The left panel also 
displays the
quasihole contribution and the results obtained with various energy
cut-offs in the integration of the spectral functions.
The right panel compares the momentum distribution obtained with the
Green's function (total) with that of other approaches: local density 
approximation (LDA)~\cite{bffs94}, Fermi hypernetted chain (FHNC)~\cite{cff94}
 and Variational Monte Carlo (VMC)~\cite{pwp92}.
It should be noted that different normalizations were used for the
presentation of these panels.
}
\end{minipage}
\end{center}
\end{figure}

The results for the spectral function $S_{\ell j}^{c} (k; \omega)$
for the $p_{1/2}$ quantum numbers are shown 
at three different energies in Fig.~\ref{fig:npe12}.
The long-dashed curve corresponds to -50 MeV, the full curve
to -150 MeV, and finally the short-dashed curve to -250 MeV. 
From these results it is clear that an important change in
the momentum content of the sp strength occurs with
increasing excitation energy in the $A = 15$ system.
At higher excitation energy one finds more high-momentum
components. 
Moreover, these high-momentum components are not observed
in the quasihole states.
This can be concluded by considering Fig.~\ref{fig:nto1}
where the total momentum distribution, including the contribution from the
quasihole states, is shown in the left panel.
This requires the energy integration of the continuum hole strength
for each $k$ according to Eq.~(\ref{eq:occ_numb}).
This distribution is
presented 
for various energy cut-offs. The quasihole part reflects
the cross section for knockout reactions with small energy transfer, i.e. 
leading to the ground state of the final nucleus and excited states
up to $\approx 20$ MeV. The curve denoted by $E>$ -100 MeV reflects
the momentum distribution including all states of the final
nucleus up to around 80 MeV, etc. 
This result for $^{16}{\rm O}$ is very similar to the observation
for $^3{\rm He}$ made in Ref.~\cite{tadok}, where the contribution
of the ground state to ground state transition exhibits also
very few high-momentum components.
 As a consequence the high-momentum components of the momentum
distribution due to short-range correlations can be
observed mainly in
knockout experiments with an energy transfer of the order of 100 MeV
or more.
The right panel of Fig.~\ref{fig:nto1} compares the total momentum
distribution
for $^{16}{\rm O}$ obtained with different approaches, and different
interactions, all of which  properly include the effects of short-range
correlations~\cite{pmd2}.

To understand this result, it is important to recall
that the appearance of high-momentum components at a certain
energy in the $A-1$ system is related to the self-energy contribution
containing 2h1p states at this energy.
From energy conservation it is then clear that at low energy
it is much harder to find such states with a high-momentum particle
than at high energy.  
This same feature is observed in nuclear matter where the peak
of the sp spectral function for momenta above $k_F$
increases in energy as $k^2$~\cite{ciofb}.
As a result, the hole strength in nuclear matter as a function
of momentum shows the same tendency as the result shown in
Fig.~\ref{fig:npe12}~\cite{gdpr},
{\it i.e.} higher momenta become more dominant
at higher excitation energy. 

In order to show the importance of the continuum part of the spectral
functions as compared to the quasihole contribution and to visualize
the effects of correlations, we have included in Tab.~\ref{tab:noc} 
the particle numbers for each partial wave including the degeneracy of the
states 
\begin{equation}
{\hat n_{\ell j}} = 2 (2j+1) \int_{-\infty}^{\varepsilon_F}
dE\, \int_0^{\infty} dk\,k^2
S_{\ell j}(k,E) \;, \label{eq:nocn}
\end{equation}
also separating the contributions originating from the quasihole states 
and those due to the continuum~[as in Eq.~(\ref{eq:renor})]. 
As already noticed above only 14.025 out
of the 16 nucleons of $^{16}$O occupy the quasihole states in this
calculation (while the experimental data suggests a much 
smaller number).
 Another 1.13
nucleons are found in the 2h1p continuum with partial wave quantum
numbers of the $s$ and $p$ shell, while an additional 0.687 nucleons
are obtained from the continuum with orbital quantum numbers of the $d$
and $f$ shells. The distinction between quasihole and continuum
contributions is somewhat artificial for the $s_{1/2}$ orbital since
the coupling to low-lying 2h1p states leads to a strong fragmentation of the
strength~\cite{domit}, which is also observed experimentally~\cite{moug}.
It should be noted that the depletion of the occupation 
probabilities of the hole states, indicated in Tab~\ref{tab:noc}, 
is larger for the $s_{1/2}$ orbit. 
This feature can be ascribed to the closeness of the $s_{1/2}$
Hartree-Fock energy to the 2h1p continuum which yields more leakage of
strength to the continuum than for the $p_{1/2}$ and $p_{3/2}$ quasihole
states.
The sum of the particle numbers listed in Tab.~\ref{tab:noc} is 
slightly smaller (15.841) than the number particle in $^{16}$O
this due to the fact that only  partial waves
up to $\ell=3$ were taken into account~\cite{mkp95}. 
One must also keep in mind that the approach to the
sp Green's function discussed here is not number-conserving,
as the Green's functions used to evaluate the self-energy
are not determined in a self-consistent way, as dicussed in
Sec.~\ref{sec:SCGF}.

\begin{table}[!tb]
\begin{center}
\begin{tabular}{c|rrrr|rr}
 \hline
 \hline
&&&&&\\
$lj$&\multicolumn{1}{c}{$\hat n^{qh}$}&\multicolumn{1}{c}{$\hat n^c(E<-150)
$}&\multicolumn{1}{c}{$\hat n^c(E<-100)$}&
\multicolumn{1}{c|}{$\hat n^c$}&\multicolumn{1}{c}{$\hat n$}&
\multicolumn{1}{c}{$\hat n/(2(2j+1))$}
 \\
&&&&&&\\ \hline
&&&&&&\\
$s_{1/2}$ & 3.120 & 0.033 & 0.244 & 0.624 & 3.744 & 0.936\\
$p_{3/2}$ & 7.314 & 0.032 & 0.133 & 0.332 & 7.646 & 0.956\\
$p_{1/2}$ & 3.592 & 0.026 & 0.086 & 0.173 & 3.764 & 0.941\\
&&&&&&\\
$d_{5/2}$ & 0.0 & 0.033 & 0.106 & 0.234 & 0.234 & 0.020\\
$d_{3/2}$ & 0.0 & 0.036 & 0.108 & 0.196 & 0.196 & 0.025\\
$f_{7/2}$ & 0.0 & 0.025 & 0.063 & 0.117 & 0.117 & 0.007\\
$f_{5/2}$ & 0.0 & 0.032 & 0.084 & 0.140 & 0.140 & 0.012\\
&&&&&&\\
$\sum$ & 14.025 & 0.217 & 0.824 & 1.816 & 15.841 & \\
&&&&&&\\
 \hline
 \hline
\end{tabular}
\caption{\label{tab:noc}
Distribution of nucleons in $^{16}$O. 
Listed are the total occupation number $\hat n$ for various
partial waves [see Eq.~(\ref{eq:nocn})]
but also the contributions from the quasihole ($\hat n^{qh}$) and the 
continuum part ($\hat n^c$) of the spectral function, separately.
The continuum part is split further into contributions originating 
from energies $E$ below -150 MeV ($n^c(E<-150)$) and from energies 
below -100 MeV.
The last line shows the sum of particle numbers for all partial waves
listed.
}
\end{center}
\end{table}

\begin{table}[|tb]
\begin{center}
\begin{tabular}{c|rrr|rrr|rrr}
 \hline
 \hline
&&&&&&&&&\\
&\multicolumn{3}{c|}{HF} &\multicolumn{3}{c|}{BHF} &\multicolumn{3}{c}
{Total} \\
$lj$&\multicolumn{1}{c}{$\epsilon$}&\multicolumn{1}{c}{$t$}&
\multicolumn{1}{c|}{$\Delta E$}
&\multicolumn{1}{c}{$\epsilon$}&\multicolumn{1}{c}{$t$}&
\multicolumn{1}{c|}{$\Delta E$}
&\multicolumn{1}{c}{$\epsilon$}&\multicolumn{1}{c}{$t$}&
\multicolumn{1}{c}{$\Delta E$}\\
&&&&&&&&&\\
\hline
&&&&&&&&&\\
$s_{1/2}$ qh & -36.91 & 11.77 & -50.28 & -42.56 & 11.91 & -61.30 & -34.30
& 11.23 & -35.98 \\
$s_{1/2}$ c & &&&&&& -90.36 & 17.09 & -22.89 \\
$p_{3/2}$ qh & -15.35 & 17.62 & 9.08 & -20.34 & 18.95 & -5.59 & -17.90 &
18.06 & 0.37 \\
$p_{3/2}$ c  & & & & & & & -95.19 &
35.19 & -9.96\\
$p_{1/2}$ qh & -11.46 & 16.63 & 10.34& -17.07 & 18.46 &  2.76 & -14.14 &
17.19 & 5.47 \\
$p_{1/2}$ c  & & & & & & & -103.62&
35.94 & -5.84\\
$l>1$ c  & & & & & & & -98.87 &
63.17 & -12.27 \\
&&&&&&&&&\\
\hline
&&&&&&&&&\\
$E/A$(MeV)&\multicolumn{3}{c|}{-1.93} &\multicolumn{3}{c|}{-4.01} &\multicolumn{3}{c}
{-5.12} \\
$\left\langle r \right\rangle$(fm) &\multicolumn{3}{c|}{2.59}
&\multicolumn{3}{c|}{2.49} &\multicolumn{3}{c}
{2.55} \\
&&&&&&&&&\\
 \hline
 \hline
\end{tabular}
\end{center}
\caption{\label{tab:kol}
Ground-state properties of $^{16}$O. Listed are the energies
$\epsilon$ and kinetic energies $t$ of the quasihole states (qh) and the
corresponding mean values for the continuum contribution (c), normalized to 1,
for the various partial waves. Multiplying the
sum: $\frac{1}{2}(t+\epsilon )$ of these
mean values with the corresponding particle numbers of Tab.I, one obtains
the contribution $\Delta E$ to the energy of the ground state [as given
by the Migdal-Galitski-Koltun sumrule, Eq.~(\ref{eq:MGKsum})].
 Summing up all these contributions and dividing 
by the nucleon number yields the
energy per nucleon $E/A$. Furthermore, the {\em rms} radius
for nucleon distribution is shown. Results are presented for
the Hartree-Fock (HF), Brueckner-Hartree-Fock (BHF) and the complete
calculation of Secs.~\ref{sec:finiteSRC} and \ref{sec:finiteSRC2}
(Total). The particle numbers for the qh states in HF and BHF
are equal to the degeneracy of the states, all other occupation numbers
are zero. All the energies are given in MeV.
}
\end{table}

The contributions to the total energy, as derived from the 
Migdal-Galitski-Koltun sume rule~(\ref{eq:MGKsum}), are
shown in Tab.~\ref{tab:kol} for different angular momenta.
The first two columns give the analogous results obtained from the
solution of the Hartree-Fock (HF) and Brueckner-Hartree-Fock (BHF)
terms. The latter result includes the 2p1h
correction to the nuclear-matter $G$-matrix (see Ref.~\cite{mpd1}).
The BHF result 
continues to describe the nucleus in terms of fully occupied sp
 states as in HF. 
 However, as the sp states in BHF are more bound, the
gain in binding energy from HF to BHF is accompanied by a reduction
of the calculated radius of the nucleon distribution.
The inclusion of the 2h1p contributions to the self-energy in the complete
calculation reduces the absolute values of the quasihole energies (compare
BHF and ``Total'' in Tab.~\ref{tab:kol}). Despite this reduction of the
quasihole energies, however, the total binding energy is increased as
compared to BHF. This increase of the binding energy is mainly due to the
continuum part of the spectral function. Comparing various contributions
to the intergral in Eq.~(\ref{eq:MGKsum}), one finds that only 37\%
of the total energy is due to the quasiholes Eq.~(\ref{eq:skoqh}).
The dominating part (63\%)
results from the continuum part of the spectral functions
although this continuum part only represents 1.8 nucleons (that is
11\% 
of the total, see Tab.~\ref{tab:noc}).

Summarizing, the calculation of the complete energy dependence 
of the hole
spectral function demonstrates that the presence of
high-momentum components in the nuclear ground state will only show up
unambiguously at high excitation energy when probed by $(e,e'p)$ reactions.
 These deeply bound nucleons, not only generate the enhancement of the momentum
distribution for momenta $> 400$Mev/c, depicted in Fig.~\ref{fig:nto1},
but they are essential in understanding the binding of nuclear systems.
More discussion of these high-momentum components is given in 
Sec.~\ref{sec:Pb208deep}.

\subsection{Faddeev approach to the treatment of collective excitations
\label{sec:Faddform} }

The results for the spectroscopic factors of Sec.~\ref{sec:48Ca90Zr}
demonstrate that
the understanding of the occupation of quasihole states, can only be
achieved with a complete description of the coupling to the low-energy
collective motion of the system.
In this section, we review a method that has been recently developed
in Refs.~\cite{badi01,badi02},
with the aim of pursuing a complete description of these features.

 The degrees of freedom that contribute to long-range correlations
can be very complicated. However, when one is interested in the
dynamics at small missing energies, close to the Fermi level,
the problem can be simplified by focusing only on the excitations
at low energy, generally a small number.
 If the other phonons away from the Fermi surface are high in energy 
they can be expected to mix weakly or not at all with
the low-energy configurations.
It is important to observe that this picture can be too optimistic
if one works in terms of particle and hole orbitals as generated by
an IPM. For example the direct diagonalization of the nuclear
hamiltonian in a shell-model basis requires the inclusion of
configurations at rather high energy (several $\hbar\omega$),
while still obtaining sizable mixing of strength in the lowest excited 
states (see for example Ref.~\cite{warb92}).
In this respect, the Green's function approach appears to
account more easily for the shift in energy of the nuclear strength,
 as discussed in relation to the
results of Fig.~\ref{fig:rijs_Sh}.
The coupling to 2p1h fragments is responsible for moving 
sizable amounts of strength into the particle domain,
even at of several tens of MeVs from its original location.
 This mixing comes into the
sp propagator through the solution of the Dyson equation and 
therefore it is  included in the description of the low-energy
excitations.
 One thus expects that the dressed propagator represents a more
accurate approximation of the sp excitations. 
As a result,
the decoupling between  
low-energy excitations and those at higher energy
is more effectice when the full fragmentation
of the spectral strength is taken into account, as in the SCGF
theory. 
 To some extent, the use of fragmented propagators corresponds to 
expanding the wave function in terms of an already correlated single
particle basis, as in the Monte Carlo shell model of Ref.~\cite{Otsuka}.
From a more practical point of view, the energy distribution of 
the spectral function and the quenching of the spectroscopic factors
act as to renormalize the interaction between quasiparticles and
phonons (the real degrees of freedom in the system) allowing for a
better convergence of the results.
A second motivation to follow the approach of SCGF comes from the 
observation that a fully self-consistent calculation guarantees the 
fulfillment of the basic conservation laws and sum rules (see 
Sec.~\ref{sec:SCGF})~\cite{kaba2,baym62}.

\begin{figure}
 \begin{center}
    \parbox[b]{.4\linewidth}{
\caption[]{Example of diagrams that are summed to all orders by means of 
the Faddeev equations.
 \vspace{1in} 
\label{fig:FaddSum} }  } 
    \parbox[b]{7mm}{~}
    \parbox[b]{.45\linewidth}{
 \includegraphics[width=.7\linewidth]{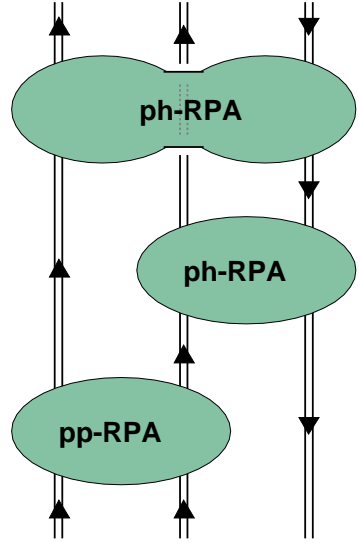} }
\end{center}
\end{figure}

In general, one may need to allow the sp motion to couple
to both pp(hh) and ph motion and at the
same time achieve a satisfactory
description of these collective phonons.
 It has been known for a long time that the excitation spectrum of $^{16}$O,
as obtained in TDA approximation, is particularly inadequate.
In order to account for the coupling to collective excitations
that are actually
observed in this nucleus it is necessary to at least consider 
an RPA description of the isoscalar negative parity states~\cite{cz2}.
To account for the low-lying isoscalar positive parity states an even more
complicated treatment will be required.
Sizable collective effects are also present in the pp and hh
excitations involving tensor correlations for isoscalar and pair
correlations for isovector states.
Each of the diagrams $c$) to $f$) of Fig.~\ref{fig:rijs_selfen} represents
one of these couplings in a separate way. Moreover
one must also remember that the diagrams coupling the
ph and sp channel are affected by a violation of the Pauli principle
already at the 2p1h/2h1p level,
due to neglect of the exchange between the freely propagating
hole or particle with the one propagating inside the ph phonon.
 More serious inconsistencies show up if one naively
sums the above diagrams, with the intent of accounting for both 
pp(hh) and ph collective modes. 
 In this case a double counting of the second order diagrams,
Fig.~\ref{fig:rijs_selfen}$a$) and $b$), would come into play and
there is no simple way to correct for this except to introduce
spurious poles in the
self-energy~\cite{badi01}.
Therefore the inclusion of both the pp~(hh) and ph collective phonons 
does require a complete treatment by summing these excitations
to all orders. Such an expansion sums a series of diagrams like
the one shown in Fig.~\ref{fig:FaddSum}. The interplay between the
various phonons
would ultimately generate an expansion of the 2p1h/2h1p propagator
$R(\omega)$ [see Sec.~\ref{sec:EOM}] 
that automaically resolves the above issues.
In Ref.~\cite{badi01}, it was shown that this all order 
summation can by achieved by extending the technique of
the Faddeev equations for the three-body problem~\cite{Fadd1,JoachBk,glock}
to the case of three interacting quasiparticle excitations.
 A corresponding approximation for the self-energy can then
be obtained from Eq.~(\ref{eq:selfen_R}).

 To describe the formalism of the Faddeev equations as applied to
the many-body case, we consider the Bethe-Salpeter equation for the 2p1h
propagator [which is analogous to Eq.~(\ref{eq:Pi-3t_BSE}) for the ph case].
 In the time formulation, one has
\begin{eqnarray}
\lefteqn{
 R_{\mu \nu \lambda , \alpha \beta \gamma}(t_1 , t_2 , t_3 ; t')}
    \hspace{.5in} & &
\nonumber \\
  &=&  g_{\mu \alpha}(t_1 - t') g_{\nu \beta}(t_2 - t') g_{\gamma \lambda}(t' - t_3)
     ~-~ g_{\nu \alpha}(t_2 - t') g_{\mu \beta}(t_1 - t') g_{\gamma \lambda}(t' - t_3)
\nonumber \\
  & & + g_{\mu \mu'}(t_1 - t_1') g_{\nu \nu'}(t_2 - t_2') g_{\lambda' \lambda}(t_3' - t_3)
\nonumber \\  
  & & ~ \times \; K_{\mu' \nu' \lambda' , \alpha' \beta' \gamma'}(t_1' , t_2' , t_3' ; t_4' , t_5' , t_6')
    \;  R_{\alpha' \beta' \gamma' , \alpha \beta \gamma}(t_4' , t_5' , t_6' ; t') \; ,
\label{eq:Rbt}
\end{eqnarray}
where here and in the rest of this section we use the convention
of summing over
all repeated indices and integrate from $-\infty$
to +$\infty$ over all repeated time variables, unless specified otherwise.
%
%
The interaction vertex is given by
\begin{eqnarray}
\lefteqn{
 K_{\mu \nu \lambda , \alpha \beta \gamma}(t_1 , t_2 , t_3 ; t_4 , t_5 , t_6) }
    \hspace{.5in} & &
\nonumber \\
  &=& K^{(ph)}_{\nu \lambda , \beta \gamma}(t_2 , t_3 ; t_5 , t_6) g^{-1}_{\mu \alpha}(t_1 - t_4) ~+~
  K^{(ph)}_{\mu \lambda , \alpha \gamma}(t_1 , t_3 ; t_4 , t_6) g^{-1}_{\nu \beta}(t_2 - t_5)
\nonumber \\
 & & ~+~ K^{(pp)}_{\mu \nu , \alpha \beta}(t_1 , t_2 ; t_4 , t_5) g^{-1}_{\gamma \lambda}(t_6 - t_3)
 ~+~K^{(pph)}_{\mu \nu \lambda , \alpha \beta \gamma}(t_1 , t_2 , t_3 ; t_4 , t_5 , t_6)  \; .
\label{eq:Rbtvertex}
\end{eqnarray}
In Eq.~(\ref{eq:Rbtvertex}),$K^{(ph)}$ represent the ph irreducible
vertex that appears in Eq.~(\ref{eq:Pi-3t_BSE}) while
$K^{(pp)}$ and $K^{(pph)}$ are the pp and 2p1h irreducible vertices.
It should be noted that in Eq.~(\ref{eq:selfen_R}) the propagator
$R_{\mu \nu \lambda , \alpha \beta \gamma}$ is only required at
two times and therefore the complete knowledge of its time dependence
is not necessary to obtain the self-energy.
On the other hand, the dependence on the time variables $t_1$, $t_2$ and $t_3$
is employed in the Bethe-Salpeter Eq.~(\ref{eq:Rbt}), thus 
an exact solution of the 2p1h motion would require at
least a 4-time object.

Equation (\ref{eq:Rbt}) can be reduced to a set of coupled equations in a way
similar to the method proposed by Faddeev to solve the three-body
problem~\cite{Fadd1,JoachBk}.
The inclusion of pp and ph RPA phonons in a consistent way requires
an approach that provides a natural framework for correctly
iterating in the self-energy quantities that have already been summed
to all orders, like pp and ph RPA phonons.
The calculation reviewed in Sec.~\ref{sec:Faddres}
neglect the contribution of the irreducible $K^{(pph)}$ term
in Eq.~(\ref{eq:Rbtvertex}) and
therefore require only three Faddeev components.
Following standard notation in the literature~\cite{glock},
the component  
$R^{(i)}_{\mu \nu \lambda , \alpha \beta \gamma}$ is related to
the sum of all diagrams ending with a vertex between legs $j$
and $k$, with $(i,j,k)$ cyclic permutations of $(1,2,3)$.
We will employ the convention in which the third leg propagates in the opposite
direction with respect to the first two.
The Faddeev components $R^{(i)}$ can be written in terms of the 2p1h
propagator $R$ and the contribution of the three dressed but noninteracting
sp propagators.
This definition is given in detail here for all three components, omitting
explicit reference to the time variables for convenience of notation
\begin{subequations}
\label{eq:faddcmp}
\label{eq:faddcmpall}
\begin{eqnarray}
    R^{(1)}_{\mu \nu \lambda , \alpha \beta \gamma} ~=~
     g_{\nu \epsilon} g_{\rho \lambda} ~
        K^{(ph)}_{\epsilon \rho,\eta \sigma} ~
          R_{\mu \eta \sigma , \alpha \beta \gamma}
     ~+~ \frac{1}{2} \left(
                       g_{\mu \alpha} \; g_{\nu \beta} \; g_{\gamma \lambda}
                     - g_{\nu \alpha} \; g_{\mu \beta} \; g_{\gamma \lambda}
                     \right) \; ,
\label{eq:faddcmp1}
\end{eqnarray}
\begin{eqnarray}
    R^{(2)}_{\mu \nu \lambda , \alpha \beta \gamma} ~=~
     g_{\mu \epsilon} g_{\rho \lambda} ~
       K^{(ph)}_{\epsilon \rho,\eta\sigma} ~
         R_{\eta \nu \sigma , \alpha \beta \gamma}
     ~+~ \frac{1}{2} \left(
                       g_{\mu \alpha} \; g_{\nu \beta} \; g_{\gamma \lambda}
                     - g_{\nu \alpha} \; g_{\mu \beta} \; g_{\gamma \lambda}
                     \right) \; ,
\label{eq:faddcmp2}
\end{eqnarray}
\begin{eqnarray}
    R^{(3)}_{\mu \nu \lambda , \alpha \beta \gamma} ~=~
     g_{\mu \epsilon} g_{\nu \rho} ~
       K^{(pp)}_{\epsilon \rho,\eta\sigma} ~
         R_{\eta \sigma \lambda , \alpha \beta \gamma}
     ~+~ \frac{1}{2} \left(
                       g_{\mu \alpha} \; g_{\nu \beta} \; g_{\gamma \lambda}
                     - g_{\nu \alpha} \; g_{\mu \beta} \; g_{\gamma \lambda}
                     \right) \; ,
\label{eq:faddcmp3}
\end{eqnarray}
%
\end{subequations}
where $\frac{1}{2}$ is a symmetry factor.
 With these definitions the full propagator $R(\omega)$ is given by
\begin{equation}
    R_{\mu \nu \lambda , \alpha \beta \gamma} ~=~
   \sum_{i=1,2,3}     R^{(i)}_{\mu \nu \lambda , \alpha \beta \gamma}
     ~-~ \frac{1}{2} \left( g_{\mu \alpha} \; g_{\nu \beta} \; g_{\gamma \lambda}
        - g_{\nu \alpha} \; g_{\mu \beta}  \; g_{\gamma \lambda} \right) \; .
\label{eq:faddfullR}
\end{equation}
The Faddeev equations now take the following form
\begin{eqnarray}
    R^{(i)}_{\mu \nu \lambda , \alpha \beta \gamma} &=&
          \frac{1}{2} \left( g_{\mu \alpha} \; g_{\nu \beta} \; g_{\gamma \lambda}
               - g_{\nu \alpha} \; g_{\mu \beta}  \; g_{\gamma \lambda} \right)
\nonumber \\ 
   & & +~  g_{\mu \mu'} \; g_{\nu \nu'} \; g_{\lambda' \lambda} ~
            \Gamma^{(i)}_{\mu' \nu' \lambda' , \mu'' \nu'' \lambda''} ~
            (  R^{(j)}_{\mu'' \nu'' \lambda'' , \alpha \beta \gamma} ~+~
               R^{(k)}_{\mu'' \nu'' \lambda'' , \alpha \beta \gamma} ) \; ,
                ~ ~ i = 1, 2, 3
\label{eq:BSFaddeq}
\end{eqnarray}
where the $\Gamma^{(i)}_{\mu \nu \lambda , \alpha \beta \gamma}$ vertices
obey the following symmetry relations and are defined by 
\begin{subequations}
\label{eq:G2vsBS}
\begin{eqnarray}
   \Gamma^{(1)}_{\mu \nu \lambda , \alpha \beta \gamma}
                                       (t_1 , t_2 , t_3 ; t_4 , t_5 , t_6) =
   \Gamma^{(2)}_{\nu \mu \lambda , \beta \alpha \gamma}
                                       (t_2 , t_1 , t_3 ; t_5 , t_4 , t_6) =
  g^{-1}_{\mu \alpha}(t_1 - t_4) ~
 \tilde{\Gamma}^{(ph)}_{\nu \lambda , \beta \gamma}(t_2 , t_3 ; t_5 , t_6) \; ,
\label{eq:G2vsBS_ph}
\end{eqnarray}
\begin{eqnarray}
   \Gamma^{(3)}_{\mu \nu \lambda , \alpha \beta \gamma}
                                       (t_1 , t_2 , t_3 ; t_4 , t_5 , t_6) =
   \Gamma^{(3)}_{\nu \mu \lambda , \beta \alpha \gamma}
                                       (t_2 , t_1 , t_3 ; t_5 , t_4 , t_6) =
    g^{-1}_{\gamma \lambda}(t_6 - t_3) ~
     \tilde{\Gamma}^{(pp)}_{\mu \nu , \alpha \beta}(t_1 , t_2 ; t_4 , t_5) \; .
\label{eq:G2vsBS_pp}
\end{eqnarray}
\end{subequations}
In Eqs.~(\ref{eq:G2vsBS}), $g^{-1}(\tau)$ is the inverse sp propagator
and matrices $\tilde{\Gamma}^{(pp)}$ and
$\tilde{\Gamma}^{(ph)}$ correspond to the proper time orderings of the
four-point reducible vertex $\Gamma^{4-pt}$ that appears in
Eq.~(\ref{eq:Gamma_4pt}). These 
quantities solve the Bethe-Salpeter equation for the
pp and ph motion, respectively
\begin{subequations}
\label{eq:BetheSalpeter}
\begin{eqnarray}
 \lefteqn{
  \tilde{\Gamma}^{(pp)}_{\gamma \delta , \alpha \beta}(t_1 , t_2 ; t_3 , t_4) ~=~
   K^{(pp)}_{\gamma \delta , \alpha \beta}(t_1 , t_2 ; t_3 , t_4) }
     \hspace{.5in} & &
\nonumber \\
  &+&
    \tilde{\Gamma}^{(pp)}_{\gamma \delta , \mu \nu}(t_1 , t_2 ; t_1' , t_2') ~
     g_{\mu \eta}(t_1' - t_3') \; g_{\nu \sigma}(t_2' - t_4') ~
      K^{(pp)}_{\eta \sigma , \alpha \beta}(t_3' , t_4' ; t_3 , t_4) \; ,
\label{eq:ppBS}
\end{eqnarray}
\begin{eqnarray}
 \lefteqn{
  \tilde{\Gamma}^{(ph)}_{\gamma \delta , \alpha \beta}(t_1 , t_2 ; t_3 , t_4) ~=~
   K^{(ph)}_{\gamma \delta , \alpha \beta}(t_1 , t_2 ; t_3 , t_4) }
     \hspace{.5in} & &
\nonumber \\ 
  &+&
    \tilde{\Gamma}^{(ph)}_{\gamma \delta , \mu \nu}(t_1 , t_2 ; t_1' , t_2') ~
     g_{\mu \eta}(t_1' - t_3') \; g_{\sigma \nu}(t_4' - t_2') ~
      K^{(ph)}_{\eta \sigma , \alpha \beta}(t_3' , t_4' ; t_3 , t_4) \; .
\label{eq:phBS}
\end{eqnarray}
\end{subequations}

Apart from neglecting the $K^{(pph)}$ vertex,
Eq.~(\ref{eq:BSFaddeq}) is otherwise an exact
equation for the 2p1h propagator, that involves quantities
which depend on several times and for this reason still
intractable.
In order to construct a manageable approximation scheme
one first requires a reduction to a set of equations
that involve only two-time quantities but still include the
relevant physical ingredients of interest.
Here we give only a brief overview of this issue. More
datails can be found in Refs.~\cite{badi01,carloth}

\begin{figure}
 \begin{center}
\includegraphics[width=0.80\textwidth]{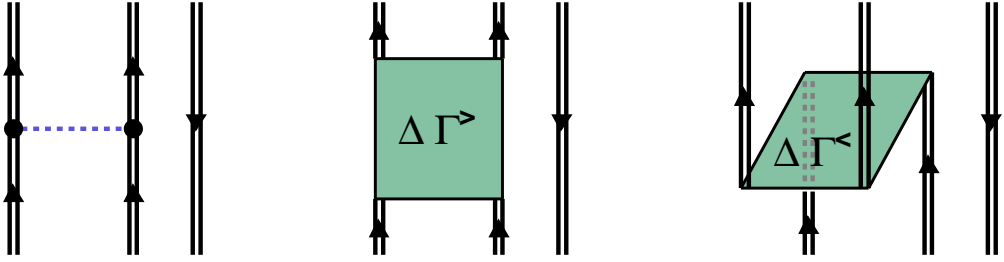}
\vspace*{-.5cm}       
\begin{minipage}[t]{16.5 cm}
\caption[]{Diagrams that are included in the definition of the vertex
for the pp channel.
 Here $\Delta \Gamma^>$ and $\Delta \Gamma^<$ are the forward- and
backward-going part of the energy dependent contribution to the pp DRPA
vertex~(\ref{eq:ppBS}).
  The contribution of these three diagrams can be 
factorized in an expression of the form ${G^0}^> \; \Gamma^{(3)} \; {G^0}^>$ 
only after having redefined the propagators ${G^0}^>$ and $\Gamma^{(3)}$
to depend also on the particle and hole
fragmentation indices ($n$,$n'$,$k$).
\label{fig:Gammai_vs_DRPAS} }
\end{minipage}
 \end{center}
\end{figure}

The calculations of Sec.~\ref{sec:Faddres} employ a bare interaction
$V_{\alpha \beta , \gamma \delta}$ for the vertices
$K^{(pp)}$ and $K^{(ph)}$ so that
the Bethe-Salpeter equations~(\ref{eq:BetheSalpeter}) reduce to the usual
dressed RPA (DRPA) equations~\cite{schuckbook,DRPApaper}. The solutions
of these equations depend only on two times.
These pp and ph phonons correspond to the dressed version of the phonons
that are considered in Ref.~\cite{rijs} (see also
Fig.~\ref{fig:rijs_selfen}) and represent 
the minimum step that maintains the simultaneous inclusion
of both pp and ph collective low-lying excitations in the self-energy.
 However, it is possible to extend these vertex to include
correlations that go beyond the RPA level [see Sec.~\ref{sec:BSEO16}].
A second approximation requires to construct the 
Faddeev equations~(\ref{eq:BSFaddeq}) for only two time
variables. In doing this, the sp line that propagates freely
in the Faddeev vertices~(\ref{eq:G2vsBS}) remains blocked 
and can propagate only in one time direction. This means that the 
set of Eqs.~(\ref{eq:BSFaddeq}) split up in two
separate expansions for the 2p1h and the 2h1p components.
For this reason, the pp and ph phonons will be summed only in
one time direction in a TDA way contributing separately to the 2p1h
and 2h1p components of the self-energy.
However,  the collective RPA correlations in the pp and ph channels have
already been computed through Eqs.~(\ref{eq:BetheSalpeter}) and therefore 
remain properly included in the this approach. An example of the 
diagrams that are included in the pp channel, Eq.~(\ref{eq:G2vsBS_pp}),
is shown in Fig.~\ref{fig:Gammai_vs_DRPAS}.

 The remaining complication, related to the use of dressed propagators,
concerns the interactions vertices~(\ref{eq:G2vsBS}).
In order to keep track of which quasiparticle or quasihole
 is coupled to the DRPA phonons, the $\Gamma^{(i)}$ and the
propagators $R^{(i)}$ need to be redefined in such a way that their matrix
elements also depend on the indices ($n$, $n'$, $k$), which label the
fragments of the sp propagator, Eq.~(\ref{eq:fullg_Leh}).
This implies that the eigenvalue equations will involve
summations on both the sp indices ($\alpha$, $\beta$, $\gamma$) and the
ones corresponding to the fragmentation, ($n_\alpha$, $n_\beta$, $k_\gamma$).
 The particular expression to be employed for the vertices $\Gamma^{(i)}$
depends on the approximation chosen for the pp and ph phonons
and it can be crucial to control spurious solutions
in the Faddeev equations. Here we only note that the diagrams of 
Fig.~\ref{fig:Gammai_vs_DRPAS} are those required for the DRPA, as
discussed in Ref.~\cite{badi01}. 
 The 2p1h propagator and its Faddeev components, as defined
in Eqs.~(\ref{eq:faddcmpall}), are recovered only at
the end by summing the solutions over all values
of ($n_\alpha$, $n_\beta$, $k_\gamma$) and~($n_\mu$,~$n_\nu$,~$k_\lambda$).

 Putting together all the above considerations, the resulting approximation to
the Faddeev equations~(\ref{eq:BSFaddeq}) can be rewritten in a 
way where all the propagators involved depend only on one energy variable
(or two time variables).
 The forward-going part of this expansion can be written as follows
(after Fourier transformation to the energy representation)
\begin{eqnarray}
  \lefteqn{
  R^{(i)}_{\mu    n_\mu    \nu   n_\nu   \lambda k_\lambda , 
           \alpha n_\alpha \beta n_\beta \gamma  k_\gamma}(\omega) }
        \hspace{.5in} & &
\nonumber  \\
   &=&  \frac{1}{2}\left( 
      {G^0}^>_{\mu    n_\mu    \nu   n_\nu   \lambda k_\lambda , 
               \alpha n_\alpha \beta n_\beta \gamma  k_\gamma}(\omega)
    - {G^0}^>_{\nu   n_\nu   \mu    n_\mu    \lambda k_\lambda ,
               \alpha n_\alpha \beta n_\beta \gamma  k_\gamma}(\omega) \right)
\nonumber  \\
  & & +~ {G^0}^>_{\nu  n_\nu   \mu  n_\mu   \lambda  k_\lambda ,
                  \mu' n_\mu'  \nu' n_\nu'  \lambda' k_\lambda'}(\omega) ~
    \Gamma^{(i)}_{\nu'  n_\nu'  \mu'  n_\mu'   \lambda'  k_\lambda' ,
                  \mu'' n_\mu'' \nu'' n_\nu''  \lambda'' k_\lambda''}(\omega) ~
\nonumber  \\
  & & ~ ~ ~ ~\times ~
  \left( R^{(j)}_{\mu''    n_\mu''    \nu''   n_\nu ''  \lambda'' k_\lambda'' , 
                  \alpha n_\alpha \beta n_\beta \gamma  k_\gamma}(\omega) ~+~
         R^{(k)}_{\mu''    n_\mu''    \nu''   n_\nu ''  \lambda'' k_\lambda'' , 
                  \alpha n_\alpha \beta n_\beta \gamma  k_\gamma}(\omega)
       \right)  \; , ~ ~ i=1,2,3 ~ .
\label{eq:FaddTDA}
\end{eqnarray}
 In Eq.~(\ref{eq:FaddTDA}), ${G^0}^>$ is the forward-going part of the 2p1h
propagator for three dressed but noninteracting lines. This quantity is 
therefore defined
as follows (without implicit summation):
\begin{eqnarray}
 \lefteqn{
  {G^0}^>_{\mu    n_\mu    \nu   n_\nu   \lambda k_\lambda , 
           \alpha n_\alpha \beta n_\beta \gamma  k_\gamma}(\omega) ~=~
        \delta_{n_\mu , n_\alpha} \;
        \delta_{n_\nu , n_\beta} \;
        \delta_{k_\lambda , k_\gamma} }
    \hspace*{0.5cm} & &
\nonumber \\
  & & \times   \frac{ 
   \langle \Psi^A_0 \vert c_\mu            \vert \Psi^{A+1}_{n_\mu}     \rangle
   \langle \Psi^A_0 \vert c_\nu            \vert \Psi^{A+1}_{n_\nu}     \rangle
   \langle \Psi^A_0 \vert c^{\dag}_\lambda \vert \Psi^{A-1}_{k_\lambda} \rangle
            \; 
   \langle \Psi^{A+1}_{n_\alpha} \vert c^{\dag}_\alpha \vert \Psi^A_0 \rangle
   \langle \Psi^{A+1}_{n_\beta}  \vert c^{\dag}_\beta  \vert \Psi^A_0 \rangle
   \langle \Psi^{A-1}_{k_\gamma} \vert c_\gamma        \vert \Psi^A_0 \rangle
         }
    { \omega - ( \varepsilon^+_{n_\alpha} + \varepsilon^+_{n_\beta} -
                                     \varepsilon^-_{k\gamma} ) + i \eta } \; .
\label{eq:G0fw}
\end{eqnarray}
A completely  analogous set of equations apply for the backward-going
(2h1p) expansion of $R(\omega)$.

 Equations (\ref{eq:FaddTDA}) reduce the general 
``Faddeev-Bethe-Salpeter'' expansion, Eq.~(\ref{eq:BSFaddeq}),
to a tractable set of equations involving only two-time objects.
It is important to note that these equations are still expressed in terms
of the self-consistent solution $g_{\alpha \beta}(\omega)$ and include
both the collective pp and ph phonons in a correct way. 
 Thus they maintain all the features relevant for the physics that have been
discussed at the beginning of this subsection.

\subsection{Single-particle spectral function of ${}^{16}{\rm O}$
\label{sec:Faddres} }

A self-consistent solution of the spectral function of ${}^{16}{\rm O}$
was obtained in Ref.~\cite{badi02} by solving the set of 
Faddeev equations~(\ref{eq:FaddTDA}). The Dyson equation was solved within
a model space consisting of harmonic oscillator sp states.
An oscillator parameter $b =$ 1.76~fm was chosen (corresponding to
$\hbar\omega =$ 13.4~MeV) and all the first four major
shells (from $1s$ to $2p1f$) plus the $1g_{9/2}$ where included.
 Inside this model space, the interaction used was a Brueckner
$G$-matrix~\cite{CALGM} derived from the Bonn-C potential~\cite{bonnABC}.
%
%
%
For the solution of the Faddeev equations a $G$-matrix evaluated at a
fixed starting energy $\omega=$-25~MeV was chosen as a suitable
average of the energy the most important 2h1p states that couple to
the experimentally observed quasiholes~\cite{leus}.
However, the energy dependence of $G(\omega)$ was taken into account
in the calculation of the mean field part of the self-energy, resulting
in the Brueckner-Hartree-Fock approximation (this can be compared
 to the HF case in
Eq.~(\ref{eq:selfen_HF})~),
\begin{equation}
 \Sigma^{BHF}_{\alpha \beta}(\omega) =
   i \sum_{\gamma \delta} \int \frac{d \omega '}{2 \pi} 
   G_{\alpha \gamma, \delta \beta}(\omega + \omega ')
      g_{\gamma \delta}(\omega ') 
\label{eq:SigmaBHF}
\end{equation}
The self-energy was then computed as in Eq.~(\ref{eq:selfen_R}),
with Eq.~(\ref{eq:SigmaBHF}) instead of $\Sigma^{HF}$.

The normalization (spectroscopic factor) of each sp fragment was 
computed by means of Eq.~(\ref{eq:2nd21}), which we rewrite here in the form
\begin{equation}
 Z_k = \sum_{\alpha}
\left| {\cal Y}^{k}_{\alpha} \right|^2
    =  1 + 
     \sum_{\alpha , \beta}   \left( {\cal Y}^{k}_{\alpha} \right)^* 
       \left.  \frac{ \partial \Sigma^\star_{\alpha \beta}(\omega) }
                    { \partial \omega}
       \right|_{\omega = \varepsilon^{-}_k }
	       {\cal Y}^{k}_{\beta}  \; \; .
\label{eq:normBHF}
\end{equation}
where ${\cal Y}^{k}_{\alpha} \sim  {\mbox{$\langle {\Psi^{A-1}_k} \vert $}}
c_\alpha {\mbox{$\vert {\Psi^A_0} \rangle$}}$ represent the $k$-th solution
for the spectrocopic amlitudes and $\varepsilon^{-}_k \sim E^A_0 - E^{A-1}_k$
the corresponding eigenvalue.
It is important to note that the 
$\Sigma^{BHF}(\omega)$ component of the self-energy also 
contributes to
reduction the spectroscopic factor. This energy dependence, associated with the
use of the $G$-matrix in Eq.~(\ref{eq:SigmaBHF}), is not 
present in a standard HF self-energy and originates from the
states at high energy that are excluded from the model space.
%
 In Ref.~\cite{GeurtsO16}, it was found that this procedure properly
accounts 
for the depletion of spectroscopic factors that is induced by SRC,
at least for the normally occupied shells in the IPM.
 
The solution of RPA equations may lead to collective
instabilities whenever interactions are employed that are too attractive.
A particular sensitivity to the strength of the $G$-matrix interaction
was found in the calculation of~\cite{badi02,badi03}, only for the lowest
isoscalar $0^+$ solution. For this reason an artificial
shift of its eigenvalue to the experimental energy of
first excited state, 6.049~MeV, was maintained 
during the successive iterations.
 Note that the screening due the redistribution of the sp
strength tends
to reduce the instability problem in DRPA.
Moreover, the origin of the instability for this particular solution
has been identified and corrected later, in Ref.~\cite{badi03}.

\begin{table}[!tb]
 \begin{center}
 \begin{tabular}{ccccccccccccc}
  \hline 
  \hline 
    Shell      & ~~ & TDA    &~~ & RPA    & ~~& 1st itr. &~ ~& 2nd itr. &~~ & 3rd
    itr. &~ & 4th itr. \\
  \hline 
 $Z_{p_{1/2}}$ & & 0.775  & & 0.745  & & 0.775    & &  0.777   & & 0.774    & & 0.776 \\ \\
 $Z_{p_{3/2}}$ & & 0.766  & & 0.725  & & 0.725    & &  0.727   & & 0.722    & & 0.724    \\
               & &        & &        & & 0.015    & &  0.027   & & 0.026    & & 0.026 \\ \\
  \hline  \\
%
%
%
  Total occ.   & & 14.56  & & 14.56  & & 14.56    & &  14.57   & & 14.58    & & 14.63   \\
  \hline 
  \hline 
 \end{tabular}
  \end{center}
    \parbox[t]{1.\linewidth}{
      \caption[]{\label{tab:SpectFact}
  Hole spectroscopic factors ($Z_{\alpha}$) for knockout of a $\ell =1$
      proton from ${}^{16}{\rm O}$.
       These results refer to the initial IPM calculation in TDA and RPA
      and to the first
      four iterations of the DRPA equations.
       All the values are given as a fraction of the corresponding IPM value.
       Also included is the total number of nucleons, as deduced from the
       complete one-hole spectral function, for each iteration.}
      }
\end{table}

The Faddeev equations were solved first in terms of a IPM input
propagator in both TDA and RPA approximations.
 For an IPM ansatz, the TDA calculation is equivalent to the one of 
Ref.~\cite{GeurtsO16} and yields the same results for the spectroscopic 
factors of the main particle and hole shells
close to the Fermi energy.
  These correspond to 0.775 for $p_{1/2}$ and 0.766
for $p_{3/2}$ as reported in Tab.~\ref{tab:SpectFact}.
 The introduction of correlations reduces these values and brings them
down to 0.745 and 0.725, respectively.
This result reduces the discrepancy with the experiment by about 4\% and
shows that collectivity beyond the TDA level is relevant to
explain the quenching of spectroscopic factors.
 Since the present formalism does not account for center-of-mass
effects, the above quantities need to be increased by about 7\%
before they are compared with the experiment~\cite{diedef,O16com}.

\begin{figure}[!tb]
 \begin{center}
 \includegraphics[height=4in]{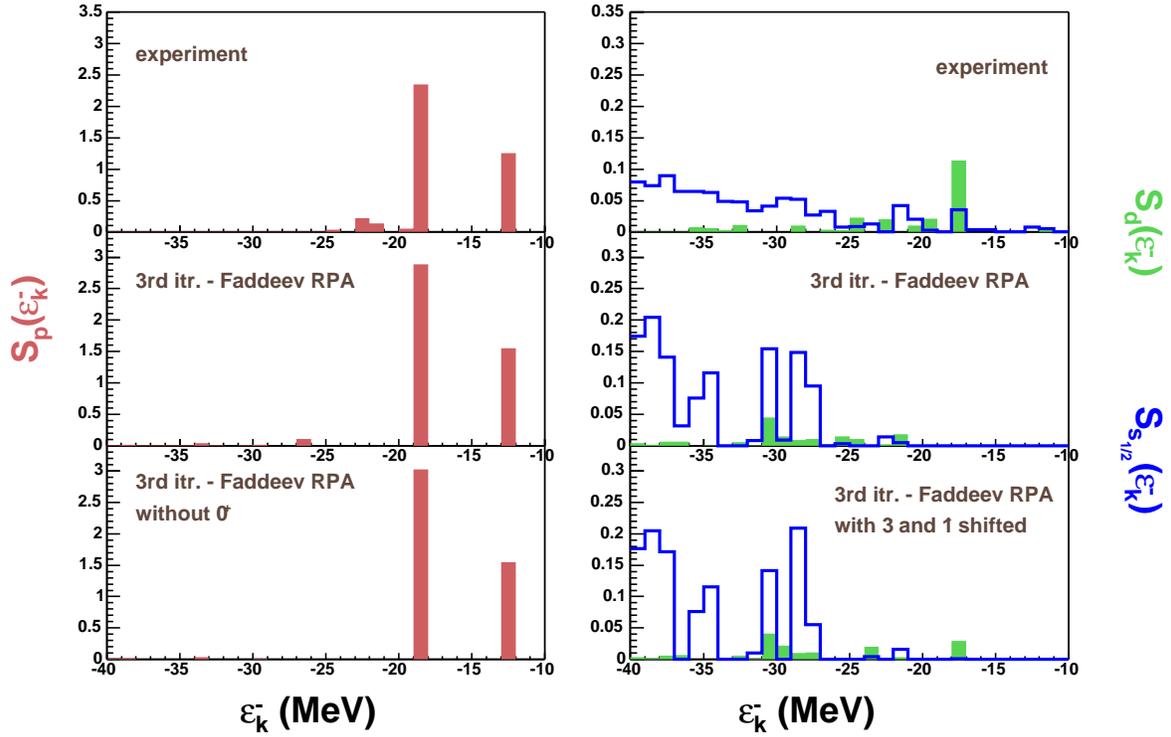}
\begin{minipage}[t]{16.5 cm}
\caption[]{One-proton removal strength as a function of the 
   hole sp energy $\varepsilon^{-}_k = E^A_0 - E^{A-1}_k$
    for ${}^{16}{\rm O}$ for angular momentum
  $\ell =1$ (left) and $\ell =0,2$ (right).
   For the positive parity
   states, the solid bars correspond
   to results for $d_{5/2}$ and $d_{3/2}$ orbitals, while the
   thick lines refer to $s_{1/2}$.  
   The top panels show the experimental values taken from~\cite{leus}.
   The middle panels give the theoretical results for the self-consistent
   spectral function.
   The bottom panels show the results obtained by repeating the 3$^{\rm rd}$
   iteration with a modified ph-DRPA spectrum, in which
   the lowest eigenstates have been either removed or
   shifted to the corresponding
   experimental values.
\label{fig:p+sd_hsf} }
\end{minipage}
 \end{center}
\end{figure}

The RPA results have been iterated several times, with the aim of studying
the effects of fragmentation on the RPA phonons and, subsequently,
on the spectral strength. 
Since one is mainly interested in the low-energy excitations,
it is sufficient to keep track only of the strongest 
fragments that appear ---close to the Fermi energy--- in the
(dressed) sp propagator, Eq.~(\ref{eq:fullg_Leh}),
while the residual strength is collected in two effective
poles, above and below the Fermi level, for each orbital~\cite{jyuan,badi02}.
As shown in Tab.~\ref{tab:SpectFact}, only 
a few iterations are required to reach convergence for the values
of the spectroscopic factors. The converged 
distribution of one-hole strength is compared with the experimental one
in Fig.~\ref{fig:p+sd_hsf}. 
 The main difference between these results and the one obtained by using
an IPM input is the appearance of a second smaller $p_{3/2}$ fragment at 
-26.3~MeV. This peak arises in the first two iterations and appears
to become stable in the last one, with a spectroscopic factor of 2.6\%.
 This can be interpreted as a peak that describes the fragments seen
experimentally at slightly higher energy.
  This result corresponds to the first time that such a fragment is
obtained in calculations of the spectral strength.

 A second effect of including fragmentation in the construction of the RPA 
phonons is to increase the strength of the main hole
peaks.
  The $p_{1/2}$ 
strength increases from the 0.745 obtained with IPM input to 0.776,
essentially canceling the ``improvement'' gained by 
the introduction of RPA correlations over the TDA ones.
 The main peak of the $p_{3/2}$ remains at 0.722 but the appearance of the
secondary fragment slightly increases the overall strength for this orbital
as well.
 This behavior is due to the fact that redistribution of the 
strength (and principally, the reduction of the main spectroscopic
factors) tends to screen the nuclear interaction, with respect to
the IPM case.
 The consequence is a reduced effect of RPA correlations
when fragmentation is included in the construction of the phonons.
This feature has also been observed in other self-consistent calculations
of the sp spectral strength, for example in nuclear matter~\cite{libth}.
 Obviously, this makes the disagreement with experiments a little worse
and additional work is needed to resolve the disagreement with the data.
Nevertheless, it is clear 
that fragmentation is a relevant feature of nuclear systems and that
it has to be properly taken into account.

 Together with the main fragments, the Dyson equation produces also a
large number of solutions with small spectroscopic factors.
 This strength extends down to about -130~MeV for the one-hole spectral
function and up to about 100~MeV for the one-particle case. 
This background represents 
the strength that is removed from the main peaks
and shifted up or down to medium missing energies.
 Another 10\% of strength is moved to very high energies
due to SRC~\cite{vond1}. Its location cannot be explicitly
calculated in the present approach but the effects on the
reduction of spectroscopic factors at low energy is accounted
through the
energy dependence of the $G$-matrix.
Accordingly the total number of 
nucleons deduced from the fully self-consistent spectral function is
14.6 as reported in Tab.~\ref{tab:SpectFact}.
 This gives an estimate for the overall occupancy of high-momentum states of
about 10\%, in agreement with direct calculations~\cite{md94,mpd1}
as discussed in Secs.~\ref{sec:finiteSRC} and \ref{sec:finiteSRC2}.

 We note that the present approach also yields predictions  for the strength
distribution of quasiparticle excitations. Although little experimental
information is available for the addition of a nucleon (as well as for
the removal of a neutron).
Other knockout and nucleon transfer reactions, employing nuclear probes,
have been recently been developed or reanalyzed to extract
 hole spectroscopic factors of both stable
and radioactive nuclei~\cite{kbl01,hato03}. 
 These tools may yield better
accuracies than in the past and can, in principle, be used to extract
information on the one-particle spectral distribution.

 A deeper insight into the mechanisms that generate the fragmentation
pattern can be
gained by investigating directly the connection between
the spectral function and some specific collective states.
 To clarify this point we repeated the third iteration using exactly the same
input but without keeping the lowest ph $0^+$ state at its 
experimental energy, it was discarded instead.
 The resulting $p$ hole spectral function is shown in the lower-left panel
of Fig.~\ref{fig:p+sd_hsf}.
 In this calculation no breaking of the $p_{3/2}$ strength is obtained but 
only a single peak is found with a spectroscopic factor equal to 0.75, which
is the sum of the two fragments that are obtained when the 
$0_2^+$ state is taken into account.
 This result can be interpreted by considering the $p_{3/2}$ fragments as a
hole in the ground state and in the first excited $0^+$ state of the
${}^{16}{\rm O}$ core, respectively.
 The correct fragmentation pattern can be reproduced only when the latter
two levels are close enough to one another in energy, so that the
two configurations can mix.

The other two low-lying states of ${}^{16}{\rm O}$ that may be of some
relevance are the isoscalar $1^-$ and $3^-$. 
These excitations are reproduced reasonably well
by RPA type calculations~\cite{cz2} but are typically found at
$\sim$3~MeV above the experimental results.
 The lower panels of Fig.~\ref{fig:p+sd_hsf} show the results for the 
even parity spectral functions that are obtained when both
the $3^-$ and the $1^-$ ph-DRPA solutions are shifted to
match their experimental values. 
In this case, a $d_{5/2}$ hole peak is obtained at a missing energy
-17.7~MeV, in agreement with 
experiment,  and with a spectroscopic strength of~0.5\%.
 It is interesting to note that the shifting of the $3^-$ and $1^-$
collective states
does not produce any other noticeable change in the theoretical spectral
function.

\subsection{Spectrum of ${}^{16}{\rm O}$ and extension of RPA
\label{sec:BSEO16} }

\begin{figure}[!tb]
 \begin{center}
 \includegraphics[width=.65\linewidth]{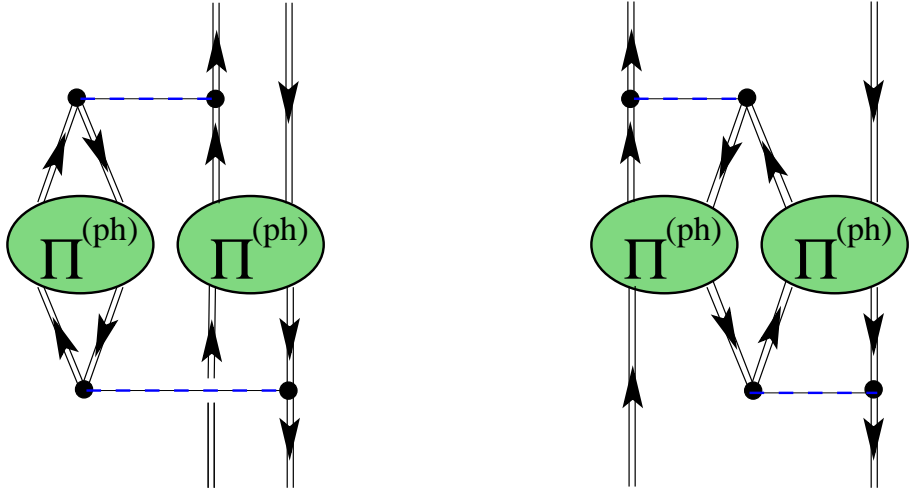} 
\begin{minipage}[t]{16.5 cm}
  \caption[]{\label{fig:2PiERPA}
 Examples of contributions involving the coupling of two independent
ph phonons. 
 In total, there are sixteen possible diagrams of this type, obtained
by considering all the possible couplings to a ph state.
The two-phonon ERPA equations sum all of these
contributions in terms of dressed sp propagators.
 }
\end{minipage}
 \end{center}
\end{figure}

The above results suggest that the main impediment for further improvements
of the description of the experimental spectral strength is associated
with the deficiencies of the RPA (DRPA) description of the excited states.
One important problem is the first excited state, with quantum numbers $0^+$,
which is seen to produce important effects on the fragmentation
of the $p$ shell. The DRPA result do find this state 
at low energy but can describe
at most one collective phonon for a given angular momentum, parity, and isospin
combination $J^{\pi},T$ whereas several low-lying isoscalar $0^+$ and $2^+$
excited states are observed at low energy in ${}^{16}{\rm O}$, as well
as additional $3^-$ and $1^-$ states.
A possible way to proceed would be to first concentrate on an improved 
description of the collective phonons by extending the RPA to explicitly
include the coupling to two-particle$-$two-hole (2p2h) states.
Such an extended RPA procedure has been applied in heavier nuclei with
considerable success~\cite{brand88,brand90}.
In order to be relevant for ${}^{16}{\rm O}$, this approach requires
an extension in which the coherence of the 2p2h states is included
in the form of the presence of two-phonon excitations~\cite{wim2ph}.
This corresponds to  adding  two-phonon configurations to the kernel
$K^{ph}$ of the BSE for the ph propagator~(\ref{eq:Pi-3t_BSE}),
as shown in Fig.~\ref{fig:2PiERPA}~\cite{badi03}. 

\begin{figure}[!tb]
 \begin{center}
 \includegraphics[width=0.65\linewidth]{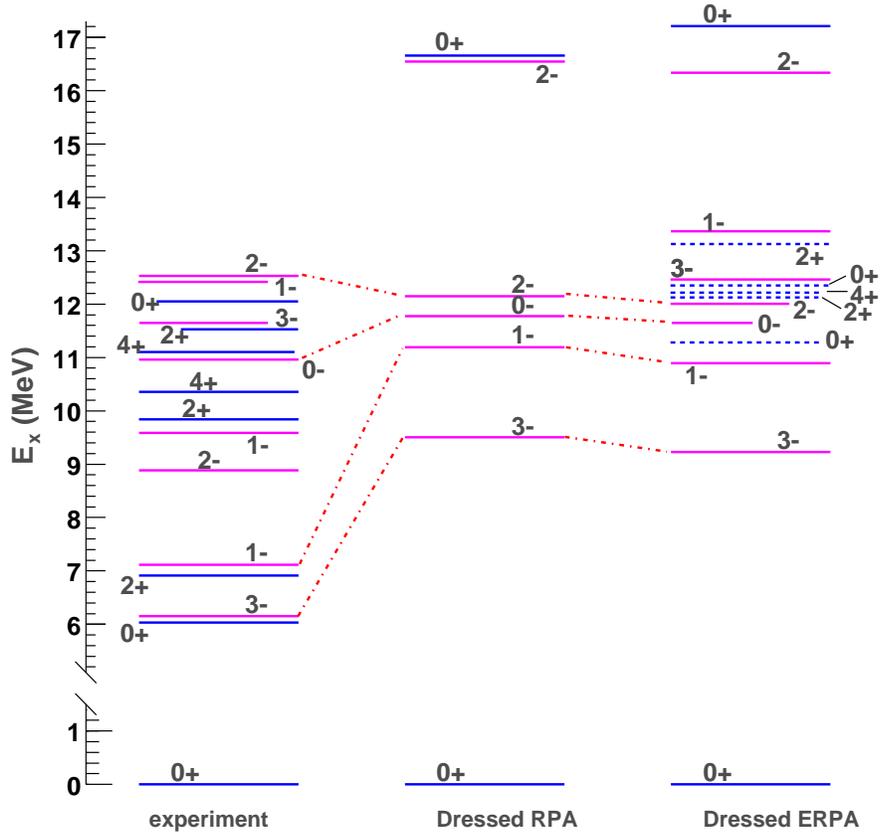}
\begin{minipage}[t]{16.5 cm}
\caption{ \label{fig:2PiERPA_3g}
 Results for the DRPA and the two-phonon ERPA propagator of ${}^{16}\textrm{O}$
 with the dressed input propagator computed in Ref.~\cite{badi02}, middle and
 last column respectively.
  In solving the ERPA equation, the lowest $3^-$, $1^-$ and $0^+$ levels
 of the DRPA propagator where shifted to their experimental energies. All
 other DRPA solutions were left unchanged.
 The excited states indicated by dashed lines are those for which the (E)RPA
 equation predicts a total spectral strength $Z_{n_\pi}$ lower than 10\%.
  The first column reports the experimental results~\cite{azj82}.
 }
\end{minipage}
 \end{center}
\end{figure}

The formalism to include two-phonon configurations has been presented
in Ref.~\cite{badi03}, where it is referred as two-phonon extended RPA
(ERPA).
In that work the ph-DRPA equation has been solved first,
using the self-consistent sp propagator discussed in Sec.~\ref{sec:Faddres}.
The lowest DRPA solutions for both the $0^+$, $3^-$ and $1^-$ channels
were shifted down to their respective experimental energies and
then employed to generate the two-phonon contributions
for the ERPA calculation.
The spectrum obtained for  ${}^{16}{\rm O}$ is
displayed in Fig.~\ref{fig:2PiERPA_3g} and Tab.~\ref{tab:2PiERPA_3g}
together the total ph strength $Z_{n_\pi}$ of each state,
\begin{equation}
 Z_{n_\pi} ~=~  \sum_{\alpha \beta} ~
      \left| 
       {\mbox{$\langle {\Psi^A_{n_\pi} } \vert $}}
            c^{\dag}_\alpha c_\beta {\mbox{$\vert {\Psi^A_0} \rangle$}} 
      \right|^2   \; .
\label{eq:Z_tot_ph}
\end{equation}
 Tab.~\ref{tab:2PiERPA_3g} also
reports the relative strength of ph and two-phonon admixtures in
the wave function for each solution of the ERPA equation.
  Due to the screening
effects associated with dressing the sp propagators, the solution for the first
isoscalar $0^+$ state in DRPA is found much higher in energy
at $\sim$17~MeV.
 It is important to realize that this solution has a sharp ph character
and therefore it cannot be identified with the experimental $0_2^+$ 
state.
The latter state is known to be dominated by 4$\hbar\omega$
configurations~\cite{Brown0+,Haxton0+,warb92}.
On the other hand inelastic electron scattering experiments~\cite{buti86}
clearly excite this state. 
From the point of view of the SCGF approach, 
the one-body response 
is completely described by the polarization propagator, Eq.~(\ref{eq:Pi_Leh}),
and therefore, 
the total one-body strength must be represented
by $Z_{n_\pi}$~(\ref{eq:Z_tot_ph}). This indicates a strong
coupling to the (dressed) ph propagator~(\ref{eq:Pi_Leh}). 
 On the basis of this similarity, it still seems
plausible to shift the ph-DRPA solution down in energy in order to study its
mixing with other configurations.

 The corresponding results yield an isoscalar $0^+$ 
state with a predominant ph character at $\sim$17~MeV,
as it was found in DRPA. Although this solution is now characterized
by a partial mixing with two-phonon configurations. 
 Tab.~\ref{tab:2PiERPA_3g} shows that this is the result
of mixing with 
the lowest solution in the same channel, which ends up at $\sim$11~MeV.
 The latter is predominately a two-phonon state.
  It is also seen that in both cases the
relevant configuration comes from the coupling of two $0^+$ phonons
themselves.
 in Ref.~\cite{badi03} it was found that the wave functions for these two
states contain several relevant ph configurations, obtained
from different quasiparticle fragments in the $pf$~($sd$) shells combined 
with quasihole fragments of the $p$~($s$) shells. 
 Therefore, the situation is more complicated than the simple picture
of only two levels interacting with each other.

 The calculations of Ref.~\cite{iach1} have shown that the bulk of
the 4p4h contributions to the first excited state of ${}^{16}\textrm{O}$
may come from the coupling of four different phonons with negative
parity ($3^-$ and $1^-$). However,
the self-consistent role of  coupling positive parity states was
not considered in that work. Both this effect,
the RPA correlations and the inclusion of the nuclear fragmentation allow
for the --at least partial-- inclusion of configurations beyond the 2p2h case
even when no more than two-phonon coupling is considered, as discussed
here.
 Still it would be helpful to further study the importance
of three- and four-phonon configurations within this apporach.

 The low-lying negative parity isoscalar states ($3^-$ and $1^-$) are
only slightly
 affected by two-phonon contributions and remain substantially
above the experimental energy at 9.23 and 10.90~MeV, respectively,
and more correlations will be needed in order to lower their energy.
We note that these wave functions contain a two-phonon admixture obtained by
coupling the low-lying $0^+$ excitation to either the $3^-$ or $1^-$ phonons.
 According to the above interpretation of the first $0^+$ excited state
this corresponds to the inclusion of 3p3h and beyond.
 At the same time two additional negative parity solutions, with two-phonon
character, are found at higher energy in accordance with experiment.
 These configurations are not reproduced by the simple DRPA.

\begin{table}[t]
 \begin{center}
 \begin{tabular}{c|ccc|cccccc|cccccc}
  \hline 
  \hline 
 $T=0$
   &  & \multicolumn{2}{c|}{dressed/DRPA} 
   &  & \multicolumn{5}{c|}{dressed/ERPA}
       &  & $(0_2^+)^2$   & $(3_1^-)^2$
          & $(0_2^+,3_1^-)$ & $(0_2^+,1_1^-)$
          & \\
 $J^\pi$  &  & $\varepsilon^\pi_n$ & $Z_{n_\pi}$ 
          &  & $\varepsilon^\pi_n$ & $Z_{n_\pi}$
          & \hspace{.05truein} & $ph$(\%) & $2\Pi$(\%) 
      & ~ & (\%)   & (\%)  & (\%)   & (\%)
          & \\
  \hline 
  \hline 
 $1^-$ &  &        &         &   &  13.37 &  0.148 &   & 21   & 79   & &      &      &      &  79   & \\
 $3^-$ &  &        &         &   &  12.35 &  0.113 &   & 16   & 84   & &      &      & 84    \\
 &&&&&&&&&&&&&&& \\
 $0^+$ &  &        &         &   &  12.15 &  0.001 &   &  1   & 99   & &  3   & 96   \\
 $4^+$ &  &        &         &   &  12.14 &  0.007 &   &  1   & 99   & &      & 99   \\
 $2^+$ &  &        &         &   &  12.12 &  0.008 &   &  1   & 99   & &      & 98   \\
 &&&&&&&&&&&&&&& \\
 $0^+$ &  &  16.62 &   0.717 &   &  17.21 &  0.633 &   & 88   & 12   & &  10  &  0.5 \\
 &&&&&&&&&&&&&&& \\
 $0^+$ &  &        &         &   &  11.28 &  0.092 &   & 12   & 88   & &  85  &  2   \\
 &&&&&&&&&&&&&&& \\
 $1^-$ &  &  11.19 &   0.720 &   &  10.90 &  0.680 &   & 94.1 &  5.9 & &      &      &      &  5.8  & \\
 $3^-$ &  &   9.50 &   0.762 &   &   9.23 &  0.735 &   & 95.9 &  4.1 & &      &      &  4.0  \\
  \hline 
  \hline 
 \end{tabular}
  \end{center}
    \parbox[t]{1.\linewidth}{
      \caption[Two-phonon ERPA spectrum and strengths; dressed input]
  {  \small 
    Excitation energy and total spectral strengths obtained for the 
   principal solutions of DRPA and two-phonon ERPA equations.
    The total contribution of ph and two-phonon states
   of the ERPA solutions are shown.
    For states below 15~MeV, the columns on the right side give the 
individual contributions of
all the relevant two-phonon contributions.
    \label{tab:2PiERPA_3g}  }
    }
\end{table}

The two-phonon ERPA approach also generates a triplet of states at 
about 12~MeV, with quantum numbers $0^+$,  $2^+$ and $4^+$.
A similar triplet is found experimentally at 12.05, 11.52
and 11.10~MeV, which also corresponds to
twice the experimental energy of the first $3^-$ phonon.
 The ERPA solutions for this triplet are almost 
exclusively obtained by $3^- \otimes 3^-$ states and therefore have a 2p2h
character, in accordance
with Refs.~\cite{buti86,zis70,loba72}.
Further interaction in the pp and hh channels will be needed in order
to reproduce the correct experimental splitting and experimental
strength~\cite{iach1,iach2}.


\subsection{Results for the ${}^{16}{\rm O}(e,e'pp)$ cross section
\label{sec:ee1-2N} }

Data from two-nucleon knockout reactions have been measured recently at
NIKHEF~\cite{KHK95,ond1,ond2,groep} and MAMI~\cite{blom99,rosner} facilities
for different targets.
 Theoretically, these cross sections can be computed with the models
 developed by the Pavia~\cite{giupa98,giupa99}
and Ghent~\cite{Ry97} groups for the emission of both two protons
or a proton-neutron pair.
 Calculation for the ${}^{16}{\rm O}(e,e'pp)$ case have shown that the 
transition to the ground state of ${}^{14}{\rm C}$ is dominated by
the high-momentum components in the two-nucleon overlap function. 
Comparison with the data from NIKHEF for this reaction have
provided a clear signature of SRC effects~\cite{ond2,star00}.

An important ingredient in the description of two-nucleon
knockout reactions is the two-hole spectral function. This quantity can be 
extracted from the imaginary part of the two-body
propagator $g^{II}(\omega)$, compare Eqs.~(\ref{eq:g2})
and~(\ref{eq:SpecFunc}), and contains the nuclear structure information
on the correlated pair before the reaction takes place.
 The propagator $g^{II}(\omega)$ can be obtained by 
solving pp(hh) RPA equations as done in Ref.~\cite{geu2h}.
 However, typical calculations of this kind yield results 
within a model space that is too small to contain the short-range effects.
In these cases one needs to compute the high-momentum components separately.
In the following we outline the approach taken in Refs.~\cite{geu2h,giupa98}
and show a few results for the ${}^{16}{\rm O}(e,e'pp)$ cross section.

The sp spectral function of ${}^{16}{\rm O}$
obtained in Ref.~\cite{GeurtsO16} was employed as the starting point
of a hhDRPA calculation (see Eq.~(\ref{eq:g2DRPA})~), eventually obtaining 
a two-hole spectral function of the form~\cite{geu2h},
\begin{eqnarray}
  \lefteqn{
     S^{hh}_{J T}( {\bf p}_1, {\bf p}_2, {\bf p}_{1'}, {\bf p}_{2'}, \omega ) =
          }
 \hspace{0.5cm} & &
\nonumber \\
& & 
        \sum_{k}
        \sum_{a b c d \in {\cal P}} \sum_{M T_3}
        \Phi^{\ast JM}_{cd}( {\bf p}_{1'}, {\bf p}_{2'} )
        \left( X^{k}_{cd J T} \right)^* X^{k}_{ab J T} 
        \Phi^{JM}_{ab}( {\bf p}_1, {\bf p}_2 )
        \;
        \delta(\omega - (E^{A}_{0} -E^{J T,A-2}_{k}) )
\; ,
\label{eq:SpecFuncW}
\end{eqnarray}
where $J$ and $T$ refer to the total angular momentum and isospin of the
residual nucleus and $M$ and$T_3$ are their projections.
In Eq.~(\ref{eq:SpecFuncW}), the energies $E^{J T,A-2}_{k}$ are the 
the DRPA solutions for excitation energies of the residual nucleus
and the quantities
\begin{equation}
 X^{k}_{ab J T} \sim  \langle \Psi^{k,A-2}_{J} ||
			\tilde{(a_{\beta} a_{\alpha})}_J ||
		     \Psi^A_0 \rangle
\label{eq:XacJT}
\end{equation}
represent the components of the solutions
for the two-body overlap amplitudes.
The sums
extend only over those orbitals that are occupied in the model space
${\cal P}$ and the spectral function $S^{hh}(\omega)$ is
transformed into the momentum representation by employing two-body harmonic
oscillator wave functions~$\Phi^{JM}_{ab}( {\bf p}_1, {\bf p}_2 )$.

 The description of 
the high-momentum components due to SRC requires 
the inclusion of a very large number of basis states, at least up to 
$100\hbar\omega$ in a harmonic oscillator basis~\cite{CALGM}.
However, the description of long-range correlations by solving
a Bethe-Salpeter
or an RPA equation~(\ref{eq:g2DRPA}) is not feasible within such a large space.
For this reason the complete basis is split into a model space 
${\cal P}$, and a complementary space ${\cal Q}={\bf 1}-{\cal P}$.
For $^{16}$O the space ${\cal P}$ was chosen to contain the orbitals up to
the $pf$ shell only,
which is large enough to accommodate long-range effects.
The justification for this procedure is that SRC are caused by close encounters
of two nucleons, which mainly depend on the nuclear density and therefore are 
not very sensitive to details of the long-range structure.
The latter, on the other hand, may be calculated within the space ${\cal P}$
with a suitable effective interaction in which the SRC are incorporated at
least in ladder approximation.
This effective interaction can then be obtained by following Brueckner's
individual pair approach~\cite{BGW58,BLR61} by solving the Bethe-Goldstone 
equation (BGE). Using the technique of Ref.~\cite{CALGM} the equation for the 
correlated pair wave function 
        \begin{equation}
                |\Psi_{ab}\rangle
        =
                |\Phi_{ab}\rangle
        +
                \frac{\hat{Q}}
                {\omega - \hat{H}_0}
                \hat{V}
                |\Psi_{ab}\rangle
        \label{eq:BGEP}
        \;,
        \end{equation}
is solved in the finite nucleus. In this equation
the Pauli operator $\hat{Q}$ prohibits scattering into orbits of the
finite shell-model space ${\cal P}$, in which long-range correlations are
trated in an DRPA approach.
In Eq.\ (\ref{eq:BGEP}),
$\Phi_{ab}$ represents the same uncorrelated shell model wave functions
appearing
in Eq.~(\ref{eq:SpecFuncW}), $\omega$ is the propagation energy
of the pair and $\hat{H}_0$
is the Hamiltonian without residual interaction.
In this case, the energy $\omega$ refers to the propagation of two holes which
precludes the vanishing of the denominator in Eq.\ (\ref{eq:BGEP}).

From the solution of Eq.\ (\ref{eq:BGEP}) one obtains the defect
wave function
as the difference between the correlated and uncorrelated pair wave functions
\begin{equation}
        |\chi\rangle
=         
        |\Psi\rangle - |\Phi\rangle  \; ,
\label{eq:def2}
\end{equation}
which represent the components of the two-body wave function
that are orthogonal to the space ${\cal P}$, 
and the $G$ matrix as the effective interaction in ${\cal P}$ 
according to
\begin{equation}
        \langle\Phi^{JM}_{cd}| G |\Phi^{JM}_{ab}\rangle
=
        \langle\Phi^{JM}_{cd}| V |\Psi^{JM}_{ab}\rangle
\;.
\end{equation}
%

The essential step taken here, is to approximate the spectral
function (\ref{eq:SpecFuncW}) by the expression
\begin{eqnarray}
  \lefteqn{
     S^{hh}_{J T}( {\bf p}_1, {\bf p}_2, {\bf p}_{1'}, {\bf p}_{2'}, \omega ) =
          }
 \hspace{0.5cm} & &
\nonumber \\
& & 
        \sum_{k}
        \sum_{a b c d \in {\cal P}} \sum_{M T_3}
        \Psi^{* \; JM}_{cd}( {\bf p}_{1'}, {\bf p}_{2'} )
        \left( X^{k}_{cd J T} \right)^* X^{k}_{ab J T} 
        \Psi^{JM}_{ab}( {\bf p}_1, {\bf p}_2 )
        \;
        \delta(\omega - (E^{A}_{0} -E^{J T,A-2}_{k}) )
\; ,
\label{eq:SpecFuncC}
\end{eqnarray}
where the summation over orbits is still 
limited to the finite shell-model space 
${\cal P}$
but the uncorrelated wave functions are replaced by the correlated ones
(see Eq.\ (\ref{eq:BGEP})~). 
This is in line with the argument just given,
that hard binary collisions, treated in the BGE and giving rise to 
high-momentum components, proceed independently of the long-range correlations.
The latter are taken into account in the shell model amplitudes $X$ within the
limited space ${\cal P}$.

\begin{figure}[!tb]
\begin{center}
\vspace*{1.cm}       
 \parbox[b]{.45\linewidth}{ \includegraphics[width=.95\linewidth]{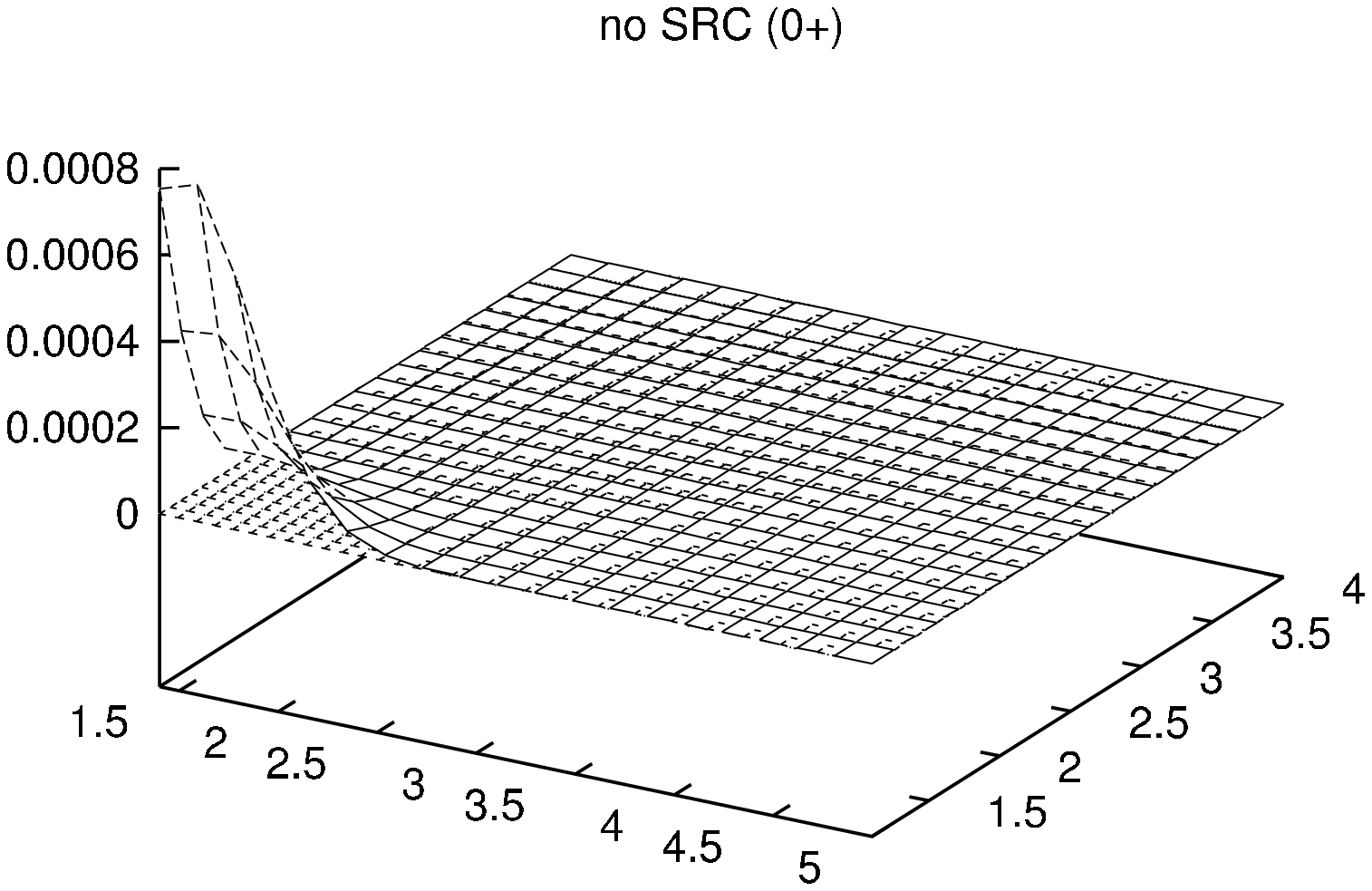} }
 \parbox[b]{.05\linewidth}{~}
 \parbox[b]{.45\linewidth}{ \includegraphics[width=.95\linewidth]{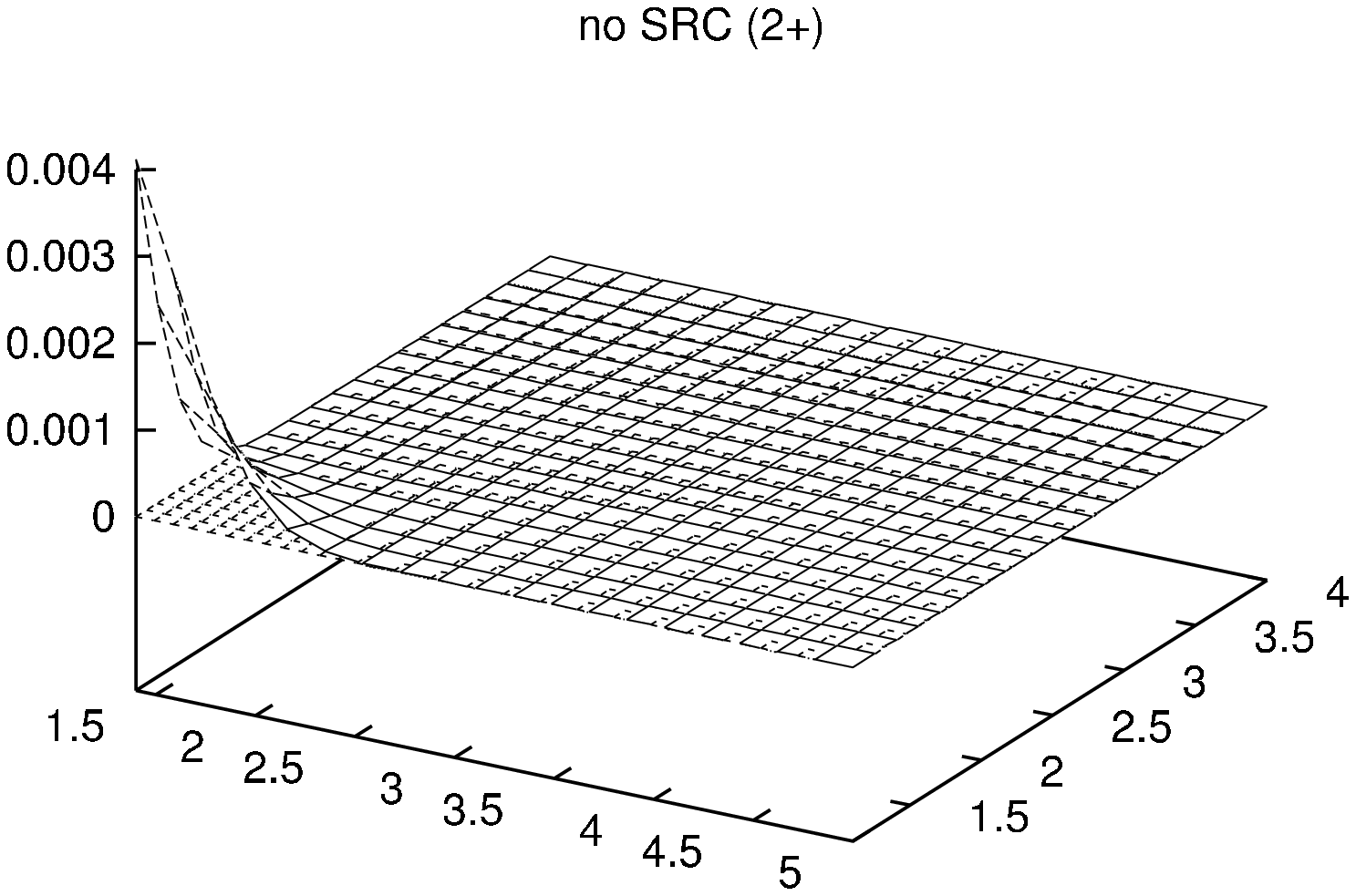} }
\end{center}
\begin{center}
 \parbox[b]{.45\linewidth}{ \includegraphics[width=.95\linewidth]{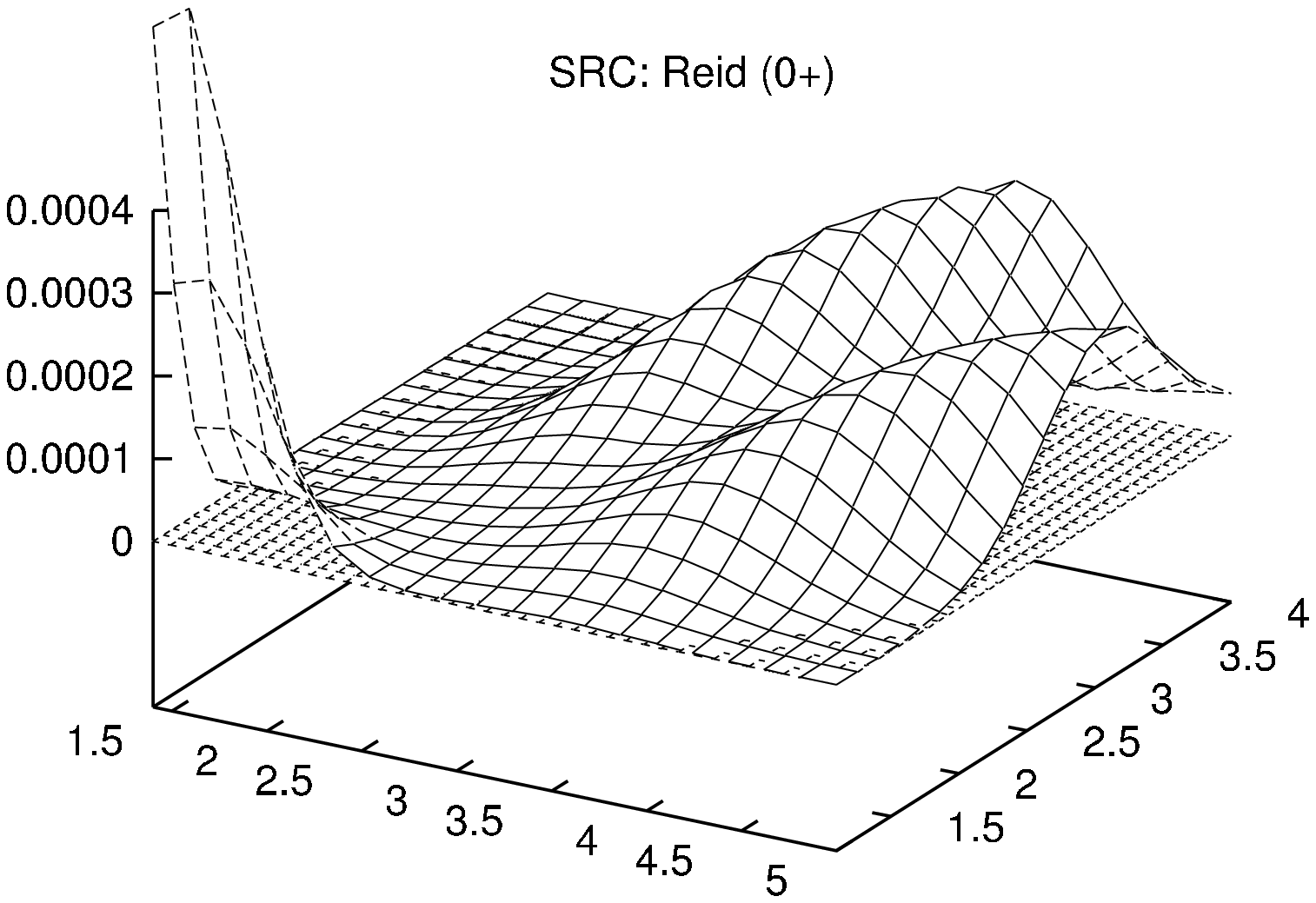} }
 \parbox[b]{.05\linewidth}{~}
 \parbox[b]{.45\linewidth}{ \includegraphics[width=.95\linewidth]{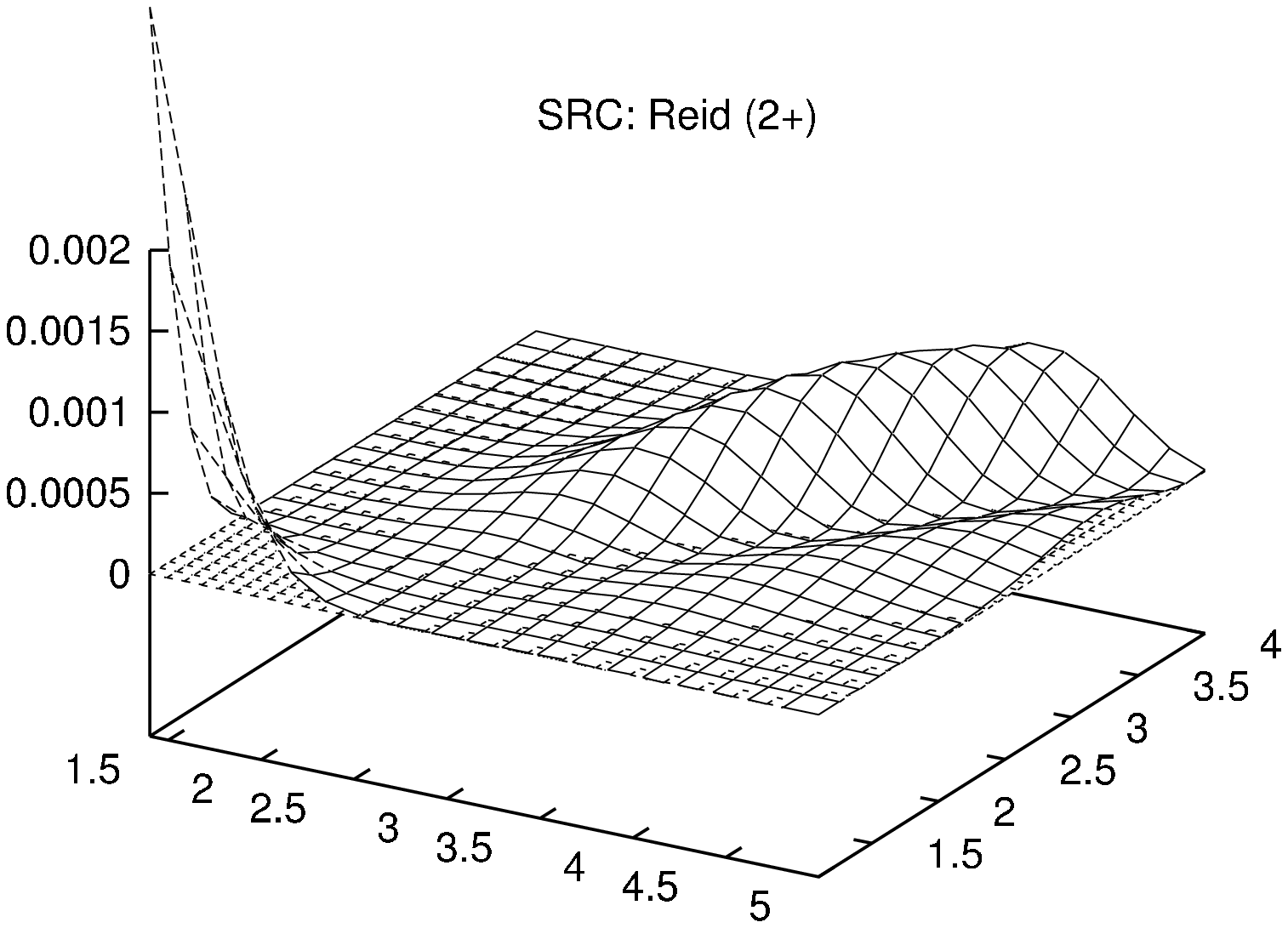} }
\end{center}
\begin{center}
 \parbox[b]{.45\linewidth}{ \includegraphics[width=.95\linewidth]{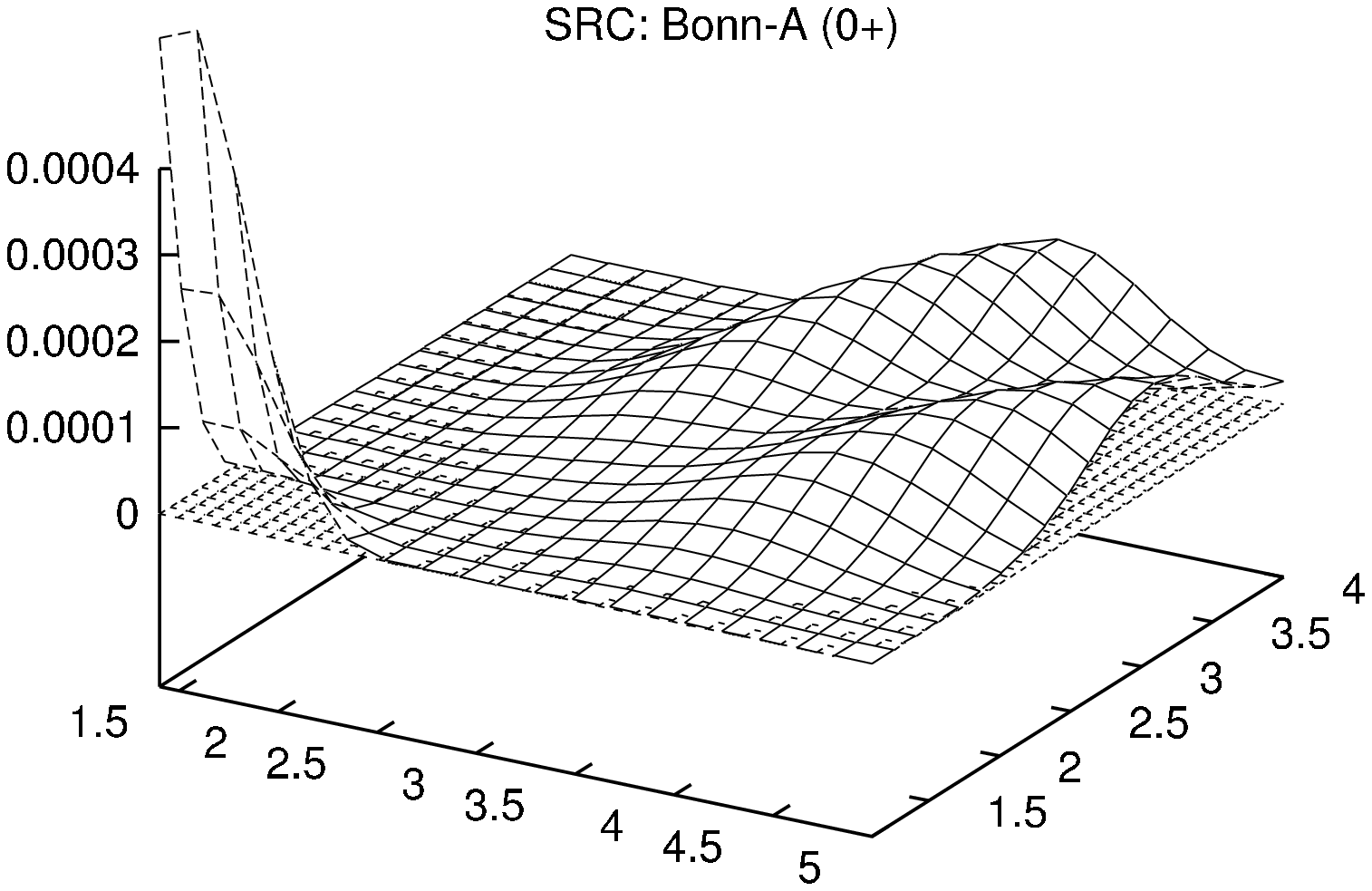} }
 \parbox[b]{.05\linewidth}{~}
 \parbox[b]{.45\linewidth}{ \includegraphics[width=.95\linewidth]{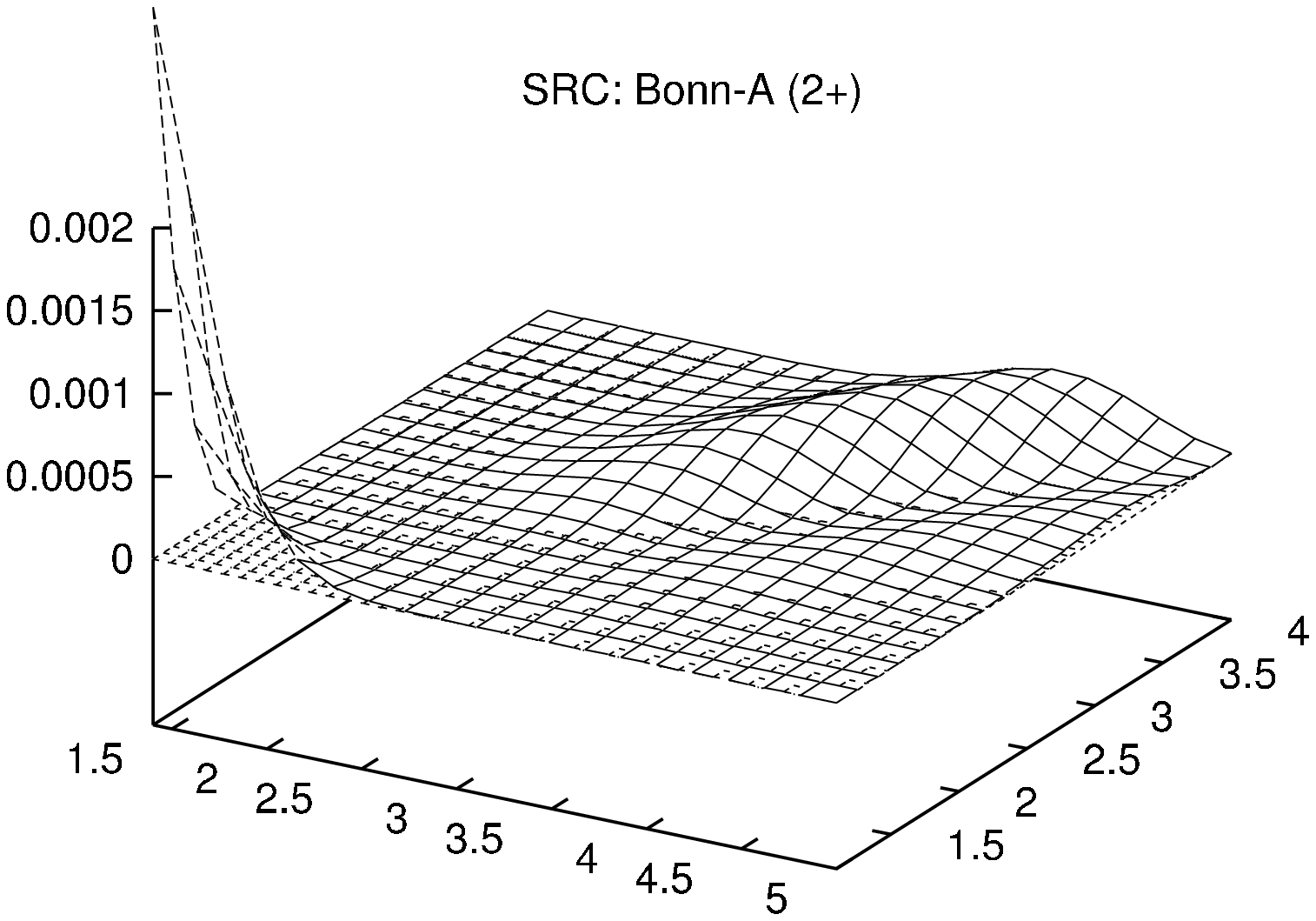} }
\begin{minipage}[t]{16.5 cm} 
\caption[]{%
Superposition of spectral functions~(\ref{eq:Shat}) appropriate
for the removal of two protons 
with final momenta ${\bf p}_1$ and ${\bf p}_2$ from $^{16}$O leading to the 
final $0^+$ ground state or first excited $2^+$ state in $^{14}$C. The plots 
are given for a kinematic setting used in experiments at NIKHEF~\cite{ond1}.
The momentum vector ${\bf q}$ is fixed along the $z$-axis,
with length $313$~MeV/c.
The momenta ${\bf p}_1$ and ${\bf p}_2$ are in the same plane with ${\bf q}$ 
at $-49$ and $123$ degrees angles with respect to this transferred momentum,
respectively.
The upper plots correspond to harmonic oscillator wave functions without the 
inclusion of short-range correlations (see Eq.~(\ref{eq:SpecFuncW})~).
 In the lower plots these SRC are 
incorporated by the defect functions computed the Reid 
and the Bonn-A potential (Eqs.~(\ref{eq:def2}) and~(\ref{eq:SpecFuncC})~).
\label{fig:sfC1}}
\end{minipage}
\end{center}
\end{figure}

In the spectator model with a plane-wave approximation for the
outgoing protons, the contribution of the spectral function to the
cross section is given by the superposition
$\hat{S}(\bm{p}_1, \bm{p}_2, E)$, Eq.~(\ref{eq:Shat}).
This function is plotted in Fig.~\ref{fig:sfC1}, at the
kinematics corresponding to the ${}^{16}{\rm O}(e,e'pp)$ 
measurements at NIKHEF~\cite{ond1}, for the 
transitions to the $0^+$ ground state and to the excited $2^+$
levels of ${}^{14}{\rm C}$.
 In these kinematics the momenta $\bm{p}_1$ and $\bm{p}_2$ of the 
protons are almost antiparallel to each other and coplanar with the
momentum $\bm{q}$ transferred by the electron. As a consequence, high and
equal values of their magnitudes correspond to high
relative momenta in the plots of Fig.~\ref{fig:sfC1}.
 The effects of including the SRC is shown in the lower plots, where
the correlation functions~(\ref{eq:def2}) have been computed with both the
Reid-Soft-Core and Bonn-A potentials.
Figure \ref{fig:sfC1}, also shows the plots labeled `No SRC' that correspond
to neglecting the correlation functions and computing the 
spectral functions with Eq.~(\ref{eq:SpecFuncW}).
As expected, one observes that the latter gives no
contribution at higher momenta.
The high-momentum part of the spectral function is about a factor of two 
larger 
for the Reid-Soft-Core potential than for the Bonn-A potential
in the momentum range around $3$--$5$~fm$^{-1}$.
The short-range correlations give rise to a spectral 
function which is clearly distinct from the one without SRC.

\begin{figure}[!tb]
\vspace*{.5cm}       
\begin{center}
\includegraphics[width=0.8\textwidth]{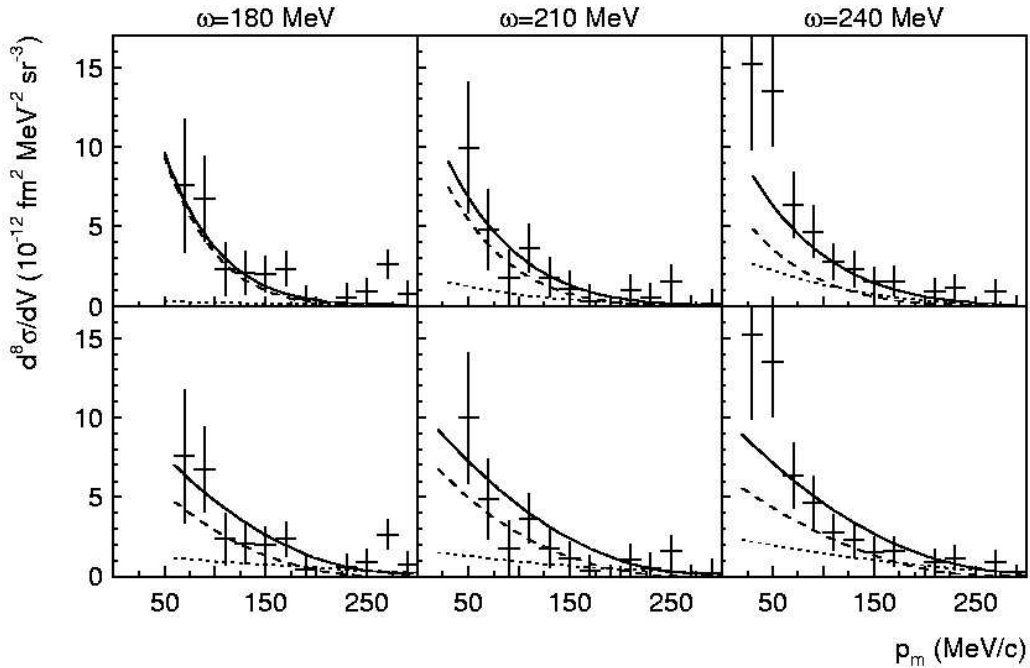}
\begin{minipage}[t]{16.5 cm}
\caption{ \label{fig:plb5}
The cross sections for the transition to the ground state in ${}^{14}{\rm C}$
as a function of the missing momentum, measured for the reaction
${}^{16}{\rm O}(e,e'pp){}^{14}{\rm C}$  at three values of $\omega$.
 The curves are obtained from calculations of Refs.~\cite{giupa98}
(top) and~\cite{Ry97} (bottom). The solid curves
represent the calculated cross sections; the dashed and dotted curves
correspond to the contributions of the one- and two-body hadronic currents,
respectively.
}
\end{minipage}
\end{center}
\end{figure}

The Ghent-model~\cite{Ry97} accounts for SRC in a different way. There,
single-particle wave functions obtained
from Hartree-Fock calculations are used and the many-body wave function,
is approximated
to first order in the correlation function,
\begin{equation}
\Psi({\bf r}_1, \ldots {\bf r}_A) =
 \prod^A_{i < j = 1} \left( 1-g(r_{ij}) \right)
   \Phi({\bf r}_1, \ldots {\bf r}_A) / \sqrt{N}
   \; ,
\end{equation}
where $N$ is a normalisation factor. The effects of
SRC are included in the factor $(1-g(r_{ij}))$ by means of the
correlation function $g(r_{ij})$ (which zero in the absence of
correlations). 
 The calculations presented here employed the $g(r_{ij})$ of 
Ref.~\cite{christh}, also shown in Fig.~\ref{fig:corrfunc}.

Figure \ref{fig:plb5} compares the cross sections for the calulations of
Ref.~\cite{giupa98,Ry97} with the experimental data,
as a function of the missing
momentum at the three values of $\omega$~\cite{star00}. 
For this calculation final state interactions for the outgoing protons were
taken into account by using distorted waves but neglecting their mutual
interaction.
 A pair of $(p_{1/2})^2$ or $(p_{3/2})^2$
protons can be coupled to total angular momentum $J$=0,
leaving the ${}^{14}{\rm C}$ nucleus in the ground state, 
through either a ${}^1S_0$ or a ${}^3P_1$  relative state.
 The coupling scheme used in the calculation is as follows.
 The ${}^1S_0$  state is always associated with an
 angular momentum of the center-of-mass motion $L=0$, and the
${}^3P_1$ state always with $L=1$.
 From Fig.~\ref{fig:plb5} it is clear that for all three values of the energy
the
missing-momentum dependence of the measured cross sections is similar. In
Refs.~\cite{ond1,ond2}, it has been pointed out that such a momentum
distribution
reflects an angular momentum $L=0$ for the center-of-mass motion of the pair,
and thus suggests a dominant role for the knockout of a ${}^1S_0$
pair driven by SRC. 
The theoretical cross sections are represented in Fig.~\ref{fig:plb5} 
by the solid curves;
the contributions of the one- and two-body currents are given by the dashed
and dotted curves, respectively. The calculated cross sections agree well
with the data at all three values of $\omega$.
 The curves, representing the
contributions of the one- and two-body currents to the ground-state
transition, indicate that at $\omega$=180 and 210 MeV the reaction is dominated by
one-body currents, and that the contribution of two-body currents increases
with increasing energy transfer. Conceptually both models are quite similar,
and neither of the two contains free parameters.
The effects due to SRC, though, are accounted for in different ways 
as discussed above.
 The Bonn-A and Reid
Soft Core potentials, adopted in the calculations, are both realistic
NN-potentials and successfully used in many nuclear-structure calculations.
Hence, the data agree with the theoretical results obtained independently
with the two models. Furthermore, both models predict that the largest
contribution to the cross section stems from one-body hadronic currents
driven by SRC. This justifies the conclusion that evidence is obtained for
SRC although the data do not distinguish details of the actual form of
the SRC. 
Future experiments will include the removal of proton-neutron pairs and
therefore may also yield information on tensor correlations~
\cite{giupa99,bffilb}.
An improved description of the mutual interaction between the outgoing 
protons~\cite{knmu01,sbgp03}
will help in further developing the usefulness of this reaction
to elucidate details of short-range nuclear dynamics.

\subsection{Extraction of spectroscopic information from $(e,e'p)$
reactions at high ${\cal Q}^2$
\label{sec:highQ} }

The analysis of the (e,e$'$p) data has relied on the Distorted Wave Impulse
Approximation (DWIA) for both the Coulomb distortion of the electron waves
(in heavy nuclei) and the outgoing proton~\cite{frmo,Paviabook,bgprep,kelly}.
Some aspects of this analysis have been discussed in Sec.~\ref{sec:exp}.
The proton distortion is described in terms of an optical potential required
to describe elastic proton scattering data at relevant energies~\cite{lap93}.
There is some uncertainty related to this treatment since elastic proton
scattering is considered to be a surface reaction and no detailed information
is obtained related to the interior of the nucleus.
This uncertainty gives rise to an estimated error of about 10\%.
Such an estimate may be inferred by considering the difference between
the relativistic and nonrelativistic treatment of the proton distortion.
It is shown in Ref.~\cite{udias} that this difference is essentially
due to the reduction of the interior wave function in the relativistic
case. This feature can also be generated by including a reasonable amount of
nonlocality in the optical potential~\cite{udias}.

A serious challenge to the interpretation of (e,e$'$p) experiments
was recently published in Refs.~\cite{lapy,fsz}.
This challenge consists in questioning the validity of the constancy of the
spectroscopic factor as a function of the four-momentum (squared), $Q^2$,
transferred by the virtual photon to the knocked-out nucleon.
In Ref.~\cite{lapy} a conventional analysis of the world's data for (e,e$'$p)
experiments on ${}^{12}{\rm C}$ at low $Q^2$ generated results 
consistent with previous expectations.
Data at higher values of $Q^2$ were then analyzed within the framework
of a theoretical model which employs Skyrme-Hartree-Fock bound-state
wave functions for the initial proton, a Glauber-type description of the
final-state interaction of the outgoing proton, and a factorization
approximation for the electromagnetic vertex~\cite{lapy,fsz}.
Within the framework of this theoretical description, spectroscopic factors
were obtained which increase substantially with increasing $Q^2$ for the
${}^{12}{\rm C}$ nucleus.

The spectroscopic factor defined in Sec.~\ref{sec:form} is a many-body
quantity defined without reference to a probe.
The discussion in Sec.~\ref{sec:exp} shows that this spectroscopic factor
is obtained by comparing a calculation of the cross section with the actual
data.
The wave function of the removed nucleon is the solution of a Woods-Saxon 
potential at the appropriate removal energy.
This potential is adjusted to generate an optimum description of the
shape of the experimental cross section.
This resulting theoretical representation of the cross
section is then multiplied by a constant factor to coincide
with the experimental cross section.
This constant factor is then interpreted as the spectroscopic factor.
Another important ingredient in this analysis is the choice of the 
electron-proton cross section which must be considered off-shell~\cite{defor}.
This leads to a small additional uncertainty in the analysis of low $Q^2$ data
as discussed recently in Ref.~\cite{udiasa}.

Further clarification of this intriguing situation with different 
spectroscopic factors at different $Q^2$ has been obtained in 
Ref.~\cite{rdrs02}.
In this paper a study of recently
published (e,e$'$p) data for ${}^{16}{\rm O}$ at $Q^2 = 0.8$ 
(GeV/$c)^2$~\cite{e89003} has been presented. 
An unfactorized approach was used, as it is required at low 
$Q^2$~\cite{Paviabook} and as it
 has been recently advocated also for high $Q^2$
reactions in Ref.~\cite{jes}. 
However, the higher energy 
of the outgoing proton requires 
a description which contains different elements than the conventional 
low $Q^2$ analysis. 
For the electromagnetic current operator 
the approach of Ref.~\cite{kelly1} was followed, where a relativistic current 
operator was used in a Schr\"odinger-based calculation, avoiding any 
nonrelativistic reduction and including the effect of spinor distortion 
by the Dirac scalar and vector potentials. As for the final state, 
a recently developed eikonal description of the final-state
interaction (FSI) of the proton with the nucleus that has been tested 
against DWIA solutions of a complex spin-dependent optical 
potential~\cite{br2,br3} was employed. 
The absorption of the proton was described theoretically by linking it to
the corresponding absorption of a nucleon propagating through nuclear
matter. The relevant quantity is the nucleon self-energy which is obtained from
SCGF calculations of nucleon spectral functions discussed in 
Sec.~\ref{sec:selfnmr}.
This description of FSI was combined with previous results for the bound-state
wave functions of the $p$-shell quasihole states in
${}^{16}{\rm O}$~\cite{mpd1} discussed in
Sec.~\ref{sec:finiteSRC}.
In Ref.~\cite{rad97} these wave functions have been used to analyze low $Q^2$
data for the ${}^{16}$O(e,e$'$p) reaction~\cite{leus}. 
These very same wave functions produce a good description of the 
shape of the coincidence cross section for the $p$-shell quasihole states 
at high $Q^2$~\cite{e89003} using the same spectroscopic factors obtained from 
the low $Q^2$ data~\cite{leus}.

\begin{figure}[tb]
\begin{center}
\includegraphics[angle=0,width=0.6\textwidth]{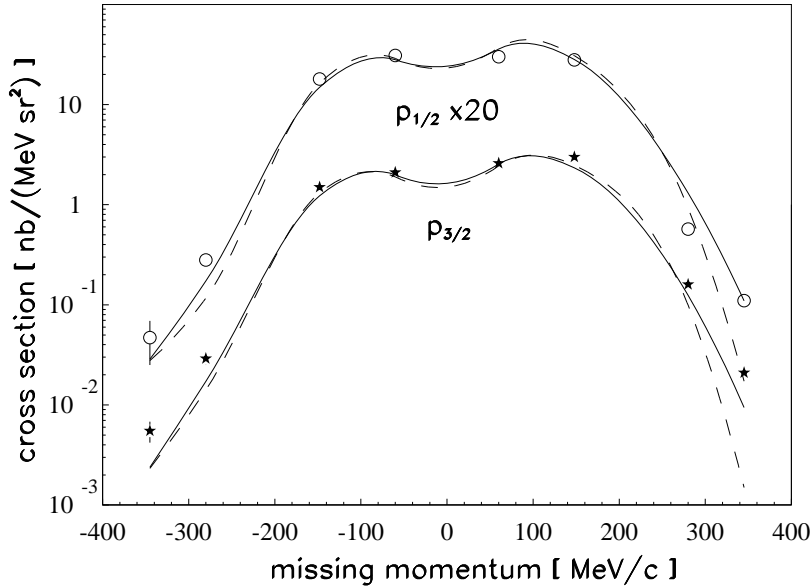}
\begin{minipage}[t]{16.5 cm}
\caption{Cross section for the $^{16}$O(e,e$'$p)$^{15}$N 
reaction at $Q^2=0.8$ (GeV/$c)^2$ in perpendicular 
kinematics~\cite{e89003}. Data for the $p1/2$ state have been multiplied 
by 20. The solid lines represent 
the result of the calculation~\cite{rdrs02}. 
The dashed lines are obtained by replacing 
the quasihole states with the bound state wave functions of 
Ref.~\cite{Zhalov}. 
In all cases, the results have been 
rescaled by the same spectroscopic factors as used for the NIKHEF kinematics
(see Fig.~\ref{fig:N15p32} for example), 
namely Z=0.644 and Z=0.537 for the $p1/2$ and 
$p3/2$ states, respectively.}
\label{fig:figIV-2}
\end{minipage}
\end{center}
\end{figure}

We first establish the quality of the original assertion pertaining to
low $Q^2$ by referring to Fig.~\ref{fig:N15p32} for the $p3/2$ state
in ${}^{16}$O. A similar result is obtained for the $p1/2$ state~\cite{rdrs02}.
The corresponding spectroscopic factors to obtain agreement with
these low $Q^2$ data are given by 0.644 and 0.537 for the $p1/2$ and $p3/2$
states, respectively.
In Fig.~\ref{fig:figIV-2} the same reaction is considered in a very different 
kinematical regime, namely at constant $(\bm{q}, \omega)$ with $Q^2=0.8$ 
(GeV$/c)^2$~\cite{e89003}. The data here refer to a five-fold differential 
cross section, avoiding any ambiguity in modelling the half off-shell 
elementary cross section $\sigma_{ep}$.
Results for the transition to the ground state $p_{1/2}$ have 
been multiplied by 20.
 The theoretical calculations are displayed with solid lines for the results 
with the quasihole 
bound state and dashed lines by employing the wave function of 
Ref.~\cite{Zhalov}. 
The preference for the first choice is evident. In any case, it is 
remarkable that the calculations reproduce the data by using the same 
spectroscopic factors as in the previous kinematics, i.e. $Z_{0p1/2}=0.644$ 
and $Z_{0p3/2}=0.537$. 
Therefore, contrary to the findings of Ref.~\cite{fsz}, 
no need for a $Q^2$ dependence of the spectroscopic factors over 
a wide kinematical range has been established.
This is also in agreement with the results obtained by a mean-field 
description in the context of relativistic DWIA~\cite{mgp}. 
This outcome is reassuring and is further confirmed in Ref.~\cite{rmd03}.
In this paper it is also shown that a similar statement holds when the
analysis of the data is performed using a relativistic description
of the distorted proton waves combined with a corresponding mf description
of the bound states.
Also in this case, the slightly larger spectroscopic factors of 0.708 and
0.602 for the $p1/2$ and $p3/2$, respectively describe both the low
\textbf{and} high $Q^2$ data.
The difference between the spectroscopic factors for these different analyses
is about 5\% 
which may be considered as a measure of the uncertainties associated with
the assignment of spectroscopic factors until an improved microscopic
desciption of the FSI becomes available.
 
 Finally, one may conclude that the treatment of the bound-state wave 
function is not responsible 
for the $Q^2$ dependence found in Ref.~\cite{fsz}. 
From this observation one may 
infer that it is useful to extend the analysis of the high $Q^2$ data to other 
nuclei using the eikonal description supplemented with the nuclear matter 
damping description while foregoing the use of microscopic quasihole wave 
functions. Clearly an analysis of the data considered in Refs.~\cite{lapy} and 
\cite{fsz} will further clarify the validity of the present analysis. It is 
already encouraging to note that the damping employed in the 
 calculations is adequate to describe the observed absorption in the NE18 
experiment~\cite{ne18a} as discussed in Ref.~\cite{rdrs02}.

\subsection{Recent experimental data at higher missing energies
\label{sec:Pb208deep} }

The bulk of experimental data obtained during the last decades by means of
the (e,e$'$p) reaction have focussed
almost exclusively on the shells in the immediate vicinity of
 the Fermi energy, especially for heavier nuclei.
As a consequence the amount of protons seen experimentally
has been considerably smaller than the total number of
protons $Z$ inside the nucleus.

\begin{figure}[!tb]
\vspace*{.5cm}       
\begin{center}
\includegraphics[height=0.6\textwidth,width=0.55\textwidth]{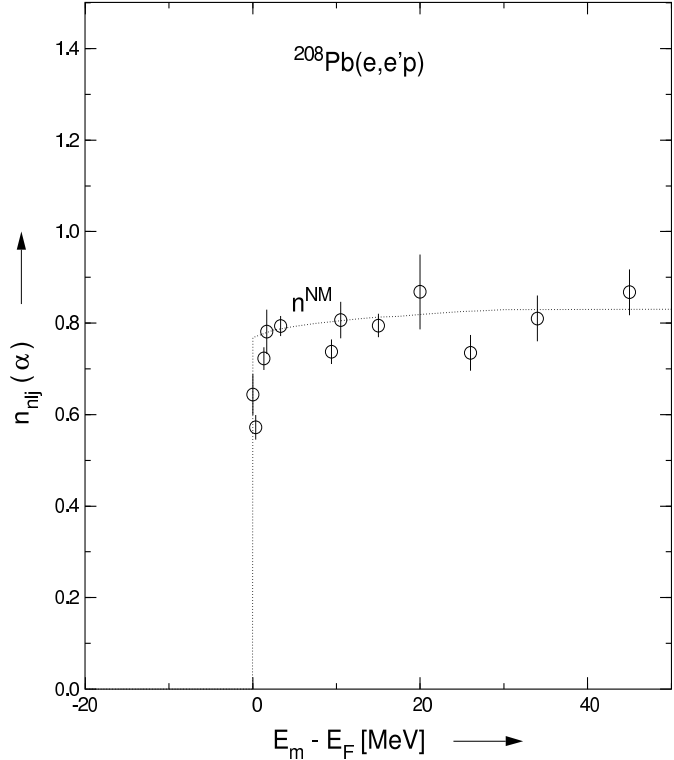}
\begin{minipage}[t]{16.5 cm}   
\caption{ \label{fig:pb208deep}
 Experimental occupation number for the all the mean field shells
of ${}^{208}{\rm Pb}$~\cite{lap2}. The dotted line gives the
value of occupation probability $n^{NM}$ computed
for nuclear matter~\cite{benf90}.
}
\end{minipage}
\end{center}
\end{figure}

More recently, 
a welcome addition to our knowledge of the sp strength distribution has been
provided by the analysis of one of the last $(e,e'p)$ experiments performed
at NIKHEF.
In this experiment the spectroscopic strength in a
large range of missing momentum (up to 270 MeV/c) and
missing energy (up to 100 MeV) was measured for 
${}^{208}\textrm{Pb}$~\cite{lap2,bat}.
The analysis of this experiment for the first time identifies the occupation
numbers of all the mean-field protons in this nucleus.
These results
are shown in Fig.~\ref{fig:pb208deep} and
compared with a theoretical calculation
for nuclear matter by Benhar et al.~\cite{benf90}.
The analysis for these deeply bound levels
yields a global depletion
of 15\% 
for all these proton orbits.
This confirms the expectation that nuclear matter results
yield a reliable estimate of the depletion due to SRC~\cite{vond1}.
Recent theoretical calculations of second generation spectral 
functions in nuclear matter~\cite{libth}
confirm the robustness of this observation as shown in Fig.~\ref{fig:nkself}. 
These new results have been obtained from self-consistent Green's functions
which incorporate full off-shell propagation and include the effects of
short-range and tensor correlations as discussed in Sec.~\ref{sec:selfnmr}.

\begin{figure}[!tb]
\vspace*{.5cm}       
\begin{center}
\includegraphics[width=0.9\textwidth]{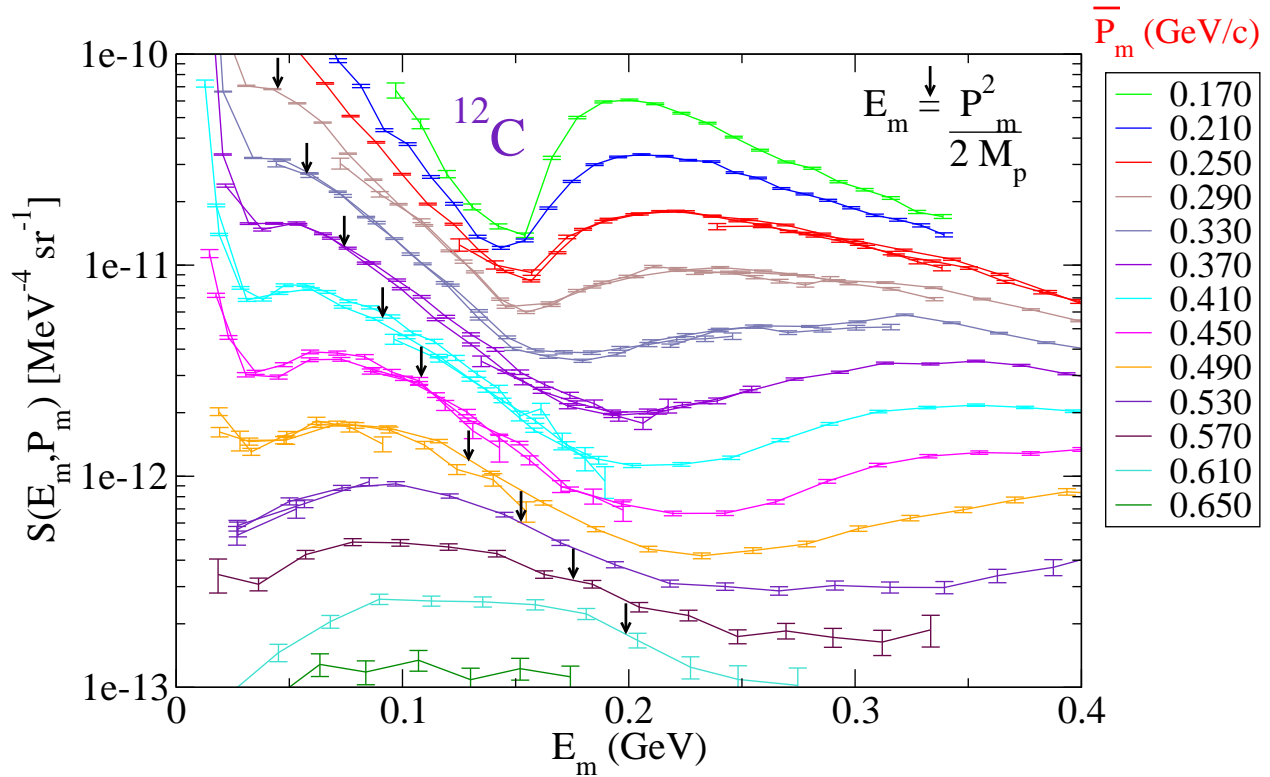}
\begin{minipage}[t]{16.5 cm} 
\caption{ \label{fig:daniela}
 Reduced spectral function for ${}^{12}\textrm{C}$ in the short-range
correlated region obtained with $(e,e'p)$ reaction~\cite{danielaPavia}.
 At missing energies above 150~MeV, sizable contributions
 from the excitation of a $\Delta$ resonance are seen.
 The arrows indicate indicate the location of the maximum of the 
spectral function expected according to a simplified picture.
  }
\end{minipage}
\end{center}
\end{figure}

The discussion
 in Secs.~\ref{sec:finiteSRC} and~\ref{sec:finiteSRC2} for the case of
finite nuclei, emphasized
another consequence associated of the presence
of short-range correlations in nuclear systems.
Besides the sizable amount of
strength that is removed from the mean-field region, these same 
correlations also supply a corresponding amount of strength in
the ground state in the form of high-momentum components.
A recent experiment at JLab, more fully discussed in
Refs.~\cite{danielaPavia,rescatt},
has been able to identify these correlations for different nuclei at high
missing energy~\cite{danielaElba}.
Figure~\ref{fig:daniela} shows first results from this experiment
for the spectral function of ${}^{12}\textrm{C}$.
The location of these high-momentum components conforms with the mechanism
that admixes these correlations with 2h1p states.
This can only be accommodated at such high missing energies due to the 
following observation.
Two-hole states in the nucleus tend to have a small total momentum. 
Momentum conservation then
requires that a high momentum admixture is accompanied by a particle
with equal and opposite momentum inside the corresponding self-energy
contribution.
The resulting energy for the presence of these high-momentum components
is thus given by $\langle \varepsilon_{2h} \rangle
- \bm{p}^2/2m$.
These missing energies can therefore easily be as large as 100 to 200 MeV.
As one can see from the picture, the maximum
of the theoretical spectral function follows this expectation 
(indicated by the arrows) but the experiment
clearly finds the maximum at slightly smaller energy with respect to the
naive application of the above argument.
A spreading of the peak in energy when moving to high 
momenta appears to be in agreement with the trend shown by 
Fig.~\ref{fig:selfsf}.

The spectral function at high momenta has been computed in finite nuclei
with the approach of Secs.~\ref{sec:finiteSRC} and~\ref{sec:finiteSRC2}%
~\cite{md94,mpd1,mkp95} and in local density approximation~\cite{bffs94}.
The experimental momentum distribution obtained for carbon is found
to be in between these theoretical predictions~\cite{danielaPavia}.
 We note that one limiting issue in taking measurements at higher
missing energies
comes from the necessity of transferring a large amount of energy from
the probe to the struck nucleon. The outgoing proton in an
experiment that probes this high-energy region can
have kinetic energies of several hundreds of MeV, leading to
strong rescattering effects, excitation of nucleon resonances,
etc.
  In certain kinematical conditions the associated final state effects
can overwhelm the direct process. This has been the case for some
previous experiments~\cite{oldhighk1,oldhighk2,oldhighk3}.
The issue of how to minimize the FSI has been addressed
in Ref.~\cite{E97proposal}. There, it is shown that FSI
in exclusive $(e,e'p)$
cross sections are dominated by two-step processes that become particularly 
relevant when perpendicular kinematics are employed to probe
the regions of small spectral strength.
A study of the kinematic conditions shows that for  perpendicular
kinematics the rescattered nucleons move spectral strength in the
$k$-$E$ plane, from the top of the ridge toward regions where the
correlated strength is small, therefore submerging the direct signal
in a large background noise.
 In Ref.~\cite{E97proposal}, it was suggested that the 
contribution of rescattering can be diminished using parallel kinematics.
 The Jefferson Lab data discussed here was
therefore taken in these conditions~\cite{danielaElba,danielaPavia}.
First calculations of the rescattering process appear to confirm the
expected trend with respect to the kinematics~\cite{rescatt1,rescatt}

We also note that other important contributions to the yield of
this experiment are expected to come from $(e,e'\Delta)$
reactions followed by the decay of the $\Delta$ resonance. The broad 
peak due to the excitation of the $\Delta$ is clearly visible on the right
hand side of Fig.~\ref{fig:daniela}. For heavier nuclei or when perpendicular
kinematics are used, the above rescattering effects are seen to become more and more
relevant and at the same time the strength from the delta region tends to
come at low missing energy, filling the gap that separates it from
correlated region. Therefore a proper treatment of final state interactions
becomes
necessary to study the behavior of the strength in the  short-range
region when the size of the nucleus is increased~\cite{rescatt}.
First calculations in this direction have been reported in
Refs.~\cite{rescatt1,rescatt}.

The relevance of the experiments discussed in this section
lies in the fact that they extend the experimental knowledge of
correlations to the regions of the momentum-energy plane
that were previously inaccessible.
Being able to identify high-momentum components~\cite{danielaPavia}
in addition to locating all
the sp strength associated with the mean-field orbits~\cite{lap2,bat}
completes the search for the missing protons.


\section{What a proton does in the nucleus and comparison with other fermion 
systems\label{sec:other}}

From the results discussed sofar it is clear that a crucial role in
obtaining pertinent information about the properties of protons in the
nucleus has been played by coincidence experiments involving electron beams.
In the following we will therefore summarize on what has been
learned in general terms about the properties of these protons.
Before embarking on this overview, it is important to note that recent work
on single-nucleon knockout with fast radioactive beams~\cite{hato03}
proposes to extend this information for valence protons to neutrons and
unstable nuclei.
Some promising work in this direction has been reported in 
Refs.~\cite{bhst02,ender03}.
Indeed, there is a long tradition with hadronic probes to study spectroscopic
factors in nuclei.
Due to the inherent complexity of hadron-induced reactions it has been more
difficult to establish absolute spectroscopic factors from corresponding 
experimental data.
Nevertheless, it is possible to generate a consistent analysis of (e,e$'$p)
and (d,${}^3$He) experiments as shown in Ref.~\cite{kbl01}.
The original discrepancies between these different experiments disappear
when a new analyis of the (d,${}^3$He) experiments is performed
using finite-range DWBA procedures combined with bound-state wave functions
obtained from the analysis of (e,e$'$p) reactions.
The resulting spectroscopic factors then appear to be quite consistent with
the spectroscopic factors obtained from the (e,e$'$p) reaction~\cite{kbl01}.
This observation demonstrates that with proper care it is possible to generate
valuable information from hadron-induced nucleon knockout experiments.
It may therefore be possible to extend the extraction of spectroscopic factors
to nuclei far off stability.
This exciting new development will allow the study of the properties
of nucleons in the nuclear medium in different regions of the periodic table
and may provide new challenges for our theoretical understanding.

The present situation concerning the knowledge of the properties of protons
in the nucleus may be summarized as follows.
The consequences of short-range correlations in nuclear systems on the
properties of protons appear to be theoretically well understood and
becoming available for experimental scrutiny as discussed in
Sec.~\ref{sec:finnuc}.
These results show that the effect of SRC is two-fold.
First, it involves the depletion
of spectroscopic strength from the mean-field domain as discussed in 
Sec.~\ref{sec:Pb208deep} for the experimental data obtained for
${}^{208}$Pb.
These data show that this depletion in heavy nuclei corresponds to about 15\%
for all the deeply bound proton levels.
This result was predicted more than
ten years ago based on nuclear-matter calculations
briefly reviewed in Sec.~\ref{sec:brian}.
For lighter nuclei all theoretical work suggests that this amount may be
closer to 10\%
as discussed in Sec.~\ref{sec:finiteSRC}.
Accompanying this information is the realization that valence shells
near the Fermi energy will not contain substantial amounts of high-momentum
components.
This has been experimentally confirmed and clarifies the other role played by 
SRC in nuclei, \textit{i. e.} the admixtures of high-momentum components
at high missing energy to account for the missing protons removed from the
mf location.
Recent attempts to quantify the amount of this strength and the corresponding
location have been presented in Sec.~\ref{sec:Pb208deep}.
It now appears that about the right amount of high-momentum strength can
be identified in the expected region although additional work is needed
to complete the analysis.
The location of these high-momentum components nevertheless broadly
conforms with the mechanism
that admixes these correlations with 2h1p states
at large missing energies as discussed in Sec.~\ref{sec:Pb208deep}.

\begin{figure}[!pbt]
 \begin{center}
    \includegraphics[height=0.70\textheight]{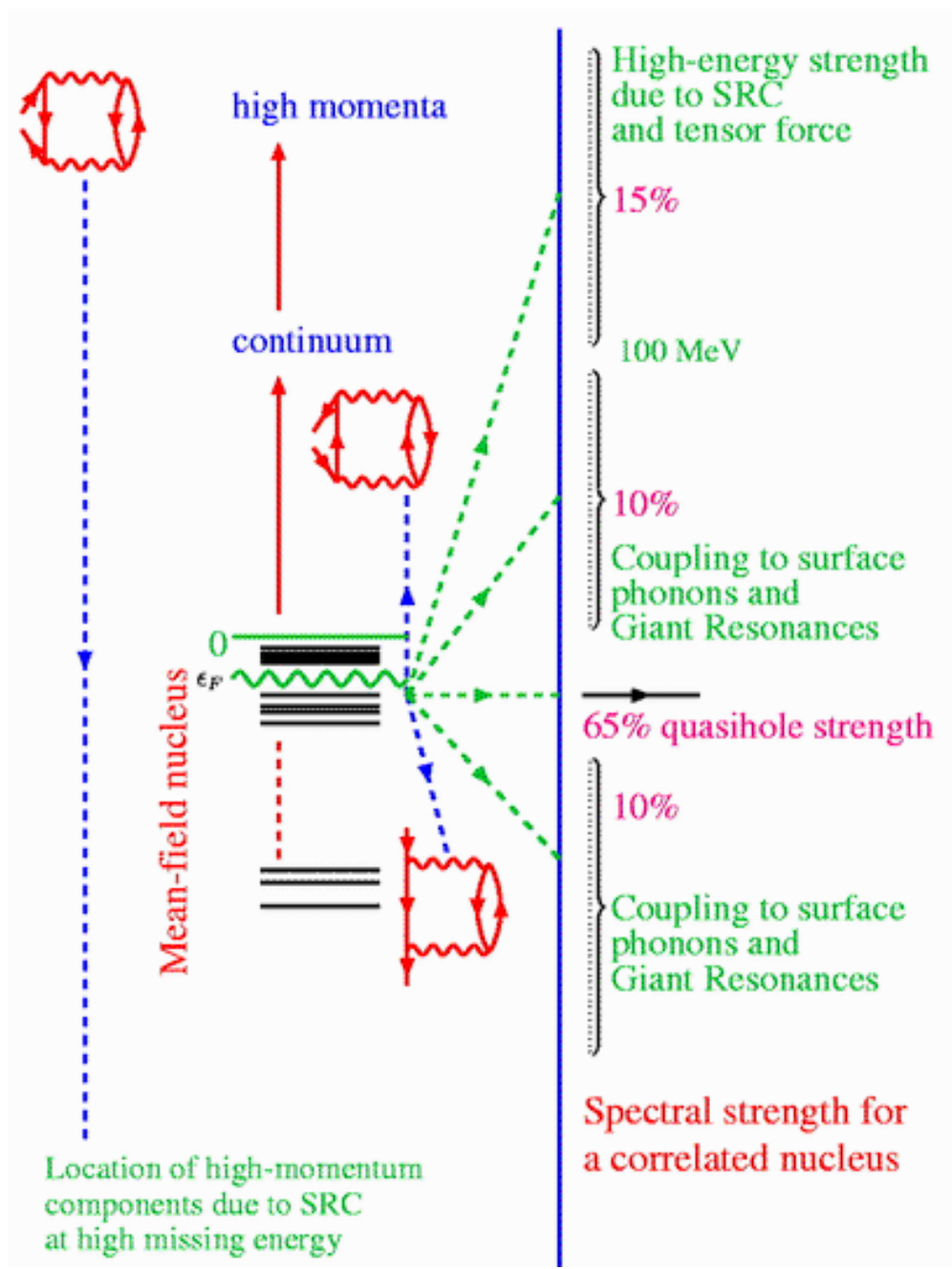}
\begin{minipage}[t]{16.5 cm}
\caption{The distribution of single-particle strength in a nucleus like
${}^{208}$Pb. The present summary is a synthesis of experimental and
theoretical work discussed in this review.
A slight reduction (from 15\% 
to 10\%)
of the depletion effect due to SRC must be considered
for light nuclei like ${}^{16}$O. 
\label{fig:strx} }
\end{minipage}
\end{center}
\end{figure}

Being able to identify high-momentum components in addition to locating all
the sp strength associated with the mean-field orbits~\cite{lap2,bat}
completes the identification of the properties of protons in the
ground state of the nucleus.
The latter understanding is illustrated in Fig.~\ref{fig:strx}.
Several generic diagrams are identified in Fig.~\ref{fig:strx}
which have unique physical
consequences for the redistribution of the single-particle strength when
they are taken into account in the solution of the Dyson equation as discussed
in this paper.
The middle column of the figure characterizes the mean-field
picture that is used as a starting point of the theoretical description.
The right column identifies the location of the sp
strength of the orbits just below the Fermi energy when correlations
are included.
One may apply this picture for example to the $3s_{1/2}$ proton
orbit in ${}^{208}\textrm{Pb}$.
The physical mechanisms responsible for the correlated strength distribution
are also identified.
The strength of this orbit remaining at the quasihole energy is
about 65\%.
Long-range correlations are responsible for the loss of 20\% 
of the strength
due to the coupling to nearby 2p1h and 2h1p states.
This loss is symmetrically distributed above and below the Fermi energy
and is physically represented by the coupling to low-lying surface modes
and higher-lying giant resonances.
The resulting occupation number of this orbit therefore corresponds to 75\%.
More deeply bound nucleons have higher occupation numbers corresponding
to 85\%.
As discussed in Sec.~\ref{sec:Pb208deep} this is true for all these
deep-lying orbits and is consistent with a global depletion due to SRC of 15\%.
The corresponding location of this strength is identified
at very high energy in the particle domain
and is due to the short-range and tensor correlations induced by a realistic
nucleon-nucleon interaction as discussed in Sec.~\ref{sec:nucmat}.
In the left column of Fig.~\ref{fig:strx} 
the generic diagram that is responsible for the admixture
of high-momentum components in the ground state is depicted.
The energy domain of these high-momentum nucleons is at large missing
energies as discussed above.

This rather complete picture of the properties of a strongly 
interacting proton in 
the nucleus signals a unique position for the field of nuclear physics.
Indeed, unlike other fields with strong correlations effects, like particle or
condensed matter physics, it is now possible in nuclear physics to 
state that the properties of the constituent protons inside the nucleus
are identified experimentally and understood in global terms theoretically.
For atoms and molecules it is also possible to extract this kind of information
by employing the corresponding (e,2e) reaction~\cite{mcwe91,comd94}.
This reaction generates the best possible information on the properties
of individual electrons in these systems.
Indeed, it was shown in 1981 that one can ``measure''  the square
of the $1s$ wave function of the Hydrogen atom in momentum 
space~\cite{lowe81}.
Similar results for electron wave functions in the medium have been obtained
for a wide range of atoms and molecules~\cite{mcwe91}.
Similar wave function results for nuclei are given in
Fig.~\ref{fig:louk1} of Sec.~\ref{sec:exp}.
This technique may also become successful in identiying the properties of
electrons in solids~\cite{vomc95}.
In nuclear physics the improved analysis of two-nucleon knockout
reactions is expected to provide detailed information about the
short-range interaction of nucleons in the nuclear medium.
With the additional improvement of the analysis of 
experiments that probe high-momentum nucleons
in the nucleus, one may therefore look forward to an even deeper understanding
of nucleon properties and their interactions in the future.

\section{Summary and Outlook\label{sec:sumout}}

Several recent developments have been reported that are of general
importance to the understanding of nuclei.
We single out here the improved understanding of the nuclear-matter
saturation problem and in particular the essentially complete understanding
of the properties of protons in nuclei.
These achievements have been accomplished by applying the
method of self-consistent Green's function.
This method has been presented in this overview together with a summary
of the
calculations that have been performed in nuclear physics.
One of the motivations to employ this method is that the quantities that
form the ingredients of the calculations have a very clear relationship
with experimental data.
This is true for the one-body propagator which contains information
about one-nucleon removal (and addition) quantities but also
for other propagators discussed in this work.
These include the polarization propagator which yields information about
excited states and corresponding transition probabilities but
also the pp(hh) propagator which for the hh part can be related to two-nucleon
knockout data.
The appearance of all these quantities as ingredients of self-consistent
calculations provides us with a multitude of opportunities to
compare theory with experimental data.
One of the strengths of the method is therefore its ability to identify
relevant physical correlations that need to be included in the calculation
based on this comparison with experiment.
As an example, the fragmentation of the removal strength observed in nuclei
immediately identifies the need for a dynamic content of the self-energy
of a nucleon below the Fermi energy.
The comparison of the theoretical sp strength distribution obtained
with the second-order self-energy with experimental
data then immediately points to the need for a better description of the
intermediate states in the self-energy.
This in turn leads to the consideration of the coupling to collective
motion at low energy.
The process is illustrated by the development of the Faddeev technique
to the many-body problem
discussed in Sec.~\ref{sec:finnuc} which allows the inclusion of these
collective features of both ph as well as pp(hh) character.
In turn, these microscopic phonons yield information that can be directly
compared with the excitation spectra in the corresponding nuclei with 
$A$ and $A \pm 2$ nucleons.
This comparison suggests further improvements of these phonons which
in turn can be taken into account in describing the sp properties and so on.

Another important advantage of the SCGF method is associated with its
ability to handle the relevant physical correlations.
As has been discussed throughout this review the treatment of SRC can
be handled by the summation of ladder diagrams whereas for LRC
it appears that the Faddeev technique is necessary in finite nuclei.
The treatment of SRC is most advanced for nuclear matter where the method
is applied by several groups but for finite nuclei improvements should
be studied in the future.
The discussion of the results for the saturation properties of nuclear
matter with the inclusion of SRC as given in Sec.~\ref{sec:satprop}
suggests that it may be important to actually establish the validity
of one of the underlying assumptions of the nuclear-matter problem.
This assumption is that the best equation of state for a given
Hamiltonian will automatically yield information relevant for finite nuclei.
This assumption has been questioned in this review since
it is unclear that long-range pionic modes which contribute significantly
to nuclear-matter binding actually play a significant role in finite
nuclei.
We therefore urgently suggest investigations which clearly identify
whether pion-exchange in ${}^{208}$Pb is really similar to
the corresponding process in nuclear matter.
If the answer is negative, nuclear physicists may have worried too much in
the past about their inability to describe the empirical saturation properties
of nuclear matter.

It has been argued in this review that it has been possible to track the
properties of \textbf{all} the protons in closed-shell nuclei.
We emphasize that this achievement reflects the combined effort of experiment
and theory.
The arrival of new facilities which aim to probe the properties of nuclei
far off stability will also provide us with new challenges to theory.
The inclusion of the continuum in SCGF calculations is a technical challenge
but the method appears sufficiently flexible to include this difficulty.
Knowledge of the sp properties of these exotic nuclei will widen the scope
of our understanding of the properties of nucleons in nuclei.
Additional work is also necessary to improve the description of the
spectrum and sp strength in ${}^{16}$O.
It appears likely that the Faddeev method can be applied to study the
low-energy properties associated with particle addition (elastic
scattering) which also has relevance to astrophysical problems.
The flexibility of this method is not restricted to nuclear physics
and may also be applied to other many-body systems like the electron gas.

\section*{Acknowledgments}
The authors wish to acknowledge the numerous collaborations that have
been instrumental in obtaining many of the results reviewed in this paper.
In particular we would like to thank Klaas Allaart, Mario Brand, 
Sigfrido Boffi, Yves Dewulf,
Chris Gearhart, Wouter Geurts,
Carlotta Giusti, Willem Hesselink,
Byron Jennings, Louk Lapik\'{a}s, Andrea Meucci,
Herbert M\"{u}ther, Franco Pacati, Artur Polls, Marco Radici,
Angels Ramos, Neil Robertson, Daniela Rohe, G\"{u}stl Rijsdijk,
Jean-Marc Sparenberg,
Libby Roth Stoddard, Dimitri Van Neck, Brian Vonderfecht, and
Michel Waroquier for important insights and fruitful collaborations.
This work was supported in part by the U.S. National Science Foundation
under Grant No.~PHY-0140316 and in part by the Natural
Sciences and Engineering Research Council of Canada (NSERC).

\end{document}